\documentclass{ar-1col-S2O}

\usepackage{amssymb}
\usepackage{url}
\usepackage{subfig}
\usepackage{wrapfig}
\usepackage{enumerate} 
\usepackage[hidelinks]{hyperref}
\hypersetup{
    colorlinks,
    citecolor={blue},
}
\usepackage[numbers]{natbib}
\setcitestyle{square}
\usepackage[many]{tcolorbox}    	% for COLORED BOXES (tikz and xcolor included)

\definecolor{main}{HTML}{5989cf}    % setting main color to be used
\definecolor{sub}{HTML}{cde4ff}     % setting sub color to be used

\tcbset{
    sharp corners,
    colback = white,
    before skip = 0.2cm,    % add extra space before the box
    after skip = 0.5cm      % add extra space after the box
}                           % setting global options for tcolorbox

\newtcolorbox{boxA}{
    fontupper = \bf,
    boxrule = 1.5pt,
    colframe = black % frame color
}

\newtcolorbox{boxB}{
    fontupper = \bf\color{main}, % font color
    boxrule = 1.5pt,
    colframe = main,
    rounded corners,
    arc = 5pt   % corners roundness
}

\newtcolorbox{boxC}{
    colback = sub, % background color
    boxrule = 0pt  % no borders
}

\newtcolorbox{boxD}{
    colback = sub, 
    colframe = main, 
    boxrule = 0pt, 
    toprule = 3pt, % top rule weight
    bottomrule = 3pt % bottom rule weight
}

\newtcolorbox{boxE}{
    enhanced, % for a fancier setting,
    boxrule = 0pt, % clearing the default rule
    borderline = {0.75pt}{0pt}{main}, % outer line
    borderline = {0.75pt}{2pt}{sub} % inner line
}

\newtcolorbox{boxF}{
    colback = sub,
    enhanced,
    boxrule = 1.5pt, 
    colframe = white, % making the base for dash line
    borderline = {1.5pt}{0pt}{main, dashed} % add "dashed" for dashed line
}

\newtcolorbox{boxG}{
    enhanced,
    boxrule = 0pt,
    colback = sub,
    borderline west = {1pt}{0pt}{main}, 
    borderline west = {0.75pt}{2pt}{main}, 
    borderline east = {1pt}{0pt}{main}, 
    borderline east = {0.75pt}{2pt}{main}
}

\newtcolorbox{boxH}{
    colback = sub, 
    colframe = main, 
    boxrule = 0pt, 
    leftrule = 6pt % left rule weight
}

\newtcolorbox{boxI}{
    colback = sub, 
    colframe = main, 
    boxrule = 0pt, 
    toprule = 6pt % top rule weight
}

\newtcolorbox{boxJ}{
    sharpish corners, % better drop shadow
    colback = sub, 
    colframe = main, 
    boxrule = 0pt, 
    toprule = 4.5pt, % top rule weight
    enhanced,
    fuzzy shadow = {0pt}{-2pt}{-0.5pt}{0.5pt}{black!35} % {xshift}{yshift}{offset}{step}{options} 
}

\newtcolorbox{boxK}{
    sharpish corners, % better drop shadow
    boxrule = 0pt,
    toprule = 4.5pt, % top rule weight
    enhanced,
    fuzzy shadow = {0pt}{-2pt}{-0.5pt}{0.5pt}{black!35} % {xshift}{yshift}{offset}{step}{options} 
}

\newtcolorbox{boxL}{
    fontupper = \color{main},
    rounded corners,
    arc = 6pt,
    colback = sub, 
    colframe = main!50, 
    boxrule = 0pt, 
    bottomrule = 4.5pt 
}

\newtcolorbox{boxM}{
    fontupper = \color{white},
    rounded corners,
    arc = 6pt,
    colback = main!80, 
    colframe = main, 
    boxrule = 0pt, 
    bottomrule = 4.5pt,
    enhanced,
    fuzzy shadow = {0pt}{-3pt}{-0.5pt}{0.5pt}{black!35}
}
\newcommand{\msun}{{\rm M}_\odot}
\newcommand{\diff}{{\rm d}}

\begin{document}
% Page header
\markboth{Daniel G. Figueroa \& Alberto Sesana}{NanoHz Gravitational Waves}

% Title
%\title{Origin of the nanohertz gravitational wave signal}
%\title{On the nature of the nanohertz gravitational wave signal}
%\title{Origin of the nanohertz gravitational wave signal}
%\title{On the nanohertz gravitational wave signal}
%\title{On the origin of the nano-Hz GW signal observed by PTAs}
%\title{On the origin of the nano-Hz GW signal}
%\title{On nanohertz GW signal observed by PTAs}
%\title{On nanohertz GW signal candidates}
%\title{On nanohertz gravitational wave signals}
\title{Nanohertz gravitational waves}

%Authors, affiliations address.
\author{Alberto Sesana$^{1,2,3}$ and  Daniel G. Figueroa$^4$ 
\affil{$^1$ Dipartimento di Fisica ``G. Occhialini'', Università degli Studi di Milano-Bicocca, Piazza della Scienza 3, 20126 Milano, Italy}
 \affil{$^2$ INFN - Sezione di Milano-Bicocca, Piazza della Scienza 3, 20126 Milano, Italy}
 \affil{$^3$ INAF - Osservatorio Astronomico di Brera, Via Brera 20, 20121 Milano, Italy\\{\bf Email}: alberto.sesana@unimib.it}
\affil{$^4$Instituto de F\'isica Corpuscular (IFIC), Consejo Superior de Investigaciones Cient\'ificas (CSIC) and Universitat de Val\`{e}ncia, 46980, Valencia, Spain \\{\bf Email}: daniel.figueroa@ific.uv.es}
}

%Abstract
\begin{abstract}
Evidence of a gravitational wave (GW) signal has emerged in pulsar timing array (PTA) data, opening a new window into the nanoHz GW Universe. We explore the physics of GW signals potentially explaining the data, with a primary focus on GW backgrounds (GWBs), considering both astrophysical and cosmological origins. We describe how:\vspace*{0.1cm}

\,\,$\bullet$ An astrophysical nanoHz GWB emerges as the superposition of individual signals from inspiralling massive black-hole binaries (MBHBs).\vspace*{0.1cm}

\,\,$\bullet$\, Environment coupling, eccentricity, and sparse sampling, %affect the SMBH signal, %spectrum and statistical properties, 
cause great uncertainty in the theoretical prediction of the SMBH signal, but offer simultaneously a handle to discriminate the %potential astrophysical 
origin of the signal.\vspace*{0.1cm}

\,\,$\bullet$ PTA data offers unprecedented opportunities to constrain high-energy physics beyond the standard model (BSM), via probing early Universe GWBs, originated during or after inflation.\vspace*{0.1cm} 

\,\,$\bullet$ Different early Universe GWBs, typically created by non-linear and out-of-equilibrium dynamics, can explain the PTA data, as {\it e.g.}~from inflation scenarios, first order phase transitions, or topological defects.\vspace*{0.1cm} 

\,\,$\bullet$ The PTA detection of GWs opens a new window to explore the Universe, with profound implications for astrophysics and particle physics, probing {\it e.g.}~the equation of state of the early universe, the origin of the cosmological perturbations, the nature of the dark matter, or whether exotic objects like primordial black holes or cosmic strings exit.
%\end{itemize}
 
\end{abstract}

%Keywords, etc.
\maketitle

%\section*{Review summary and plan}

%Table of Contents

\vspace{2.0cm}

\begingroup
\hypersetup{linkcolor=blue}
\tableofcontents
%\listoffigures (e.g.)
\endgroup

%\tableofcontents

%\section*{Plan/Summary}

%Final deadline: Sep 2025

%\noindent Length: 40 pages (18000 words / 200 references)

% Heading 1
\section{Introduction}
\label{sec:Intro}

~~~~Introduced as ``{\it The discovery that shook the world}\," at the 2017 Nobel Price ceremony, the direct detection of gravitational waves (GWs) has brought an unprecedented revolution in physics, providing a new cosmic messenger to explore the Universe~\cite{2016PhRvL.116f1102A}. In less than three years of effective data taking, up to the {\it O4a} catalog, the LIGO and Virgo interferometers have observed over 200 compact binary mergers~\cite{LIGOScientific:2025slb}, piercing for the first time into the {\tt Gravitational Universe}: a realm of dynamical phenomena driven by gravity, often inaccessible to other means like electromagnetic (EM) radiation. Those observations unveiled a restless dark side of the Universe, where black hole binaries (BHBs) whirling at relativistic speeds and crushing into each other, offer unique avenues to probe %constrain astrophysical models of 
stellar evolution \cite{KAGRA:2021duu}, the strong field regime of general relativity, and the underlying theoretical framework itself of gravity \cite{KAGRA:2025oiz}. When GW and electromagnetic (EM) signals met for the first time in the detection of the merging neutron star binary (NSB) GW170817 \cite{2017PhRvL.119p1101A}, countless new scientific opportunities unfolded: from the testing the Gamma Ray Burst-NSB merger connection and the equation of state of nuclear matter, to the understanding the heavy metal enrichment of the Universe and the measurament of its expansion. Multimessenger astronomy showcased its full potential. 

%%%%%%%%%%%%%%%%%%%%%%%%%%%%%%
%\begin{figure}[h!]
%    \vspace{-2mm}
%    \centering
%    \includegraphics[width=0.9\textwidth]{The_spectrum_of_gravitational_waves_pillars.png}%[scale=0.08]{path_to_coalescence.png}
%    \vspace{-3mm}
%    \caption{\small The GW spectrum, showing sources at the bottom and probes at the top. Courtesy of ESA, source: \url{www.esa.int/ESA_Multimedia/Images/2021/09/}}
%    \label{fig:fig_gwspec}
%    \vspace{-2mm}
%\end{figure}
%%%%%%%%%%%%%%%%%%%%%%%%%%%%%%%

\begin{figure*}[t]
    \centering
    \includegraphics[width=\linewidth]{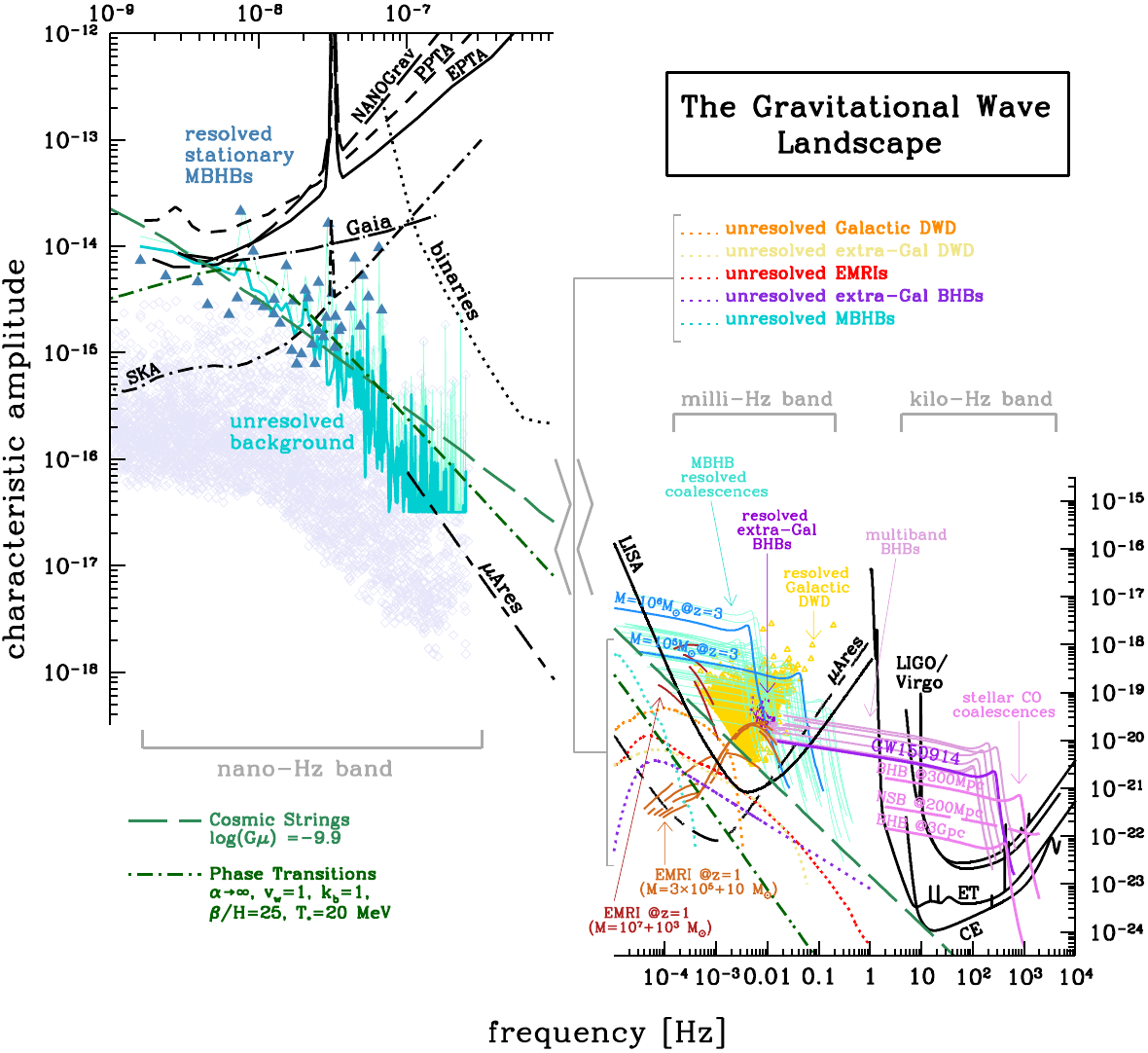}
    \caption{The gravitational wave landscape. In the nano-Hz band zoom-in, each lavander diamond is an individual MBHB. The overall GW signal is marked by the light-turquoise jagged line, whereas the residual GWB following subtraction of the loudest source (dark-blue triangles) at each frequency is marked by the dark-turquoise jagged line. Sensitivity curves of various experiments are shown in black, as labeled. We also show the milli-Hz and kilo-Hz band for completeness. In each band, we plot in black GW detectors and with colored lines and marks a variety of resolved GW signals and unresolved GWBs as labeled.}
    \label{fig:GWlandscape}
\end{figure*}

~~~~Because of inherent length and seismic noise limitations, ground based detectors can access only the GW `audio band' that covers the Hz-kHz frequency range: a narrow window within an otherwise broad frequency spectrum, which potentially extends over 20 decades, from the inverse age of the Universe up to MHz and potentially above~\cite{Lasky:2015lej}. %To make a parallelism with EM observations, it is like if today we were able to observe only the local Universe (out to distances $\lesssim 1$~Gpc) in optical wavelength. Our knowledge of galaxies and structure formation would be very different and severely incomplete. 
Similarly as to how different EM wavelengths unveil different astrophysical phenomena, % (i.e. violent non thermal physical processes emitting in X-rays and $\Gamma$-rays), 
different GW frequencies probe different mass and energy scales, bringing up the possibility to  potentially revolutionize our understanding of the Universe. For this reason, a great deal of effort is put into opening new frequency windows in the ``GW sky", schematically represented by Figure~\ref{fig:GWlandscape}. As the sensitivity of ground based detectors keeps improving, the deployment of LISA, the Laser Interferometer Space Antenna~\cite{LISA:2017pwj,2024arXiv240207571C}, is well on track for the mid 2030s' and is expected to probe the mHz frequency window. Simultaneously, observational cosmologists have plans to search~\cite{LiteBIRD:2023iei,Wolz:2023lzb} for the imprint of primordial GWs at frequencies below $10^{-15}$\,Hz, in the B-mode polarization of the Cosmic Microwave Background (CMB). 

~~~~Bridging the gap between the previous two bands, the nano-Hz frequency region has been the stage of the latest breakthrough in GW astronomy, as several pulsar timing array (PTA) collaborations reported evidence on a GW background (GWB) of unknown origin~\cite{NANOGrav:2023gor, Antoniadis:2023ott, Reardon:2023gzh, Xu:2023wog}. While a GWB from the incoherent superposition of GWs emitted by inspiralling massive black hole binaries (MBHBs) is naturally expected at PTA frequencies \citep[{\it e.g}\,][]{Rajagopal:1994zj,Jaffe:2002rt,Sesana:2008mz}, GWBs can also be produced by a variety of phenomena potentially occurring in the early Universe, like inflationary scenarios, first-order phase transitions, topological defects, and others~\cite{Caprini:2018mtu}. 

~~~~In this review, we consider mostly {\it stochastic} GW signals, %{\it i.e.}~%in the nanoHz window, 
focusing on GWBs that might explain PTA observations, but delving also briefly into deterministic signals. %Assuming %We assume 
%General Relativity, %{\it i.e.}~spin-2 gravity with only two tensor modes representing GWs. We 
We start by introducing the basic description of stochastic GW signals %in General Relativity (GR) %relevant to the nano-Hz frequency band  (Sec.~\ref{sec:gwb_basics}), 
and their main detection techniques. We put a specific focus on PTAs (Sec.~\ref{sec:pta}), but also mention astrometry and the potential of combining the two techniques (Sec.~\ref{sec:astrometry}). We summarize the current evidence of a nano-Hz signal in PTA data (Sec.~\ref{sec:PTA_evidence}). We then move onto the description of the main astrophysicals sources of nano-Hz GWs, MBHBs (Sec.~\ref{sec:astroGWB}). We describe in detail the build-up of a GWB from  unresolved individual binaries, its main features (Secs.~\ref{sec:MBHB_GWbasics} and~\ref{sec:GWB_circular}) and the role played by MBHB properties (eccentricity and environmental coupling) in shaping its spectrum and statistical properties (Secs.~\ref{sec:binary_dynamics} and~\ref{sec:SMBH_totalGWsignal}). We also describe deterministic GWs stemming from MBHBs, including resolvable continuous GWs (CGWs), eccentric bursts, and bursts with memory (Sec.~\ref{sec:MBHB_resolvable}). Finally, %before finishing by commenting 
we comment on the astrophysical interpretation of current PTA observations (Sec.~\ref{sec:MBHB_interpretation}). In Sec.~\ref{sec:cosmo_GWsignals} we describe relevant GWBs arising from hypothetical processes in the early Universe. These are referred to %these signals 
as early Universe backgrounds (EUB), and we classify them into signals coming from inflation (Sec.~\ref{subsec:EUBinflation}), first-order phase transitions (Sec.~\ref{subsec:EUBphaseTransitions}), cosmic defects (Sec.~\ref{subsec:EUBcosmicDefects}), and others (Sec.~\ref{subsec:EUBothers}). We also discuss cosmological deterministic signals (Sec.~\ref{subsec:EUBdeterministic}), albeit more briefly. In Sec.~\ref{sec:discriminating}, we complete this review by exploring future prospects for disentangling %astrophysical from early universe 
the origin of the signal, before making our final remarks in Sec.~\ref{sec:conclusions}.

~~~~Before proceeding, we warn the reader of what this review is not supposed to be. This is not a review about PTAs, an excellent long review on the subject can be found in \cite{Taylor:2021yjx}, while for more concise readings, we suggest \cite{2018ASSL..457...95P,2025arXiv250500797K}. Neither this is a review on data analysis methods, we briefly touch on signal to noise ratio calculations and their scalings in Sec.~\ref{sec:gwb_basics}, without entering in the details of frequentist and Bayesian analysis, the definition of PTA likelihood functions, advanced sampling techniques etc. The reader is referred to the PTA analysis papers ~\cite{NANOGrav:2023gor, Antoniadis:2023ott, Reardon:2023gzh, Xu:2023wog} and references therein as a useful starting point for exploring PTA analysis methods and algorithms. This review rather presents a basic introduction to GW signals in the nano-Hz band and current evidence of their existence, and above all it represents a detailed overview of all leading astrophysical and cosmological processes that could potentially explain the signal observed in current PTA data. %{\color{blue} DGF: Alberto, you ok with the last sentence?}

\section{Nano-Hz GW basics: signals, detectors and observations}
\label{sec:nHzGWbasics}

We begin with the characterization of %GW signals, focusing on the standard quantities defining 
a GWB. Following, %We then move to the 
we discuss on nano-Hz GW detection techniques, with a focus on PTAs, including detection principles, noise sources, definition of the {\it Hellings} \& {\it Downs} correlation, and a survey of current PTA experiments. We provide a dictionary between %for navigating the 
different quantities %used to represent a %an unresolved GW signal GWB: 
such as the characteristic strain $h_c$, energy density fraction $\Omega_{\rm GW}$, power spectra density $S_h$, and RMS residual, which is particularly relevant for understanding the PTA results. We also provide simple scalings for estimating the signal-to-noise ratio of nano-Hz GWs with different detectors. We conclude by discussing the current evidence for GWs in PTA data.

\subsection{Basic Concepts defining a GWB}
\label{sec:gwb_basics}

In General Relativity (GR), a GW propagating in %a given 
direction $\hat\Omega$, has two polarization modes, $h_{+}$ and $h_{\times}$, which in full generality can be written as a metric perturbation of the form
\begin{equation}
h_{ij}(t,\hat\Omega) = e_{ij}^+(\hat\Omega)h_+(t,\hat\Omega) + 
e_{ij}^{\times}(\hat\Omega)h_\times(t,\hat\Omega),
\label{hij}
\end{equation}
where the polarization tensors $ e_{ij}^A(\hat\Omega)$ (with $A=+,\times$) are defined as
%\begin{subequations}
\begin{align}
e_{ij}^+(\hat\Omega) =  \hat{m}_i \hat{m}_j - \hat{n}_i \hat{n}_j\,,~~~e_{ij}^{\times}(\hat\Omega) &= \hat{m}_i \hat{n}_j + \hat{n}_i \hat{m}_j\,.
\label{eq:polarizations}
\end{align}
%\end{subequations}
Here $\hat{m}$ and $\hat{n}$ are the principal axes that, together with $\hat\Omega$, define the GW propagation orthonormal basis. The characterization of $h_{+}$ and $h_{\times}$ depends on the physics of source of the GW, either of astrophysical or cosmological origin.

~~~~By construction, a GWB can be expanded -- at any point in space -- as an incoherent superposition of plane GWs, with different amplitudes, phases, and frequencies. Placing the observer in the origin at ${\bf x}=0$, 
%and adopting the TT gauge, 
we can write~\cite{Maggiore:2007ulw}
\begin{equation}
    h_{ij}(t)=\sum_{A=+,\times}\int_{-\infty}^{\infty}{\rm d}f\int_{4\pi}{\rm d}\hat\Omega\,\tilde{h}(f,\hat\Omega)e^{-2\pi ift}e^A_{ij}(\hat\Omega)\,.
\label{eq:SGWB_hij}
\end{equation}
If the GWB is isotropic, stationary and unpolarized, then 
\begin{equation}
  \langle \tilde{h}^*_A(f,\hat{\Omega}) \tilde{h}_A'(f',\hat{\Omega}') \rangle=\delta(f-f')\frac{\delta^2(\hat\Omega,\hat\Omega')}{4\pi}\delta_{AA'}\frac{1}{2}S_h(f).
\label{eq:Sh_def}
\end{equation}
%$\langle h_{ij}(t)\,h_{ij}(t) \rangle$
so that using Eq.~(\ref{eq:Sh_def}), we obtain the expectation value\footnote{Throughout this  review we assume a perturbation of the space components of the metric over a flat background. For this reason, we do not differentiate between covariant and contravariant indices and we always use subscripts.} (variance) of the GW field as
\begin{equation}
\langle h_{ij}(t)\,h_{ij}(t) \rangle=4\int_0^\infty{\rm d}f\, S_h(f)\equiv 2\int_0^\infty{\rm d\,ln}f\, h^2_c(f),
\label{eq:hc1}
\end{equation}
where we used the fact that $S(-f)=S(f)$ to limit the integral over positive ({\it i.e.}~physical) frequencies. Eq.~(\ref{eq:hc1}) defines the {\it characteristic strain} of the GWB, which is then related to its {\it power spectral density} (PSD) simply as 
\begin{equation}
h^2_c(f)=2f\,S_h(f).
\label{eq:hc}
\end{equation}

~~~~A simple characterization of a GWB within a narrow frequency interval, is given by a power law (PL) spectrum
\begin{equation}
    h_c(f)=A_{\rm GWB}\left(\frac{f}{f_*}\right)^{\alpha},
    \label{eq:hc_analytical}
\end{equation}
with $A_{\rm GWB}$ the strain amplitude at the reference frequency $f_*$, which in the  PTA band is typically fixed to $f_* = 1\,{\rm yr}^{-1}$. As we will see below, the strain of the GWB from a cosmic population of circular, GW-driven binaries, can be described as a PL within an  extended frequency range (which encompasses the PTA window), with tilt $\alpha=-2/3$. 

~~~~We can now connect $h_c$ and $S_h$ to the energy density of a GWB, which is given by the 00 component of its stress-energy tensor~\cite{Maggiore:2007ulw} 
\begin{equation}
\epsilon_{\rm GW} \equiv \rho_{\rm GW} c^2 = T_{00}= \frac{c^2}{32\pi G}\langle \dot{h}_{ij}(t)\,\dot{h}_{ij}(t) \rangle,
\label{eq:rho_gw}
\end{equation}
where $\dot{h}$ means derivative with respect to the time coordinate. Since in Eq.~(\ref{eq:SGWB_hij}) time appears only in the exponent, it is straightforward to demonstrate that
\begin{equation}
\epsilon_{\rm GW} \equiv c^2\int_0^\infty{\rm d\,ln}f\left(\frac{{\rm d}\rho_{\rm GW}}{{\rm d\,ln}f}\right)=\frac{\pi c^2}{4G}\int_0^\infty{\rm d\,ln}f\,f^2h_c^2(f)=\frac{\pi c^2}{2 G}\int_0^\infty{\rm d\,ln}f\,f^3\,S_h(f),
\label{eq:epsilon_gw}
\end{equation}
from where it follows that
\begin{equation}
\Omega_{\rm GW}(f)\equiv\frac{1}{\rho_c}\frac{{\rm d}\rho_{\rm GW}}{{\rm d\,ln}f}\equiv \frac{8\pi G}{3H_0^2}\frac{{\rm d}\rho_{\rm GW}}{{\rm d\,ln}f}=\frac{2\pi^2}{3H_0^2}f^2h_c^2(f)=\frac{4\pi^2}{3H_0^2}f^3S_h(f).
\label{eq:omega_gw}
\end{equation}
Here $\Omega_{\rm GW}(f)$ defines the energy content of the GWB per logaritmic frequency, normalized to the critical energy density of the Universe $\rho_c=3H_0^2/(8\pi G)$, where $H_0=100~h_0$~km~s$^{-1}$~Mpc$^{-1}$ is the Hubble expansion rate and $h_0\approx0.67$~\cite{Planck:2018vyg}. Therefore, Eq.~(\ref{eq:omega_gw}) connects the (normalized) spectral energy density spectrum $\Omega_{\rm GW}$ to the PSD $S_h$ and characteristic strain $h_c$. %Integrating over all frequencies we get %this leads to the total energy density of the GWB 
%\begin{equation}
%\epsilon_{\rm GW}=\rho_cc^2\int_0^\infty{\rm d\,ln}f\,\Omega_{\rm GW}(f)=\frac{\pi c^2}{4G}\int_0^\infty{\rm d\,ln}f\,f^2h_c^2(f)=\frac{\pi c^2}{2G}\int_0^\infty{\rm d\,ln}f\,f^3S_h(f)\,.
%\label{eq:epsilon_gw}
%\end{equation}

~~~~The above derivation assumed a GWB statistically homogeneous and isotropic, stationary and unpolarised, %and Gaussian, 
as these circumstances apply (at least to a large degree) to the majority of cases, see Ref.~\cite{Caprini:2018mtu} for a discussion on this. The possibility of sizeable anisotropies might be however a key discriminating factor to pin down the origin of the PTA signal as astrophysical, as opposed to cosmological, %\footnote{Cosmological signals might also display small anisotropies~\cite{Dimastrogiovanni:2021mfs,LISACosmologyWorkingGroup:2022kbp}, as well as non-Gaussian statistics, non-stationary bursts, or chiral features, see {\it e.g.}~\cite{Bartolo:2018qqn,Damour:2000wa,Sorbo:2011rz}. We will not dwell, however, into these aspects, as they are not generic features, and it is uncertain that they will play any role in discriminating the origin of the PTA signal.}, 
see Sec.~\ref{sec:discriminating}.

%\begin{marginnote}
%\entry{Aerosol}{a fine solid or liquid particle in the atmosphere}
%\entry{XANES}{X-ray absorption near-edge structure}
%\end{marginnote}

\subsection{Pulsar timing arrays (PTA)} %detection techniques}
\label{sec:pta}

Millisecond pulsars (MSPs) are magnetized neutron stars emitting beamed radio waves \cite{2019MNRAS.483.1731L}. They spin hundreds of times per second along a rotation axis, typically misaligned with their magnetic axis, making them cosmic lighthouses. If %at any rotational phase 
the radio beam intercepts the line of sight to the Earth, we see an extremely precise pulsation: we essentially have a clock in geodesic motion within the gravitational potential of our galaxy. Any GW passing between the MSP and the Earth perturbs the null geodesic along which the radio waves travel, causing a change in their time of arrival (TOA). 
MSPs have been monitored for $T\approx\,$decades with weekly cadence $\Delta{t}$, and therefore their TOA series are sensitive to GWs with $1/T\approx{\rm nHz}\lesssim f\lesssim 1/(2\Delta t)\approx \mu{\rm Hz}$. This makes them {\it ideal nano-Hz GW detectors} ({\it c.f.} Fig.~\ref{fig:GWlandscape}) when an ensemble of them is monitored to form a pulsar timing array \citep[PTA,][]{1990ApJ...361..300F}. 
\begin{marginnote}[]
\entry{Pulsar Timing \\Array}
\,Systematic monitoring of an ensemble of MSPs over decades to detect GWs via correlation of their pulses' TOAs.
\end{marginnote}

%Unlike GW interfereometers, the PTA observable is not directly the strain $h$ but a induced time delay $\Delta t$, known as {\it timing residual}, $R$.

\subsubsection{Response of PTA to a passing GW}
\label{sec:ptaresponse}
Radio waves propagate along the GW-perturbed null geodesic, which leads to a characteristic shift \footnote{\footnotesize We denote the frequency shift by $z(t)$, which is standard in the PTA literature and should not be confused with the cosmological redshift $z$.} $z
(t)=(\nu-\nu(t))/\nu=-\Delta\nu/\nu$. Originally computed for spacecraft tracking \cite{1975GReGr...6..439E}, $z(t)$ takes the form
\begin{equation}
\footnotesize{ z(t,\hat\Omega)  =  \frac{1}{2} \frac{\hat{p}^i\hat{p}^j}{1+\hat{p}^i\hat\Omega_i}\left\{e_{ij}^+(\hat\Omega)\left[h_+(t,0)-h_+(t-\frac{L}{c},{\bf x_p})\right]-e_{ij}^\times(\hat\Omega)\left[h_\times(t,0)-h_\times(t-\frac{L}{c},{\bf x_p})\right]\right\}},
\label{eq:ztgeneral}
\end{equation}
where we placed the Earth\footnote{In reality, in PTA measurements, the pulse TOAs are referred to the solar system baricenter (SSB) by applying an appropriate coordinate transformation that requires the knowlege of the solar system ephemeris. While this is irrelevant for our discussion above, it is however crucial to take it into account in models for timing and GW data analysis.} in ${\bf x}=0$, $L$ is the distance to the pulsar and ${\bf x_p}=L\hat{p}$ its 3-D position vector, $\hat{p}$ is the unit vector pointing from Earth to %the direction of 
the pulsar, and $\hat\Omega$ is the direction of the incoming wave. Eq.~(\ref{eq:ztgeneral}) %quantifies the $\sim$ in Eq.~(\ref{eq:deltaP}) by 
shows that the induced shift is due to the difference between the GW strain at the Earth at the TOA of the pulse (known as the Earth term) and the strain at the location of the pulsar at the time of emission of the pulse (known as pulsar term), multiplied by a geometric factor depending on the relative orientation between the pulsar and the Earth. It is also useful to note that the pulsar term can be equivalently written by considering the strain at the Earth location at an appropriate time $\tau$, $h_A(t-\tau,0)$, where $\tau=(L/c)(1+\hat{p}\cdot\hat\Omega)$. Eq.~(\ref{eq:ztgeneral}) can be written in compact form as 
\begin{equation}
z(t,\hat\Omega) = \sum_A F^A(\hat\Omega)\Delta h_{A}(t)\,,
\label{eq:z1}
\end{equation}
where $\Delta h_{A}(t)=h_A(t,0)-h_A(t-L/c,L\hat{p})$ and $F^A(\hat\Omega) = \hat{p}_i\hat{p}_je_{ij}^A(\hat\Omega)/[2({1+\hat{p}\cdot\hat\Omega})]$
%\begin{equation}
%$F^A(\hat\Omega) = \frac{1}{2} \frac{\hat{p}^i\hat{p}^j}{1+\hat{p}^i\hat\Omega_i} e_{ij}^A(\hat\Omega)$
%\label{e:FA}
%\end{equation}
are the antenna pattern functions representing the geometrical response of the pulsar-Earth system to each GW polarization. This notation separates the physics of the GW from the geometric response due to the projection of the radiation frame defined by the orthonormal unit vectors ($\hat{m}, \hat{n}, \hat\Omega$) onto the pulsar-Earth baseline \citep[see][for the full equations of the coordinate transformations]{2013LRR....16....9Y}. The explicit form of the response functions (or antenna patterns) is universal, i.e. it does not depend on the specific GW signal\footnote{This is true so long as only GR tensor polarizations are considered. In alternative theories of gravity, scalar and vector polarizations might also arise, requiring different response functions \cite{2013LRR....16....9Y}.}, and is given by
\begin{equation}
F^+(\hat\Omega) = \frac{1}{2} \frac{(\hat{m} \cdot \hat{p})^2 - (\hat{n} \cdot \hat{p})^2}{1 + \hat\Omega \cdot \,\,\hat{p}}\,;\,\,\,\,\,\,\,\,\,\,\,\,\,\,\,\,\,\,\,
F^\times(\hat\Omega) = \frac{(\hat{m} \cdot \hat{p})\,(\hat{n} \cdot \hat{p})}{1 + \hat\Omega \cdot \hat{p}}\,.
\end{equation}
The {\it timing observable} is the residual $R$, defined as the difference between the expected TOA according to some MSP timing model with and without the presence of the GW. This is simply the integral of $z(t)$ over an observation time $T$,
\begin{equation}
  R(T)=\int_0^Tz(t)\,{\rm d}t\equiv-\int_0^T\frac{\Delta\nu(t)}{\nu}{\rm d}t ~\sim \int_0^Th(t)\,{\rm d}t.  
  \label{eq:deltaP}
\end{equation}

The final $\sim$ ignores geometric factors, highlighting that for a sinusoidal wave of the form $h\sim h_0{\rm cos}(2\pi f t)$, the induced residual is of order $R\approx h_0/(2\pi f)$. 
%\begin{marginnote}
%    \entry{PTA sensitivity} \, 
    Since the best MSPs are monitored with  $\approx 100$ ns precision, assuming a 10 nHz fiducial frequency, PTAs are sensitive to perturbations $h_0\sim 10^{-14}-10^{-15}$ ({\it c.f.} Fig.~\ref{fig:GWlandscape}).   
%\end{marginnote}

%\begin{subequations}
%\begin{align}
%F^+(\hat\Omega) & = \frac{1}{2} \frac{(\hat{m} \cdot \hat{p})^2 - (\hat{n} \cdot \hat{p})^2}{1 + \hat\Omega \cdot \hat{p}}\,,\\
%\nonumber
%F^\times(\hat\Omega) & = \frac{(\hat{m} \cdot \hat{p})\,(\hat{n} \cdot \hat{p})}{1 + \hat\Omega \cdot \hat{p}}\,.
%\nonumber
%\end{align}
%\end{subequations}

\subsubsection{Response of a PTA to a GWB} If we now consider a GWB expanded as in Eq.~(\ref{eq:SGWB_hij}) and the fact that each component in the expansion produces a frequency shift given by Eq.~(\ref{eq:z1}), it is straightforward to write the overall shift as [see definition of $\tau$ above Eq.~(\ref{eq:z1})]
\begin{equation}
z(t)=\sum_{A=+,\times}\int_{-\infty}^\infty \diff f\int \diff ^2\hat{\Omega}\,F^A(\hat{\Omega})\tilde{h}_A(f,\hat{\Omega})e^{-2\pi i f t}\left[1-e^{-2\pi i f \tau}\right],
\label{stochred}
\end{equation}
where we recognize in the two exponentials the Earth and the pulsar terms. 
%If we now enforce the assumptions of an isotropic, stationary and unpolarized GWB as per
Using Eq.~(\ref{eq:Sh_def}), the expectation value of the frequency shift correlation  between MSPs $a$ and $b$ is
\begin{equation}
  \langle z_a(t)z_b(t) \rangle= \frac{1}{2}\int_{-\infty}^\infty \diff f S_h(f)\int \frac{\diff ^2\hat{\Omega}}{4\pi}\sum_{A=+,\times}F_a^A(\hat{\Omega})F_b^A(\hat{\Omega}),
\label{crossredshift}
\end{equation}
where all of the terms involving $e^{-2\pi i f \tau}$ have been neglected by assuming the short wavelength limit $\lambda_{\rm GW}\ll {\rm min}(L_a,L_b, {\bf x}_a-{\bf x}_b)$. This is apprioriate for PTAs that are sensitive to GW wavelengths of about a parsec ($\lambda=c/f$, with $f\sim$ several nano-Hz), whereas the closest known MSP is at about 150 pc from Earth and typical inter-MSP distances are in the kpc range.\footnote{Note that this assumption might break down if multiple MSPs of a PTA are located within the same globular cluster, where their typical distances is of the order of the cluster core size, which can be smaller than a parsec. In this case, the correlation of the pulsar terms cannot be neglected, causing an extra correlation in the frequency shifts of the pulsars. Given typical globular cluster distances, this can have an impact only for pulsars separated by an angle $\theta_{ab}\ll 1\,$deg.}

~~~~The integral over ${\diff }^2\hat{\Omega}$ in Eq.~(\ref{crossredshift}) was first computed by \cite{1983ApJ...265L..39H}, and takes the form
\begin{eqnarray}
C(\zeta_{ab}) \equiv \sum_{A=+,\times}\int  \frac{\diff ^2\hat{\Omega}}{4\pi}F_a^A(\hat{\Omega})F_b^A(\hat{\Omega})=\frac{1}{4}\left[1+\frac{\cos\zeta_{ab}}{3}+4(1-\cos\zeta_{ab})\ln\left(\sin\frac{\zeta_{ab}}{2}\right) \right],
\label{e:hdcurve}
\end{eqnarray}
where $\zeta_{ab}$ is the angle between pulsars $a$ and $b$ on the sky. Finally, the observable quantity in PTAs is the %ensemble average 
cross-correlation of the timing residuals between two pulsars $r_{ab}=\langle R_a(t)R_b(t)\rangle$, where $R_x(t)$ is defined %as the integral of the redshift as per {\it c.f.}~
Eq.~(\ref{eq:deltaP}). % for the definiotion of $R_x(t)$. 
Integrating Eq.~(\ref{crossredshift}) over time, we get
\begin{equation}
r_{ab}=\Gamma(\zeta_{ab})\int_0^\infty \diff f P_h(f),
\label{e:rabPh}
\end{equation}
where we re-normalized the Hellings \& Downs (HD) correlation of Eq.~(\ref{e:hdcurve}) as
\begin{equation}
\Gamma(\zeta_{ab})=\frac{3}{2}C(\zeta_{ab})(1+\delta_{ab}).
\label{eq:HD} 
\end{equation}
The $3/2$ renormalization and the $\delta_{ab}$ term ensure the correlation coefficient to be $\Gamma(\zeta_{ab})=1$ when $a=b$ (i.e., the GWB has perfect autocorrelation). Note that when $a \neq b$ and $\zeta_{ab}=0$, $\Gamma_{ab}=1/2$; this is because for pulsars at the same location, but at different distances, only the Earth terms are phase correlated, whereas the pulsar terms act as an additional source of noise. %We will see below that this has important implications for GWB detection with PTAs. 
In Eq.~(\ref{e:rabPh}) we introduced the timing residual PSD as
\begin{equation}
  P_h(f)=\frac{h^2_c(f)}{12\pi^2f^3}=\frac{A_{\rm GWB}^2}{12\pi^2}\frac{f^{\gamma}}{(1{\rm yr}^{-1})^{2\alpha}}, \,\,\,\,\,\,\,\,\,\,\,\,\,\,\gamma=3-2\alpha\,,
\label{e:Ph}
\end{equation}
where in the last equality we have substituted $h_c$ from Eq.~(\ref{eq:hc_analytical}). Note that $P_h(f)$ has dimensions of $[s^{3}]$, as expected for a spectral density of a time series. 

\begin{boxC}
    {\bf A note on PTA results.} When presenting their results, PTAs usually quote the strain amplitude $A_{\rm GWB}$ at $f=1{\rm yr}^{-1}$, and the slope $\gamma$ of the residual PSD $P_h$. Since for a cosmic population of circular GW driven MBHB $\gamma=13/3$, quite often $A_{\rm GWB}$ is also quoted after fixing $\gamma$ to this reference value.
\end{boxC}

Finally, the total root-mean-square (RMS) timing residual is obtained by integrating the PSD over frequency, ${\rm RMS}^2=\int {\diff f} P_h(f)$. It is standard in PTA analysis to divide the frequency domain in bins $\Delta{f_i}=f_{i+1}-f_i$, where $f_i=i/T$ such that $\Delta f_i=1/T,\,\forall\, i$, and to define the spectrum of the RMS residual as a function of frequency as 

\begin{equation}
    {\rm RMS}^2(f)_i=\int_{\Delta f_i}{\diff f} P_h(f)\approx P_h(f_i)\Delta{f_i}=\frac{P_h(f_i)}{T},
    \label{eq:rms_spectrum}
\end{equation}
which, as expected, has units of time. The RMS spectrum can then be converted into $h_c$ using Eq.~(\ref{e:Ph}) and then into  $S_h$ and $\Omega_{\rm GW}$ via Eq.~(\ref{eq:omega_gw}).

\begin{marginnote}
    \entry{Energy density}\, $\Omega_{\rm GW}(f)\equiv\frac{1}{\rho_c}\frac{{\rm d}\rho_{\rm GW}}{{\rm d\,ln}f}$
    \entry{Strain PSD}\, $S_h(f)=\frac{3H_0^2\Omega_{\rm GW}(f)}{4\pi^2f^3}$
    \entry{Characteristic strain} {$h_c(f)=\sqrt{2fS_h(f)}$}
    \entry{Timing residual PSD}\, $P_h(f)=\frac{h^2_c(f)}{12\pi^2f^3}$
    \entry{RMS residual spectrum}\, RMS$^2(f)=\frac{P_h(f)}{T}$
\end{marginnote}

\subsubsection{PTA signal-to-noise ratio computations and scalings}

The detectability of a signal is generally assessed through the construction of a detection statistic and the definition of a signal-to-noise (S/N) ratio $\rho$. An in-depth discussion of PTA data analysis techniques is beyond the scope of this review and an excellent account can be found in \cite{Taylor:2021yjx}. Here we just provide simple S/N estimates that stem from contrasting the GWB amplitude %GW signature in the array 
within the PTA window versus the potential noises affecting the TOAs of the pulses. In order to do this, we briefly introduce first such noise sources.

\paragraph{Main noise sources in PTAs} The noise PSD in the pulse TOAs is given by
\begin{equation}
P_n(f)=P_{\rm wn}+P_{\rm rn,ac}(f)+P_{\rm rn,c}(f),
\label{eqrho}
\end{equation}
where:
\begin{itemize}
    \item $P_{\rm wn}=2\sigma^2\Delta{t}$ is a frequency-independent white noise, determined by the cadence of the observations $\Delta{t}$ (typically weeks) and by the RMS TOA uncertainty  $\sigma$. The latter is mainly due to radiometer and jitter noise and has typical values of tens to hundreds of ns for high quality MSPs \cite{Cordes:2013iea};
    \item $P_{\rm rn,c}(f)$ is a chromatic 
    red noise depending %, by definition, 
    on the observed radio-photon frequency $\nu_\gamma$. It originates from dispersion ($\propto \nu_\gamma^{-2}$) and scattering ($\propto \nu_\gamma^{-4}$) effects in the ionized interstellar medium \citep[IISM,][]{Stinebring:2013yza}. Because of its frequency dependence, the chromatic 
    red noise contribution can be separated by using wide band receivers;
    \item $P_{\rm rn,ac}(f)$ is achromatic red noise due to intrinsic pulsar spin variations \cite{Shannon:2010bv}. These likely arise from complex crust-superfluid interactions and have been observed in several pulsars. Although this noise usually has a steep red spectrum that can mimic a GWB, it is %generally 
    expected to be uncorrelated among different MSPs.  
\end{itemize}

Besides the aforementioned noise sources %which are  
uncorrelated among pulsars, there are additional ones featuring specific correlation patterns. Most notably, those include clock offsets that affect all pulsars by inducing a monopole correlated signal, and errors in the determination of the solar system ephemeris, which result in a dipole correlation pattern \cite{2016MNRAS.455.4339T}. It can be shown, however, that the GWB-induced HD correlation has a prominent quadrupolar pattern that is orthogonal to both monopole and dipole \citep[e.g.][]{Taylor:2021yjx}.

\paragraph{S/N for deterministic signals}
For deterministic signals, we can apply the match filtering theory and define the S/N as \citep[e.g.][]{Sesana:2010mx}

\begin{equation}
\rho^2=\sum_a \rho_a^2,\,\,\,\,\,\,\,\,\rho_a^2 = (R(t)|R(t)) \approx \frac{2}{P_{0,a}}\int_0^{T_a}\diff t\, R^2_a(t),
\label{e:innerxyapprox}
\end{equation}
where the sum runs over all pulsars in the PTA, $(\cdot|\cdot)$ is the weighted inner product, and in the last passage we assumed a monochromatic sine wave.
%The total S/N is defined as the root-squared sum of $\rho_a$, which denotes the S/N in the $a$-th pulsar of the array.  $(\cdot|\cdot)$ is the weighted inner product, and in the last passage we assumed a monochromatic sine wave. 
$R_a$ is the residual imprinted in the $a$-th pulsar, and $P_0,a$ is the PSD of the noise in the $a$-th pulsar at the frequency of the GW. Eq.~(\ref{e:innerxyapprox}) also applies to a generic signal by decomposing it into its harmonics $h_n$, summing  over them, and considering $P(f_n)$ at the observed frequencies $f_n$ of each harmonic \citep[see][for an application to eccentric MBHBs]{Truant:2024aci}.

~~~~The integral Eq.~(\ref{e:innerxyapprox}) can be easily computed getting $R_a$ from Eqs.~(\ref{eq:ztgeneral}) and~(\ref{eq:deltaP}). If we consider an array of $N_p$ equal pulsars dominated by white noise, drop the pulsar term and  average over $F^+, F^{\times}$ and wave polarizations, we can solve 
%we obtain
%\begin{equation}
%\rho \approx \frac{1}{\sqrt{15}\pi}\frac{h}{\sigma f}(N_{\rm obs}N_p)^{1/2},
%\label{rhosingle}
%\end{equation}
%where $h$ is the inclination averaged polarization strain amplitude and $N_{\rm obs}=T/\Delta{t}$ is the number of observations for eahc pulsar. We see that the S/N for a monochromatic wave is proportional to the square root of the number of pulsars in the array and of the number of observations (i.e. the total observation time $T$, for a uniform observation cadence), and is inversely proportional to the rms residual $\sigma$. Equation  (\ref{rhosingle}) can be inverted to obtain 
for the minimum strain $h$ observable at a given S/N threshold, as
\begin{equation}
h\approx 6\times 10^{-15}\,\frac{\rho}{10}\,\frac{\sigma}{1\mu{\rm s}}\,\frac{f}{10^{-8} {\rm Hz}}\left(\frac{N_{\rm obs}}{500}\right)^{-1/2}\left(\frac{N_p}{100}\right)^{-1/2},
\label{aminCGW}
\end{equation}
where $N_{\rm obs}=500$ is approximately the number of TOAs  of each MSP under the assumption of weekly observations for 10 years. Assuming uniform sampling, $N_{\rm obs}\propto T$, we %can therefore 
see that the PTA sensitivity to a deterministic GW grows with $\sqrt{N_p}$, $\sqrt{T}$ and the RMS precision timing $\sigma$. As anticipated, we need strain amplitudes $h>10^{-15}$ to have a chance of detection. For comparison, typical strains detected in ground based detectors %inspiralling binaries are of 
are $h\sim 10^{-22}-10^{-21}$.

\paragraph{S/N for a GWB}
For stochastic signals, the strategy is to detect the HD cross-correlated power in the array. The S/N can be written as \cite{Allen:1997ad,2015MNRAS.451.2417R}
\begin{equation}
\rho^2=2\sum_{a=1,M}\sum_{b>a}T_{ab}\int \frac{\Gamma_{ab}^2P_h^2}{P_{n,\hat{ab}}^2}df,
\label{eqrho}
\end{equation}
where the sums run over all pulsar pairs, $T_{ab}$ is the timespan of overlapping observations for pulsars $a$ and $b$,\footnote{Note that, in general, MSPs have different observation coverage, depending on when they were discovered, the schedule requirement at observatories, etc.} and $\Gamma_{ab}$ is defined by Eq.~(\ref{eq:HD}). The correlated noise term is given by
\begin{equation}
  P_{n,\hat{ab}}^2=P_{n,a}P_{n,b}+P_h[P_{n,a}+P_{n,b}]+P_h^2(1+\Gamma_{ab})^2,
    \label{eqsn}
\end{equation}
We assume that all pulsars are equal so that $P_{n,x}=P_n$ for each pulsar $x$. Eq.~(\ref{eqrho}) has a different scaling depending on the relative magnitude of $P_h$ and $P_n$, i.e. on whether we are in the weak or strong signal regime \cite{2013CQGra..30v4015S}. So long as $P_h<P_n$ one can ignore $P_h$ in the denominator of Eq.~(\ref{eqrho}). By assuming a red spectrum of the form of Eq.~(\ref{eq:hc_analytical}), and using an average value $\sqrt{\langle\Gamma^2\rangle}=1/(4\sqrt{3})$, the integral is trivial, and one can solve for $A_{\rm GWB}$ to obtain the scaling for the minimum detectable amplitude at $f={1{\rm yr}^{-1}}$:
\begin{equation}
A_{\rm GWB}\approx 9\times10^{-14} \,\left(\frac{\rho}{10}\right)^{1/2}\,\frac{\sigma}{1\mu{\rm s}}\,\left(\frac{T}{1{\rm yr}}\right)^{\alpha-1}\left(\frac{N_{\rm obs}}{50}\right)^{-1/2}\left(\frac{N_p}{100}\right)^{-1/2}.
\label{eqaweak}
\end{equation}
Here we assumed weekly observations for 1 year. We see that by keeping observing with the same cadence, $A_{\rm GWB}\propto T^{\alpha - 3/2}$. Assuming $\alpha=2/3$, a typical GWB amplitude of $A_{\rm GWB}\approx 10^{-15}$ can be detected on a $T=10$ yr timescale.
Once $P_h \gtrsim P_{n}$, it dominates the noise budget and a crude approximation to the S/N becomes: 
\begin{equation}
\rho \approx 0.15\,T^{1/2}N_p(f_{\rm max}-f_{\rm min})^{1/2}\approx 0.15\, N_p \Sigma^{1/2},
\label{eq:snrGWBstrong}
\end{equation}
where $f_{\rm min}=1/T$ and $f_{\rm max}$ is the maximum frequency for which the condition $P_h> P_{n}$ is satisfied. For the last $\approx$, we divided the frequency domain in resolution bins $\Delta{f}=1/T$, and took $\Sigma$ as the number of frequency bins for which $P_h> P_{n}$ (usually a few). While the S/N grows linearly with $N_p$, it now only grows as $\sqrt{T}$, due to the uncorrelated part of the GWB acting as an effective common red noise source. Note also that so long as $P_h> P_{n}$, the RMS residual $\sigma$ plays a marginal role, by weakly affecting the $f_{\rm max}$ limit (or alternatively $\Sigma$). We note that the scaling in Eq.~(\ref{eq:snrGWBstrong}) applies regardless of the GWB spectral shape.

\subsubsection{Current PTA efforts}
There exist three main historical PTA collaborations: the European PTA \citep[EPTA,][]{EPTA:2023sfo}, the Parkes PTA \citep[PPTA,][]{Zic:2023gta} and the North American Nanohertz Observatory for gravitational waves \citep[NANOGrav,][]{NANOGrav:2023hde}. They are taking data for over 100 MSP combined for about two decades and signed an agreement in 2010 to form the International PTA \citep[IPTA,][]{Perera:2019sca}, a Consortium of Consortia %framework 
aimed at enhancing the experiment sensitivity by combining regional PTA data. The IPTA effort was recently joined by the Indian PTA \citep[InPTA,][]{Rana:2025ano} and by the MeerKAT PTA \citep[MPTA,][]{Miles:2022lkg}, the PTA experiment conducted at the MeerKAT radio telescope in South Africa and taking data since 2018. To complete the picture, the Chinese PTA \citep[CPTA,][]{Chen:2025qqb} also began its observational campaign in 2019 and is now joining the IPTA. We discuss the latest PTA results in Sec.~\ref{sec:PTA_evidence}. The sensitivity of current and future PTAs is shown in Fig.~\ref{fig:GWlandscape}.

\subsection{Other nano-Hz GW detection techniques}
\label{sec:astrometry}

\subsubsection{Astrometry}

The same way GWs alter the path of radio photons emitted by MSPs, they also perturb the path of light coming from stars, causing measurable shifts in their apparent positions in the sky. The derivation of the effect is similar to the frequency shift induced in MSPs TOAs.\begin{marginnote}[]
\entry{Astrometry}
\,precise measurement of stellar position and proper motion, can be used to detect GWs due to the induced perturbation in the stars apparent position.
\end{marginnote} If ${\hat n}$ is the incoming photon wave vector, ${\hat\Omega}$ is the propagation direction of a plane wave GW $h_{ij}$, and the distance to the star is much larger than the GW wavelength, the astrometric deflection is then~\cite{1996ApJ...465..566P,Book:2010pf}
\begin{equation}
\delta n^i = \left(\frac{n^i  +\Omega^i}{2 (1 + \hat{n} \cdot \hat\Omega)} n^j n^k - \frac{1}{2} \delta^{ik}n^j\right)  h_{jk}\,,
\label{eq:astrometric_shift}
\end{equation}
with $h_{jk}$ %is %the spatial part of 
%the GW %tensor 
evaluated at the Earth location.     The Gaia mission \cite{Gaia:2016zol} is releasing precise astrometric measurements of billions of stars in the Milky Way over 10 years with average monthly cadence, which makes this dataset sensitive to GWs in the 3 nHz - 100 nHz frequency range, overlapping with PTAs ({\it c.f.} Fig.~\ref{fig:GWlandscape}).    

%\begin{boxC}
%The Gaia mission \cite{Gaia:2016zol} is releasing precise astrometric measurements of billions of stars in the Milky Way over 10 years with average monthly cadence, which makes this dataset sensitive to GWs in the [3 nHz - 100 nHz] frequency range, overlapping with PTAs.    
%\end{boxC}
The $\delta n_i$ in astrometry are equivalent to $z(t)=-\delta\nu/\nu$ in PTA. However, there are two degrees of freedom on the celestial sphere, and the expectation value of the correlation between astrometric deflections from two stars $a$ and $b$ is not a scalar function, taking the form
%\begin{marginnote}
%\entry{Astrometry band} \,
%\end{marginnote}

\begin{equation}
\langle \delta n_a^i \delta n_b^j\rangle \propto  \mathcal{T}(\theta)U^{ij}(\hat{n}_a,\hat{n}_b)\times S_h(f)\,,
\label{eq:Mihaylov}
\end{equation}
where $\mathcal{T}(\theta)$ is the astrometric analog of the HD curve, only depending on the separation angle between the two stars $\theta$, while $U^{ij}(\hat{n}_a,\hat{n}_b)$ is a function of the unperturbed star directions that depends on the coordinate choice \cite{Mihaylov:2018uqm}. 

A limitation of Eq.~(\ref{eq:Mihaylov}) is the reliance of a fixed reference frame that is practically extremely hard to determine. For this reason \cite{Pardo:2023cag} introduced the concept of {\it relative} astrometry, where the positions of stars within a field-of-view are evaluated relative to each other. A similar concept was developed by \cite{Vaglio:2025tex}, who investigated the use of pairs of close stars as {\it differential} astrometric rulers. 
Given a pair $a$ of two stars separated by an angle $\psi^a_{12}$ and a pair $b$ of stars separated by another angle $\psi^b_{12}$, both $\psi^a_{12}$ and  $\psi^b_{12}$ experience a small change due to the GW passage, and one can evaluate the expectation value of the correlation of those changes, $\bigl\langle \delta \!\cos\psi^a_{12}\,\delta\!\cos\psi^b_{12} \bigr\rangle$. 
Noticeably, when averaged over the relative orientations of the two star pairs, this correlation is identical to the HD curve. We also note that both the {\it absolute} $\delta n_i$ and {\it relative} $\delta\!\cos\psi^b_{12}$ astrometric shifts can be cross-correlated with the frequency shift $z(t)$, to obtain the observables $\langle z\,\delta n_i\rangle$ and $\langle z\, \delta\!\cos\psi^b_{12}\rangle$, allowing to combine astrometric and PTA measurement in a coherent framework 
\cite{Qin:2018yhy,2024JCAP...05..030C,2024PhRvD.110f3547I,Cruz:2024diu,Vaglio:2025tex}.

\paragraph{S/N computation for astrometry and astrometry+PTAs.}

The S/N of a GWB in an astrometric experiment has the same form as Eq.~(\ref{eqrho}) but: (i) the sums run over all possible pairs of stars (or pairs of stellar pairs if we consider relative astrometry); (ii) the HD curved is replaced by the appropriate astrometric correlation (being always a ${\cal O}(1)$ factor); (iii) $P_n=2\sigma^2_\star\Delta{t_\star}$, being $\sigma_\star$ and $\Delta{t_\star}$ the astrometric precision and cadence of the measurements, respectively; (iv) $P_h$ is replaced by $S_h=h_c^2/(2f)$ since we are dealing with angular rather than time measurement. By solving for $A_{\rm GWB}$ as before, we obtain:
\begin{equation}
A_{\rm GWB}\approx 9\times10^{-14} \,\left(\frac{\rho}{10}\right)^{1/2}\,\frac{\sigma_\star}{0.2{\rm mas}}\,\left(\frac{T}{1{\rm yr}}\right)^{\alpha}\left(\frac{N_{\rm obs}}{20}\right)^{-1/2}\left(\frac{N_\star}{10^9}\right)^{-1/2},
\label{eqaweak}
\end{equation}
normalized to Gaia mission performance \cite{Gaia:2016zol}. % %\begin{marginnote}
    %\entry{Astrometry sensitivity}\, 
    A striking feature to note is that, with the current precision, astrometry needs to monitor ${\cal O}(10^9)$ stars to achieve performaces similar to PTAs.   
%\end{marginnote}
%\begin{boxC}
% A striking feature to note is that astrometry needs to monitor ${\cal O}(10^9)$ stars to achieve performaces similar to PTAs.   
%\end{boxC}
 We also note that, by keeping the same observing cadence, $A_{\rm GWB}\propto T^{\alpha - 1/2}$, which means that astrometry benefits less than PTAs from longer surveys. This is because the latter are sensitive to temporal rather than angular perturbations, resulting in a steeper sensitivity as a function of frequency ($h_c\propto f^{3/2}$ for PTA vs $h_c\propto f^{1/2}$ for astrometry for $f>1/T$, see Fig. ~\ref{fig:GWlandscape}). 

~~~~When combining PTAs and astrometry, the total $S/N$ is $\rho^2_{\rm tot}=\rho^2_{\rm PTA}+\rho^2_\star+\rho^2_{\star/{\rm PTA}}$, where the latter is given by the correlation of the two measurements and can be also written by appropriately modifying Eq.~(\ref{eqrho}). Beyond the simple scalings derived here, the sensitivity of PTAs, astrometric measurements and their correlations have been carefully explored by a number of Authors %and tools exist for proper S/N computation and parameter estimation forecasts 
~\citep[see, {\it e.g.},][]{Qin:2018yhy,Cruz:2024diu,Vaglio:2025tex}.

\subsubsection{Laser interferometry} Laser interferometry is the most successful technique to detect kilo-Hz GWs, and is anticipated to open the milli-Hz frequency band in the mid 2030's with LISA 
\cite{LISA:2017pwj,2024arXiv240207571C}, TianQin \cite{TianQin:2015yph} and Taiji \cite{Hu:2017mde}. Going to lower frequencies would require building a solar system scale detector. Although futuristic, mission concepts for satellite constellations forming a triangle of the size of the Earth or Mars orbit have been studied \cite{Sesana:2019vho}. Pushing the interferometry detection window deep in the $\mu$Hz range, these detectors might achieve a sensitivity of $h_c\approx 10^{-16}$ around 100 nHz, complementing PTAs at high frequencies ({\it c.f.} Fig.~\ref{fig:GWlandscape}).

\subsubsection{Binary resonances}

GWs can resonantly interact with astrophysical binaries, causing small deviations in their orbital period \cite{Hui:2012yp,Foster:2025nzf}. The order of magnitude effect that can be detected depends on the precision of the determination of the binary period, of its periastron passage, and on the number of observed passages \cite{Hui:2012yp}. A projected sensitivity this technique might achieve by the end of the 30s' by monitoring currently known MSPs in binaries is shown in Fig.~\ref{fig:GWlandscape} \citep[from][]{Foster:2025nzf}.

\subsection{Current evidence of nano-Hz GW signal from PTA data}
\label{sec:PTA_evidence}

\begin{figure*}[t]
    \centering
    \includegraphics[width=\linewidth]{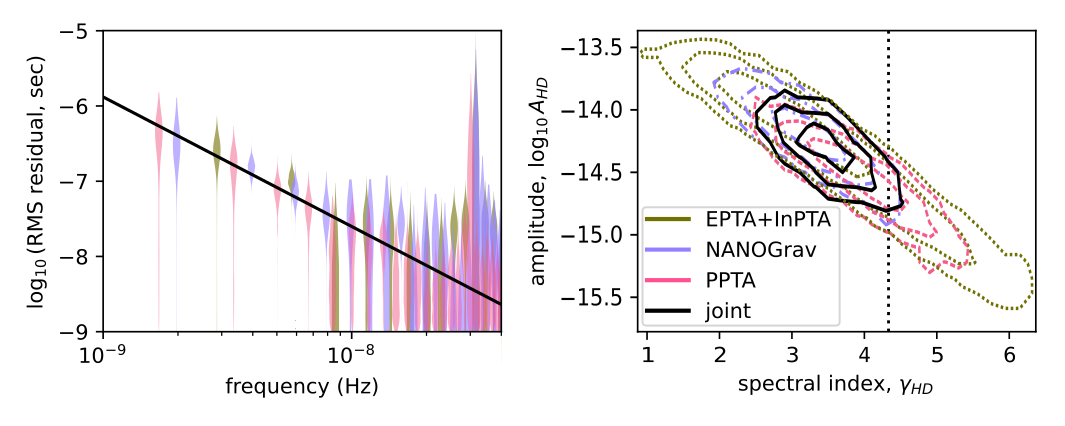}
    \caption{Left panel: free spectrum showing the PSD of the RMS residuals measured by EPTA+InPTA, NANOGrav and PPTA. Right Panel: 2-D $A_{\rm GWB}$ - $\gamma$ posterior distribution of the GWB parameters as determined from the data (68\%, 95\% and 99.7\% confidence intervals) together with the joint posterior. The vertical dotted line represents $\gamma=13/3$ for reference. Adapted from \cite{InternationalPulsarTimingArray:2023mzf}.}
    \label{fig:SGWB_evidence} 
\end{figure*}

On June 29th 2023, the major PTAs forming the IPTA (EPTA+InPTA,\footnote{For this analysis EPTA and InPTA worked together by joining their data into a common EPTA+InPTA dataset.} NANOGrav and PPTA) together with CPTA, coordinated four independent publications reporting evidence on a HD correlated signal in their respective datasets~\cite{NANOGrav:2023gor, Antoniadis:2023ott, Reardon:2023gzh, Xu:2023wog}. MPTA followed in early 2025, also reporting evidence of a correlated signal in their data~\cite{Miles:2024seg}. The signal is consistent with an isotropic GWB, with evidence ranging from $2\sigma$ to $4\sigma$, depending on the dataset, see %Measured values of $A_{\rm GWB}$ and $\gamma$, {\it c.f.}~Eq.~(\ref{e:Ph}), are reported in 
Tab.~\ref{tab:pta_results}. While EPTA+InPTA, NANOGrav and PPTA spectra are well constrained and broadly consistent (as also shown in Fig.~\ref{fig:SGWB_evidence}), CPTA and MPTA data are  much less constraining, mostly due to the reduced timespan of these datasets (roughly 3 and 4.5 years respectively, compared to 10-to-20 years of other PTAs). While the FAST (CPTA) and MeerKAT (MPTA) telescopes have a superior sensitivity allowing them to detect a correlated signal with a much shorter observing baseline, more observational time is still required for a properly spectral signal characterization. %this prevents a proper characterization of the spectral properties of the  signal. 

\begin{table*}[t]
\begin{center}
    ~~~\begin{tabular}{cccc}
\hline\\[-0.75em]
 PTA & ~ ~  $A_{\rm GWB}/10^{15}$ ~ ~ & ~ ~ $\gamma$ ~ ~ & ~ ~ $A_{\rm GWB}(\gamma=13/3)/10^{15}$ ~ ~ \\[0.25em] \hline\\[-0.5em]
 EPTA+InPTA & $11.48^{+8.02}_{-7.68}$ & $2.71^{+1.18}_{-0.73}$ & $2.5^{+0.7}_{-0.7}\,(90\%)$\\
 [+0.3em]
 NANOGRAV & $6.4^{+4.2}_{-2.7}$ & $3.2^{+0.6}_{-0.6}$ & $2.4^{+0.7}_{-0.6}\,(90\%)$\\
  [+0.3em]
  PPTA & $3.1^{+1.3}_{-0.9}$ & $3.9^{+0.4}_{-0.4}$& $2.0^{+0.3}_{-0.1}\,(68\%)$\\
 \\[-0.5em] \hline \\[-0.5em]
 CPTA & $3.98^{+33.82}_{-3.97}$ & - & $1.99^{+13.85}_{-1.97}\,(95\%)$\\
 [+0.3em]
 MPTA & $7.5^{+0.8}_{-0.9}$ & $3.52 ^{+1.12}_{-0.9}$ & $4.8^{+0.8}_{-0.9}\,(68\%)$ \\ [0.5em] \hline
\end{tabular}
\vspace{0.5cm}
\caption{Amplitude $A_{\rm GWB}$ and slope $\gamma$ of the HD correlated common red signal measured by several PTAs ({\it c.f.} Eq.~\eqref{e:Ph}). Reported are both the best fit $A_{\rm GWB}$ and $\gamma$ (col. 2 and 3) and the $A_{\rm GWB}$ measured by fixing $\gamma=13/3$ (col. 4).}
\label{tab:pta_results}
\end{center}
\end{table*}

~~~~The PSD of the residuals [{\it c.f.} Eq.~(\ref{eq:rms_spectrum})] and the 2-D $A_{\rm GWB}-\gamma$ posterior distributions for EPTA+InPTA, NANOGrav and PPTA are depicted in Fig.~\ref{fig:SGWB_evidence}, together with a 'face value combination' obtained by stacking them,\footnote{This implies taking a flat prior across the three PTAs and assuming the three experiments are independent, which is not the case since several pulsars are shared among different arrays.}. %while an example of HD reconstruction is shown in the upper left panel of Fig.~\ref{fig:epta_interpretation}. 
The emerging picture is a GWB characterized by $A_{\rm GWB}\in[2\times10^{-15} - 3\times10^{-15}]$ (for a reference $\gamma=13/3$) and $\gamma\in[2.5-4]$ (1$\sigma$ confidence). At the current sensitivity level, the signal is broadly consistent with an isotropic, Gaussian GWB and searches for spectral features \cite{Agazie:2024jbf}, anisotropies \cite{NANOGrav:2023tcn} and resolved deterministic signals \cite{NANOGrav:2023pdq,EPTA:2023gyr} did not yield any significant result. Both EPTA and NANOGrav published comprehensive papers dissecting these results and their possible interpretations \cite{EPTA:2023xxk,NANOGrav:2023hvm,NANOGrav:2023hfp}, we will comment on that in the next two Sections.

\section{Astrophysical nano-Hz GW signals}
\label{sec:astroGWB}

Primary astrophysical GW sources are compact object binaries \cite{1963PhRv..131..435P}. The GW emission causes energy loss from the pair and the binary orbit shrinks until the two compact objects eventually coalesce. For a circular binary of BHs of total mass $M$, the inspiraling phase lasts up to the innermost stable circular orbit (ISCO), corresponding to an emitted GW frequency
\begin{equation}
    f_{\rm ISCO}\approx \left(6^{3/2}\pi\frac{GM}{c^3}\right)^{-1}\approx 4\times10^{-6}\left(\frac{M}{10^9\msun}\right)^{-1}{\rm Hz}.
\end{equation}
GWs emitted during the inspiral have $f < f_{\rm ISCO}$. The most massive black holes (MBHs) observed in the Universe have masses of about $10^{10}\msun$ \cite{Kormendy:2013dxa}, for which %Therefore, even for the most massive compact binaries 
$f_{\rm ISCO} \gtrsim 10^{-7}$Hz. Since we are interested in nano-Hz GWs, and since both PTAs and astrometric techniques are most sensitive at $f<1{\rm yr}^{-1}\approx 3\times10^{-8}$Hz, for all the astrophysically relevant mass range the signals of interest are adiabatically inspiralling binaries, moving at sub-relativistic speeds $v \ll c$, hence allowing the use of the quadrupolar approximation \cite{1963PhRv..131..435P}.\footnote{This assumption breaks down for systems with $M>10^{10}\msun$ at $z > 10$. As we shall see below, the observed signal is redshifted by the expansion of the Universe, and the observed $f_{\rm ISCO}$ is $\approx10$ nHz, right in the band of interest. Despite the wealth of MBHs at high $z$ found by JWST \citep[e.g.][]{2024A&A...691A.145M}, there is no evidence of the existence of $M>10^{10}\msun$ MBHs at $z \gtrsim10$.} 

\subsection{Quadrupolar GW emission from a generic eccentric binary}
\label{sec:MBHB_GWbasics}

The most general system considered in this chapter is an eccentric, non spinning binary with total mass $M=M_1+M_2$, mass ratio $q=M_2/M_1<1$, orbiting at a semimajor axis $a$ with orbital frequency $\omega_K=2\pi f_K=(GM/a^3)^{1/2}$. 
The system emits GWs at integer harmonics of the orbital frequency, and $h_{+,\times}$ can be written as \citep[e.g][]{Taylor:2021yjx}
\begin{eqnarray} \label{eq:h_plus_cross}
h_+(t) &=&  \sum_{n = 1}^{\infty} -\left(1+\cos^2{\iota}\right) \left[a_n\cos{(2\gamma(t))} - b_n\sin{(2\gamma(t))}\right] + \left(1-\cos^2{\iota}\right)c_n,\\ 
        %h_+(t) &=  \sum_{n = 1}^{n = \infty} -\left(1+\cos^2{\iota}\right) \left[a_n\cos{(2\gamma)} - b_n\sin{(2\gamma)}\right] + \left(1-\cos^2{\iota}\right)c_n,\\ 
h_\times(t) &=& \sum_{n = 1}^{\infty} 2\cos{\iota}\left[b_n \cos{(2\gamma(t))} + a_n \sin{(2\gamma(t))}\right],\\
 %   \end{aligned}
&&\hspace*{-2.4cm}{\rm where}\nonumber\\
a_n &=& - (n/2)A(f_K)\Big[J_{n-2}(ne) -2eJ_{n-1}(ne)+(2/n) J_n(ne) \nonumber\\%[-1ex] 
              && +~ 2eJ_{n+1}(ne)-J_{n+2}(ne)\Big]\cos{(nl(t))}, \nonumber\\
        b_n &=& -(n/2)A(f_K) \sqrt{1-e^2}\left[J_{n-2}(ne)-2J_n(ne) + J_{n+2}(ne)\right]\sin{(nl(t))}, \nonumber\\
        c_n &=& A(f_K)J_n(ne)\cos(nl(t)),\nonumber
\end{eqnarray}
with $n$ the harmonic number, and $J_n(x)$ Bessel functions of the first kind (of order $n{-}\rm th$). The amplitude $A(f_K)$ is given by
\begin{equation}\label{eq:Agw}
\begin{split}
A(f_K)  & = 2 \frac{{\cal M}^{5/3}}{D}\,\left(2\pi f_{K,r}\right)^{2/3} \equiv 2 \frac{{\cal M}_z^{5/3}}{D_l}\,\left(2\pi f_K\right)^{2/3}\\ & \approx
3\times 10^{-16}\left(\frac{f_{K,r}}{5\times10^{-9}{\rm Hz}}\right)^{2/3}\left(\frac{\cal M}{10^9\msun}\right)^{5/3}\left(\frac{D}{1{\rm Gpc}}\right)^{-1}.
\end{split}
\end{equation}

\begin{boxC}
\noindent{\bf Note on redshift.} Due to the  Universe expansion, GWs are redshifted just like EM waves. A GW signal can be expressed either in the source rest frame (at the source redshift $z$) or in the observer frame (at $z=0$). We adopt the following convention:
\begin{itemize}
    \item $f$ is the observed frequency, which is related to the source frame one, $f_r$, by $f=f_r/(1+z)$ (and thus the observed angular frequency $\omega=\omega_r/(1+z)$)
    \item similarly, we define the observed $t$ and source frame $t_r$, such that $t=t_r(1+z)$
    \item all plain mass symbols, i.e. ${\cal M}$, are defined in the source frame, while observed ones carry a $z$ label, i.e. ${\cal M}_z$. The relation between the two is ${\cal M}_z={\cal M}(1+z)$. Thus, in the observer frame we talk about {\it redshifted} masses;
    \item the comoving distance $D$ is adopted in the source frame, while the luminosity distance $D_l=D(1+z)$ is adopted in the observer frame.
\end{itemize} 
\end{boxC}

~~~~It is straightforward to see that the GW amplitude can be expressed in the source frame by using ($f_r,\,t_r,\,{\cal M},\,D$) or in the observer frame by using ($f,\,t,\,{\cal M}_z,\,D_l$), as in Eq.~(\ref{eq:Agw}). 
%The preference of one frame over another depends on the focus of the research. 
When dealing with GW observations it is generally preferable the observer frame; however, rest frame quantities are more intuitive when connecting GW signals with the physical properties of the sources. For example, the mass function of the population is physically more relevant than the redshifted mass function. We will generally use rest frame quantities. 
%Although with the definitions given here there should not be ambiguity.

~~~~In the above set of equations, ${\cal M}=(M_1M_2)^{3/5}/M^{1/5}$ is the {\it chirp} mass, $\iota$ is the inclination angle defined as the angle between the GW propagation direction and the binary orbital angular momentum $\hat{L}$, $l(t)$ is the binary {\it mean anomaly} $l(t)\,{\equiv}\,l_0 + 2\pi \int_{t_0}^t f_K(t') dt'$ (note that $f_K(t')$ is the {\it observed} Keplerian frequency, see above box); finally, $\gamma(t)$ is the angle that measures the direction of the pericenter with respect to the direction $\hat{x}$, defined as $\hat{x} \equiv (\hat{\Omega} + \hat{L}\cos{i})/\sqrt{1-\cos^2{i}}$, where $\hat\Omega$ is the unit vector defining the GW propagation direction \cite{2004PhRvD..69h2005B}.
%In principle, $f)$ can evolve during the observation time due to the GW emission and environmental interaction. However,  here we assume that the orbital frequency and eccentricity do not evolve during the observation time, hence, $f_k(t)$ can be treated as a constant value, namely, $f_k$ (see the discussion on this assumption in Sect.~\ref{sec:Caveats}). The orbital angular frequency is then given by $\omega\,{=}\,2\pi f_k$, while  $\gamma$ is the angle that measures the direction of the pericenter with respect to the direction $\hat{x}$, defined as $\hat{x} \equiv (\hat{\Omega} + \hat{L}\cos{i})/\sqrt{1-\cos^2{i}}$.\\ 
GWs extract energy from the binary at a rate
\begin{equation}
   \frac{{\rm d}E}{{\rm d}t_r}=L_{\rm GW}=\frac{32}{5}\frac{G^{7/3}}{c^5} (2\pi f_{K,r}{\cal M})^{10/3}F(e),
\end{equation}
causing the binary orbital elements to evolve at a rate
\begin{equation}
\begin{split}
   \frac{{\rm d} a}{{\rm d}t_r}=-\frac{64}{5}\frac{G^3}{c^5}\frac{{\cal M}^{5/3}M^{4/3}}{a^3} & {\cal F}(e) \,\,\,\,\,\,\,\,\,\, \rightarrow \,\,\,\,\,\,\,\,\,\,\frac{{\rm d}f_{K,r}}{{\rm d}t_r} =\frac{96}{5}(2\pi)^{8/3}\left(\frac{G{\cal M}}{c^3}\right)^{5/3}f_{K,r}^{11/3\,}\mathcal{F}(e),\\
   \frac{\diff e}{\diff t_r} &  = -\frac{(2\pi)^{8/3}}{15}\left(\frac{G{\cal M}}{c^3}\right)^{5/3}f_{K,r}^{8/3}\,\mathcal{G}(e),
   \label{eq:dfdt}
\end{split}
\end{equation}
Where $\mathcal{F}(e)$ and $\mathcal{G}(e)$ take the form: 
\begin{equation}\label{eq:FeGe}
%\begin{split}
\mathcal{F}(e) = \sum_{n=0}^\infty g(n,e)= \frac{1+(73/24)e^2 + (37/96)e^4}{(1-e^2)^{7/2}}\,, ~~~~~
\mathcal{G}(e) = \frac{304e + 121e^3}{(1-e^2)^{5/2}}\,, 
\end{equation}
and 
\begin{equation}
\begin{split}
\label{eq:gne}
g(n,e) &= \frac{n^{4}}{32} \bigg[ \bigg( J_{n-2}(ne)-2eJ_{n-1}(ne)+\frac{2}{n}J_n(ne) +2eJ_{n+1}(ne)-J_{n+2}(ne) \bigg)^{2} \\
&\quad +(1-e^{2})\bigg(J_{n-2}(ne)-2J_n(ne)+J_{n+2}(ne)  \bigg)^{2}\frac{4}{3n^{2}}J^{2}_n(ne) \bigg].
\end{split}
\end{equation}
Eq.~(\ref{eq:dfdt}) indicates that as the binary shrinks, increasing its frequency, it also circularizes. It is therefore expected that comparable mass binaries in the late stage of evolution are rather circular, which is why most early literature on nHz GW signals assumed circular binaries \cite{Rajagopal:1994zj,Jaffe:2002rt,Wyithe:2002ep,Sesana:2004sp}.

\subsection{Case study: the GWB from a cosmic population of GW-driven circular MBHBs}
\label{sec:GWB_circular}

We start by discussing circular binaries, $e=0$. In this case, GWs are emitted only at the second harmonic $n=2$, and therefore at $f=2f_K$. The strain in Eq.~(\ref{eq:h_plus_cross}) takes the form:

\begin{equation}\label{eq:h_circular}
%\begin{aligned}
h_+(t) = (1 + \cos^2 \iota) A \cos(\omega t+\Phi_0)\,;~~~~
h_{\times}(t) =-2 \cos\iota \, A \sin(\omega t+\Phi_0)\,,
%\end{aligned}
\end{equation}
where $\omega=2\pi f=2\omega_K$ is the wave angular frequency and  $\Phi_0$ is an irrelevant initial phase. In this case, ${\cal F}(e=0)=1$ and ${\cal G}(e=0)=0$, implying that circular binaries stay circular. 

During the adiabatic inspiral the binary emits GWs with energy spectrum
\begin{equation}
   \frac{{\rm d}E}{{\rm d}f_r}=\frac{{\rm d}E}{{\rm d}t_r}\frac{{\rm d}t_r}{{\rm d}f_r}=\frac{\pi}{3G}\frac{(G{\cal M})^{5/3}}{(\pi f_r)^{1/3}}.
    \label{eq:dEdf}
\end{equation}

We consider a cosmic population of compact binaries in circular orbits. %Let us now suppose we have a cosmic population of compact objects inspiralling in circular orbits. 
The population can be generic %(stellar mass BH binaries, neutron star binaries, white dwarf binaries, and even main sequence star binaries), 
but, as we will see, the nano-Hz GW band is dominated by very massive binaries, {\it i.e.}~MBHBs, 
\begin{marginnote}
    \entry{MBHBs}{
    With masses $M\approx 10^9 \msun$, they dominate the nano-Hz GW band with wave strains up to $h\sim 10^{-14}$.}
\end{marginnote}
so we will specialize all our calculations to this case. We define ${\rm d}^2n/({\rm d}z\,\rm{d{\cal M}})$ to be the merger rate density ({\it i.e.}~the number of mergers per unit of comoving volume, per unit of redshift, and per unit of chirp mass), and we denote with ${\rm d}E/{\rm d\,ln}f_r$ the energy emitted per unit logaritmic rest frame frequency by each source. The GW emission of this population across cosmic time generates a GWB with energy density \cite{Phinney:2001di}
\begin{equation}
    \epsilon_{\rm GW}=\int{\rm d\,ln}f\int \diff z \int \diff {\cal M} \frac{1}{1+z}\frac{\diff^2n}{\diff z\, \diff {\cal M}}\frac{\diff E}{\diff {\rm ln}f_r},
    \label{eq:egw_phinney}
\end{equation}
where $1+z$ is due to the Universe expansion. Combining Eq.~(\ref{eq:egw_phinney}) and Eq.~(\ref{eq:epsilon_gw}) we get
\begin{equation}
    h_c^2(f)=\frac{4G}{\pi c^4}\frac{1}{f^2}\int \diff z \int \diff {\cal M} \frac{\diff^2n}{\diff z\, \diff {\cal M}}\frac{\diff E}{\diff {\rm ln}f_r}.
    \label{eq:hc_phinney}
\end{equation}
%We can specialize Eq.~(\ref{eq:hc_phinney}) to a population of compact binaries by 
If we substitute $\diff E/\diff f$ from Eq.~(\ref{eq:dEdf}), assume a monochromatic mass function and write $\diff^2n/(\diff z \diff {\cal M})=(\diff n/\diff z)\delta({\cal M}-{\cal M}_0)$, by defining $n_0= \int \diff z \int \diff {\cal M}\, \diff^2n/(\diff z\, \diff {\cal M})$ we get
\begin{equation}
    h_c^2(f)=\frac{4(G{\cal M}_0)^{5/3}}{3\pi^{1/3} c^2}\frac{1}{f^{4/3}}n_0\langle(1+z)^{-1/3}\rangle.
    \label{eq:hc_phinney2}
\end{equation}
where $n_0$ can be interpreted as today's number density of merger relics, and $\langle(1+z)^{-1/3}\rangle$ is averaged over the redshift distribution of the coalescences producing those relics \cite{Phinney:2001di}. For example, if we take a coalescence rate that is constant in time, then $\langle(1+z)^{-1/3}\rangle\approx0.8$, and if we define $\chi(z)=\langle(1+z)^{-1/3}\rangle/0.8$, we can finally write 
\begin{equation}
h_c(f)\approx10^{-15}\left(\frac{f}{1{\rm yr}^{-1}}\right)^{-2/3}\left(\frac{\cal M}{10^9\msun}\right)^{5/6}\left(\frac{n_0}{10^{-4}{\rm Mpc}^{-3}}\right)^{1/2}\chi(z)^{1/2}.
    \label{eq:hc_normaliz}
\end{equation}
As anticipated in Sec.~\ref{sec:gwb_basics}, %we see that $h_c$ follows %a power law as $\propto f^{\alpha}$, {\it c.f.}~
%Eq.~(\ref{eq:hc_analytical}), with $\alpha=-2/3$. 
the GWB from a cosmic population of circular, GW driven, inspiralling compact objects is described by a single PL $h_c\propto f^{-2/3}$. The normalization $n_0=10^{-4}{\rm Mpc}^{-3}$ in Eq.~(\ref{eq:hc_normaliz}) is an estimate of the number density of MBHs of billion solar masses residing in massive elliptical galaxies today. Eq~(\ref{eq:hc_normaliz}) states that if each of them has undergone one major merger since $z=1$ (which is very reasonable, as we will see below), the expected GWB strain-amplitude at PTA frequencies is $A_{\rm GWB} \sim 10^{-15}$. 

\subsubsection{Relation between MBHB mass function and characteristic strain} 
\label{sec:satopolito}
%The connection between $h_c$ and the relic population of MBHBs was further explored by \cite{Sato-Polito:2023gym}. 
If the late growth of MBHs in the Universe is merger dominated, then the relic population must correspond to today's MBH mass function $\phi(M)=\diff n/\diff{\rm log}M$ \cite{Sato-Polito:2023gym}. 
Assuming that today's MBHs are the result of a single equal mass merger, Eq.~(\ref{eq:hc_phinney}) can be written as
\begin{equation} \label{eq:satopolito}
    h_{c,1}^2(f) =\int \diff {\rm log}M \frac{(GM)^{5/3}}{3\pi^{1/3} c^2}\frac{1}{f^{4/3}} \phi(M),
\end{equation}
where we assumed %, for simplicity, 
that mergers occur at $z\ll1$, and the subscript $_1$ refers to %the fact that we are 
considering each MBH today to be the result %product 
of a single merger event. %One can go a step further and compute the upper limit of a GWB produced by a given $\phi(M)$. This is done 
By assuming that MBHs today are the product of a series of subsequent equal mass mergers, that can be reconstructed backwards by doubling the number of involved MBHs while halving their masses at each round. In the most favorable toy scenario, %in which the sequence of mergers occurs at $z\approx0$, all MBHs are nearly maximally spinning and merge in prograde orbits,\footnote{i.e. the MBH spins are aligned with the binary orbital angular momentum, maximizing the mass-energy loss in GW at each merger round. This allows the merger progenitors at each round to be more massive and therefore enhances the overall GWB.}  
the series converges to $h_{c}^{\rm MAX}\approx 2 h_{c,1}$, which is the upper limit of a GWB produced by a given $\phi(M)$ (Ferranti et al. in preparation).

~~~~%We will see more about the implications of this analysis in Sec.~\ref{sec:MBHB_interpretation}, but we 
%We can exploit Eq.~(\ref{eq:satopolito}) to make some interesting considerations about the main contributors to the GWB. 
Since $\diff E/\diff{\rm ln}f_r\propto {\cal M}^{5/3}$ and $\phi(M)$ is generally a concave function, %(usually well described by a Schechter function),
Eq.~(\ref{eq:satopolito}) implies that the main contribution to $h^2_c$ (and thus to $h_c$) comes from masses around the point where ${\rm log}\phi(M)$ is tangent to a line with slope $-5/3$. This generally occurs at $M\gtrsim10^9\msun$, where $\phi(M)\approx 10^{-4}$Mpc$^{-3}$ \citep[{\it c.f.} Fig. 1 of][]{Sato-Polito:2023gym}. 
%This consideration relies on the fact that $\phi(M)$ today is the relic of MBHB mergers. 
By reverse engineering the argument leading to Eq.~(\ref{eq:satopolito}), we can then consider as a first order approximation, that the GWB from MBHBs is dominated by MBH mergers with ${\cal M}\sim10^9\msun$, justifying the normalizations chosen in Eq.~(\ref{eq:hc_normaliz}). %\as{(ADD SOME FIGURE WITH MBH MASS FUNCTION AND NUMBER OF MERGERS LIKE SESANA 2013?)}

%But by looking at Eqs.\eqref{eq:hc_phinney} and \eqref{eq:satopolito} we can make some interesting considerations about the main contributors to the GWB. Since $\diff E/\diff{\rm ln}f_r\propto {\cal M}^{5/3}$ and since the merger rate is generally a concave function, the main contribution to $h_c$ comes from chirp masses when $\diff^2n/(\diff z \diff {\cal M})$ is tangent to ${\cal}$ 

%From Eq.~\eqref{eq:hc_normaliz} is also clear that any other population of inspiralling binaries cannot produce a comparable GWB at nHz frequencies. (SEE WHAT TO SAY)

%Although this parametrization is extremely useful for order of magnitude calculations and provides a first order approximation to the GW signal, it has a number of shortcomings we will discuss in the next section (COMMENT ON THE FACT THAT SHOULD NOT BE USED FOR PE, etc...)

\subsubsection{The nature of the nano-Hz GW signal from MBHBs}\label{sec:gwsignalnature} 
Although a useful %to provide a first order 
approximation, %to the GW signal produced by a %cosmic 
%population of MBHBs, 
Eq.~(\ref{eq:hc_phinney}) is only an accurate description of the GWB when (i) the merger rate density $n$ is large and (ii) $\diff E/\diff{\rm ln}f_r$ is emitted on a timescale much shorter than the duration of observations. None of those conditions hold for MBHBs, which feature %merger rate densities much lower than unity
$n \ll 1$
and evolve at nano-Hz frequencies on timescales of thousands of years, while observations (PTAs in particular) have been only running for a couple of decades. 

~~~~It is more useful to think of nano-Hz GW detection as %taking a picture of the sky. PTAs capture 
capturing a `snapshot' of the sky where a finite, discrete number of MBHBs are caught at different stages of their evolution. Therefore, instead of thinking in terms of cosmic merger rate density, it is more useful to think in terms of number of systems with specific properties emitting GWs at a given frequency. In fact, early estimates of the nano-Hz GW signal were obtained in this way~\cite{Rajagopal:1994zj,Jaffe:2002rt}. The relation between this method and Eq.~(\ref{eq:hc_phinney}) was formalized in~\cite{Sesana:2008mz} by rewriting: 
\begin{equation}
    \frac{\diff^2n}{\diff z\, \diff {\cal M}}=\frac{\diff^3 N}{\diff z \diff {\cal M}\diff {\rm ln}f_r}\frac{\diff {\rm ln}f_r}{\diff t_r}\frac{\diff t_r}{\diff z}\frac{\diff z}{\diff V_c}.
    \label{eq:dN}
\end{equation}
Substituting Eq.~(\ref{eq:dN}) into Eq.~(\ref{eq:hc_phinney}), deriving $\diff{\rm ln}f_r/\diff t_r$ from Eq.~(\ref{eq:dfdt}) and using standard cosmological relations for $\diff V_c/\diff z$ and $\diff z/\diff t_r$, one arrives at the alternative formulation:
%%%%%%%%%%%%%%%%%%%%%%%%%%%%%%%%%
\begin{equation}
h_c^2(f) =\int_0^{\infty} 
dz\int_0^{\infty}d{\cal M}\, \frac{\diff^3N}{\diff z \diff{\cal M} \diff{\rm ln}f_r}\,
h^2(f_r)\equiv \int_0^{\infty} 
dz\int_0^{\infty}d{\cal M}\, \frac{\diff^3N}{\diff z \diff{\cal M} \diff t_r}\frac{\diff t_r}{\diff{\rm ln}f_r}\,
h^2(f_r)
\label{eq:hch2}
\end{equation}
%%%%%%%%%%%%%%%%%%%%%%%%%%%%%%%%%
where $h$ is the inclination-polarisation averaged strain amplitude given by \cite{1987thyg.book..330T}
%%%%%%%%%%%%%%%%%%%%%%%%%%%%%%%%%
\begin{equation}
h(f_r)=\sqrt\frac{32}{5} \frac{(G{\cal M})^{5/3}}{c^4 D}(\pi f_r)^{2/3}\,.
\label{eqthorne}
\end{equation}
%%%%%%%%%%%%%%%%%%%%%%%%%%%%%%%%%
Both ways of writing $h_c^2$ in Eq.~(\ref{eq:hch2}) are useful. The left one highlights that $h_c^2$ is just the integral over mass and redshift of the individual $h^2$s' of the sources emitting per unit logarithmic frequency. The right one separates the role of the MBHB merger rate, ${\diff^3N}/(\diff z \diff{\cal M}\diff t_r)$, from  their dynamical evolution, captured by the `{\it residence time}' $t_{\rm res}={\diff t_r}/{\diff{\rm ln}f_r}$ \cite{Haiman:2009te}. 
This separation is particularly useful since cosmological models for MBHB evolution usually output merger rates of the form ${\diff^3N}/(\diff z \diff{\cal M}\diff t_r)$, from which the GWB can be reconstructed by employing a specific model for the MBHB evolution, as we shall see in Sec.~\ref{sec:binary_dynamics}.

\subsubsection{Discretization}
\label{sec:discrete}
 In reality, Eq.~(\ref{eq:hch2}) is the continuum limit of a discrete distribution of sources. Moreover, if we collect data for a timespan $T$, the nominal frequency resolution bin is $\Delta f=1/T$; which means that the GWB spectrum is evaluated in discrete frequency bins $\Delta{f_j}=[j/T, (j+1)/T]$, centered at $f_j=(2j+1)/(2T)$ for $j=1,...,N_{\rm bin}$, where $N_{\rm bin}$ is the number of frequency bins considered. Eq.~(\ref{eq:hch2}) can be then discretized as \cite{Amaro-Seoane:2009ucl}
%%%%%%%%%%%%%%%%%%%%%%%%%%%%%%%%%
\begin{equation}
h_c^2(f_j) =\sum_z\sum_{\cal M} \frac{\Delta^3N}{\Delta z \Delta {\cal M} \Delta f_j} f_jh^2(f_r)\Theta\left(\frac{f_r}{1+z},\Delta{f_j}\right)\Delta z \Delta{\cal M},
\label{eq:hcdiscrete}
\end{equation}
%%%%%%%%%%%%%%%%%%%%%%%%%%%%%%%%%
where $\Theta=1$ if $f_r/(1+z)\in \Delta{f_j}$ and $\Theta=0$ otherwise, and we used the fact that $\diff{\rm ln}f_r=\diff{\rm ln}f$.\footnote{Also note that while we decided to write the right-hand sides of Eqs.~(\ref{eq:hch2}) and~(\ref{eq:hcdiscrete}) in terms of intrinsic, rest frame quantities, they can also be written adopting redshifted, observed quantities.} If we now label by the index $i$ each discrete MBHB contributing to the cosmic population, Eq.~(\ref{eq:hcdiscrete}) can be simplified as
\begin{equation}
h_c^2(f_j) = \sum_{f\in \Delta f_j}   \frac{f_jh_i^2(f_r)}{\Delta{f_j}}=
\sum_{f\in \Delta f_j}h_i^2(f_r)f_jT=
\sum_{f\in \Delta f_j}h_i^2(f_r)N_{\rm cyc}=
\sum_{f\in \Delta f_j}h_{c,i}^2(f_r),
\label{eq:hcdiscrete2}
\end{equation}
where we defined the $i$-th binary characteristic strain $h_{c,i}$ as the inclination-polarization average strain \footnote{When the signal is discretized, one can go beyond the inclination-polarization averaged approximation by substituting, for each individual binary, the factor $\sqrt{32/5}$ in Eq.~(\ref{eqthorne}) with $\sqrt{2(a^2+b^2)}$, where $a=1+{\rm cos}^2\iota$ and $b=-2\,{\rm cos}\,\iota$. This better captures the variance of the signal due to the specific inclinations of the loudest sources \cite{2015MNRAS.451.2417R}.} multiplied by the square root of the number of {\it observed} GW cycles within the observing time $T$. Eq.~(\ref{eq:hcdiscrete2}) simplifies the interpretation of the GWB strain produced by a cosmic population of MBHBs to the bone, highlighting a number of key facts:
\begin{itemize}
    \item The GWB strain is the sum in quadrature of the characteristic strains of all binaries that emit at a rest frame frequency $f_r=f(1+z)$ such that $f\in\Delta f_j$.
    \item Assuming multiple sources per frequency bin, $h_c(f_j)$ is {\it on average} independent on $T$. While the number of binaries in each frequency bin scales as $N_B(\Delta f)\propto\Delta f=1/T$, the contribution of each individual binary to $h_c^2$ scales %conversely 
    as $N_{\rm cyc}\propto T$. %By observing longer, fewer binaries contribute per frequency bin, but each of them contribute with a larger number of cycles. The two effects cancel out.
    \item For circular, GW driven binaries, the number of sources contributing to a frequency bin scales as $\diff N/\diff f=(\diff N/ \diff t)({\diff t}/{\diff f})\propto f^{-11/3}$ and the contribution of each source to $h_c^2$ as $\propto f^{7/3}$. %$h^2fT\propto f^{7/3}$. 
    Combining these two scalings gives $h_c^2\propto f^{-4/3}$.
    %, however this signal builds-up in very different ways at different frequencies. 
    At low frequencies, the signal is the superposition of a large number of dim sources, so a stationary unpolarized and isotropic GWB with $h_c\propto f^{-2/3}$ emerges. Conversely, at high frequencies %$h_c$ is dominated by few sparse, loud sources (see Fig.~\ref{fig:hc_MBHBcirc}), and 
    none of the assumptions made to describe a stationary unpolarized and isotropic GWB apply: sparse loud binaries dominate and can be individually resolved, the spectrum becomes spiky, and the residual GWB steeper \cite{Sesana:2008mz}.
    \item The properties of the detector play an active role in the definition of the nature of the signal. In fact, if $T$ increases and $\Delta f$ decreases, the GWB is better resolved in its individual components, there is more variance in the strain between adjacent frequency bins and eventually more sources are resolvable.\footnote{For PTAs source resolvability also depends on the number of pulsars \cite{Boyle:2010rt}}.
    
    %At low frequencies, the signal is the superposition of a large number of dim sources, so a stationary unpolarized and isotropic GWB emerges approximately. Conversely, at high frequencies $h_c$ is dominated by few sparse, loud sources (see Fig.~\ref{fig:hc_MBHBcirc}), and none of the assumptions made to describe a stationary unpolarized and isotropic GWB apply. The spectrum tends therefore to be a smooth PL as $h_c\propto f^{-2/3}$ at low frequencies, whereas it is more spiky at high frequencies, where sparse loud binaries can be resolved individually, and the GWB level drops to a steeper then -2/3 slope \cite{Sesana:2008mz}.
    %\item The properties of the detector do play an active role in the definition of the nature of the signal. In fact, if $T$ increases and $\Delta f$ decreases, the GWB is better resolved in its individual components, there is more variance in the strain between adjacent frequency bins and eventually more sources are resolvable.\footnote{For PTAs source resolvability also depends on the number of pulsars \cite{Boyle:2010rt}.}
\end{itemize}

\begin{figure}[t!]
  \begin{minipage}[c]{0.5\textwidth}
    \includegraphics[width=\textwidth]{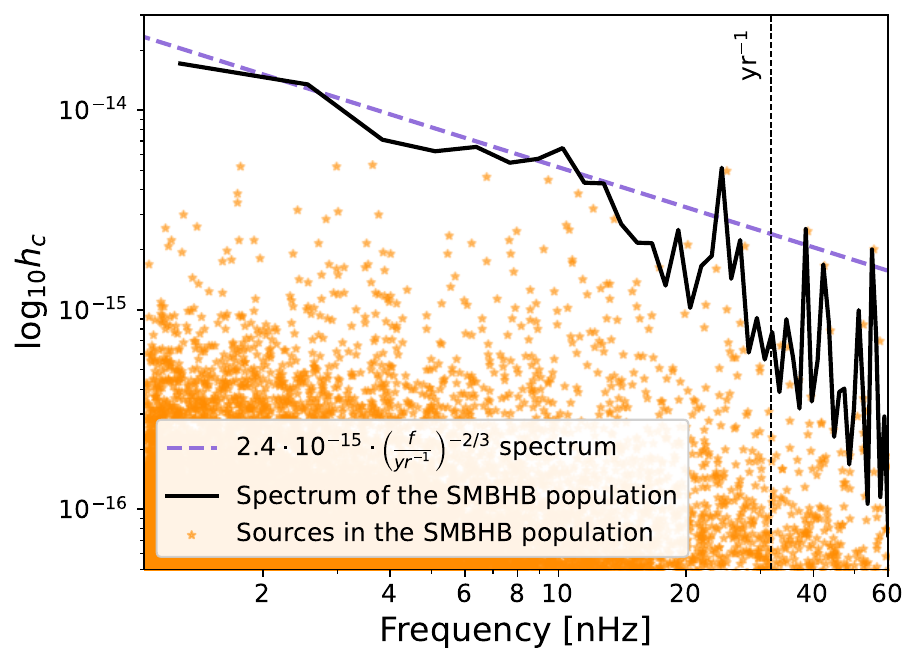} 
  \end{minipage}\hfill
  \begin{minipage}[c]{0.47\textwidth}
  \vspace{-1.0cm}
	\caption{Example of a GWB generated from a population of circular, GW driven MBHBs assuming $T\approx 25$ years. The overall GWB (black) is built by the sum of each individual source (orange stars) as described in the text. The spectrum becomes appreciably spiky and departs from the $f^{-2/3}$ PL (dashed purple line) above 10 nHz. Adapted from \cite{Ferranti:2024jsh}.}
 \label{fig:hc_MBHBcirc}
  \end{minipage}
\end{figure}
%\vspace{2.0cm}

\paragraph{Stochastic GWB vs resolved sources} As a useful order of magnitude exercise, let us consider the signal produced by a cosmic population of MBHBs with ${\cal M}=10^9\msun$, residing in large ellipticals with number density $n_0=10^{-4}$Mpc$^{-3}$ ({\it c.f.}~Eq.~(\ref{eq:hc_normaliz})). We arbitrarily limit our calculation to $z=1$.\footnote{For most astrophysical models, $\gtrsim$90\% of the GW signal originates at $z<1$ \cite{Sesana:2008mz}.} Within $z=1$ there are $\sim10^7$ massive ellipticals, if each MBH residing in their center experienced one major merger since $z=1$, approximating the lookback time to $z=1$ at about 10 Gyr, then the MBHB merger rate is $\diff N_B/\diff t=10^7/(10^{10}{\rm yr})=10^{-3}{\rm yr}^{-1}$. The number of MBHBs emitting per frequency bin is given by $N_B(\Delta f)=(\diff N_b/\diff t)(\diff t/\diff f)\Delta f$, where $\Delta f=1/T$. If we now substitute $\diff f/\diff t$ from Eq.~(\ref{eq:dfdt}) assuming ${\cal M}=10^9\msun$, we take $T=30$yr and solve for $N_B(\Delta f)=1$, we get $f_c\approx1.8\times10^{-8}$ Hz. At $f>f_c$ there is less than one source per frequency bin and the loudest signals can be resolved as deterministic continuous GWs (CGWs). Conversely, at $f<f_c$ multiple sources contribute to the signal, which can be considered stochastic. Note that since ${\diff t/\diff f}\propto f^{-11/3}$, by changing the number of sources by an order of magnitude, $f_c$ changes by less then a factor of two. Therefore it is a fairly robust estimate that {\it most resolvable sources should  arise at $f\gtrsim 10^{-8}$Hz}, consistent with Fig.~\ref{fig:hc_MBHBcirc}.  
\begin{marginnote}
    \entry{Resolvable CGWs}{Deterministic signal rising on top of the GWB, they can dominate the signal at $f\gtrsim 10^{-8}$Hz.}
\end{marginnote}

\subsubsection{On the dominant role of MBHBs at nano-Hz GW frequencies} Using Eq.~(\ref{eq:hcdiscrete2}) it is also immediate to demonstrate that no other astrophysical population of binaries can compete with MBHBs in the nano-Hz GW band. Let us assume that two  binary populations $A$ and $B$ have monochromatic mass functions and share the same distance distribution. At any given frequency, the ratio of the characteristic strain produced by them is approximately
\begin{equation} \label{eq:orderofmag}
    \frac{h_{c,A}^2}{h_{c,B}^2}=\frac{N_A}{N_B}\left( \frac{{\cal M}_A}{{\cal M}_B}\right)^{10/3}.
\end{equation}
Let $A$ be the cosmic MBHB population; at $f= 10^{-8}$Hz $N_A\approx 1$ while ${\cal M}_A=10^9\msun$. %(as we just saw above). 
Let $B$ be the population of all stars in the observable universe. If we assume %(as a hard upper limit) 
that all these stars were in binaries emitting at $f= 10^{-8}$Hz, then $N_B\approx10^{23}$ and ${\cal M}_B=1\msun$. Substituting in Eq.~(\ref{eq:orderofmag}) leads to $h_{c,A}^2/h_{c,B}^2\approx 10^9$, demonstrating that the MBHB GWB is dominant. %showing that even if all the stars in the Universe emit GWs, the GWB produced by MBHBs would still be more than four orders of magnitudes higher at $f= 10^{-8}$Hz.
\footnote{ One might argue that the distance distributions of stars and MBHBs are not comparable, primarily because there are $10^{11}$ stars within the Milky Way (i.e. at kpc distances) whereas the closest MBHB is likely several Mpc away. A similar calculation assuming all stars in our galaxy are in binaries emitting GW at $f=10^{-8}$Hz still yields a strain much smaller than that produced by MBHBs.} 

~~Other astrophysical GW sources could be relevant for particular PTA configurations. For example, a PTA featuring many precise MSPs in the MW center might detect early EMRIs around SgrA$^\star$ \cite{Kocsis:2011ch}; likewise, a PTA featuring many MSPs residing in a globular cluster might be used to detect an intermediate mass BHB (IMBHB) lurking within that same cluster \cite{Chen:2025uzf}.

%This representation clarifies a number of features of the astrophysical GWB. In each frequency resolution bin, the characteristic strain of the GWB is given by the sum of all the stra

%Eq.~\eqref{eq:SGWB_Population} is the continuum limit solution of an inherently discrete problem. In fact the GWB is made of a discrete population of SMBHB and PTAs observe for a finite amount of time $T$ and hence have a resolution limit $\Delta{f}=1/T$. This means that the GWB spectrum is evaluated in discrete frequency bins $\Delta{f_j}=[j/T, (j+1)/T]$, centered at $f_j=(2j+1)/(2T)$ for $j=1,...,N_{\rm b}$, being $N_{\rm b}$ the number of considered frequency bins. Throughout this paper we assume $T=30$ yrs. Therefore, in a real-life experiment, Eq.~\eqref{eq:SGWB_Population} takes the form \cite{2010MNRAS.402.2308A}:

\subsection{Eccentricity and environmental effects: dynamical evolution of MBHBs}
\label{sec:binary_dynamics}
When evaluating Eq.~(\ref{eq:hch2}), we considered
circular binaries driven by GWs. This simplifying assumption is likely inaccurate for astrophysical MBHBs evolving in dense galactic nuclei for two reasons: (i) MBHBs might have significant eccentricity, if anything because they are the product of galaxy mergers that usually occur on almost radial orbits \cite{Fastidio:2024crh}, and (ii) binaries exchange energy and angular momentum with the dense environment, which might significantly contribute to their dynamical evolution. In fact, the interaction with the environment {\it must} be dominant at a sufficiently low frequency, where the expected GW coalescence timescale becomes longer than the Hubble time $t_H$. By integrating Eq.~(\ref{eq:dfdt}) backwards from coalescence to a time $t_0$, we find
\begin{equation} \label{eq:fgwmin}
    f_{K,0}\approx 5\times 10^{-11} \left(\frac{t_0}{10^{10}\,{\rm yr}}\right)^{-3/8}\left(\frac{\cal M}{10^9\,\msun}\right)^{-5/8} [{\rm Hz}]
\end{equation}
which, for a reference circular equal mass binary of $M_1=M_2=10^9\msun$ corresponds to a semimajor axis $a_0\approx 0.5$pc: {\it if MBHBs did not lose energy to their environment at parsec scales, they would stall and would never coalesce because of GW emission}. To fully model the nano-Hz GWB, we therefore need to understand MBHB dynamical evolution.

MBHBs form in the aftermath of galaxy mergers. The journey of the two MBHs, from kpc separations down to final coalescence, was first laid out in~\cite{Begelman:1980vb}, who identified three evolutionary stages, which we discuss below, with a specific focus on massive systems relevant to PTA. For the sake of argument, we model the galaxy merger remnant as a {\it singular isothermal sphere} (SIS), characterized by a density profile $\rho(r)=\sigma^2/(2\pi G r^2)$, where $\sigma$ is the stellar velocity dispersion \cite{BinneyTremaineBook2ndEd2008}.\footnote{Not to be confused with the RMS residual.} We define the sphere of influence of a MBHB of total mass $M$ located at the center of the galaxy as 
\begin{equation}
    r_{\rm inf}=\frac{GM}{\sigma^2}
\end{equation}
%\as{Put in formula} $r_{\rm inf}=GM/\sigma^2$ 
and connect the MBH mass to the stellar bulge velocity dispersion through the $M-\sigma$ relation of the form $(M/10^6\msun)=(\sigma/70\,{\rm km \, s}^{-1})^4$ \cite{Kormendy:2013dxa}.\footnote{In general, these relations hold for single MBHs while here we assume that galaxy merger remnants correlate with MBHBs. For the sake of the discussion, the difference is negligible.} 

%\begin{marginnote}
%    \entry{Reference PTA system}{
% The typical PTA system we consider is composed by two MBHs of $M_1=M_2=10^9\msun$, evolving in a merger remnant with $\sigma=300\,{\rm km\,s}^{-1}$, so that $r_{\inf}=50\,$pc and $\rho_{\rm inf}=\rho(r_{\rm inf})=10^3\msun\,$pc$^{-3}$.}
%\end{marginnote}
%\begin{boxC}
The typical PTA system we consider is composed by two MBHs of $M_1=M_2=10^9\msun$, evolving in a merger remnant with $\sigma=300\,{\rm km\,s}^{-1}$, so that $r_{\inf}=50\,$pc and $\rho_{\rm inf}=\rho(r_{\rm inf})=10^3\msun\,$pc$^{-3}$.
% \end{boxC}
Although the SIS is a rough approximation for the density profile of massive galaxies, it suffices to provide an order of magnitude estimate of the relevant physical processes. We stress, however, that central density profiles of massive galaxies are extremely diverse, ranging from flat cores to steep cusps \cite{2007ApJ...664..226L}. As a result, $r_{\rm inf}$ and $\rho_{\rm inf}$ can vary significantly, with $10\,{\rm pc}< 
r_{\rm inf}<500\,{\rm pc}$ and $10\, \msun\,{\rm pc}^{-3}<\rho_{\rm inf}<10^5 \msun\,{\rm pc}^{-3}$. In general, galaxies with larger cores have larger $r_{\rm inf}$ and lower $\rho_{\rm inf}$.

\subsubsection{Phase 1: Dynamical friction: 10 kpc $\rightarrow$ 10 pc}
Right after the merger of two massive galaxies, the MBHs residing at the center of the parent galaxies are found in the outskirts of the galaxy merger remnant, 
%at several kpc distances from its center and from each other. The two MBHs thus 
and {\it independently} sink to the center because of dynamical friction (DF) on a timescale \cite{BinneyTremaineBook2ndEd2008}
\begin{equation}\label{eq:t_df}
    t_{{\rm df},i}\approx1.1 \left(\frac{{\rm ln}\Lambda}{10}\right)^{-1} 
    \left(\frac{r_{0,i}}{10\,{\rm kpc}}\right)^2
    \left(\frac{\sigma}{300\,{\rm km\,s}^{-1}}\right) 
    \left(\frac{M_i}{10^9\msun}\right)^{-1}\,{\rm Gyr},
\end{equation}
where $i=1,2$, $r_0$ is the initial distance from the center and ${\rm ln}\Lambda\approx{\cal O}(10)$. %is the Coulomb logarithm. 
%arising from integrating the interaction with the field stars over the whole range of impact parameters. 
Eq.~(\ref{eq:t_df}) is oversimplified in many ways, since: (i) it includes only stars moving slower than the MBH and it has been shown \cite{Dosopoulou:2016hbg} that faster stars do also contribute to DF;
(ii) the initial MBH orbit can be very eccentric \cite{Fastidio:2024crh}, implying a shorter $t_{\rm df}$ \cite{Colpi:1999cm};
(iii) the sinking MBHs are initially surrounded by the respective progenitor galaxy stellar nuclei, making the effective sinking masses $M_i$ bigger and $t_{\rm df}$ shorter.
%;(iv) it neglects contributions of gas and dark matter which are, however, likely subdominant in massive, late type, gas poor galaxies.
    
Note that $t_{\rm df}$ can be $>t_H$ for light BHs or large initial $r_0$. In fact, %\cite{McWilliams:2012an,Kelley:2016gse,Izquierdo-Villalba:2021prf} 
light MBHs generally fail to reach the remnant center and wander in the galaxy outskirts \citep[e.g.][]{McWilliams:2012an,2017MNRAS.464.3131K,2020MNRAS.495.4681I}. Thus, DF preferentially selects the formation of massive,  equal mass binaries. 
When the two MBHs come within $r_{\rm inf}$ of each other, they bind into a MBHB. As the binary shrinks, its circular velocity $V_c$ increases to $V_c\gg \sigma$ (i.e. the binary becomes {\it hard}), making DF ineffective \cite{BinneyTremaineBook2ndEd2008}. If there were no further physical mechanisms to extract energy and angular momentum from the MBHB, it would stall at a separation of several pc, much larger than the value at which GWs can make it coalesce within $t_H$ ({\it c.f.} Eq.~(\ref{eq:fgwmin})).

\subsubsection{Phase 2: Binary hardening: 10 pc $\rightarrow$ 0.1 pc} 
\label{sec:stellarhard}

There are at least three ways to drive a MBHB to the GW dominated inspiral. We discuss them separately in the following.

\paragraph{Stellar hardening} A hard binary loses energy and angular momentum to surrounding stars via three-body scatterings \cite{Quinlan:1996vp}, evolving according to \cite{Sesana:2010qb}:
\begin{equation}\label{eq:stellarhard}
\begin{split}
\frac{\diff a}{\diff t_r} \,{=}\, -\frac{a^2G\rho_{\rm inf}}{\sigma}H \,\,\,\,\,\,\,\,\,\, & \rightarrow  \,\,\,\,\,\,\,\,\,\, \frac{\diff f_{K,r}}{\diff t} =\frac{3G^{4/3} M^{1/3} H \rho }{2(2\pi)^{2/3} \sigma}f_{K,r}^{1/3} \\
\frac{\diff e}{\diff t_r} &\,{=}\, \frac{aG\rho_{\rm inf}}{\sigma}HK, 
\end{split}
\end{equation}
where $H$ and $K$ are dimensionless rates derived from numerical three body scattering experiments. In particular $H\approx 15-20$ depending on $q$ and $e$, while $K\rightarrow0$ for $e\rightarrow 0\,\wedge\,e\rightarrow 1$, is generally positive, and has a maximum $K\approx 0.1-0.4$ (depending on $q$) for $e\approx 0.6$ \cite{Quinlan:1996vp,Sesana:2006xw}. %The values of $H$ and $K$ have been tabulated by several authors \cite{Quinlan:1996vp,Sesana:2006xw}. 
Eq.~(\ref{eq:stellarhard}) shows that stars shrink the MBHB while increasing its eccentricity. %(as both $H$ and $K$ are positive).
Stars interacting with the MBHB are ejected from the galactic nucleus, and the evolution of the system depends on how efficiently new stars can be supplied to the binary, i.e. on the pace at which the so-called binary {\rm loss cone} can be repopulated \cite{Milosavljevic:2002bn}. The literature on this subject is vast, but  \cite{Sesana:2010qb} showed that using $\rho_{\rm inf}$ in Eq.~(\ref{eq:stellarhard}) corresponds to assuming a {\it full loss cone at the influence radius}, which is typically the case for triaxial  merger remnants \cite{Preto:2011gu,Khan:2011gi,2015ApJ...810...49V,2015MNRAS.454L..66S}.

%If we consider, for simplicity, an isothermal sphere, we substitute $\rho_i$ in equation (\ref{adotstar}), and we assume $M_{BH}\propto\sigma^5$, we get that in the stellar driven case $dt/d{\rm ln}f\propto f^{2/3}M_1^{2/3}$, which yields to a contribution of the single binary to the GW background of the form $h_c\propto M_1^2qf$. 

\paragraph{Gas hardening} Galaxy mergers trigger cold gas infall toward the center of the merger remnant \cite{Mihos:1995ri}. Part of this gas reaches the nucleus where it can form a circumbinary disk (CBD) at sub-pc scales. The MBHB-CBD  energy and angular momentum exchange is an extremely complex topic that has been recently tackled by several authors \citep[see][for the status of the field]{Duffell:2024fwy}. The outcome of the interaction depends on the detailed dynamics into the CBD cavity, on the role of self-gravity, on the binary parameters and disk aspect ratio and equation of state \citep[e.g.][]{2019ApJ...871...84M,2019ApJ...875...66M,Duffell:2019uuk,2020A&A...641A..64H,Franchini:2021uiy,2021ApJ...914L..21D,2022ApJ...929L..13F}, and on whether the disk rotation is prograde or retrograde with respect to the binary orbital angular momentum \cite{Nixon:2010by,2014MNRAS.439.3476R}. Interaction with a prograde CBD drives the MBHB towards $q=1$ \citep[due to preferential accretion onto the secondary MBH,[]{2014ApJ...783..134F}, while $e$ tends to saturate to values $0.4<e<0.7$ \cite{2011MNRAS.415.3033R,2021ApJ...914L..21D,Siwek:2023rlk}.

%While retrograde disks have been shown to consistently shrink the binary and increase their eccentricity \cite{Nixon:2010by,Roedig:2013lqa}, the physics of prograde disks is much more subtle. In particular, recent simulations showed that circular, equal mass binaries surrounded by a disk with aspect ratio $H/R=0.1$ tend to widen \cite{Munoz:2018tnj,Moody:2019nes}, which would pose a significant problem for merging MBHBs. This result has later been shown to be $H/R-$, $q-$, and $e-$dependent \cite{Duffell:2019uuk,Heath:2020chl,DOrazio:2021kob,Franchini:2022paz}. In particular, binaries interacting with more realistic disks with $H/R\lesssim0.03$ tend to shrink. This is generally the case for self-gravitating CBDs, relevant for massive PTA binaries \cite{Franchini:2021uiy}. Self gravity tends to collapse the disk vertically, resulting in an $H/R$ that is small enough to shrink the binary. In general, interaction with a prograde CBD drives the MBHB towards $q=1$ (due to preferential accretion onto the secondary MBH \cite{Farris:2013uqa}, while $e$ tends to saturate to values in the range $0.4<e<0.7$ \cite{Rodig:2011jz,DOrazio:2021kob,Siwek:2023rlk}. As a final note, we stress that the nano-Hz GW signal is dominated by MBHBs with $M\sim 10^9\msun$ at $z<1$, residing in gas poor ellipticals \cite{Izquierdo-Villalba:2022hmh}. Therefore, while extremely important for lower mass, higher redshift binaries, relevant for LISA, gas dyanamics is not expected to play a dominant role in shaping the nano-Hz GW signal.  

%To give an idea of the impact of MBHB-CBD interaction,
As an example, we consider the case of a prograde CBD, with a cavity maintained by the torque exerted by the binary onto the disk and with no mass allowed to enter the cavity. \cite{Ivanov:1998qk} showed that the binary evolution can be approximated as 

%The problem admits a self-consistent, non-stationary solution that was derived by \cite{Ivanov:1998qk}. In this case, the binary evolution rate can be approximated as %\cite{Ivanov:1998qk}
%%%%%%%%%%%%%%%%%%%%%%%%%%%%%%%%%
\begin{equation}
\frac{\diff a}{\diff t_r} = 2(aa_0)^{1/2}\frac{(1+q)^2}{q}\frac{\eta_E}{t_{\rm S}} \,\,\,\,\,\,\,\,\,\, \rightarrow  \,\,\,\,\,\,\,\,\,\, \frac{\diff f_{K,r}}{\diff t_r} = 3\pi^{1/3}a_0^{1/2}\frac{(1+q)^2}{q}\frac{\eta_E}{t_{\rm S}}(GM)^{-1/6}f_{K,r}^{4/3}
\label{adotgas}
\end{equation}
%%%%%%%%%%%%%%%%%%%%%%%%%%%%%%%%
(we neglect $\diff e/\diff t_r$ here) where $a_0$ is the semimajor axis at which the mass of the unperturbed disk equals the mass of the secondary black hole:
%%%%%%%%%%%%%%%%%%%%%%%%%%%%%%%%%
\begin{equation}
a_0= 0.05\left(\frac{\alpha}{0.1}\right)^{4/7}\left(\frac{M_1}{10^9\msun}\right)^{1/7}{\eta_E^{-3/7}q^{5/7}}\,{\rm pc}.
\end{equation}
%%%%%%%%%%%%%%%%%%%%%%%%%%%%%%%%
Here $\eta_E=\dot{M}/\dot{M}_{\rm E}$ is the accretion rate normalized to the Eddington rate, $t_S=M/{\dot M}=0.44\,$Gyr is the Salpeter time, and $\alpha$ is the disk viscosity parameter. Note that when $a=a_0$ the residence time $\diff t_r/\diff{\rm ln}a\sim qt_S/\eta_E$, which means that equal mass binaries accreting at the Eddington rate evolve on a Salpeter timescale.

%Here, $\dot{M}$ is the mass accretion rate at the outer edge of the disk, $a_0$ is the semimajor axis at which the mass of the unperturbed disk equals the mass of the secondary black hole, and $\mu$ is the reduced mass of the binary. Considering a standard geometrically thin, optically thick disk model \cite{ss73}, one finds $dt/d{\rm ln}f\propto f^{-1/3}M_1^{1/6}$, which yield to a contribution of the single binary  to the GW background of the form $h_c\propto M_1^{7/4}q^{3/2}f^{1/2}$. 

\paragraph{MBH triple interactions} In massive, gas-poor galaxies with shallow cores, the hardening timescale can exceed $t_H$, causing binary stalling. A subsequent merger may bring a third MBH to the center, where triple interactions can trigger a merger—via secular or chaotic processes—or eject one MBH, leaving a wide binary \cite{2002ApJ...578..775B}. Mergers often follow the brief formation of tight, highly eccentric binaries with short GW coalescence times, making high eccentricity a signature of triple interactions. Cosmological models show that up to 20\% of MBHBs form triplets \cite{Bonetti:2017lnj,2018MNRAS.473.3410R,Sayeb:2023vav}, and \cite{Bonetti:2017dan} found that about 40\% of these triplets merge, implying that up to $\sim$10\% of all MBHB mergers might result from triple interactions.

%In massive, gas poor galaxies with shallow cores, the hardening timescale can become several Gyrs, and even exceed $t_H$, leading to binary stalling. The stall can be solved by a third MBH brought to the center by a subsequent merger. The resulting triple interaction can either lead to a merger prompted by secular processes or chaotic encounters, or to the ejection of one of the MBHs leaving behind a binary still far from merger. Mergers typically occur following the temporary formation of tight, very eccentric binaries characterized by short GW coalescence timescales. High eccentricity is therefore expected to be a signature of triple interactions.MBH cosmological evolution models \cite{Bonetti:2017lnj,Sayeb:2023vav} have shown that up to 20\% of MBHBs can form a triplet due to a subsequent merger and \cite{Bonetti:2017dan} found that up to 40\% of these triplets are resolved by a merger. Therefore, ${\cal O}(10\%)$ of the MBHB mergers in the Universe can be the result of triple interactions.

\subsubsection{Phase 3: GW emission: 0.1 pc $\rightarrow$  a$_{\rm \bf ISCO}$} Eventually,
GW emission becomes efficient at small separations, causing the binary to evolve according to Eq.~(\ref{eq:dfdt}) until coalescence. 

To see if nano-Hz MBHBs are primarily driven by GWs or by interaction with their environment we can simply equate 
%We can now get an idea of whether MBHBs at nano-Hz frequencies are primarily driven by GWs or by interaction with their environment. We can do that by simply equating 
the GW orbital frequency evolution rate of Eq.~(\ref{eq:dfdt}) to its stellar and gas driven counterparts of Eq.~(\ref{eq:stellarhard}) and ~(\ref{adotgas}) and solve for the frequency to get
%%%%%%%%%%%%%%%%%%%%%%%%%%%%%%%%%
\begin{equation}
\begin{split}
f_{K, {{\rm \star/GW}}} & \approx 5\times10^{-10}\left(\frac{M_1}{10^9\msun}\right)^{-5/8}\left(\frac{\rho_{\rm inf}}{\rho_{\rm inf,SIS}}\right)^{3/10}q^{-3/10} {\cal F}(e)^{-3/10}\,{\rm Hz}\\
f_{K,{{\rm gas/GW}}} & \approx 5\times10^{-10}\left(\frac{\alpha}{0.1}\right)^{-6/49}\left(\frac{M_1}{10^9\msun}\right)^{-37/49}q^{-69/98}\eta_E^{-33/98}{\cal F}(e)^{-6/14}\,{\rm Hz}.
\end{split}
\label{decoup}
\end{equation}
%%%%%%%%%%%%%%%%%%%%%%%%%%%%%%%%
From the decoupling point, the lifetime of the MBHB estimated using $\diff t_r/\diff{\rm ln}f_K$ is:
%%%%%%%%%%%%%%%%%%%%%%%%%%%%%%%%%
\begin{equation}
\begin{split}
t_{{\rm \star/GW}} & \approx 10^{8}\left(\frac{\rho_{\rm inf}}{\rho_{\rm inf,SIS}}\right)^{4/5}q^{-1/5} {\cal F}(e)^{-1/5}\,{\rm yr}\\
t_{{\rm gas/GW}} & \approx 10^{8}\left(\frac{\alpha}{0.1}\right)^{-16/49}\left(\frac{M_1}{10^9\msun}\right)^{-17/49}q^{-43/49}\eta_E^{44/49}{\cal F}(e)^{1/7}\,{\rm yr},
\end{split}
\label{decoup}
\end{equation}
%%%%%%%%%%%%%%%%%%%%%%%%%%%%%%%%
where we also retain, for the stellar case, the dependence on $\rho_{\rm inf}$, keeping in mind that for our reference binary $\rho_{\rm inf, SIS}=10^3\msun\,$pc$^{-3}$. We highlight few key points ({\it c.f.} Fig.~\ref{fig:hc_MBHBdynamics}):
\begin{itemize}
    \item by substituting ${\diff f/\diff t}$ from Eqs.~(\ref{eq:stellarhard}) and~(\ref{adotgas}) into Eq.~(\ref{eq:hch2}) one finds $h_c\propto f$ and $h_c\propto f^{1/2}$ for stellar and gas driven binaries respectively. Therefore in both cases there is a low frequency turnover in the $h_c$ spectrum;
    \item for $e\approx 0$, the turnover occurs around $f\approx 1$ nHz, at the lower end of the PTA band;
    \item the turnover frequency has a very weak dependence on $\rho_{\rm inf}$ and $\eta_E$. Pushing it to 10 nHz, where PTAs today are most sensitive, would require $\rho_{\rm inf}\gtrsim 10^6\msun$pc$^{-3}$ or $\eta_E\gtrsim 10^3$, which are both extremely unlikely in massive, low redshift galaxies;
    \item $e$ decreases both the decoupling frequency and the remaining binary lifetime at such frequency. This means that when binaries are very eccentric, there are fewer of them distributed over a much larger orbital frequency range. Therefore, even without considering the redistribution of GW power at higher harmonics, high eccentricity suppresses the low frequency GW signal, as clear from Fig.~\ref{fig:hc_MBHBdynamics}.

\end{itemize}

%\as{COMMENT: (i) show f scalings for gas and stars at low f ($h_c\propto f$ and $h_c\propto f^{1/2}$) (ii) turnover quite at low freq; (iii) weak dependence on $f_E$ and $\rho$ so env effects unlikely to drive low freq turnover in PTA band; (iv) $e$ also hase a minor impact on lifetime, but it shifts $f$ equiv. at lower freq. So there are the same amount of systems over a much larger $f$ range, which is expected to cause the signal to drop. Comment on the fact that this is the main reason for the turnover and not the emission at higher harmonics.}  

\subsection{Putting everything together: the full GW signal from MBHBs}
\label{sec:SMBH_totalGWsignal}
By adding all the relevant physics, including multiple hardening mechanisms and eccentricity evolution, the characteristic strain $h_c(f)$ of the GWB can be written as:
\begin{equation} \label{eq:SGWB_Population}
  \begin{aligned}
h_c^2(f)=\int_0^{\infty} \diff z \int_0^{\infty} \diff  M_1 \int_0^1 {\diff } q \int_0^1 {\diff } e \frac{\diff ^5 N}{{\diff } z\, {\diff } M_1\, {\diff } q\, {\diff }t\, {\diff }e_0}\left[\frac{\diff t}{{\diff } \ln f_{{k}, r}}\frac{\diff e_0}{\diff e}\right]\times \\
\left. h^2(f_{K,r}) \sum_{n=1}^{+\infty} \frac{g\left(n, e\right)}{(n / 2)^2}\right|_{f_{{K}, r}=f(1+z) / n} ,
\end{aligned}
\end{equation}
where now each individual MBHB contributes a formally infinite series of harmonics of the orbital frequency $f_{{K}, r}$, adjusted by $1+z$ to account for cosmological redshift and thus observed at frequency $f=nf_{{K}, r}/(1+z)$. The normalization of each harmonic is given by $h_c^2(f_K,r)$ as per Eq.~(\ref{eqthorne}), weighted by the $g(n,e)$ factor given in Eq.~(\ref{eq:gne}). Like in the circular case, Eq.~(\ref{eq:SGWB_Population}) is the continuum limit of an inherently discrete problem. 
Using the same procedure of Sec.~\ref{sec:discrete}, the GW signal is discretized as ~\cite{Amaro-Seoane:2009ucl}:
\begin{equation}
\label{eq:hok}
h_{c}^{2} (f_{_j}) = \sum_{i=1}^{N_B}\sum_{n=1}^{+\infty} h^2_i(f_{K,r})\frac{g\left(n, e_i\right)}{(n / 2)^2}\frac{n f_{K,r_i}}{\Delta f (1 + z)}\Theta\left(\frac{n f_{K,r_i}}{1+z},\Delta{f_j}\right),
\end{equation}
where $\Theta=1$ if $n f_{K,r_i}/(1+z)\in \Delta{f_j}$ and $\Theta=0$ otherwise, the first sum is over the $N_B$ MBHBs of the population, each contributing to the signal with the sum of $n$ harmonics. For circular binaries $g(n,e=0)=1$ if $n=2$ and 0 otherwise; Eq.~(\ref{eq:hok}) simplifies to Eq.~(\ref{eq:hcdiscrete2}), which in the continuum limit, for GW driven binaries returns the well known PL of Eq.~(\ref{eq:hc_normaliz}).

\subsubsection{Connecting the GWB shape and normalization to the MBHB population}

 \begin{figure}[t!]
  \begin{minipage}[c]{0.53\textwidth}
  \hspace{-0.3cm}
    \includegraphics[width=\textwidth]{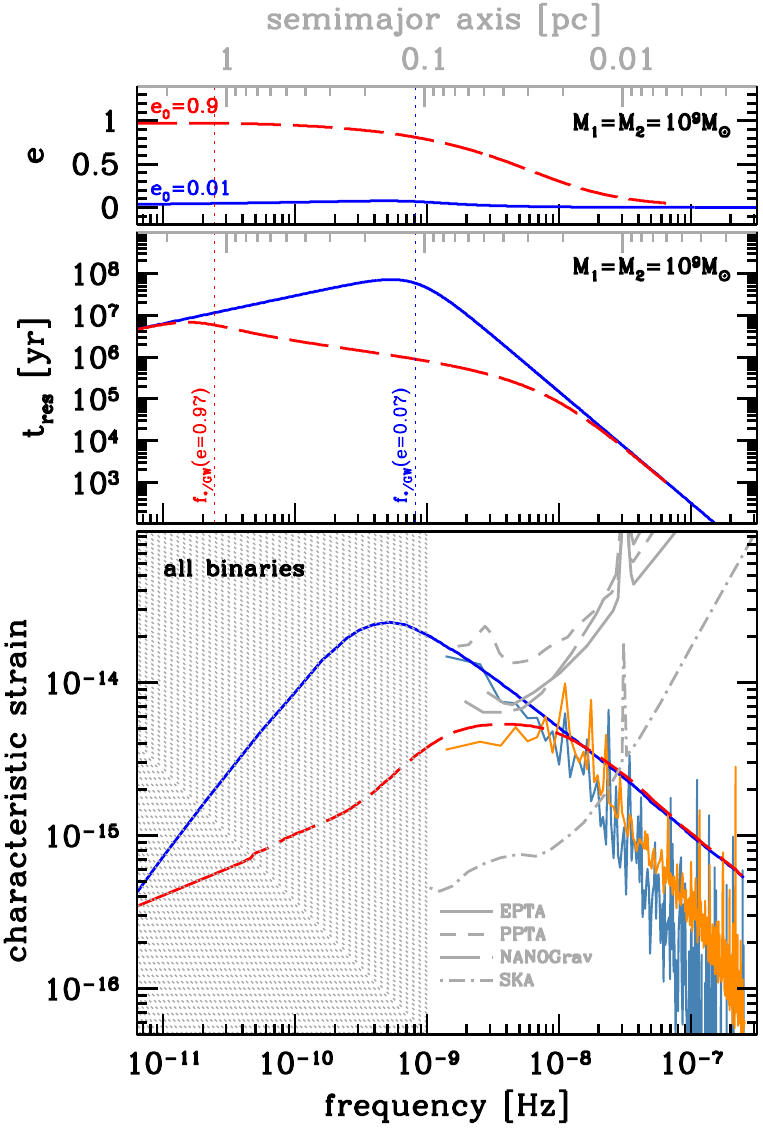} 
   \end{minipage}\hfill
   \begin{minipage}[c]{0.45\textwidth}
  \vspace{-0.2cm}
	\caption{GW signal and MBHB dynamics. Solid blue and dashed red lines are for MBHBs with $e_0=0.01$ and $e_0=0.9$ at formation. {\it Top panel}: evolution of $e$. {\it Middle panel}: Evolution of $t_{\rm res}={\diff t}/{\diff {\rm ln}f}$ of a fiducial MBHB with $M_1=M_2=10^9\msun$, as a function of $f$ (lower $x-$axes) and $a$ (gray upper $x-$axes). Binaries are initially hardened by 3-body scattering (Sec.~\ref{sec:stellarhard}) and reach to $e \approx 0.07$ (for $e_0=0.01$) and $e\approx 0.97$ (for $e_0=0.9$), until GW emission takes over leading to circularization. As discussed in Sec.~\ref{sec:gwsignalnature}, $t_{\rm res}$ peaks around $f_{\star/{\rm GW}}$, {\it c.f.} Eq.~(\ref{decoup}). {\it Bottom panel}: GW strain $h_c(f)$ from the {overall cosmic population of MBHBs} integrated over masses and redshift. Smooth lines are GWB calculations from Eq.~(\ref{eq:SGWB_Population}), whereas spiky lines are actual realizations from the discrete population according to Eq.~(\ref{eq:hok}) for $e_0=0.01$ (orange) and $e_0=0.9$ (steel blue). Also shown PTA sensitivities and the area inaccessible with $T<30$ yrs (shaded area).} 
 \label{fig:hc_MBHBdynamics}
  \end{minipage}
 \end{figure}

In Eq.~(\ref{eq:SGWB_Population}), we have separated merger rate from dynamical evolution by writing
\begin{equation}\label{eq:cosmicrate}
    \frac{\diff ^5 N}{{\diff } z\, {\diff } M_1\, {\diff } q\, {\diff } \ln f_{{k}, r}\, {\diff }e} = \frac{\diff ^5 N}{{\diff } z\, {\diff } M_1\, {\diff } q\, {\diff }t\, {\diff }e_0}\left[\frac{\diff t}{{\diff } \ln f_{{k}, r}}\frac{\diff e_0}{\diff e}\right].
\end{equation}

\paragraph{GWB normalization: cosmic merger rate} 

\begin{figure}[t]
    \centering
    \includegraphics[width=0.95\linewidth]{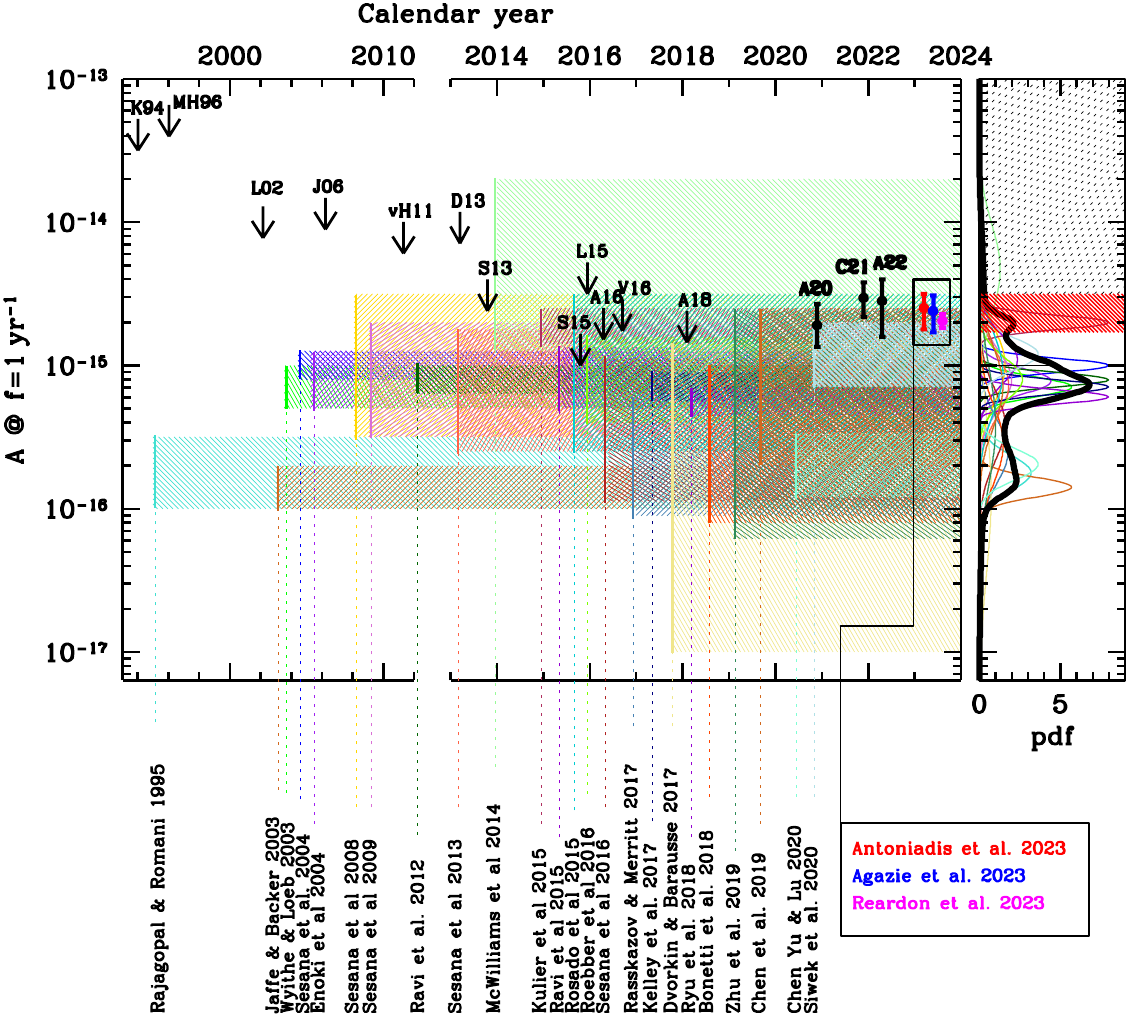}
    \caption{Progression of predictions, limits and measurements of the GWB produced by MBHBs. The main plot shows: the predicted amplitudes $A_{\rm GWB}$ assuming $h_c\propto f^{-2/3}$ (Eq.~(\ref{eq:hc_analytical}), vertical colored lines) up to 2020 \cite{Rajagopal:1994zj,Jaffe:2002rt,Wyithe:2002ep,Sesana:2004sp,Enoki:2004ew,Sesana:2008mz,Sesana:2008xk,Ravi:2012bz,Sesana:2012ak,McWilliams:2012an,Kulier:2013gda,Ravi:2014nua,2015MNRAS.451.2417R,Roebber:2015iva,Sesana:2016yky,Rasskazov:2016jjk,Kelley:2017lek,Dvorkin:2017vvm,Ryu:2018yhv,Bonetti:2017lnj,Chen:2018znx,Chen:2020qlp,Siwek:2020adv}; upper limits placed by several PTAs up to 2018 \citep[][downward pointing arrows]{Kaspi:1994hp,McHugh:1996hd,Lommen:2002je,Jenet:2006sv,EPTA:2011kjn,Demorest:2012bv,Shannon:2013wma,Shannon:2015ect,EPTA:2015qep,NANOGrav:2015aud,Verbiest:2016vem,NANOGRAV:2018hou}; evidence of common red noise since 2020 \citep[][thick black dots with errorbars]{NANOGrav:2020bcs,EPTA:2021crs,Antoniadis:2022pcn}; and evidence of a correlated GWB from EPTA+InPTA, NANOGrav and PPTA \citep[][colored marks with errorbars, shifted with respect to each other for clarity]{NANOGrav:2023gor, Antoniadis:2023ott, Reardon:2023gzh}. By giving equal credit to each prediction, the left panel shows the pdf of the theoretically expected $A_{\rm GWB}$ (thick black line). The red shaded area marks the  $A_{\rm GWB}$ range consistent with PTA measurements, whereas the black shaded area marks the excluded $A_{\rm GWB}$ region.
    }
    %\caption{pepe}
    \label{fig:ptaprogression}
\end{figure}

The first term on the right-hand side of Eq.~(\ref{eq:cosmicrate}), ${\diff ^5 N}/({{\diff } z\, {\diff } M_1\, {\diff } q\, {\diff }t\, {\diff }e_0})$, represents the differential cosmic merger rate of MBHBs as a function of the relevant binary parameters, and mainly determines the normalization $A_{\rm GWB}$ of the GWB spectrum. This quantity has been derived by several authors either theoretically from semianalytic models \cite{Rajagopal:1994zj,Jaffe:2002rt,Wyithe:2002ep,Sesana:2004sp,McWilliams:2012an} and cosmological simulations \cite{Kelley:2017lek,Siwek:2020adv,Izquierdo-Villalba:2021prf,2025ApJ...991L..19C}, or empirically by combining observations of the galaxy mass function, pair fraction, and MBH mass-host galaxy relation to obtain an estimate MBHB merger rate \cite{Sesana:2012ak,Ravi:2014nua,Simon:2023dyi}. A compilation of the estimated $A_{\rm GWB}$ from models pre-dating the first observational evidence of a signal in PTA data \cite{NANOGrav:2020bcs} is shown in Fig.~\ref{fig:ptaprogression}. The uncertainty of the underlying physics and the diversity of techniques adopted result in a wide range of $10^{-16}\lesssim A_{\rm GWB}\lesssim 3\times10^{-15}$. Note that the upper limit is essentially dictated by the MBH mass function today, as discussed in Sec.~\ref{sec:satopolito}\footnote{A notable outlier is the prediction by \cite{McWilliams:2012an}. However, their $A_{\rm GWB}$ estimate appears to be  higher than the upper limit imposed by the argument of \cite{Sato-Polito:2023gym} applied to the mass function they use.} 

\paragraph{GWB spectral shape: binary dynamics} 

The term in square brackets in Eq.~(\ref{eq:cosmicrate}) represents a map $(t,e_0)\rightarrow(f_{K.r},e)$, transforming a merger rate described by an eccentricity distribution $e_0$ into the number of systems emitting at all frequencies with their eccentricities. This map must be obtained by numerically solving the equations:

\begin{figure}
 %    \centering
 %    \begin{subfigure}[b]{0.49\textwidth}
 %        \centering
 %        \includegraphics[width=\textwidth]{plots/GWB_theo_e0.pdf}
 %        %\caption{$y=x$}
 %        %\label{fig:y equals x}
 %    \end{subfigure}
 %    \hfill
 %    \begin{subfigure}[b]{0.49\textwidth}
 %        \centering
 %        \includegraphics[width=\textwidth]{plots/GWB_theo_e9.pdf}
 %        %\caption{$y=3\sin x$}
 %        %\label{fig:three sin x}
 %    \end{subfigure}
 %    \caption{yobbroz}
 %    \label{fig:theo_GWB}
\end{figure}

\begin{equation}\label{eq:frequency_Evolution}
%\begin{split}
\frac{\diff f_{K,r}}{\diff t_r} \,{=}\, \sum_i \frac{df_{K,r}}{dt_r}\Bigg|_i\,,~~~~~~~~~~~~
\frac{\diff e}{\diff t_r} \,{=}\, \sum_i \frac{\diff e}{\diff t_r}\Bigg|_i,
%\end{split}
\end{equation} 
where the sum is over the relevant physical mechanism driving the binary: GW emission and interaction with  stars and gas, as given by Eqs.~(\ref{eq:dfdt}),~(\ref{eq:stellarhard}),~(\ref{adotgas}). Examples of the full GW signal for a quasi-circular and a very eccentric population of MBHBs are shown in Fig. ~\ref{fig:hc_MBHBdynamics}. The figure shows the connection between $e$, $t_{\rm res}$ and $h_c$ shape. Note that while the function $g(n,e)$ redistributes the GW emission at higher harmonics, the main contribution to the GWB drop comes from the much shorter $t_{\rm res}$, driven by the ${\cal F}(e)$ function of Eq.~(\ref{eq:FeGe}). Conversely, the pile-up of harmonics for multiple sources at high frequency causes a slight increase of the GWB compared to the circular case (orange vs steel blue lines in the lower panel of Fig. ~\ref{fig:hc_MBHBdynamics}) while decreasing its variance. We stress that that high freq drop is due to sparse sampling of high frequency sources. The square averaged total GW signal computed over 1000 realization of the MBHB population sits around the $h\propto f^{-2/3}$ PL at $f>10^{-8}$Hz.

%\as{EXAMPLES GIVEN IN FIGURE \ref{fig:theo_GWB}.
%Discuss: (i) variance and jugged signal, (ii) drop at high $f$, (iii) turnover at low $f$.}.

\subsection{Resolvable signals from MBHBs}
\label{sec:MBHB_resolvable}

\subsubsection{Continuous GWs (CGWs)} As mentioned in Sec.~\ref{sec:discrete}, particularly loud or nearby MBHBs can be resolved individually as deterministic signals, known as CGWs.
%which generally go under the name of continuous GWs (CGWs). 
CGWs have been studied mainly in the context of PTA detection, where the signal in the data is obtained by inserting Eq.~(\ref{eq:h_plus_cross}) into Eq.~(\ref{eq:ztgeneral}). Full expressions can be found in \cite{Taylor:2021yjx,Truant:2024aci}. 
Circular CGWs have been studied extensively \cite{2011MNRAS.414.3251L,2012ApJ...756..175E,2012PhRvD..85d4034B,2014MNRAS.444.3709Z,2022PhRvD.105l2003B,2025A&A...694A.194F,Gardiner:2025mwf}. The signal in each MSP is composed of an Earth and a pulsar terms, the latter evaluated at a time $\tau$ in the past ({\it c.f.} Eq.~\eqref{eq:ztgeneral}). During the time $\tau$ the MBHB has evolved according to Eq.~(\ref{eq:dfdt}), such that the GW frequency of the two terms is different and $f_p<f_E$:
\begin{itemize}
    \item if  $f_E-f_p>\Delta f$ pulsar and Earth terms fall at different frequencies. All Earth terms can be added coherently and properly matched with a deterministic filter;
    \item $f_E-f_p<\Delta f$ pulsar and Earth terms fall in the same frequency bin and their superposition partially destroys the coherency of the signal.
\end{itemize}
For a typical $10^9\msun$ MBHB and for typical PTA observing times ${\cal O}(10\,{\rm yr})$, the separation between the two classes of sources occurs at about $10^{-8}$Hz \cite{2015MNRAS.451.2417R}. Although CGWs have not been detected yet, it is expected that a fully operational SKA will enable to resolve tens of them \cite{Boyle:2010rt,Truant:2024aci}, with a sky localization precision of tens of deg$^2$ \cite{Sesana:2010mx}. This might still be sufficient to allow effective EM counterpart searches \cite{Burke-Spolaor:2013aba}, since CGWs are expected to be sourced by MBHBs hosted in massive and nearby elliptical galaxies \cite{2025arXiv250401074T,Zhou:2025lsk,2025arXiv251014613Q}. 
%\as{ADD A FIGURE? like properties of resolved sources by SKA? }

\subsubsection{Eccentric bursts} Another class of deterministic nano-Hz GW signals is bursts from very eccentric binaries, especially during triple interactions. At every periastron passage, the system emits a burst of GW with peak frequency \cite{Wen:2002km}
\begin{equation}
    f_b = f_K (1-e^2)^{-3/2}(1+e)^{1.1954}.
\end{equation}
Note that $f_b\approx 2f_K(1-e)^{-3/2}$, i.e. the frequency of a circular binary of the size of the periastron $r_p=a(1-e)$. The duration of the burst is $\Delta t_b\approx 1/f_b$. This means that binaries with orbital frequencies much lower than 
$1/T$ can produce localized bursts well within the PTA band. For example a MBHB with a period of 100 yrs and eccentricity 0.99, at periastron generates a burst with $f_b\approx3\times 10^{-7}$Hz of a duration $\Delta t_b\approx3\times10^6$s $\sim 1$ month. Using a rudimental model for the cosmic evolution of MBHBs and triplets \cite{Amaro-Seoane:2009ucl} found up to 50\% probability of observing GW bursts at several nanosecond levels in PTA data. %The observation of such bursts is an intriguing possibility that would shed light on the rate of MBH triplet formation and their extreme dynamics.

\subsubsection{Bursts with memory}
GW bursts can feature non-oscillatory components that leave a permanent deformation of spacetime, known as memory \cite{2009ApJ...696L.159F,1991PhRvL..67.1486C}. Such bursts, termed bursts with memory (BWM), occur because GW emission induces lasting second time derivatives in the source’s mass-energy quadrupole moments, causing the metric to relax to a new configuration. At nano-Hz, the most promising BWM sources are MBHB mergers. Assuming a circular binary inspiral, the permanent metric shift has a vanishing $h_\times$ component and approximately follows \citep[e.g.][]{2010MNRAS.401.2372V}

\begin{equation}
  h_{+}\approx \frac{1-\sqrt{8}/3}{24}\frac{\mu}{D_l}\sin^2\iota(17+\cos^2\iota)\left[1+{\cal O}(\mu^2/M^2)\right]\approx 1.5\times 10^{-15}\frac{\mu}{10^9\msun}\left(\frac{D_l}{1{\rm Gpc}}\right)^{-1},
  \label{hmem}
\end{equation}
where $\mu$ is the redshifted reduced mass and in the last passage we have averaged over $\iota$. This shift rapidly grows at merger on a timescale $t_r\approx 2\pi R_S/c\approx 1{\rm day}\,(M/10^9\msun)$,
where $R_s$ its Schwarzschild radius. The waveform is therefore described by $h(t)=h_{+}\Theta(t-t_0)$, where $\Theta(t-t_0)$ is the heaviside step function: the perturbation is null until time $t_0$, and quickly jumps to the value given by Eq.~(\ref{hmem}).
%Note that the strain modification is of ${\cal O}(10^{-15})$, potentially within PTA capabilities. 
Since the induced timing residual $R$ is the integral of a heaviside function, it takes the form of a  linear drift, similar to a pulsar glitch, which is partially absorbed by fitting a quadratic function to the pulsar spin and spin derivative, leaving behind a cusp-like burst shape similar to the one caused by the Shapiro delay \cite{1964PhRvL..13..789S}. BWM are expected to be observable if sourced by $\approx10^9\msun$ binaries at $z<1$. Their cosmic rate is equal to the merger rate of these MBHBs, which is expected to be $\ll 1$ yr$^{-1}$. Searches of BWM have been performed on several PTA data yielding null results \cite{2015MNRAS.446.1657W,Agazie:2025oug}.%\begin{equation}
% {\color{blue}R(t)}=\frac{1}{2}\cos(2\psi)(1-\cos\theta)h_+[(t-t_e)\Theta(t-t_e)-(t-t_p)\Theta(t-t_p)],
%\end{equation}
%where $t_e-t_p\equiv\tau$ defined in Sec.~\ref{sec:ptaresponse} and $\psi$ is the GW polarization angle \cite{2013LRR....16....9Y}. Note that in PTA, a linear residual is absorbed in the fit for the MSP frequency, therefore a BWM can be detected only if one of the trigger times $t_e, t_p$ occurs within the duration $T$ of the observations. Note that $t_e-t_p\ \gg T$, and if $t_e$ occurs withing $T$, $t_p$ does not. Since also for BWM only the Earth term will be correlated among all pulsars in the array, the most compelling case is when $t_e$ lies within $T$. The signature is a linear drift similar to a glitch, which is partially absorbed by fitting a quadratic function to the pulsar spin and spin derivative, leaving behind a cusp-like burst shape similar to the one caused by the Shapiro delay \cite{Demorest:2010bx}. BWM are expected to be observable if sourced by $\approx10^9\msun$ binaries at $z<1$. Their cosmic rate is equal to the merger rate of these MBHBs, which is expected to be $\ll 1$ yr$^{-1}$. Searches of BWM have been performed on several PTA data yielding null results \cite{Wang:2014zls,Agazie:2025oug}.

 \subsection{Astrophysical interpretation of PTA observations}
 \label{sec:MBHB_interpretation}
%%%%%%%%%%%%%%%%%%%%%%%%%%%%

\begin{figure}[t]
    \centering
    % Left side: stack two panels
    \begin{minipage}{0.4\textwidth}
        \centering
%        \begin{subfigure}
            \includegraphics[width=\linewidth]{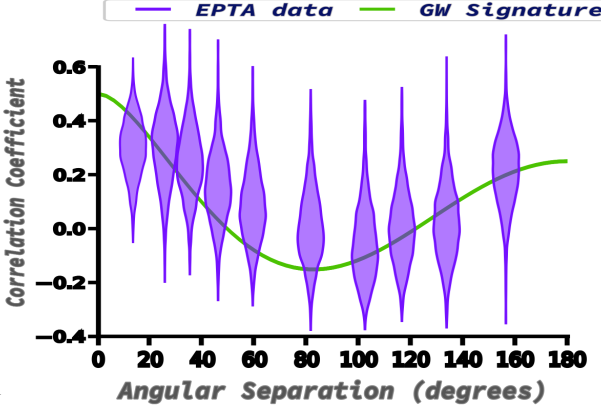}
           %\caption{Small panel 1}
%        \end{subfigure}
        
        \vspace{-0.4cm} % space between the stacked panels
        
%        \begin{subfigure}
            \includegraphics[width=\linewidth]{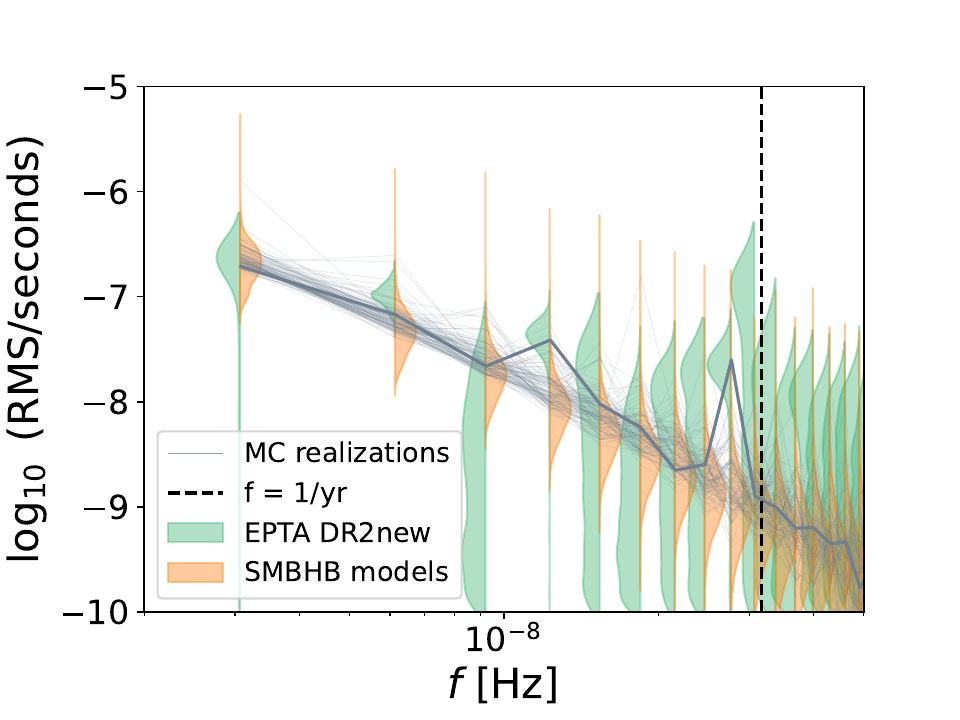}
 %           \caption{Small panel 2}
%        \end{subfigure}
    \end{minipage}
    \hfill
    % Right side: bigger panel
    \begin{minipage}{0.57\textwidth}
        \centering
%        \begin{subfigure}{1.0\textwidth}
            \includegraphics[width=\linewidth]{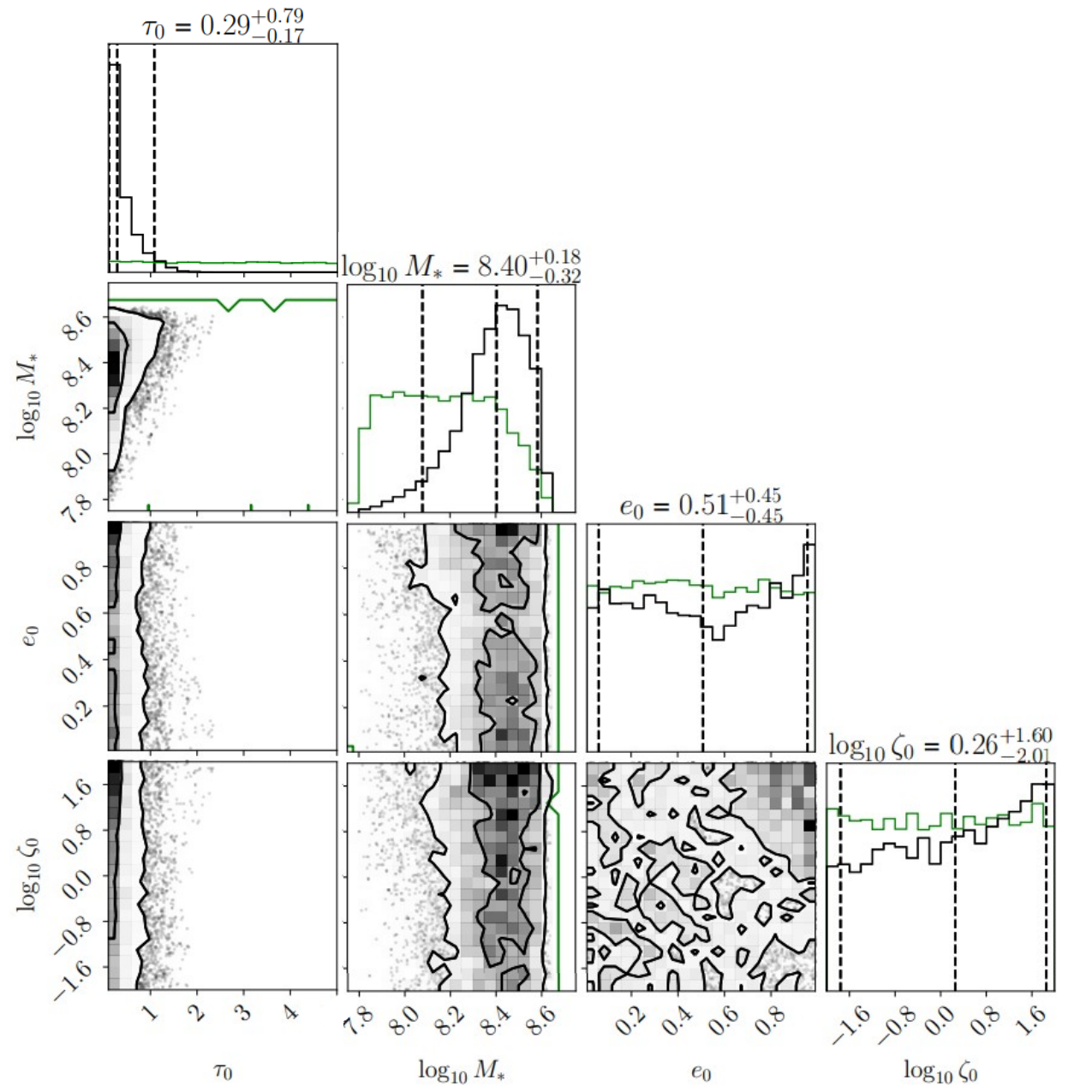}
            %\caption{Big panel}
%        \end{subfigure}
    \end{minipage}
    
    \caption{Birds eye view of the EPTA results. Top left: measured correlation (pink violins) compared to the HD expectation value (green line). Bottom left: measured RMS PSD (orange violins) compared to expected distribution from the theoretical MBHB models of \cite{Chen:2018znx} (green violins). The jagged lines show 100 realizations of the PSD from the models. Right: triangle plots of the posterior distribution (black, priors shown in green) of four key parameters of the theoretical models:
    the average coalescence time $\tau_0$ of a MBHB from 20 kpc separation; the normalization $M_*$ of the $M_{\rm BH}-M_{\rm bulge}$ relation at $M_{\rm bulge}=10^{11}\msun$ ; the eccentricity  $e_0$ at MBHB formation; the normalized density of the environment $\zeta_0=\rho_{\rm inf}/\rho_{\rm inf, SIS}$ ({\it c.f.} Eq.~(\ref{decoup}). Adapted form \cite{EPTA:2023xxk}.}
    \label{fig:epta_interpretation}
\end{figure}

%PTA observations summarized in Sec.~\ref{sec:PTA_evidence} can be interpreted in the context of the MBHB hypothesis. 
Although $A_{\rm GWB}$ lies at the upper end of the range shown by Fig.~\ref{fig:ptaprogression} and $\gamma<13/3$, a cosmic population of MBHBs remains  as a natural explanation of the observe PTA signal presented in Sec.~\ref{sec:PTA_evidence}. Fig.~\ref{fig:epta_interpretation} shows few highlights from the EPTA intepretation paper \cite{EPTA:2023xxk} as an example. The RMS residual PSD aligns quite well with the expected distribution from empirical models (bottom left violin plot). as expected, When fitting the model presented by \cite{Chen:2018znx} to the data, the high value of $A_{\rm GWB}$ translates into a short average MBHB coalescence time ($\tau_0<1$ Gyr with 90\% confidence) and a high normalization $M_*$, implying $M_{\rm BH}/M_{\rm bulge}=2.5^{+1.3}_{-1.3}\times10^{-3}$ for $M_{\rm bulge}=10^{11}$. Likewise, adding evolution of the $M_{\rm BH}-M_{\rm bulge}$ into the picture, favors positive redshift evolution, implying more massive MBHs for a given galaxy mass in the past \cite{Matt:2025bao}. Although \cite{Sato-Polito:2023gym} argued that such a high amplitude is difficult to reconcile with the observed local MBH mass function, this observation has been disputed by \cite{Liepold:2024woa}. It should be noticed that the maximum $A_{\rm GWB}$ is very sensitive to the high mass end of the local MBH mass function \cite{Izquierdo-Villalba:2021prf}, which is determined by a handful of sparse measurements and suffers of considerable uncertainties. The low $\gamma$ would naturally arise from either high eccentricity or strong environmental coupling ({\it c.f.} Sec.~\ref{sec:binary_dynamics}), the latter expressed by the parameter $\zeta_0$. Although there is a slight preference for the high $e-\zeta_0$ corner, the marginalized posteriors of both parameters only show a tentative preference for such high values, leaving the analysis inconclusive. 

Finally, the results discussed above are conditioned to the astrphysical priors included in the \cite{Chen:2018znx} model. Assuming less informative priors, might result in differences in the inferred MBHB observables \cite{NANOGrav:2023hfp}. This is a hint that astrophysical inference is still dominated by the prior choice, meaning that current PTA observations are not yet very informative for constraining underlying signal models (Ferranti et al. in prep.)

\section{Cosmological GW Signals}
\label{sec:cosmo_GWsignals}

~~~~The evidence~\cite{NANOGrav:2023gor, Antoniadis:2023ott, Reardon:2023gzh, Xu:2023wog} for a gravitational wave signal in PTA data, offers unprecedented opportunities for exploring high-energy physics {\it beyond the standard model} (BSM), embedded in early Universe cosmology~\cite{Caprini:2018mtu}. The key point is to note that early universe dynamics, typically operating at very high energies, create GWBs that carry information about their source(s). These are referred to as {\it cosmological} backgrounds, in contrast to the {\it astrophysical} ones introduced in Sec.~\ref{sec:astroGWB}. There is potentially a plethora of cosmological backgrounds permeating the universe~\cite{Caprini:2018mtu},  including signals possibly originated during or after inflation, typically created by non-linear and out-of-equilibrium field dynamics, as {\it e.g.}~in first order phase transitions or topological defect evolution.

~~~~While the PTA signal is likely to be due to MBHBs, {\it c.f.}~Sec.~\ref{sec:MBHB_interpretation}, cosmological backgrounds also represent viable candidates to explain the data. In fact, the parameter space of models producing a GWB which can either fit, significantly exceed, or simply fail to explain the PTA signal, can be clearly identified. Remarkably, these parameter constraints are independent of the origin of the PTA signal. For example, parameter regions in certain models may lead to signals in conflict with the PTA data, leading to excluded regions and upper limits on the individual parameters. In other scenarios, PTA constraints can be complementary to existing bounds, probing previously unexplored parameter regions. 

~~~~The constraining power of PTA data in the search for new physics, marks the dawn of {\it early universe gravitational wave (GW) cosmology} as a field. With the advent of PTA astronomy, we can start addressing fundamental questions, such as: {\it Q1: What is the parameter space region(s)
of BSM models that can explain the observed data? Q2: Is there any preference between astrophysical and cosmological explanations of the signal? Q3: Could the observed signal arise simultaneously from a combination of both MBHBs and a cosmological source ?}, etc. Here in Sec.~\ref{sec:cosmo_GWsignals}, we review the status of the art of the answer(s) to $Q1$, as current PTA data already allows for a quantitative and rich analysis in this respect. We focus mainly in cosmological GWB signals, though for completeness we also discuss deterministic signals from dark matter models, albeit more briefly. 
We postpone however any discussion on $Q2, Q3$, and other questions, to Sec.~\ref{sec:discriminating}, as current data is not precise enough to address these aspects yet, so only qualitative answers (and future prospects) can now be laid out.  

\subsection{Early universe backgrounds (EUB)}
\label{subsec:EUB}

The present Universe my harbor a large variety of cosmological backgrounds, including signals potentially originated during inflation, {\it e.g.}~from vacuum fluctuations~\cite{Grishchuk:1974ny,Starobinsky:1979ty, Rubakov:1982df,Fabbri:1983us}, strong particle production~\cite{Anber:2006xt,Sorbo:2011rz,Barnaby:2011qe,Adshead:2013qp,Namba:2015gja,Maleknejad:2016qjz,Dimastrogiovanni:2016fuu,Domcke:2016bkh,Guzzetti:2016mkm,Bartolo:2016ami,Garcia-Bellido:2017aan,Garcia-Bellido:2023ser,Figueroa:2023oxc,Figueroa:2024rkr,Sharma:2024nfu}, or induced by large scalar fluctuations~\cite{Matarrese:1992rp,Matarrese:1993zf, Matarrese:1997ay,Nakamura:2004rm,Ananda:2006af,Baumann:2007zm,Unal:2018yaa,Adshead:2021hnm,Domenech:2021ztg}; signals that could have been generated after inflation but are still related to the inflationary set-up,  {\it e.g.}~from preheating dynamics~\cite{Easther:2006gt,GarciaBellido:2007dg,GarciaBellido:2007af,Dufaux:2007pt,Dufaux:2008dn,Dufaux:2010cf,Bethke:2013aba,Bethke:2013vca,Figueroa:2017vfa,Adshead:2019igv,Cosme:2022htl,Antusch:2025ewc} or enhanced by kination-domination~\cite{Giovannini:1998bp,Giovannini:1999bh,Boyle:2007zx,Li:2016mmc,Li:2021htg,Figueroa:2018twl,Figueroa:2019paj,Li:2021htg,Gouttenoire:2021wzu,Co:2021lkc,Gouttenoire:2021jhk,Oikonomou:2023qfz}; and signals potentially generated after inflation (typically unrelated to the inflationary set-up), {\it e.g.}~from thermal plasma motions~\cite{Ghiglieri:2015nfa, Ghiglieri:2020mhm, Ringwald:2020ist, Ghiglieri:2022rfp}, oscillon dynamics~\cite{Zhou:2013tsa,Antusch:2016con,Antusch:2017vga,Liu:2017hua,Amin:2018xfe}, first order phase transitions (1stO-PhTs)~\cite{Kamionkowski:1993fg,Caprini:2007xq,Huber:2008hg,Hindmarsh:2013xza,Hindmarsh:2015qta,Caprini:2015zlo,Hindmarsh:2017gnf,Cutting:2018tjt,Cutting:2018tjt,Cutting:2019zws,Pol:2019yex,Caprini:2019egz,Cutting:2020nla,Han:2023olf,Ashoorioon:2022raz, Athron:2023mer,Li:2023yaj}, or cosmic defects~\cite{Vachaspati:1984gt,Sakellariadou:1990ne,Damour:2000wa,Damour:2001bk,Damour:2004kw,Figueroa:2012kw,Hiramatsu:2013qaa,Blanco-Pillado:2017oxo,Auclair:2019wcv,Gouttenoire:2019kij,Figueroa:2020lvo,Gorghetto:2021fsn,Chang:2021afa,Yamada:2022aax,Yamada:2022imq,Kitajima:2023cek,Baeza-Ballesteros:2023say,Dankovsky:2024ipq,Ferreira:2024eru,Baeza-Ballesteros:2024otj,Notari:2025kqq,Babichev:2025stm}.  We refer to all these signals as early Universe backgrounds (EUBs). For a comprehensive review on EUBs see~\cite{Caprini:2018mtu}, and references therein.  Below we discuss how leading EUB models may succeed or fail to explain current PTA data, reviewing current constraints on their parameter space. We mostly  follow Ref.s~\cite{EPTA:2023xxk,NANOGrav:2023hvm,Figueroa:2023zhu,Ellis:2023oxs}, as these consider an extensive list of EUB signals fit against PTA data, enabling us to compare their respective analysis on commonly considered signals. 

\subsection{EUB: Inflation}
\label{subsec:EUBinflation}

Today's spectrum of a GWB from inflation can be written as~\cite{Caprini:2018mtu}
\begin{eqnarray}
    h^2 \Omega_{\rm GW}^{(0)}(k) = \mathcal{T}_{\rm RH}(k)\mathcal{G}_{\rm RD}\frac{h^2 \Omega_{\rm rad}^{(0)}}{24} \mathcal{P}_t(k) 
    \simeq 1.19 \cdot 10^{-16} \left(\frac{r}{0.032}\right) %\left(\frac{A_s}{2.1 \cdot 10^{-9}}\right)
    \left(\frac{f}{f_{\rm pivot}}\right)^{n_t},
    \label{eq:GWB_inflation_SR}
\end{eqnarray}
with the first equality representing an exact expression, where $\mathcal{P}_t(k)$ is the inflationary tensor power spectrum, $\mathcal{T}_{\rm RH}(k)$ the {\it reheating} transfer function connecting the end of inflation with the onset of radiation-domination (RD), and $\mathcal{G}_{\rm RD} \equiv (g_{\rm RD}/g_{\rm 0})(g_{s,0}/g_{s,{\rm RD}})^{4/3}$ a factor indicating the changes in the relativistic degrees of freedom from RD till today. The second equality, however, is only an approximation, valid for {\it vanilla} inflationary models (canonically-normalized, single-field, slow-roll models, assuming general relativity), where a primordial tensor spectrum is expected in the form of a PL. The latter is parametrized around CMB scales as $ \mathcal{P}_t(k) = r A_s (k/k_{\rm pivot})^{n_t}$, with $k_{\rm pivot} = 0.05 ~{\rm Mpc^{-1}}$, $n_t$ the spectral tilt, $r$ the tensor-to-scalar ratio bounded as $r \leq 0.032$~\cite{Tristram:2021tvh}, and $A_s$ the primordial scalar perturbation measured as $A_s \simeq 2.1 \cdot 10^{-9}$~\cite{Planck:2018jri}. For convenience, we assumed instant reheating ($\mathcal{T}_{\rm RH} = 1$) and only SM degrees of freedom during RD ($\mathcal{G}_{\rm RD} \simeq 0.39$), and converted $k_{\rm pivot}$ into a frequency today $f_{\rm pivot} \simeq 8 \cdot 10^{-17}$ Hz. 

~~~~As the inflationary signal can only be at most
$h^2\Omega_{\rm GW} \simeq 10^{-16}$ at CMB scales, {\it c.f.}~Eq.~(\ref{eq:GWB_inflation_SR}), explaining a signal with higher amplitude at PTA frequencies, clearly requires a {\it blue} tilt $n_{\rm t} > 0$ (the smaller is $r$, the larger the tilt required).  Vanilla scenarios must however verify the so called {\it consistency condition},  $n_t = -r/8$, and hence current CMB constraints allows them to admit at most a tiny {\it red} tilt $-n_{\rm t} \lesssim 0.0035$. Therefore, to explain PTA data, more complex scenarios are needed, in particular incorporating some mechanism leading to a large positive tilt at small scales. This is the case, for example, of models with exponential particle production during inflation, see {\it e.g.}~\cite{Bartolo:2016ami,Guzzetti:2016mkm,Caprini:2018mtu} and references therein. In such scenarios, a `sizeable' running of $n_{\rm t}$ is also expected, so Eq.~(\ref{eq:GWB_inflation_SR}) represents only a simplifying parametrization, which helps to understand the `averaged' positive tilt required to explain the PTA data by an inflationary blue-tilted signal.

~~~~Fitting $\Omega_{\rm GW}^{(0)} = \mathcal{A}_{\rm yr}(f/f_{\rm yr})^{n_{\rm t}}$ to PTA data, with free amplitude $\mathcal{A}_{\rm yr}$ and tilt $n_{\rm t}$, leads straightforwardly to a direct constraint on $r$ (at the CMB scales), via Eq.~(\ref{eq:GWB_inflation_SR}). For example, Ref.~\cite{EPTA:2023xxk} obtains
$\log_{10} r = -12.18^{+8.81}_{-7.00}$ and $n_{\rm t} = 2.29^{+0.87}_{-1.11}$ at 90\% CL using EPTA data, whereas Ref.~\cite{Figueroa:2023zhu} finds $r_{0.05} \in [4.37\cdot 10^{-14},3.47\cdot 10^{-9}]$ and $n_t = 2.08_{-0.30}^{+0.32}$ at 68\% CL, when combining NG15 and EPTA data.  Ref's~\cite{NANOGrav:2023hvm,Ellis:2023oxs} obtain similar contour plots in the plane $(n_{\rm t},\log_{10}r)$ with a reheating transfer function $\mathcal{T}_{\rm RH}(k) \neq 1$ taken from \cite{Kuroyanagi:2014nba,Kuroyanagi:2020sfw} and using NG15, finding a strong correlation as $n_{\rm t} \simeq -0.14 \log_{10}r + 0.58$. Ref.~\cite{Ellis:2023oxs} highlights $n_t \simeq 2.6$ and $\log_{10}r \simeq -14$ as their best fit. 

\begin{figure}[t]
  \begin{minipage}[c]{0.35\textwidth}   \includegraphics[width=\textwidth,angle=90]{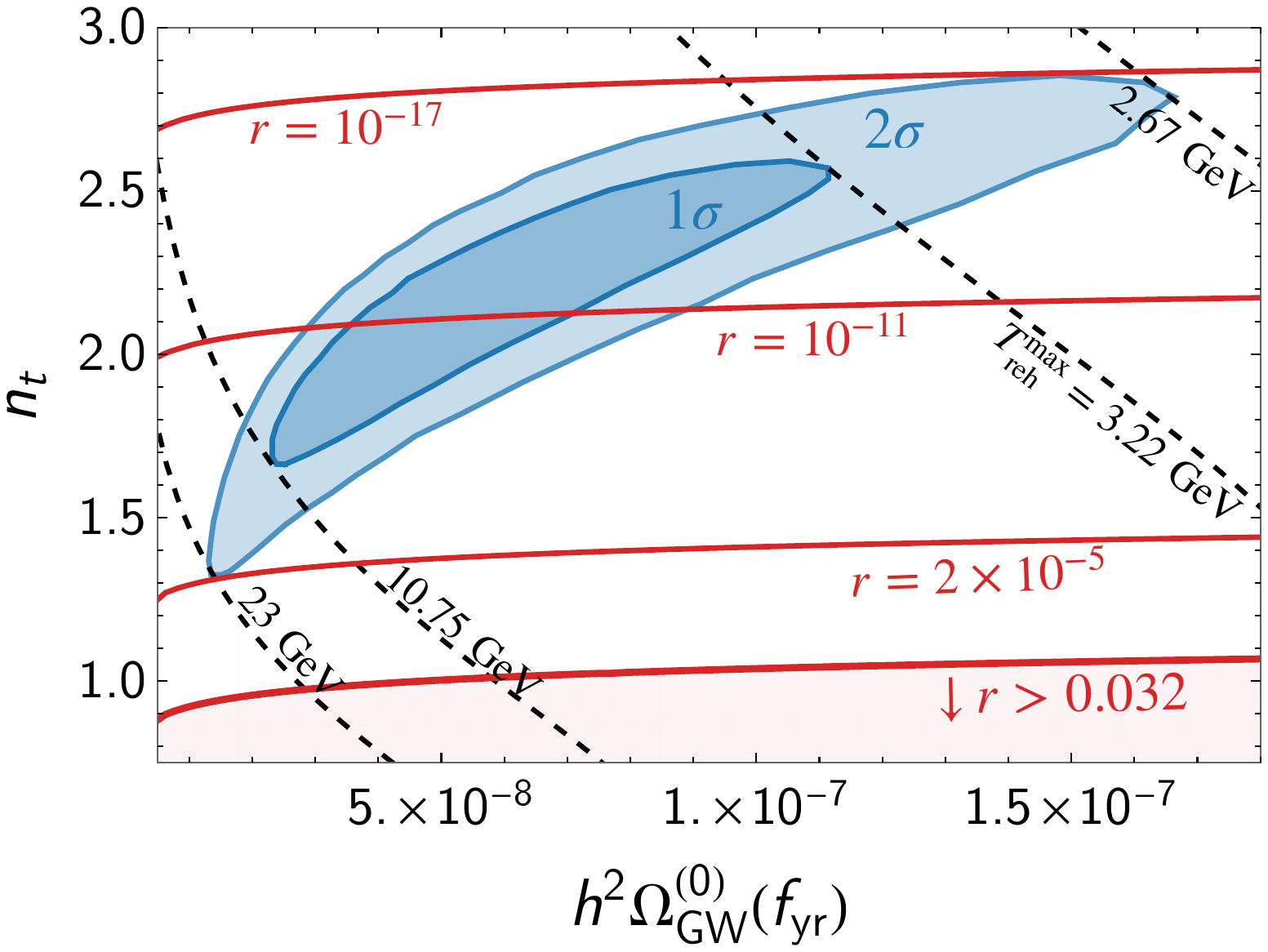} 
  \end{minipage}\hfill
  \begin{minipage}[c]{0.5\textwidth}
	\caption{Parameter space $n_{\rm t}$ vs $h^2\Omega_{\rm GW}(f_{\rm yr})$, required to explain EPTA+NG15 data together, under the assumption of an inflationary signal as in Eq.~(\ref{eq:GWB_inflation_SR}). The CMB bound on the tensor-to-scalar ratio (in red) excludes the region with $r > 0.032$. The $\Delta N_{\rm eff}$ bound requires a low-scale reheating, $T_{\rm rh} < 23 $ GeV, at 95\% CL. This figure is adapted from~\cite{Figueroa:2023zhu}. 
    }
 \label{fig:inf_cmb_cons}
  \end{minipage}
\end{figure}

~~~~The highly blue-tilt required can be in tension with extra-radiation bounds, typically parametrized by the number of extra neutrino species $\Delta N_\nu$~\cite{Allen:1997ad}.  This bound translates into a constraint on the reheating temperature as~\cite{Figueroa:2023zhu}
\begin{align}
    T_{\rm reh} \lesssim  {0.68 ~ \rm MeV}\times\left[1 + (4.3 \cdot 10^{9}) \, n_t \left(\frac{f_{\rm pivot}}{f_{\rm BBN}}\right)^{n_t} \left(\frac{0.032}{r}\right) \right]^{\frac{1}{n_t}}\,,
    \label{eq:infl_bound_Treh}
\end{align}
where we have assumed only SM fields during RD, fixed $\Delta N_\nu \simeq 0.2$ and $A_s = 2.1\cdot 10^{-9}$~\cite{Planck:2018vyg}, and introduced $f_{\rm BBN} \simeq 1.84 \cdot 10^{-11}$ Hz as the frequency today corresponding to the Hubble scale at the onset of BBN. Using Eq.~(\ref{eq:infl_bound_Treh}), Ref.~\cite{Figueroa:2023zhu} finds that a low-reheating temperature $2.65~{\rm GeV} \lesssim T_{\rm rh} \lesssim 23.0$ GeV at 95\% CL, must be imposed to respect $\Delta N_{\rm eff}$ constraints. This can be appreciated in Fig.~\ref{fig:inf_cmb_cons}, where the 2D posterior of $\lbrace n_{\rm t},h^2\Omega_{\rm GW}(f_{\rm yr})\rbrace$ is obtained from combining EPTA+NG15 data. 

~~~~Overall, all analyses point blue spectral tilts $n_{\rm t} \gtrsim 1$ and tiny tensor-to-scalar ratios, orders of magnitude below current CMB constraints $r < 0.032$~\cite{Tristram:2021tvh}.  An inflationary signal strongly violating the consistency relation $n_t = -r/8$, with a blue tilt as large as required by PTA data, $n_t \simeq 2$, is not impossible to conceive. Ref.~\cite{Vagnozzi:2023lwo} discusses in fact a number of inflationary (and non-inflationary) scenarios leading to blue tensor spectra. However, such scenarios either do not predict a sufficiently large amplitude at PTA scales, or a sufficiently blue tilt, or are simply ruled out due to an excessive amplitude or non-Gaussianity level of scalar perturbations. Adding on top the requisite of a very low reheating temperature, makes it even more challenging to construct a viable model. Thus, an inflationary explanation of the PTA signal looks rather implausible; see however \cite{Unal:2023srk} for an explicit model claiming otherwise.  

\subsubsection{Kination domination after inflation} We note that Eqs.~(\ref{eq:GWB_inflation_SR}) and  (\ref{eq:infl_bound_Treh}) above were obtained under the assumption of instant reheating. If there is however an extended reheating period between the end of inflation and the onset of RD, the GWB from inflation today is affected through the corresponding transfer function of this period. If the latter is modeled by a dominant  energy density characterized by a constant equation-of-state $\omega$, the spectrum receives a tilt as~\cite{Giovannini:1998bp,Boyle:2007zx,Figueroa:2019paj}
\begin{align}
\Omega_{\rm GW}^{(0)}(f) \propto  f^{n_t + \Delta n_{\rm t}(w)}\,,~~~{\rm with}~~~\Delta n_{\rm t} \equiv -2{(1-3\omega)\over(1+3\omega)}\,, 
    \label{eq:eos_bound}
\end{align}
becoming more blue for a period of kination-domination\footnote{Kination here refers to the fact that the energy budget of the universe is dominated by the kinetic energy of a scalar field, which is much larger than its potential energy. This leads to an equation-of-state $1/3 < \omega \leq 1$, see {\it e.g.}~\cite{Figueroa:2019paj}.} (KD), or more red for a stage of matter domination (MD). Assuming that between the end of inflation and RD there is first a MD phase with $w \simeq 0$, then followed by a KD period with $w \simeq 1$, PTA data can be explained by the resulting broken PL GWB signal, which grows as $\Omega_{\rm GW}^{(0)}(f) \propto f^{n_{\rm t}+1}$ in the PTA frequency range, but decays as $\propto f^{n_{\rm t}-2}$ at higher frequencies, so that the BBN constraint is fully respected. Ref.~\cite{Li:2025udi} has analyzed this model, finding a spectral tilt as $n_{\rm t} = 1.29_{-0.42}^{+0.13}$ at 68\% CL and a correlation as $n_{\rm t} \simeq -0.13 \log_{10}r + 0.7$, while  allowing for higher reheating temperatures than in the case of instant reheating. 

\subsubsection{Second order induced GWs due to large scalar fluctuations}
\label{subsubsec:2ndOIGW}

Primordial scalar curvature fluctuations $\zeta$ can generate a GWB at second order in perturbation theory~\cite{Tomita:1975kj,Matarrese:1993zf,Acquaviva:2002ud,Mollerach:2003nq,Ananda:2006af,Baumann:2007zm}, for a review see~\cite{Domenech:2021ztg}. Detectable scalar-induced GWs (SIGWs) at PTA scales from this mechanism, require enhanced curvature perturbations entering the horizon near the QCD phase transition, see {\it e.g.}~\cite{Franciolini:2023pbf,Franciolini:2023wjm,Inomata:2023zup,Ebadi:2023xhq,Figueroa:2023zhu,Yi:2023mbm,Firouzjahi:2023lzg,You:2023rmn,Balaji:2023ehk,Zhao:2023joc,Yi:2023tdk}. If $\zeta$ is  Gaussian, the spectrum today reads 
\begin{eqnarray}
\Omega^{(0)}_{\rm GW}(f) &=& \frac{\Omega_{\rm rad}^{(0)}\mathcal{G}(\eta_c)}{24} \left(\frac{2\pi f}{a(\eta_c)H(\eta_c)}\right)^2 \overline{\mathcal{P}_h^{\rm ind}(\eta_c,2 \pi f)}\,,\label{eq:transfer}\\
\overline{\mathcal{P}_h^{\rm ind}(\eta,k)} &=& 2 \int_0^\infty \hspace*{-3mm}dt \int_{-1}^{1}\hspace*{-3mm}ds \left[\frac{t(2+t)(s^2-1)}{(1-s+t)(1+s+t)}\right]^2\, \overline{I^2(u,v,k,\eta)} \,\mathcal{P}_{\zeta}(ku) \,\mathcal{P}_{\zeta}(kv)\,,
\label{eq:guassian_induced_GW}
\end{eqnarray}
where ``c" indicates {\it horizon crossing}, $\eta$ is the conformal time,  $\mathcal{P}_{\zeta}(k)$ is
the power spectra of the curvature perturbation, 
and $u = (1+s+t)/2$ and $v = (1-s+t)/2$. Here $\overline{I^2}$ is given by Eq.~(27) of Ref.~\cite{Kohri:2018awv}. 

~~~~A broad class of primordial curvature spectra can be characterized by a single peak ({\it bump})~\cite{Pi:2020otn,Vaskonen:2020lbd,Kohri:2020qqd}. This is the case of ultra slow-roll scenarios (USR) [see {\it e.g.}~\cite{Faraoni:2000vg, Kinney:2005vj, Martin:2012pe, Garcia-Bellido:2017mdw, Ezquiaga:2017fvi,Kannike:2017bxn, Germani:2017bcs, Motohashi:2017kbs, Dimopoulos:2017ged, Gong:2017qlj,Ballesteros:2017fsr, Hertzberg:2017dkh, Pattison:2018bct, Biagetti:2018pjj, Ezquiaga:2018gbw, Rasanen:2018fom,Pattison:2021oen,Figueroa:2020jkf,Figueroa:2021zah,Franciolini:2023agm}], which can also lead to the formation of primordial black holes (PBHs)~\cite{Hawking:1971ei, Chapline:1975ojl, Dolgov:1992pu, Ivanov:1994pa, Yokoyama:1995ex, GarciaBellido:1996qt, Jedamzik:1996mr, Ivanov:1997ia, Blais:2002nd} that could constitute all (or a fraction) of the dark matter in the Universe, or provide the seed for massive black holes~\cite{Carr:2020gox, Carr:2020xqk, Green:2020jor}. As bump-like spectra are naturally generated by USR models, we have naturally included SIGW signals in the section of Inflation. We note, however, that SIGW signals simply require a large scalar power spectrum at scales smaller than those of the CMB, independently of whether its origin is inflationary or not. USR inflation is simply a natural scenario in which the necessary ingredients occur.  

~~~A bump-like power spectrum around a characteristic scale $k_*$, can be parametrized as
\begin{eqnarray}
 \mathcal{P}_{\zeta}(k) = 
    \frac{\mathcal{A}_{R}}{\sqrt{2 \pi }\sigma}e^{-\frac{(\ln{k/k_*})^2}{2\sigma^2}}\,, 
 \label{eq:gaussian_scalar}
\end{eqnarray}
with $\mathcal{A}_{R}$ controlling the peak amplitude, and $\sigma$ the width of the bump. Using NG15 data, Ref.~\cite{Ellis:2023oxs} finds as best fit $\log_{10}({\rm Mpc}k_*) = 7.7$, $\mathcal{A}_{R} = 0.63$, and $\sigma = 0.21$, whereas Ref.~\cite{NANOGrav:2023hvm} reports $\mathcal{A}_{R} \gtrsim 0.037$. Ref.~\cite{Figueroa:2023zhu}, using NG15+EPTA data, reports a peak location $k_{*} \in [4.57, 88]\cdot 10^{7} $ Mpc$\rm ^{-1}$, $A_R \in [0.10, 2.82]$ and $\sigma \in [0.11,0.82]$, all at 68\% CL. We note that the 2D posteriors of the parameter space $\lbrace \mathcal{A}_{R}, \sigma, k_* \rbrace$ of Refs.~\cite{Ellis:2023oxs,NANOGrav:2023hvm,Figueroa:2023zhu}, can be found in their respective Fig.s~10, Fig.~7 and Fig.~1. 

\begin{figure*}[t]
    \centering
    {\bf 2nd Order Scalar Induced GWB (log-normal bump)}\\
    \includegraphics[width=1.0\linewidth,angle=00]{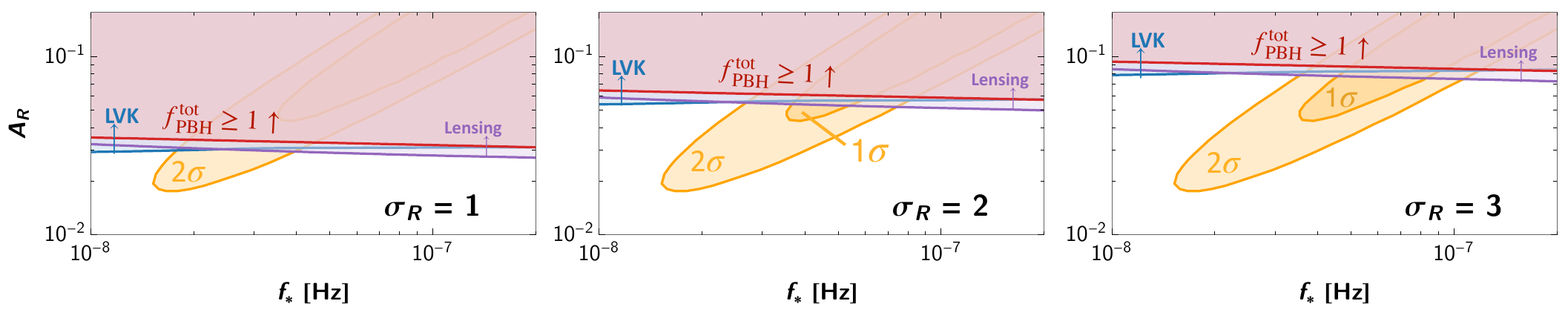}
    \vspace{0.1cm}\caption{Constraints on the Gaussian scalar-perturbation bumps in Eq.~(\ref{eq:gaussian_scalar}) from PBH abundance (over-abundance, lensing, LVK mergers) and other constraints ($\mu$-distortion). The parameters are $\lbrace \sigma$, $f_{*}$, $A_{\mathcal{R}}\rbrace$.
    The 2OSI GW interpretation of the PTA signal (in yellow) survives but cannot account for the dark matter totality. This figure is adapted from~\cite{Figueroa:2023zhu}.}
    \label{fig:SIGW_constraint}
\end{figure*}

~~- {\bf PBH Constraints}. Sufficiently large scalar perturbations can form PBHs, with different populations expected to be formed when a broad scalar power spectrum as Eq.~(\ref{eq:gaussian_scalar}) is considered. For a typical scale $k_*\sim 10^8~{\rm Mpc}^{-1}$ of SIGW in the PTA band ($f_* \sim 1-10$ nHz), PBHs are expected in the mass range $M_{H}/M_\odot \sim 0.4-40 $. While the most conservative constraint is the PBH over-abundance $f_{\rm PBH}^{\rm tot} > 1$, many observations also exclude the PBH abundance for $f_{\rm PBH}^{\rm tot} < 1$ over a wide range of masses~\cite{Green:2020jor}. Furthermore, there are also two relevant constraints for the PBH mass-range compatible with PTA data, LVK mergers (both individual events and stochastic background), and gravitational lensing. Fig.~\ref{fig:SIGW_constraint} shows these three constraints on the log-normal scalar perturbation, as obtained by Ref.~\cite{Figueroa:2023zhu}, where the bounds on the monochromatic PBH mass from Ref.~\cite{Green:2020jor} were translated for a PBH mass distribution, using the method outlined in Ref.~\cite{Carr:2017jsz}. 

~~- \textbf{CMB and $\mu$-distortions.} While the scalar perturbation spectrum considered has a peak at PTA scales, its low-momentum tail at scales larger than PTA ($k \lesssim 10^6 ~ {\rm Mpc}^{-1}$) can be also constrained via $\mu$-distortions in the CMB spectrum~\cite{Chluba:2012we,Kohri:2014lza}, or directly via CMB temperature anisotropies when the spectrum is too broad~\cite{Hunt:2015iua}. The latter constraint 
sets an upper bound for the width $\sigma \lesssim 3$ in Eq.~(\ref{eq:gaussian_scalar}), see Fig.~\ref{fig:SIGW_constraint}.

~~~~~Fitting NG15+EPTA data with SIGWs and choosing $\sigma <3$~\cite{Inomata:2018epa}, Ref.~\cite{Figueroa:2023zhu} finds a peak frequency $f_* \in [0.07, 1.44] ~ {\mu \rm Hz}$ at 68\% CL, which leads to a PBH distribution peaked around masses $M_H/M_{\odot} \in [1.5\cdot 10^{-5},5.6\cdot 10^{-3}]$. Fig.~\ref{fig:SIGW_constraint} shows how the region of PTA best-fit is cut by the PBH over-abundance criterion, as well as by gravitational lensing, and LKV bounds~\cite{Green:2020jor}. In light of Fig.~\ref{fig:SIGW_constraint}, the PTA signal cannot correspond to an interpretation of PBH as the DM totality for $\sigma \lesssim 1$. We note however that to obtain the total fraction of PBH today $f_{\rm PBH}^{\rm tot}$, different ingredients need to be accounted for, such as collapse threshold considerations, nonlinearities between $\zeta$ and the density contrast $\delta_m$, and other effects, see Ref.~\cite{Ellis:2023oxs} for a discussion and references therein. Works that have analyzed PTA data in light of SIGW signals with a bump spectrum~\cite{NANOGrav:2023hvm,Franciolini:2023pbf, Franciolini:2023wjm, Inomata:2023zup, Ebadi:2023xhq, Figueroa:2023zhu, Yi:2023mbm, Firouzjahi:2023lzg, You:2023rmn,Balaji:2023ehk,Zhao:2023joc,Yi:2023tdk,Ellis:2023oxs} have incorporated different ingredients, so the (in)compatibility of PBH bounds with a bump signal fitting PTA data, varies from work to work. For example, Fig.~\ref{fig:SIGW_constraint} is based on the considerations of Ref.~\cite{Figueroa:2023zhu}, which used the $\delta_m$-$\zeta$ non-linear relation, critical collapse, and~\emph{peak theory}~\cite{Young:2019yug} (with a \emph{real-space top-hat} window function and fixed threshold). Ref.~\cite{Ellis:2023oxs}, however, further incorporated effects of the QCD phase transition and state-of-the-art threshold criteria, reaching the conclusion that a SIGW signal from a bump spectrum, is ruled out at (almost) 3$\sigma$ by PBH over-abundance, see their Fig.~10. Future improvement of PBH abundance calculations (incorporating {\it e.g.}~effects beyond gradient expansion at super-horizon scales, non-linear transfer functions, etc), and certainly more data from LVK, lensing and PTA observations, will help clarifying whether a SIGW signal explaining the origin of the PTA signal, can also explain the abundance of DM in the universe, or at least a fraction of it. 

~~- {\bf Primordial non-Gaussianity (pNG)}. For broader generality, some works considered also a curvature perturbation with local pNG as $\zeta({\bf x}) = \zeta_{\rm G}({\bf x}) + f_{\tt nl}(\zeta_{\rm G}^2({\bf x})- \langle\zeta_{\rm G}^2({\bf x})\rangle)$, with $\zeta_{\rm G}$~Gaussian. In this case, the resulting GWB spectrum is the sum of a series of terms~\cite{Cai:2018dig,Unal:2018yaa}, a Gaussian contribution as in Eq.~(\ref{eq:gaussian_scalar}), and non-Gaussian contributions $\propto f_{\tt nl}^2$ and $\propto f_{\tt nl}^4$, which can be found explicitly in Eqs.~(2.33)-(2.39) of Ref.~\cite{Adshead:2021hnm}. We note that the calculation of the PBH abundance in the presence of pNG needs also to be modified~\cite{Young:2022phe}. Fitting NG15 data with a SIGW signal with local-pNG and Gaussian component with bump-spectrum Ref.~\cite{Ellis:2023oxs} concludes that current data are not able to constrain $f_{\rm nl}$ at PTA scales, whereas Ref.~\cite{Wang:2023ost} finds -- using the same data -- a stringent constraint as $|f_{\rm nl}| \lesssim 4.1$ at 68\% CL. Using the same modeling but fitting to NG15+EPTA+PPTA data, Ref.~\cite{Liu:2023ymk} finds that $|f_{\rm nl}| \lesssim 13.9$. Ref.~\cite{Figueroa:2023zhu} claims however that pNG can only be constrained at PTA scales if the perturbativity condition $A_{\mathcal{R}}f_{\tt nl}^2 < 1$ is imposed in the analysis, in which case NG15+EPTA data provides a stringent constrain on the pNG parameter as $|f_{\tt nl}| \lesssim 2.34$ at 95\% CL (and otherwise remains unconstrained). While more work is needed to reach a clear conclusion, it is remarkable that pNG could be potentially constrained at PTA scales, that are $\sim 8$ orders of magnitude smaller than CMB scales~\cite{Planck:2019kim}.

\subsubsection{Post-inflationary Preheating}

In many models of particle physics, the reheating of the universe can be initially driven by {\tt preheating} effects,  typically characterized by exponential particle production~\cite{Traschen:1990sw,Kofman:1994rk,Shtanov:1994ce,Khlebnikov:1996mc,Prokopec:1996rr,Kofman:1997yn,Greene:1997fu}. This leads to  very efficient production of GWs, though the details depend strongly on the considered scenario, see {\it e.g.}~\cite{Khlebnikov:1997di,Easther:2006gt,Easther:2006vd,GarciaBellido:2007dg,GarciaBellido:2007af,Dufaux:2007pt,Dufaux:2008dn,Figueroa:2011ye,Bethke:2013aba,Bethke:2013vca,Figueroa:2016ojl,Figueroa:2017vfa,Cosme:2022htl}. 
In the case of {\it parametric resonance}~\cite{Kofman:1994rk,Kofman:1997yn,Greene:1997fu}, the inflaton is assumed to have a potential $V(\phi) \propto |\phi|^p$ after inflation, and to be coupled via $g^2 \phi^2 \chi^2$ with a secondary {\it daughter} (massless scalar) field $\chi$, with $g^2$ the coupling constant. If the \emph{resonance parameter} $q_{\star} \equiv g^2 \phi_{\star}^2 /\omega_{\star}^2$ is $\gg 1$, with $\phi_{\star}$ the amplitude at the onset of oscillations and $\omega_{\star} \equiv V''(\phi_*)$, fluctuations of the daughter field grow exponentially via broad resonances, for modes $k \lesssim k_{\star} \sim q_*^{1/4} \omega_{\star}$. The spectrum of the GWB produced during this process has a peak at approximately the frequency~\cite{Figueroa:2017vfa},
\begin{eqnarray}
f_{\rm peak}\simeq 8\cdot 10^{-9}{\rm Hz} \times \bigg( \frac{\omega_\star}{\rho_\star^{1/4}} \bigg) \, \epsilon_\star^{\frac{1}{4}} q_\star^{\frac{1}{4} + \eta}\label{eq:preheating-1} \,, 
\end{eqnarray}
where $\rho_\star$ is the total initial energy density in the universe, $\eta \sim 0.3$--0.4 
is a parameter accounting for non-linear effects, and $\epsilon_\star$ parametrizes the period between the end of inflation and the onset of RD. Ref.~\cite{Figueroa:2017vfa} finds that for $V(\phi) \propto \phi^2$ and $q_{\star} \in (10^4,10^6)$ (assuming $\epsilon_{\star} = 1$), a GWB peak amplitude today as large as $\Omega_{\text{GW},0} \simeq 10^{-12} - 10^{-11}$, at a frequency $f_{\rm peak} \simeq 10^8 \, \rm{Hz} - 10^9 \, \rm{Hz}$. Similarly, for $V(\phi) \propto \phi^4$ with $q_\star \in (1,10^4)$, Ref.~\cite{Figueroa:2017vfa} finds a multi-peak GWB spectrum with leading peak amplitude $\Omega_{\text{GW},0} \simeq 10^{-13} - 10^{-11}$ and frequency today $f_{\rm peak} \simeq 10^7 \, \rm{Hz} - 10^8\,\rm{Hz}$. Clearly, given the high frequencies involved, the PTA data cannot be explained by these phenomena.

~~~~We note that GWs can be also efficiently produced when preheating leads to {\it oscillons}, which are long-lived compact scalar field configurations~\cite{Gleiser:1993pt, Amin:2010jq, Amin:2010dc, Amin:2011hj, Zhou:2013tsa, Amin:2013ika, Lozanov:2014zfa, Antusch:2015vna, Antusch:2015ziz, Antusch:2016con, Antusch:2017flz, Antusch:2017vga, Lozanov:2017hjm, Amin:2018xfe, Antusch:2019qrr, Sang:2019ndv, Lozanov:2019ylm, Fodor:2019ftc, Hiramatsu:2020obh}. Their dynamics can efficiently source GWs due to field interactions and collisions among each other~\cite{Helfer:2018vtq}, but the GWB peak today is typically well above LVK frequencies~\cite{Zhou:2013tsa, Antusch:2017flz, Lozanov:2019ylm} (see however \cite{Antusch:2016con, Liu:2017hua, Kitajima:2018zco} for models that lead to a GWB peaked at lower frequencies, though not as low as the PTA frequencies). GWs can be also efficiently produced in preheating by fields with spin $\neq 0$. For example, GWs can be generated {\it e.g.}~during the out-of-equilibrium production of fermions after inflation, for both spin-1/2~\cite{Enqvist:2012im, Figueroa:2013vif, Figueroa:2014aya} and spin-3/2~\cite{Benakli:2018xdi} fields. Similarly, GWs can be generated when the produced particles are Abelian or non-Abelian gauge fields, when coupled to charged scalar fields~\cite{Dufaux:2010cf, Figueroa:2016ojl, Tranberg:2017lrx}, or to a pseudo-scalar field via an axial coupling~\cite{Adshead:2018doq,Adshead:2019igv,Adshead:2019lbr}. Preheating can be remarkably efficient in the second case, with a resulting GW energy density as large as $\Omega_{\rm GW} \lesssim \mathcal{O}(10^{-6})$ for certain couplings~\cite{Adshead:2019igv, Adshead:2019lbr}, but still peaked at very high frequencies. Neither of these cases can produce a GWB signal within PTA scales, despite the fact that they constitute very efficient early Universe GW generation mechanisms.

\subsection{EUB: Phase transitions}
\label{subsec:EUBphaseTransitions}

First-order phase transitions (1stO-PhT) in the early Universe, occur whenever a potential barrier emerges separating higher ({\tt true} vacuum) and lower ({\tt false} vacuum) minima in a scalar field potential. A phase transition is triggered when quantum tunneling or thermal fluctuations take the scalar field amplitude over the potential barrier in some localized region. This leads to the nucleation of spherical {\it bubble} configurations, inside of which the scalar field reaches the true vacuum, whereas outside the scalar field still resides in the false vacuum. If the nucleated bubbles have a radius above a critical scale, they expand and eventually collide with each other. The expansion and collision of the bubbles~\cite{Kosowsky:1991ua,Kosowsky:1992rz,Kosowsky:1992vn,Kamionkowski:1993fg,Caprini:2007xq, Huber:2008hg}, the sound waves generated in the plasma due to its coupling to the scalar field~\cite{Hindmarsh:2013xza,Giblin:2013kea,Giblin:2014qia,Hindmarsh:2015qta}, and possible turbulent motions of the plasma~\cite{Caprini:2019egz,RoperPol:2019wvy}, can all source very efficiently GWs, leading to a large and potentially observable GWB, see Ref.~\cite{Hindmarsh:2020hop} for a review, and Ref.'s~\cite{Witten:1984rs,Hogan:1986qda} for seminal works. 

~~~~The resulting GWB produced during a phase transition can be therefore a superposition of the bubble-collision, sound-wave and turbulence contributions~\cite{Hindmarsh:2013xza,Hindmarsh:2015qta,Hindmarsh:2017gnf,Weir:2017wfa,Caprini:2019egz,Cutting:2018tjt,Cutting:2019zws,Pol:2019yex,Cutting:2020nla,Cai:2023guc}. In the case of magneto-hydrodynamic (MHD) turbulence at the QCD scale, GWs are generated exactly around the PTA $\sim${\it nHz} frequencies~\cite{Neronov:2020qrl}, whereas in the case of a dark sector, the GWB could be peaked across a wide frequency range~\cite{Schwaller:2015tja}. The spectrum today of the GWB from a 1stO-PhT can be written as 
\begin{eqnarray}\label{eq:spectrumPhT}
\Omega_{\rm GW}^{(0)} = \Omega_{\rm rad}^{(0)}\mathcal{G}(T_*)\sum_i \Omega_{{\rm GW},i}(f)\,,~~\Omega_{{\rm GW},i}(f) \equiv \mathcal N_i \, \Delta_i(v_{\rm w}) \left({\kappa_i \, \alpha \over 1+\alpha}\right)^{p_i}\left({H\over\beta}\right)^{q_i}  s_i(f)\,,
\end{eqnarray}
where $\Omega_{\rm rad}^{(0)} \simeq 9.2 \cdot 10^{-5}$, $\mathcal{G}(T) \equiv ({g(T)/ g_{0}})({g_{s,0}/g_{s}(T)})^{4/3}$ accounts for the number of relativistic degrees of freedom at temperature $T$, the index $i$ stands for $i = \rm b$ (bubbles), $i = \rm sw$ (sound waves) and $i = \rm tb$ (turbulence), and $\alpha$, $\beta$ and $T_*$, denote the strength, the (inverse) duration, and the temperature of the PhT. The normalization 
factors $\mathcal N_i$, exponents $p_i,q_i$, efficiency factor $\kappa_i$, and the function $\Delta_i(v_{\rm w})$ depending on the wall velocity $v_{\rm w}$, are summarized in Table~\ref{tab:PhT}. The spectral shapes $s_i(f)$, also given in Table~\ref{tab:PhT}, are essentially a {\it broken PL} ($i$ = b, sw) or a {\it broken double-PL} ($i$ = tb). The total spectrum can contain various peaks, due to the various contributions. We will focus however on scenarios dominated by a single contribution, with peak frequency today  $f_*$ given in Tab.~\ref{tab:PhT}. Using both the peak position and amplitude of each contribution, we can convert parameters extracted from a broken (double) PL fit to PTA data, into PhT parameters from different scenarios, considering either strong or weak-mild 1stO-PhT's. See Fig.~\ref{fig:GWlandscape} for an example of a single-peaked signal corresponding to a strong 1stO-PhT, {\it c.f.}~Sec.~\ref{subsubsec:StrongPhT}.

\begin{table}[t]
~~~\begin{tabular}{cccc}
\hline\\[-0.75em]
 & ~ ~  Bubbles ~ ~ & ~ ~ Sound-wave ~ ~ & ~ ~ Turbulence ~ ~ \\[0.25em] \hline\\[-0.5em]
$\mathcal{N}_i$ & 1 & 0.16 & 20 \\[0.5em]
$\Delta_i(v_{\rm w})$ & $\frac{0.11 \, v_{\rm w}^3}{0.42 + v_{\rm w}^2}$ & $v_{\rm w}$ & $v_{\rm w}$ \\[0.5em]
$p_i$ & 2 & 2 & 3/2 \\[0.5em]
$q_i$ & 2 & 1 & 1 \\[0.5em]
$s_i(f, f_*)$ & $\frac{3.8 (f/f_*)^{2.8}}{1 + 2.8 (f/f_*)^{3.8}}$ & $\left(\frac{f}{f_*}\right)^{3} \left[\frac{7}{4 + 3 (f/f_*)^{2}}\right]^{\frac{7}{2}}$ & $\frac{(f/f_*)^3}{(1 + f/f_*)^{11/3}\left(1 + 8 \pi f/\tilde{H}_*\right)}$ \\[0.5em]
$f_{*}$ & $11 {\rm nHz}  \left(\frac{0.62}{1.8 - 0.1 v_{\rm w} + v_{\rm w}^2}\right) \mathcal{F}$ & $13  {\rm nHz} ~ v_{\rm w}^{-1} \mathcal{F}$ & $19  {\rm nHz} ~  v_{\rm w}^{-1} \mathcal{F}$ \\ [0.5em] \hline
\end{tabular}
\vspace*{2mm}\\
\hspace*{1.5cm}{\rm where}~~
%\begin{align}
$\left\lbrace
\begin{array}{rl}
    \mathcal{F} &\equiv \left(\frac{\beta}{H_*}\right) \left(\frac{T_*}{0.1 ~ {\rm GeV}}\right) \left(\frac{g_*(T_*)}{10.75}\right)^{\frac{1}{2}} \left(\frac{g_{*,s}(T_*)}{10.75}\right)^{-\frac{1}{3}}\\
    \tilde{H}_* &\equiv H_* \left(\frac{a_*}{a_0}\right) \simeq 11 ~ {\rm nHz} \left(\frac{T_*}{0.1 ~ {\rm GeV}}\right) \left(\frac{g_*(T_*)}{10.75}\right)^{\frac{1}{2}} \left(\frac{g_{*,s}(T_*)}{10.75}\right)^{-\frac{1}{3}}
\end{array}\right.$
\vspace*{1mm}\\
%\end{align}
\hspace*{0.3cm}\rule{12.2cm}{0.4pt}
\vspace*{0.2cm}\\
\caption{Parameters and functions characterizing the bubble~\cite{Huber:2008hg}, sound-wave~\cite{Hindmarsh:2015qta} and turbulence~\cite{Caprini:2009yp} contributions to the GWB spectrum from a 1stO-PhT.}
\label{tab:PhT}
\end{table}

\subsubsection{Strong 1stO-PhT}
\label{subsubsec:StrongPhT}

If there is a \emph{super-cooling} phase with $\alpha \gg 1$, a strong first-order PhT takes place in vacuum, and the bubble collisions are the dominant contribution to the GWB. According to the \emph{envelope approximation}~\cite{Huber:2008hg}, the shape of the IR tail of the spectrum scales as $\Omega_{\rm GW} \propto f^{2.8}$, and the UV tail as $\Omega_{\rm GW} \propto f^{-1}$, see Table~\ref{tab:PhT}. While other studies have found different slope values (ranging as $f^{[-3, -0.9]}$ for the IR tail, and as $f^{[-3, -1]}$ for the UV slopes\footnote{See Fig.~6.3.1 of Ref.~\cite{Gouttenoire:2022gwi} for a summary of the differences and appropriate list of references.}), we stick to the envelope approximation just to make a concrete choice. Using  Tab.~\ref{tab:PhT} for the peak frequency $f_*$ and Eq.~(\ref{eq:spectrumPhT}) for the GWB amplitude $h^2\Omega_{\rm GW}^{(0)}(f_*)$, one can translate a broken-PL fit into the duration and temperature of a strong 1stO-PhT as~\cite{Figueroa:2023zhu}
\begin{align}
    \left(\frac{\beta}{H_*}\right)_{\rm b} &\simeq 206 \left[\Delta_{\rm b}(v_{\rm w}) \kappa_{\rm b}^2 \mathcal{G}(T_*) \left(\frac{10^{-9}}{h^2 \Omega_{\rm GW}^{(0)}(f_*)}\right)\right]^{1/2},\\
    \left(T_*\right)_{\rm b} &\simeq 0.48 ~ {\rm MeV} \left(\frac{f_*}{10 ~\rm nHZ}\right) \left(\frac{h^2 \Omega_{\rm GW}^{(0)}(f_*)}{10^{-9}}\right)^{1 \over 2} \left(\frac{g_{*,s}(T_*)}{g_*(T_*)}\right) \left(\frac{1.8 - 0.1 v_{\rm w} + v_{\rm w}^2}{0.62\kappa_{\rm b}\Delta_{\rm b}^{1/2}}\right).
\end{align}
In this case, the vacuum energy is efficiently  converted into energy of the bubbles, {\it i.e.}~$\kappa_{\rm b} = 1$, the wall velocity reaches the speed of light $v_{\rm w} \to 1$, and then $\Delta_{\rm b} \approx 0.08$.  We note that for relativistic bubble velocities, the average bubble separation in
units of the Hubble radius at the time of percolation is related to the bubble nucleation rate $\beta$, through the relation $H_*R_*=(8\pi)^{1/3}H_*/\beta$, so constraints on $\beta/H_*$ or $H_*R_*$ can alternatively be provided.

~~~~Fitting NG15+EPTA for bubble collisions with $\alpha \gg 1$, Ref.~\cite{Figueroa:2023zhu} finds $T_* \in [6.2 \cdot 10^{-3},9.0]$ GeV and $\beta/H_* \in [2.2,500]$, at 68\% CL, with best fit temperature $T_* \simeq 20$ MeV. Fitting NG15 data only for the same case, Ref.~\cite{Ellis:2023oxs} finds a best fit $T_* = 0.34 \, {\rm GeV}, \beta/H = 6.0$. Fitting NG15 but allowing $\alpha$ to vary, Ref.~\cite{NANOGrav:2023hvm} finds $\alpha > 1.1$ ($0.29$), $H_*R_* > 0.28$ ($0.14$), and $T_*\in [0.047,0.41]$ $([0.023,1.75])$ GeV at the $68\%$ ($95\%$) CL. A strong PhT with low temperature $T_* \sim 10-100$ MeV, as required by PTA data, could however affect BBN~\cite{Bringmann:2023opz}, and would be in tension with lattice QCD, which predicts a cross-over~\cite{Aoki:2006we}, not a first order PhT.   

\subsubsection{Weak-to-moderate 1stO-PhT}

When a PhT is not too strong, if sound-waves are created, they typically dominate GW production over other contributions. This happens when bubbles expand in a thermal medium, reaching a terminal velocity thanks to their interaction with the plasma (\emph{non-runaway} scenario). \cite{Hindmarsh:2015qta} finds the shape of the GWB as a PL with $\propto f^{3}$ and $\propto f^{-4}$ IR and UV tails. While recent simulations have found an intermediate slope associated to the width of the sound shell, behaving as $\propto f$ close to the peak, we stick to the above broken-PL for simplicity. The peak amplitude and position in Eq.~(\ref{eq:spectrumPhT}) [Tab.~\ref{tab:PhT}] can be translated into PhT parameters via
\begin{align}
    \left(\frac{\beta}{H_*}\right)_{\rm sw} &\simeq 68  \left(\frac{10^{-9}}{h^2 \Omega_{\rm GW}^{(0)}(f_*)}\right) \mathcal{G}(T_*) v_{\rm w} \left[\frac{(1+\alpha)^2\alpha^{-2} \kappa^{-2}_{\rm sw}(\alpha)}{100}\right]^{-1},\\
    \left(T_*\right)_{\rm sw} &\simeq 1.5 ~ {\rm MeV} \left(\frac{f_*}{10 ~\rm nHZ}\right) \left(\frac{h^2 \Omega_{\rm GW}^{(0)}(f_*)}{10^{-9}}\right) \left[\frac{10.75}{g_*(T_*)}\right]^{\frac{3}{2}} \left[\frac{g_{*,s}(T_*)}{10.75}\right]^{\frac{5}{3}} \left[\frac{(1+\alpha)}{10\alpha \kappa_{\rm sw}(\alpha)}\right]^2,
\end{align}
where the efficiency factor is $\kappa_{\rm sw}(\alpha) \simeq \alpha(0.73 + 0.083 \sqrt{\alpha} + \alpha)^{-1}$~\cite{Espinosa:2010hh}. While the degeneracy among $\alpha$, $\beta/H_*$, and $T_*$, can be broken by considering UV completions of a low-$T$ first-order PhT, PTA data by itself cannot break such degeneracy.

~~~~~During a weak-to-mild 1stO-PhT, turbulent motions can also be a relevant GW contribution. It is therefore common to expect both sound-wave and turbulence in this case, typically with $\kappa_{\rm turb} < \kappa_{\rm sw}$. In the case of magneto-hydrodinamic (MHD) turbulence, ``MHD eddies" are formed during the PhT, though we note that MHD turbulence can also arise due to other unrelated mechanisms~\cite{Kamada:2019uxp}. Either way, GWs in this case are sourced by the magnetic field present, characterized by an energy density fraction at the time of production $t_*$, scaling as $\Omega_{\rm GW}^* \sim 3 (\Omega_B^*)^2 (H_* \mathcal{T})^2$~\cite{Neronov:2020qrl}, where $\Omega_B^* = B_*^2/(2 \rho_{c,*})$ is the magnetic energy-density fraction, and 
$\mathcal{T} \sim \lambda_*/v_A \lesssim H_*^{-1}$ is the turbulence-sourcing time scale, with $v_A \simeq \sqrt{2 \Omega_B^*}$ the Alfv{\'e}n speed, and $\lambda_*$ the characteristic scale of turbulence. The GWB amplitude scales as $\Omega_{\rm GW}^* \sim (\Omega_B^*)^n (H_* \lambda_*)^2$, with $n \in [1,2]$ depending on the details. The GWB spectrum has therefore a peak's amplitude and frequency today as $\Omega_{\rm GW}^{(0)} \sim \Omega_{\rm rad}^{(0)} \mathcal{G}(T_*) (\Omega_B^*)^n (H_* \lambda_*)^2$ and $f_{*} = \frac{2}{l_*}\cdot \frac{a_*}{a_0}$~\cite{Neronov:2020qrl}. Recent numerical simulations of GW production from MHD turbulence~\cite{RoperPol:2018sap,RoperPol:2019wvy,Brandenburg:2021tmp,RoperPol:2022iel}, in particular Ref~\cite{RoperPol:2022iel}, have shown that the envelope of the produced GWB can be estimated analytically assuming the total
anisotropic stress to vary more slowly than the production of GWs itself [see also Ref.~\cite{Auclair:2022jod} for validation of this in simulations of purely kinetic turbulence]. Ref.~\cite{Caprini:2009yp} determines the GWB spectrum in the case of turbulence dynamics as a double-broken PL, with causality tail in the IR as $\Omega_{\rm GW} \propto f^{3}$, intermediate slope as $\Omega_{\rm GW} \propto f^{2}$ within scales $\tilde{H}_* < f <f_*$, and UV tail as $\Omega_{\rm GW} \propto f^{-5/3}$. Ref.~\cite{RoperPol:2022iel} estimates that the UV tail rather scales as $\Omega_{\rm GW} \propto f^{-8/3}$. 

~~~~For the  largest processed eddies (LPE), {\it i.e.}~those with maximum length scale $l_* = v_A /H_*$, the peak's amplitude and frequency of the GWB spectrum are related to the magnetic energy fraction and temperature at the time of GW production, as~\cite{Neronov:2020qrl,Figueroa:2023zhu}
\begin{align}
    \Omega_B^{*} &\simeq \left[1.4 \cdot 10^{-5} \left(\frac{h^2\Omega_{\rm GW}^{(0)}(f_*)}{10^{-9}}\right)\left(\frac{10.75}{g_*(T_*)}\right)\left(\frac{g_{*,s}(T_*)}{10.75}\right)^{\frac{4}{3}}\right]^{1/(n+1)},\\
    f_* &\simeq 16.1 ~{\rm nHz} \left(\frac{T_*}{\rm MeV}\right) \left(\frac{10^{-2}}{\Omega_B^{*}}\right)^{1 \over 2},
\end{align}
Alternatively, today's magnetic field strength $B_0$ and correlation length scale $l_B \equiv l_*(a_0/a_*)$, are given by~\cite{Neronov:2020qrl,Figueroa:2023zhu}
\begin{align}
    B_0 = 0.4 ~ {\rm \mu G} \left(\frac{\Omega_B^{*}}{10^{-2}}\right)^{\frac{1}{2}} \left(\frac{g_*(T_*)}{10.75}\right)^{\frac{1}{2}}\left(\frac{10.75}{g_{*,s}(T_*)}\right)^{\frac{2}{3}}, ~ ~ l_B \simeq 1.95 ~ {\rm pc} \left(\frac{10 ~ {\rm nHz}}{f_*}\right).
\end{align}

~~~~Fitting a signal from sound-waves + turbulence to NG15+EPTA data, 
Ref.~\cite{Figueroa:2023zhu} finds $T_* \in [5.5, 360]$ MeV,  $\beta/H_* \in [1.0, 15]$ and $\alpha \in [0.017,0.060]$, at 68\% CL, with best fit temperature $T_* \simeq 94$ MeV. Fitting to NG15+EPTA data only a contribution from MHD turbulence, Ref.~\cite{Figueroa:2023zhu} finds $T_* \in [10^{-3},3.2]$ GeV at 68\% CL, and translating the best-fit peak position into a primordial magnetic field~\cite{Neronov:2020qrl}, it leads to a strength $B_0 \simeq 23.3 \mu$G and correlation length $l_B \simeq 0.25 ~ {\rm pc}$,  which is close to current constraints on primordial magnetic fields~\cite{Planck:2015zrl, Jedamzik:2018itu, Neronov:2021xua, MAGIC:2022piy}. Fitting NG15 data alone to a signal from only sound-waves, Ref.~\cite{NANOGrav:2023hvm} finds $\alpha_* > 0.42$ ($0.37$), $H_*R_*\in[0.053,0.27]$ ($[0.046,0.89]$), and $T_*\in[4.7,33]$ $([2.7,93])$ MeV at the $68\%$ ($95\%$) CL. Fitting EPTA data alone to a signal from MHD turbulence only, Ref.~\cite{EPTA:2023xxk}
shows 2D-posteriors for $\lambda_*H_*,\Omega_{\rm B}^*,T_*$ in their Fig.~18, commenting that MHD turbulence provides a good fit to the data only in the limit of large $\Omega_{\rm B}$ (close to the upper bound of their prior $\Omega_{\rm B} \leq 1$), observing a clear degeneracy between  $\lambda_* H_*$ and $\Omega_{\rm B}$, as expected from a signal scaling as $\Omega_{\rm GW}^* \sim (\Omega_B^*)^n (H_* \lambda_*)^2$.

\subsection{EUB: Topological Defects}
\label{subsec:EUBcosmicDefects}

Topological defects are stable configurations that may form upon spontaneous symmetry breaking during a PhT~\cite{Kibble:1976sj}. Depending on the topology of the associated vacuum manifold (characterized by the homotopy groups $\Pi_n,~n = 1, 2, 3, ...$), different type of defects may appear (or no defects at all).  Independently of the type, defects can be global or local, depending on whether the symmetry broken is global or gauge. Either way, in all cosmologically viable cases ({\it i.e.}~those that do not overclose the universe), defects appear everywhere in the universe upon the completion of the PhT, forming a {\it network}. The defect network reaches soon enough a {\it scaling} regime, where the mean separation between defects tracks the horizon during cosmic evolution. Cosmic strings, corresponding to one-dimensional configurations, are topological solutions that appear in a PhT when $\Pi_1 \neq \mathcal{I}$~\cite{Kibble:1976sj} (alternatively, they can also be cosmologically stretched fundamental strings in String theory scenarios~\cite{Dvali:2003zj,Copeland:2003bj}). Domains walls, on the other hand, emerge in PhT's driven by the breaking of a discrete symmetry, and correspond to 2-dimensional topological solutions in the form of ``sheets".  Other type of defects can emerge, depending on the topology of the vacuum manifold. 

~~~~Independently of the type of defect, as the network's energy-momentum tensor adapts itself to maintaining scaling, GWs are emitted by the network~\cite{Figueroa:2012kw,Figueroa:2020lvo}. This leads to a scale invariant GWB with amplitude $\Omega_{\rm GW} \propto (v/m_p)^4$ for scaling networks (except in the case of domain walls), with $v$ the vacuum expectation of the scalar field within $\mathcal{M}$. Current CMB constraints set $v \lesssim 10^{-3}$~\cite{Lizarraga:2016onn,Lopez-Eiguren:2017dmc}, making this GWB too small to explain the PTA signal. In the case of cosmic strings, however, on top of the GWs emitted by the network (infinite) strings during scaling, there is expected a much stronger emission from the loops~\cite{Vilenkin:1981bx,Vachaspati:1984gt,Hogan:1984is,Auclair:2019wcv}, which can potentially explain PTA data. See Ref.~\cite{Gouttenoire:2019kij} for a review on GWB signals from cosmic-string networks, and Ref.~\cite{Dimitriou:2025bvq} for a classification of different string-network scenarios. Domain walls, on the other hand, are not viable defects unless some mechanism is in place so that they annihilate at a given energy scale, before their energy density overtakes the energy budget of the universe. In the viable cases, domain wall networks emit GWs until their disappearance at an annihilation temperature $T = T_{\rm ann}$~\cite{Gelmini:1988sf,Coulson:1995nv,Larsson:1996sp,Preskill:1991kd}. The resulting GWB emitted before annihilation, can also potentially explain PTA data. For recent simulations on the GWB from domain wall networks, see~\cite{Dankovsky:2024zvs,Dankovsky:2024ipq,Babichev:2025stm,Notari:2025kqq}.

\subsubsection{Local strings and superstrings} Local strings may emerge after symmetry breaking in gauge theories. They have a finite core's width, corresponding to the inverse mass scale of the particles involved (both scalar and gauge fields). As this width is much smaller than the typical length scale of the strings, the motion of the strings is expected to be well approximated by the Nambu-Goto (NG) dynamics of infinitely-thin strings~\cite{Vilenkin:2000jqa}. Under this assumption, the spectrum of the GWB emitted by a network of local CSs, in particular by the loop distribution along cosmic history, is given by~\cite{Blanco-Pillado:2017oxo,Blanco-Pillado:2017rnf,Auclair:2019wcv,Dimitriou:2025bvq}
\begin{eqnarray}
\Omega_{\rm GW}^{(0)}(f) \equiv {16\pi\Gamma (G\mu)^2\over 3p\zeta(q)H_0^2}{1\over f}
\sum_{j=1}^{\infty}{\mathcal{C}_{j}(f)\over j^{q-1}}\,,~~~
\mathcal{C}_{j}(f) 
= \int{dz'\over H(z')(1+z')^6}\,n\hspace*{-1mm}\left({2j\over (1+z')f},t_e(z')\right),\nonumber\\
\label{eq:strings}
\end{eqnarray}
with $G$ Newton's constant, $G\mu$ the string tension in Planck units, $\zeta(q)$ the {\it Riemann zeta} function, $p$ the {\it intercommutation} probability, $H(z)$ and $H_0$ the Hubble rate at {\it redshift} $z$ and today, and $n(l_e(z),t_e(z))$ the {\it number density} of loops of size $l_e(z)$ at the emission time $t_e(z)$. As Eq.~(\ref{eq:strings}) is valid for NG strings, the formulation also applies to superstrings. One considers therefore $p = 1$ for field theory local strings, and $10^{-3} \leq p < 1$ for superstrings.

~~~~To evaluate the signal, \cite{Figueroa:2023zhu}  considers a VOS modeling, with loop birth length $l = 0.1/H$~\cite{Blanco-Pillado:2013qja}, $\Gamma = 50$~\cite{Blanco-Pillado:2017oxo} and $q = 4/3$ corresponding to cusps. Fitting NG15+EPTA data to local strings represented by NG strings with $p = 1$ and the above parameters, leads to $\log_{10}(G\mu) = -9.90_{-0.19}^{+0.11}$ at 68\% CL (this signal is actually plot in  Fig.~\ref{fig:GWlandscape}). This translates into a very tight constraint on the associated energy scale as $\sqrt{\mu} \simeq 1.32^{+0.20}_{-0.24} \cdot 10^{14} ~{\rm GeV}$, which can be accommodated in a variety of particle-physics models~\cite{Vilenkin:2000jqa}, despite being $\mathcal{O}(10^{-2})$ smaller than the standard Grand Unification Theory scale $\sim 10^{16}~\rm GeV$. This GWB spectrum, if viable, would be actually accessible to higher frequency experiments allowing us to probe the cosmological history above the BBN $\sim$MeV scale~\cite{Cui:2017ufi, Gouttenoire:2021jhk, Auclair:2021jud}. This signal however, does not fit well PTA data. Fitting the NG15+EPTA instead to a superstring signal~\cite{Damour:2004kw}, represented by NG strings with $p<1$, Ref.~\cite{Figueroa:2023zhu} finds $\log_{10}(G\mu) = -11.83_{-0.15}^{+0.27}$ and $\log_{10} p = -2.63_{-0.31}^{+0.49}$ at 68\% CL. This case actually fits very well current PTA data, thanks to accommodating a smaller energy scale $\sqrt{\mu} \simeq 1.56^{+0.48}_{-0.39} \cdot 10^{13} ~{\rm GeV}$, while suppressing the intercommutation probability. Ref.~\cite{NANOGrav:2023hvm} considers more CS modelings, including NG strings with only cusps ($q=4/3$) or only kinks ($q = 5/3$) in loops, NG strings emitting GWs only at the fundamental frequency $f = 2/l_e$ of loops, and strings and GW emission as calibrated from NG simulations~\cite{Blanco-Pillado:2011egf,Blanco-Pillado:2015ana}. Fitting all these modelings against NG15 data, they obtain peaked distributions centered around values $\log_{10}G\mu \sim -\left(10.5\cdots10.0\right)$. Extending the analysis to superstrings, they find a posterior density maximized at small intercommutation probabilities $\log_{10} p \sim -3$ (at the edge of their prior range) and cosmic-string tensions $\log_{10} G\mu \sim -12$. This parameter region is however in tension with the LVK bounds~\cite{KAGRA:2021kbb}. They also analyze metastable strings, which are string networks unstable against the nucleation of GUT monopoles~\cite{Buchmuller:2021mbb}. Fitting NG15 data, they find values of the decay parameter of around $\sqrt{\kappa} \sim 8$, together with a large tension $\log_{10}G\mu \sim -\left(7\cdots4\right)$.

~~~The above treatment of field theory local strings as NG strings, ignores however that particles of the scalar and gauge fields the strings are made of, are actually emitted very efficiently, as shown by numerical simulation of the whole network of Abelian-Higgs field theory strings~\cite{Vincent:1997cx,Moore:2001px,Bevis:2006mj,Bevis:2010gj,Daverio:2015nva,Correia:2019bdl,Correia:2020gkj,Correia:2020yqg}, and more recently by isolating string loop configurations~\cite{Matsunami:2019fss,Hindmarsh:2021mnl,Baeza-Ballesteros:2024otj}. The extra energy loss due to particle production by local loops extracted from simulations of Abelian-Higgs field theory networks in scaling, would not only scale down the GWB amplitude (for given tension), but also shift the spectrum frequency profile, as it affects how string loops shrink~\cite{Baeza-Ballesteros:2024otj}. While a frequency template for such field-theory strings is still missing, Ref.~\cite{Kume:2024adn} fits against NG15 data a model that accommodates only a fraction $f_{\rm NG}$ of loops following NG dynamics, with no particle emission, and as a result it loosens the constraint of local field theory strings to $\log_{10}G \mu \lesssim -7$. This restores the compatibility of the bound from PTA observations on the tension of local strings to energy scales up to GUT scales, with $v \lesssim 10^{-3} m_p$, similarly as in CMB bounds~\cite{Planck:2013mgr}.

\subsubsection{Global Strings}

Global strings are formed after the breaking of a global symmetry with non-trivial fundamental homotopy group  ({\it e.g.}~breaking of a global $U(1)$), and    contrary to local strings, they have a spatially extended core profile, with a tension scaling radially as $\propto \log[r/\eta^{-1}]$. As a consequence, they naturally sustain long-range interactions, and Goldstone-boson emission is subjected to a strong $\Delta N_{\rm eff}$ bound~\cite{Chang:2021afa,Gorghetto:2021fsn}. 

~~~~Global strings~\cite{Gouttenoire:2019kij, Gorghetto:2021fsn} are naturally generated in BSM models featuring an additional global $U(1)$ symmetry, which gets spontaneously broken by the vacuum expectation value of a complex scalar field, providing in this way a Nambu-Goldstone Boson. The Peccei-Quinn $U(1)$ symmetry advocated to solve the strong CP problem and its associated axion particle~\cite{Peccei:1977hh, Peccei:1977ur, Weinberg:1977ma, Wilczek:1977pj}, is precisely the paradigmatic example of this. As the $U(1)$ symmetry gets also broken explicitly at later times, the axion acquires a mass, and at that moment domain walls are also formed~\cite{Sikivie:1982qv}. If the $U(1)$ is broken after inflation, cosmic strings generate axion particles throughout the cosmic history~\cite{Davis:1986xc, Gorghetto:2020qws, Buschmann:2021sdq}. In this case, the resulting population of loops generate a GWB, while the domain walls provide an additional GW production, which we discuss later in Sec.~\ref{subsubsec:DW}. 

~~~~Ref.~\cite{Servant:2023mwt} has fit NG15 data to the GWB predictions from a post-inflationary axion model that naturally leads to global strings. They find bounds both for low and large axion mass ranges, which are respectively associated with signals from axionic global strings ($N_{\rm DW} = 1$) or domain walls ($N_{\rm DW} > 1$)~\cite{Gouttenoire:2019kij}, where $N_{\rm DW}$ is the number of domain walls. In the low-axion-mass region, they find the strongest constraint on the axion decay constant as $f_a < 2.8 \times 10^{15}$ GeV, for an axion mass $m_a \ll 10^{-17}$ eV. This is competitive with the $\Delta N_{\rm eff}$ bound. At the high masses $0.1 \mbox{ GeV}\lesssim m_a\lesssim 10^3$ TeV, a substantial region, corresponding to $m_a (f_a/N_{\rm DW})^2\gtrsim 2 \times 10^{11}$ GeV$^3$, can be excluded when DW decays in the temperature $T_*\propto \sqrt{V_{\rm bias}}\sim$ $1-300$ MeV range. Below we comment more on direct fits of the data to domain walls. 

\subsubsection{Domain walls}
\label{subsubsec:DW}

Domain walls (DWs)~\cite{Vilenkin:1984ib} may emerge in the breaking of a discrete symmetry during a PhT. While they reach scaling, as any other cosmic defect, their energy density ratio to the critical energy evolves as $\rho_{\rm DW}/\rho_c = \mathcal{O}(1)\sigma H / m_p^2 H^2 \propto t$~\cite{Hiramatsu:2013qaa, Kibble:1976sj}, with $\sigma$ the wall tension (= surface mass density). If they remain in scaling for too long, they would overclose the universe. Hence, a bias term in the Lagrangian is needed so they annihilate at a given temperature $T = T_{\rm ann}$~\cite{Gelmini:1988sf,Coulson:1995nv,Larsson:1996sp,Preskill:1991kd}, before they dominate the energy budget of the universe. DWs emit gravitational waves (GWs) until they annihilate, with peak frequency $f_p = f_H(T_{\rm ann})$ and high-frequency tail $\propto f^{-1}$. The GW spectrum at annihilation can be approximated by~\cite{Hiramatsu:2013qaa,Ferreira:2022zzo}
\begin{eqnarray}\label{eq:DW}
\Omega_{\rm DW}(f,T_{\rm ann}) = \frac{3 \epsilon \alpha_*^2}{8\pi} \left( \frac{1}{4} \left[ \frac{\Omega_{\rm CT}(f_p)}{\Omega_{\rm CT}(f)}\right]^{1/\delta} + \frac{3}{4} \left[\frac{f}{f_p}\right]^{1/\delta} \right)^{-\delta} ,
\end{eqnarray}
with $\epsilon = 0.7$, $\delta = 1$, and $\alpha_ \equiv \rho_{\rm DW}(T_{\rm ann})/\rho_{\rm rad}(T_{\rm ann})$ the DW-to-radiation energy density ratio.

If DWs annihilate into dark radiation, one obtains $\Delta N_{\rm eff} = 13.6, g_*(T_{\rm ann})^{-1/3} \alpha_*$~\cite{Ferreira:2022zzo}, which must respect BBN and CMB constraints. If DWs decay into SM particles, BBN and CMB impose $T_{\rm ann} \gtrsim  5,{\rm MeV}~$\cite{Hasegawa:2019jsa,deSalas:2015glj,Allahverdi:2020bys}. Ref.~\cite{Ellis:2023oxs} has fit the DW model, {\it c.f.}~Eq.~(\ref{eq:DW}), to NG15 data, obtaining as best fits $T_{\rm ann} = 0.85,{\rm GeV}$ and $\alpha_* = 0.11$. While this is consistent with annihilation into SM particles, it is in tension with $\Delta N_{\rm eff}$ bounds if DWs decay into dark radiation (though smaller $\alpha_*$ values within $1\sigma$ remain viable). Like first-order phase transitions, DW annihilation may produce PBHs~\cite{Ferrer:2018uiu,Gelmini:2023ngs}. If efficient, PBH formation could exclude regions favored by NG15~\cite{Gouttenoire:2023ftk}, though further analysis is needed.

\subsection{EUB: Other signals} 
\label{subsec:EUBothers}

Besides GWBs from inflation, phase transitions and cosmic defects, other backgrounds may have been generated in the early Universe. For instance, the thermal plasma formed after reheating, is guaranteed to produce a GWB. The spectrum of this background is determined by the particle content and the maximum temperature $T_{\rm max}$ reached by the plasma. The GW energy density spectrum scales as $\Omega_{\rm GW} \propto f^3$ in the IR, and reaches a maximum at a peak frequency~\cite{Ghiglieri:2015nfa, Ghiglieri:2020mhm, Ringwald:2020ist} $f_{\rm peak}^{\Omega_{\rm CGMB}} \sim 80\, {\rm GHz}$. While the amplitude of this GWB can be very large at such high-frequency peak, it is however extremely suppressed at smaller frequencies, due to the $\propto f^3$ scaling. It is therefore clear that under no circumstance this background can explain a signal at PTA frequencies.

~~~~Another possibility is the production of a dark-$U(1)$ gauge field coupled to an ALP particle via ${\alpha\over 4f_a}\phi F_{\mu\nu}\tilde{F}^{\mu\nu}$~\cite{Machado:2018nqk}. This is a mechanism similar to particle production during preheating, but in a different context. In particular, way after the end of inflation and reheating, the oscillations of this axion particle lead to a chiral instability of the gauge field, which sources a large GWB. This scenario is referred to as {\it audible} axions. The peak frequency and amplitude of the resulting GWB depends on the axion's mass $m_a$ and decay constant $f_a$~\cite{Machado:2018nqk,Guo:2023hyp}. The spectrum today can be fitted as~\cite{Machado:2018nqk}
\begin{eqnarray}
\Omega_{\rm GW}^{(0)}(f) = \frac{6.3 \, \mathcal{A}_*}{\left[\frac{f}{2f_{*}}\right]^{-3/2}+\exp\left[12.9\left(\frac{f}{2 f_{*}} - 1\right)\right]},
    \label{eq:audible_axion_shape}%\\
\end{eqnarray}
where $f_{*} \simeq 3.8 \cdot 10^{-9} \, {\rm Hz} \left({\alpha\theta\over 10}\right)^{\frac{2}{3}} \left({m_a\over 10^{-12} \, {\rm meV}}\right)^{\frac{1}{2}}$, and 
$\mathcal{A}_* \simeq 3.41 \cdot 10^{-6} \left({f_a\over m_p}\right)^{4} \theta^{8/3}\left({10\over\alpha}\right)^{\frac{4}{3}}$, with $\alpha \gtrsim \mathcal{O}(10)$ and $\theta \sim \mathcal{O}(1)$ the initial misalignment angle~\cite{Machado:2018nqk}. The spectrum then has an IR tail $\propto f^{3/2}$, which extends to high frequencies till it dies off exponentially at $f > f_*$. 

~~~~\cite{Figueroa:2023zhu} has fitted $\lbrace f_*, \mathcal{A}_*\rbrace$ to NG15+EPTA data using Eq.~(\ref{eq:audible_axion_shape}), obtaining lower bounds on the axion mass and decay constant at 68\% CL, as $m_a \gtrsim 8.0 \cdot 10^{-11}$ meV and $f_a \gtrsim 1.3 \cdot 10^{18}$ GeV. Imposing a sub-Planckian $f_a \leq m_p$ motivated by string theory~\cite{Banks:2003sx}, turns the previous fit into a very tight constraint as $m_a = 0.32_{-0.24}^{+1.48}$ PeV. This preferred parameter region is however in tension with axion over-production and $\Delta N_{\rm eff}$ constraints~\cite{Geller:2023shn}. Ref.~\cite{Ellis:2023oxs} has multiplied the profile in Eq.~(\ref{eq:audible_axion_shape}) by a spectral function  that accounts for the causality tail of the spectrum at superhorizon frequencies~\cite{Caprini:2009fx}. Fitting to the NG15 data, they find best-fit values as $m_a= 3.1\times10^{-11} {\rm eV}$ and $f_a  =0.87\, M_{\rm P}$. They find a larger range for the axion mass below super-Planckian $f_a$ values, though part of such region is ruled out by the super-radiance constraints\cite{Zhang:2021mks,Baryakhtar:2020gao,Mehta:2020kwu}. The discrepancy between~\cite{Ellis:2023oxs} and~\cite{Figueroa:2023zhu} is likely due the use of NG15 vs NG15+EPTA data, and to the use of the causality tail.

\subsection{Deterministic Signals}
\label{subsec:EUBdeterministic}

In addition to the GWB signals discussed in Secs.~\ref{subsec:EUBinflation},~\ref{subsec:EUBphaseTransitions},~\ref{subsec:EUBcosmicDefects},~\ref{subsec:EUBothers}, there are also BSM scenarios that can lead to deterministic signals in PTAs. A paradigmatic example is ultralight dark matter (ULDM) -- also known as {\it fuzzy} DM --, which must be a bosonic species~\cite{Tremaine:1979we} with a tiny mass, bounded from below by the CMB as $m_\phi > 10^{-24} \, \text{eV}$~\cite{Hlozek:2017zzf}. ULDM models suppress generically structures on small scales, potentially solving the so called {\it small-scale structure} problem of the standard $\Lambda$CDM paradigm~\cite{Bullock:2017xww}. PTAs can probe their tiny mass because the frequency of the PTA signal they cause is proportional to the ULDM mass, $f_* \sim m_\phi/2 \pi$, so that the sensitivity window of PTA observations falls within $10^{-23} \, \text{eV} \lesssim m_\phi \lesssim 10^{-20} \, \text{eV}$. PTA searches for ULDM in such mass range, see {\it e.g.}~\cite{Porayko:2018sfa}, can be actually viewed as complementary to other astrophysical searches, such as measurements of the Lyman-$\alpha$ forest, galactic subhalo mass functions, stellar kinematics, or the lack of observation of superradiance at SMBHs. 

~~~~ULDM models can create various effects in PTA data. In a seminal work, Ref.~\cite{Khmelnitsky:2013lxt} showed that the travel time of a pulsar radio beam is affected by the {\it metric fluctuations} induced by ULDM particles, making PTAs excellent probes for such species. This effect is due to the oscillations of the ULDM field, which induce fluctuations in the local stress-energy tensor $T_{\mu \nu}$, and hence generate fluctuations in the metric perturbations through the Einstein's equations. Such metric perturbations, which we note they are not GWs but time-dependent gravitational potentials instead, affect nonetheless the photon geodesic on its path from the pulsar, generating a timing residual. Other effects from ULDM models, can also possibly cause timing residuals, like {\it e.g.}~{\it doppler-U(1) forces} for vector ULDM, and {\it pulsar spin fluctuations} or {reference clock shifts}, which are specific of scalar ULDM, see Ref.~\cite{NANOGrav:2023hvm} for discussion and references therein.

~~~~Furthermore, various theories of DM leave a unique fingerprints on primordial perturbations on smaller scales, giving rise to different predictions for the amount of subgalactic DM substructures. Consequently, measuring local DM substructures can be used to help determining the correct DM scenario. For example, a very simple example of such small-scale DM substructures is that of PBHs, which can be formed in inflationary set-up's with large density fluctuations on small scales, {\it c.f.}~Sec.~\ref{subsubsec:2ndOIGW}. Galactic PBH populations can be investigated by analyzing the Doppler and Shapiro signals they can leave in PTA data. In particular, the Doppler signal is due to the apparent shift in the pulsar spin frequency, generated by the acceleration induced by the gravitational pull of a passing PBH. The Shapiro signal, on the other hand,  refers to shifts in the TOAs caused by metric perturbations along the photon geodesic, produced by PBHs passing along the observer's line of sight.

~~~~Ref.~\cite{NANOGrav:2023hvm} has analyzed NG15 data searching for deterministic signals generated by models of ULDM and DM substructures. They do not find significant evidence for either of these signals, though constraints on the parameters space of both models are set. For a wide range of ULDM models, the constraints compete with (or outperform) laboratory constraints in the $10^{-23}\,{\rm eV}\lesssim m_\phi\lesssim10^{-20}\,{\rm eV}$ mass range. The signal from DM substructures is however harder to detect, so Ref.~\cite{NANOGrav:2023hvm} is only able to set very weak constraints on the local abundance of such structures. Ref.~\cite{EPTA:2023xxk} has also searched for ULDM signatures in the EPTA data. They obtain a prominent peak in the posterior  of the ULDM particle mass around ${\rm log}_{10}(m_\phi/{\rm eV})\approx -23$, which corresponds to a field oscillation frequency of ${\rm log}_{10}(f_*/{\rm Hz})\approx -8.3$, consistent with the CGW candidate they have also searched for. Performing however a joint ULDM+GWB search, Ref.~\cite{EPTA:2023xxk} finds that the peak in the ULDM mass posterior distribution vanishes, as the EPTA data strongly prefer the presence of an HD correlation in the common power. They therefore conclude that an ULDM origin of the PTA signal is disfavoured, placing a direct constraint on the abundance of ULDM. In particular they find that within our Galaxy, only about 80\% of the DM density in the solar neighbourhood could be attributed to ULDM with $-24<{\rm log}_{10}(m_\phi/{\rm eV})<-23.7$. For more stringent constraints on ULDM using EPTA data, see~\cite{EuropeanPulsarTimingArray:2023egv}.

\section{Discriminating the origin of the signal} 
\label{sec:discriminating}

Current PTA measurements are broadly consistent with a cosmic population of MBHBs as well as with a variety of cosmological signals, as detailed in Sec.~\ref{sec:astroGWB} and Sec.~\ref{sec:cosmo_GWsignals}. Countless papers have been devoted to the interpretation of those measurements, sometimes  disfavoring the MBHB interpretation because of the $\gamma=13/3$ prediction. Although caveats in these analyses are generally well spelled out,  we nonetheless warn against overinterpretation of current PTA results by noting the following points:
 \begin{enumerate}[(i)]
    \item evidence in the PTA data is only at 2-to-4$\sigma$ level and uncertainty in the spectral parameters is considerable. For example, \cite{Valtolina:2023axp} showed that the injection of a smooth $\gamma=13/3$ PL in simulated data might return offset posteriors just because of the particular realization of the MSPs stochastic noise;
    \item the signal produced by MBHBs is not smooth in general. Even in absence of eccentricity or strong environmental coupling, bright, yet unresolved, high frequency CGWs might flatten the measured spectrum to $\gamma<13/3$  \cite{Valtolina:2023axp};
    \item the algorithms employed in the analysis might be affected by the specific prior used \cite{Goncharov:2024htb} or by leakage \cite{Crisostomi:2025vue} and $\gamma$ estimates might suffer from some bias and its actual value might be closer to $13/3$. 
%\end{itemize}
\end{enumerate}

~~~~Although an astrophysical origin of the observed signal is a natural consequence of hierarchical galaxy formation, there is currently no observational evidence to substantiate this hypothesis. Going forward, besides solidifying the evidence of the detection, the primary goal of PTA experiments is to pin down the origin of the observed signal, and currently both MBHBs and a variety of cosmological signals can potentially explain the data. 

~~~~We note that as the theoretical prediction of the MBHB signal is quite uncertain (given its dependency on modeling assumptions such as eccentricity, environmental feedback, and mass distribution), fitting simultaneously MBHB and cosmological signals would make the inference on the cosmological background parameters strongly dependent on the astrophysical modeling. It is for this reason that in this review we considered the analyses on  cosmological signals in Sec.~\ref{sec:cosmo_GWsignals} isolated from the MBHB analysis in Sec.~\ref{sec:astroGWB}. We discussed in this way the constraints on either astrophysical or cosmological signals separately, always assuming that only one of them is present in the data. %explain the PTA data at once.} %Simultaneous fit of MBHB and cosmological signals to PTA data has been presented nonetheless in Refs.~\cite{....}

~~~~While cosmological GWBs are anticipated to be isotropic, Gaussian and stationary, characterized by a smooth spectrum, these properties are expected to break down for an astrophysical signal,  {\it c.f.} Sec.~\ref{sec:discrete}. Here we enumerate methods to differentiate between cosmological and astrophysical GWBs and the current status of their testing.

-- {\bf Spectral shape.} As already discussed, the spectral shape alone is likely insufficient to discriminate among astrophysical and cosmological processes, because of MBHB eccentricity, environmental coupling, and variance due to sparse sources. What can help instead is the smoothness of the spectrum. While cosmic GWBs are expected to have smooth, continuous spectra, MBHBs produce spiky features, especially at $f>10^{-8}$Hz due to loud sparse sources. Deviations from a smooth spectrum have been searched in the NANOGrav 15 years dataset \cite{Agazie:2024jbf} yielding inconclusive excursions at the $\approx 2\sigma$ level.

-- {\bf Anisotropies and hotspots.} The SMBH signal is expected to be dominated by few loud sources ({\it c.f.} Fig.~\ref{fig:hc_MBHBcirc}) which implies a Poisson rather than Gaussian power distribution across the sky. This is characterized by a flat, non-null power at all multipoles of the spherical harmonics decomposition \cite{Mingarelli:2013dsa,2013PhRvD..88h4001T}. Moreover, hot spots corresponding to the location of the brightest sources might arise \cite{Taylor:2020zpk}. When normalized to the monopole, the expected power in the higher multiples varies by orders of magnitude in the range $10^{-4}\lesssim C_l/C_0\lesssim 0.3$ for most astrophysical models \cite{Sato-Polito:2023spo,Sah:2024oyg,2024arXiv241119692R,2025ApJ...989..157Y}, increasing for higher frequencies \cite{2024ApJ...971L..10L}. Searches for anisotropies in the NANOGrav 15yr datastes constrained $C_l/C_0\lesssim 0.2$ in the five lowest frequency bins \cite{NANOGrav:2023tcn}, which is still consistent with the majority of the models. 

-- {\bf Cross correlation with the large scale structure (LSS).} MBHBs form in the center of massive galaxies, mostly found in galaxy clusters. As such, the GWB power in the sky must correlate to some extent with the LSS. Although the GWB anisotropy is dominated by Poissonian noise due to few massive sources \cite{Allen:2024mtn}, correlation with the LSS might be observable at high multipoles by an SKA-like array \cite{Semenzato:2024mtn,2025MNRAS.541.2884C,2025ApJ...993..118S}.

-- {\bf Non-Gaussianity.} Another consequence of the discreteness of the  MBHBs population is the non Gaussian distribution of the GWB power. A study of the detectability of non-Gaussianity in PTA data has been performed by \cite{Falxa:2025qxr}. They found that a Gaussian mixture model is effective at capturing non Gaussianity, both when a loud source is present in the data and when the GWB is produced by a large number of unresolved MBHBs. Searches for non-Gaussianity have not been performed on real data yet. 

-- {\bf Non stationarity.} Another feature likely arising from a cosmic population of sparse sources is non stationarity. This is especially true if the population is dominated by a few eccentric binaries, emitting loud GW bursts at periastron over timescales of months to years ({\it c.f.} Sec.~\ref{sec:binary_dynamics}). \cite{Falxa:2024ski} showed that dim eccentric MBHBs with S/N$\approx 1$ might produce detectable non stationarity in simulated PTA data. Moreover, eccentric MBHBs emit GWs across a broad frequency spectrum. If the signal is dominated by few MBHBs, the reconstucted GW power emitted across the sky at different frequencies is expected to be correlated \cite{Sah:2025dmv,Moreschi:2025qtm}. \cite{Moreschi:2025qtm} shown that an SKA-like array might detect such GW power cross-correlation at S/N$>3$.

-- {\bf Resolvable CGWs.} Finally, as more MSPs are added to the PTAs,  their RMS residual is reduced, and the observation time increases, individual CGWs will eventually emerge from the data \cite{2015MNRAS.451.2417R,2016ApJ...819L...6T,2024ApJ...965..164G}. Detection of deterministic GWBs compatible with an MBHB origin will prove beyond doubt that there is an astrophysical component to the nano-Hz detected GW signal, while identification of EM counterparts will enable multimessenger studies of MBHBs and their galactic hosts, shedding light on the dynamics of the most massive dense nuclei and on the hierarchical build-up of the most massive galaxies in the Universe. Although searches for CGWs in current PTA data so far proved inconclusive, recent analysis targeting EM selected MBHB candidates yielded marginal positive evidence for two systems \cite{Agarwal:2025cag}.

%, in particular discussing spectral techniques, the use of anisotropies and hotspots, cross correlations with large scale structure, or the use of (non-) stationarity and (non-) Gaussianity. 

% * Current status

% * Spectrum as discriminat \as{discuss why the flat spectrum is not an issue (leakage papers by Crisostomi, Boris and Hippolyte). CAUTION AGAINST OVERINTERPRETATION!!! Also discuss the high amplitude (Sato polito vs MA, IZQUIERDO, dependence on the high mass end of the mass function. Spectral discontinuity as discriminant (null detection thus far)}
  
% * Anisotropies \& hotspots
 
% * Cross-correlations with LSS
 
% * Stationarity <--> Non-gaussianity

% * \as{Eventually: Detection of individual sources and EM counterparts}

\section{Conclusions and outlook}
\label{sec:conclusions}
~~~~The nano-Hz GW window offers a new invaluable opportunity to explore the Universe, from its first instants during the inflationary phase, %, right after the Big Bang, 
to the most massive black holes populating our local cosmological neighborhood. On the one hand, the only astrophysical plausible source of the observed PTA signal is a cosmic population of MBHBs, naturally emerging in the framework of hierarchical clustering of cosmic structures. In this case, the signal amplitude, shape and statistical properties offer important information about the formation of the most massive galaxies and the merger-driven assembly of the most massive black holes in the Universe, adding an important missing piece to the puzzle of cosmic structure formation. 
On the other hand, the possibility of a signal originated in the early Universe, opens up a new door to address fundamental physics questions, such as the state of the early Universe, the origin of the cosmological perturbations, the nature of dark matter, or whether exotic objects like primordial black holes or cosmic defects may exist at all. While the current signal reconstruction does not carry enough information to characterize the GWB properties beyond an approximate estimate of the amplitude and slope of its spectrum, the parameter space of cosmological GWBs (in)compatible with PTA
data, can be clearly identified. This identification, which remarkably is independent
of the origin of the  signal, marks the dawn of {\it early universe GW-cosmology} as a research field. 

~~~~PTA collaborations are relentlessly observing night after night, amassing data of ever-increasing quality. Eventually, the IPTA, bringing together EPTA, InPTA, NANOGrav, PPTA, MPTA and CPTA, and taking advantage of the superior capabilities of the new generation of radio telescopes such as MeerKAT, FAST and eventually the SKA, will collect enough high-quality data to reveal the still hidden properties of the signal, including anisotropy, non-stationarity, non-Gaussianity, hot spots and CGWs. This will allow us to pin down its origin, heralding a new era in the exploration of the gravitational Universe.
%opening new avenues in understanding the Universe.

%Disclosure
\section*{DISCLOSURE STATEMENT}
This review covers nano-Hz GW signal generation mechanisms across cosmic time, from the early Universe to today. As such, it requires a range of expertise to which both authors contribute. The authors' ordering follows the astrophysics community standards, which is not necessarily in alphabetical order. Alberto Sesana covered the description of astrophysical GW sources (MBHBs, see Sec.~\ref{sec:astroGWB}), whereas Daniel G.~Figueroa covered the mechanisms generating GWBs in the early Universe (EUBs, Sec.~\ref{sec:cosmo_GWsignals}). The authors are not aware of any affiliations, memberships or funding that might affect the objectivity of this review. 

% Acknowledgements
\section*{ACKNOWLEDGMENTS}
%Acknowledgements, general annotations, funding.
AS thanks G. Shaifullah, I. Ferranti and A. Franchini for Fig.~\ref{fig:GWlandscape},~\ref{fig:hc_MBHBcirc} and \ref{fig:epta_interpretation}. The work of AS is supported by the European Union’s H2020 ERC Advanced Grant ``PINGU'' (Grant Agreement: 101142079).
DGF thanks Peera Simakachorn for generating customized versions of Fig.'s~\ref{fig:inf_cmb_cons} and~\ref{fig:SIGW_constraint}. The work of DGF was supported by the grants EUR2022-134028, PROMETEO/2021/083,  PID2023-148162NB-C22, PRTR-C17.I01 and ASFAE/2022/020.

%To download the appropriate bibliography style file, please see \url{https://www.annualreviews.org/page/authors/general-information}. \\

%\noindent
%Please see the Style Guide document for instructions on preparing your Literature Cited.

%The citations should be listed in alphabetical order, with no titles. For example:

%\begin{multicols}{2}
\footnotesize
  \bibliography{autoArxiv,manualArxiv}

@ARTICLE{1975GReGr...6..439E,
       author = {{Estabrook}, F.~B. and {Wahlquist}, H.~D.},
        title = "{Response of Doppler spacecraft tracking to gravitational radiation.}",
      journal = {General Relativity and Gravitation},
         year = 1975,
        month = oct,
       volume = {6},
       number = {5},
        pages = {439-447},
          doi = {10.1007/BF00762449},
       adsurl = {https://ui.adsabs.harvard.edu/abs/1975GReGr...6..439E}
}

@ARTICLE{1990ApJ...361..300F,
       author = {{Foster}, R.~S. and {Backer}, D.~C.},
        title = "{Constructing a Pulsar Timing Array}",
      journal = {Astrophys. J.},
         year = 1990,
        month = sep,
       volume = {361},
        pages = {300},
          doi = {10.1086/169195},
       adsurl = {https://ui.adsabs.harvard.edu/abs/1990ApJ...361..300F}
}

@ARTICLE{2024ApJ...965..164G,
       author = {{Gardiner}, Emiko C. and {Kelley}, Luke Zoltan and {Lemke}, Anna-Malin and {Mitridate}, Andrea},
        title = "{Beyond the Background: Gravitational-wave Anisotropy and Continuous Waves from Supermassive Black Hole Binaries}",
      journal = {Astrophys. J.},
         year = 2024,
        month = apr,
       volume = {965},
       number = {2},
          eid = {164},
        pages = {164},
          doi = {10.3847/1538-4357/ad2be8},
archivePrefix = {arXiv},
       eprint = {2309.07227},
 primaryClass = {astro-ph.HE},
       adsurl = {https://ui.adsabs.harvard.edu/abs/2024ApJ...965..164G}
}

@ARTICLE{2016ApJ...819L...6T,
       author = {{Taylor}, S.~R. and {Vallisneri}, M. and {Ellis}, J.~A. and {Mingarelli}, C.~M.~F. and {Lazio}, T.~J.~W. and {van Haasteren}, R.},
        title = "{Are We There Yet? Time to Detection of Nanohertz Gravitational Waves Based on Pulsar-timing Array Limits}",
      journal = {Astrophys. J. Letters},
         year = 2016,
        month = mar,
       volume = {819},
       number = {1},
          eid = {L6},
        pages = {L6},
          doi = {10.3847/2041-8205/819/1/L6},
archivePrefix = {arXiv},
       eprint = {1511.05564},
 primaryClass = {astro-ph.IM},
       adsurl = {https://ui.adsabs.harvard.edu/abs/2016ApJ...819L...6T}
}

@ARTICLE{2025ApJ...993..118S,
       author = {{Sah}, Mohit Raj and {Maurya}, Akash and {Mukherjee}, Suvodip and {Kumar}, Prayush and {Saeedzadeh}, Vida and {Babul}, Arif and {Mishra}, Chandra Kant and {Paul}, Kaushik and {Quinn}, Thomas R. and {Tremmel}, Michael},
        title = "{An Accurate Modeling of Nano-hertz Gravitational Wave Signal from Eccentric Supermassive Binary Black Holes: An Essential Step Toward a Robust Discovery}",
      journal = {Astrophys. J.},
         year = 2025,
        month = nov,
       volume = {993},
       number = {1},
          eid = {118},
        pages = {118},
          doi = {10.3847/1538-4357/ae0337},
archivePrefix = {arXiv},
       eprint = {2505.22745},
 primaryClass = {astro-ph.CO},
       adsurl = {https://ui.adsabs.harvard.edu/abs/2025ApJ...993..118S}
}

@ARTICLE{2025MNRAS.541.2884C,
       author = {{Cusin}, Giulia and {Pitrou}, Cyril and {Pijnenburg}, Martin and {Sesana}, Alberto},
        title = "{Measuring anisotropies in the PTA band with cross-correlations}",
      journal = {Mon. Not. Roy. Ast. Soc.},
         year = 2025,
        month = aug,
       volume = {541},
       number = {4},
        pages = {2884-2896},
          doi = {10.1093/mnras/staf1074},
archivePrefix = {arXiv},
       eprint = {2502.17401},
 primaryClass = {gr-qc},
       adsurl = {https://ui.adsabs.harvard.edu/abs/2025MNRAS.541.2884C}
}

@ARTICLE{2024ApJ...971L..10L,
       author = {{Lamb}, William G. and {Taylor}, Stephen R.},
        title = "{Spectral Variance in a Stochastic Gravitational-wave Background from a Binary Population}",
      journal = {Astrophys. J. Letters},
         year = 2024,
        month = aug,
       volume = {971},
       number = {1},
          eid = {L10},
        pages = {L10},
          doi = {10.3847/2041-8213/ad654a},
archivePrefix = {arXiv},
       eprint = {2407.06270},
 primaryClass = {gr-qc},
       adsurl = {https://ui.adsabs.harvard.edu/abs/2024ApJ...971L..10L}
}

@ARTICLE{2025ApJ...989..157Y,
       author = {{Yang}, Qing and {Guo}, Xiao and {Cao}, Zhoujian and {Shao}, Xiaoyun and {Yuan}, Xi},
        title = "{Anisotropy of Nanohertz Gravitational-wave Background and Source Clustering from Supermassive Binary Black Holes Based on Cosmological Simulation}",
      journal = {Astrophys. J.},
         year = 2025,
        month = aug,
       volume = {989},
       number = {2},
          eid = {157},
        pages = {157},
          doi = {10.3847/1538-4357/aded02},
archivePrefix = {arXiv},
       eprint = {2408.05043},
 primaryClass = {astro-ph.CO},
       adsurl = {https://ui.adsabs.harvard.edu/abs/2025ApJ...989..157Y}
}

@ARTICLE{2024arXiv241119692R,
       author = {{Raidal}, Juhan and {Urrutia}, Juan and {Vaskonen}, Ville and {Veerm{\"a}e}, Hardi},
        title = "{Statistics of the supermassive black hole gravitational wave background anisotropy}",
      journal = {arXiv e-prints},
         year = 2024,
        month = nov,
          eid = {arXiv:2411.19692},
        pages = {arXiv:2411.19692},
          doi = {10.48550/arXiv.2411.19692},
archivePrefix = {arXiv},
       eprint = {2411.19692},
 primaryClass = {astro-ph.CO},
       adsurl = {https://ui.adsabs.harvard.edu/abs/2024arXiv241119692R}
}

@ARTICLE{2013PhRvD..88h4001T,
       author = {{Taylor}, Stephen R. and {Gair}, Jonathan R.},
        title = "{Searching for anisotropic gravitational-wave backgrounds using pulsar timing arrays}",
      journal = {Phys. Rev. D.},
         year = 2013,
        month = oct,
       volume = {88},
       number = {8},
          eid = {084001},
        pages = {084001},
          doi = {10.1103/PhysRevD.88.084001},
archivePrefix = {arXiv},
       eprint = {1306.5395},
 primaryClass = {gr-qc},
       adsurl = {https://ui.adsabs.harvard.edu/abs/2013PhRvD..88h4001T}
}

@ARTICLE{1964PhRvL..13..789S,
       author = {{Shapiro}, Irwin I.},
        title = "{Fourth Test of General Relativity}",
      journal = {Phys. Rev. Letters},
         year = 1964,
        month = dec,
       volume = {13},
       number = {26},
        pages = {789-791},
          doi = {10.1103/PhysRevLett.13.789},
       adsurl = {https://ui.adsabs.harvard.edu/abs/1964PhRvL..13..789S}
}

@ARTICLE{2015MNRAS.446.1657W,
       author = {{Wang}, J.~B. and {Hobbs}, G. and {Coles}, W. and {Shannon}, R.~M. and {Zhu}, X.~J. and {Madison}, D.~R. and {Kerr}, M. and {Ravi}, V. and {Keith}, M.~J. and {Manchester}, R.~N. and {Levin}, Y. and {Bailes}, M. and {Bhat}, N.~D.~R. and {Burke-Spolaor}, S. and {Dai}, S. and {Os{\l}owski}, S. and {van Straten}, W. and {Toomey}, L. and {Wang}, N. and {Wen}, L.},
        title = "{Searching for gravitational wave memory bursts with the Parkes Pulsar Timing Array}",
      journal = {Mon. Not. Roy. Ast. Soc.},
         year = 2015,
        month = jan,
       volume = {446},
       number = {2},
        pages = {1657-1671},
          doi = {10.1093/mnras/stu2137},
archivePrefix = {arXiv},
       eprint = {1410.3323},
 primaryClass = {astro-ph.GA},
       adsurl = {https://ui.adsabs.harvard.edu/abs/2015MNRAS.446.1657W}
}

@ARTICLE{2010MNRAS.401.2372V,
       author = {{van Haasteren}, Rutger and {Levin}, Yuri},
        title = "{Gravitational-wave memory and pulsar timing arrays}",
      journal = {Mon. Not. Roy. Ast. Soc.},
         year = 2010,
        month = feb,
       volume = {401},
       number = {4},
        pages = {2372-2378},
          doi = {10.1111/j.1365-2966.2009.15885.x},
archivePrefix = {arXiv},
       eprint = {0909.0954},
 primaryClass = {astro-ph.IM},
       adsurl = {https://ui.adsabs.harvard.edu/abs/2010MNRAS.401.2372V}
}

@ARTICLE{2009ApJ...696L.159F,
       author = {{Favata}, Marc},
        title = "{Nonlinear Gravitational-Wave Memory from Binary Black Hole Mergers}",
      journal = {Astrophys. J. Letters},
         year = 2009,
        month = may,
       volume = {696},
       number = {2},
        pages = {L159-L162},
          doi = {10.1088/0004-637X/696/2/L159},
archivePrefix = {arXiv},
       eprint = {0902.3660},
 primaryClass = {astro-ph.SR},
       adsurl = {https://ui.adsabs.harvard.edu/abs/2009ApJ...696L.159F}
}

@ARTICLE{2025arXiv251014613Q,
       author = {{Quelquejay Leclere}, Hippolyte and {Li}, Kunyang and {Volonteri}, Marta and {Babak}, Stanislav and {Beckmann}, Ricarda S. and {Dubois}, Yohan and {Laigle}, Clotilde and {Webb}, Natalie A.},
        title = "{The multimessenger view of Pulsar Timing Array black holes with the Horizon-AGN simulation}",
      journal = {arXiv e-prints},
         year = 2025,
        month = oct,
          eid = {arXiv:2510.14613},
        pages = {arXiv:2510.14613},
          doi = {10.48550/arXiv.2510.14613},
archivePrefix = {arXiv},
       eprint = {2510.14613},
 primaryClass = {astro-ph.GA},
       adsurl = {https://ui.adsabs.harvard.edu/abs/2025arXiv251014613Q}
}

@ARTICLE{2025arXiv250401074T,
       author = {{Truant}, Riccardo J. and {Izquierdo-Villalba}, David and {Sesana}, Alberto and {Mohiuddin Shaifullah}, Golam and {Bonetti}, Matteo and {Spinoso}, Daniele and {Bonoli}, Silvia},
        title = "{Lighting up the nano-hertz gravitational wave sky: opportunities and challenges of multimessenger astronomy with PTA experiments}",
      journal = {arXiv e-prints},
         year = 2025,
        month = apr,
          eid = {arXiv:2504.01074},
        pages = {arXiv:2504.01074},
          doi = {10.48550/arXiv.2504.01074},
archivePrefix = {arXiv},
       eprint = {2504.01074},
 primaryClass = {astro-ph.GA},
       adsurl = {https://ui.adsabs.harvard.edu/abs/2025arXiv250401074T}
}

@ARTICLE{2025A&A...694A.194F,
       author = {{Ferranti}, Irene and {Shaifullah}, Golam and {Chalumeau}, Aurelien and {Sesana}, Alberto},
        title = "{Separating deterministic and stochastic gravitational wave signals in realistic pulsar timing array datasets}",
      journal = {Astronomy and Astrophysics},
         year = 2025,
        month = feb,
       volume = {694},
          eid = {A194},
        pages = {A194},
          doi = {10.1051/0004-6361/202451809},
archivePrefix = {arXiv},
       eprint = {2407.21105},
 primaryClass = {astro-ph.HE},
       adsurl = {https://ui.adsabs.harvard.edu/abs/2025A&A...694A.194F}
}

@ARTICLE{2022PhRvD.105l2003B,
       author = {{B{\'e}csy}, Bence and {Cornish}, Neil J. and {Digman}, Matthew C.},
        title = "{Fast Bayesian analysis of individual binaries in pulsar timing array data}",
      journal = {Phys. Rev. D},
         year = 2022,
        month = jun,
       volume = {105},
       number = {12},
          eid = {122003},
        pages = {122003},
          doi = {10.1103/PhysRevD.105.122003},
archivePrefix = {arXiv},
       eprint = {2204.07160},
 primaryClass = {gr-qc},
       adsurl = {https://ui.adsabs.harvard.edu/abs/2022PhRvD.105l2003B}
}

@ARTICLE{2014MNRAS.444.3709Z,
       author = {{Zhu}, X.-J. and {Hobbs}, G. and {Wen}, L. and {Coles}, W.~A. and {Wang}, J.-B. and {Shannon}, R.~M. and {Manchester}, R.~N. and {Bailes}, M. and {Bhat}, N.~D.~R. and {Burke-Spolaor}, S. and {Dai}, S. and {Keith}, M.~J. and {Kerr}, M. and {Levin}, Y. and {Madison}, D.~R. and {Os{\l}owski}, S. and {Ravi}, V. and {Toomey}, L. and {van Straten}, W.},
        title = "{An all-sky search for continuous gravitational waves in the Parkes Pulsar Timing Array data set}",
      journal = {Mon. Not. Roy. Ast. Soc.},
         year = 2014,
        month = nov,
       volume = {444},
       number = {4},
        pages = {3709-3720},
          doi = {10.1093/mnras/stu1717},
archivePrefix = {arXiv},
       eprint = {1408.5129},
 primaryClass = {astro-ph.GA},
       adsurl = {https://ui.adsabs.harvard.edu/abs/2014MNRAS.444.3709Z}
}

@ARTICLE{2012ApJ...756..175E,
       author = {{Ellis}, J.~A. and {Siemens}, X. and {Creighton}, J.~D.~E.},
        title = "{Optimal Strategies for Continuous Gravitational Wave Detection in Pulsar Timing Arrays}",
      journal = {Astrophys. J.},
         year = 2012,
        month = sep,
       volume = {756},
       number = {2},
          eid = {175},
        pages = {175},
          doi = {10.1088/0004-637X/756/2/175},
archivePrefix = {arXiv},
       eprint = {1204.4218},
 primaryClass = {astro-ph.IM},
       adsurl = {https://ui.adsabs.harvard.edu/abs/2012ApJ...756..175E}
}

@ARTICLE{2011MNRAS.414.3251L,
       author = {{Lee}, K.~J. and {Wex}, N. and {Kramer}, M. and {Stappers}, B.~W. and {Bassa}, C.~G. and {Janssen}, G.~H. and {Karuppusamy}, R. and {Smits}, R.},
        title = "{Gravitational wave astronomy of single sources with a pulsar timing array}",
      journal = {Mon. Not. Roy. Ast. Soc.},
         year = 2011,
        month = jul,
       volume = {414},
       number = {4},
        pages = {3251-3264},
          doi = {10.1111/j.1365-2966.2011.18622.x},
archivePrefix = {arXiv},
       eprint = {1103.0115},
 primaryClass = {astro-ph.HE},
       adsurl = {https://ui.adsabs.harvard.edu/abs/2011MNRAS.414.3251L}
}

@ARTICLE{2012PhRvD..85d4034B,
       author = {{Babak}, Stanislav and {Sesana}, Alberto},
        title = "{Resolving multiple supermassive black hole binaries with pulsar timing arrays}",
      journal = {Phys. Rev. D},
         year = 2012,
        month = feb,
       volume = {85},
       number = {4},
          eid = {044034},
        pages = {044034},
          doi = {10.1103/PhysRevD.85.044034},
archivePrefix = {arXiv},
       eprint = {1112.1075},
 primaryClass = {astro-ph.CO},
       adsurl = {https://ui.adsabs.harvard.edu/abs/2012PhRvD..85d4034B}
}

@article{EPTA:2021crs,
    author = "Chen, S. and others",
    collaboration = "EPTA",
    title = "{Common-red-signal analysis with 24-yr high-precision timing of the European Pulsar Timing Array: inferences in the stochastic gravitational-wave background search}",
    eprint = "2110.13184",
    archivePrefix = "arXiv",
    primaryClass = "astro-ph.HE",
    doi = "10.1093/mnras/stab2833",
    journal = "Mon. Not. Roy. Astron. Soc.",
    volume = "508",
    number = "4",
    pages = "4970--4993",
    year = "2021"
}

@article{Antoniadis:2022pcn,
    author = "Antoniadis, J. and others",
    title = "{The International Pulsar Timing Array second data release: Search for an isotropic gravitational wave background}",
    eprint = "2201.03980",
    archivePrefix = "arXiv",
    primaryClass = "astro-ph.HE",
    doi = "10.1093/mnras/stab3418",
    journal = "Mon. Not. Roy. Astron. Soc.",
    volume = "510",
    number = "4",
    pages = "4873--4887",
    year = "2022"
}

@article{Kaspi:1994hp,
    author = "Kaspi, V. M. and Taylor, J. H. and Ryba, M. F.",
    title = "{High - precision timing of millisecond pulsars. 3: Long - term monitoring of PSRs B1855+09 and B1937+21}",
    doi = "10.1086/174280",
    journal = "Astrophys. J.",
    volume = "428",
    pages = "713",
    year = "1994"
}

@article{McHugh:1996hd,
    author = "McHugh, M. P. and Zalamansky, G. and Vernotte, F. and Lantz, E.",
    title = "{Pulsar timing and the upper limits on a gravitational wave background: A Bayesian approach}",
    reportNumber = "PRINT-96-250 (BESANCON)",
    doi = "10.1103/PhysRevD.54.5993",
    journal = "Phys. Rev. D",
    volume = "54",
    pages = "5993--6000",
    year = "1996"
}

@inproceedings{Lommen:2002je,
    author = "Lommen, A. N.",
    title = "{New limits on gravitational radiation using pulsars}",
    booktitle = "{270th WE-Heraeus Seminar on Neutron Stars, Pulsars and Supernova Remnants}",
    eprint = "astro-ph/0208572",
    archivePrefix = "arXiv",
    reportNumber = "MPE-REPORT-278",
    pages = "114--125",
    month = "8",
    year = "2002"
}

@article{Jenet:2006sv,
    author = "Jenet, F. A. and Hobbs, G. B. and van Straten, W. and Manchester, R. N. and Bailes, M. and Verbiest, J. P. W. and Edwards, R. T. and Hotan, A. W. and Sarkissian, J. M. and Ord, S. M.",
    title = "{Upper bounds on the low-frequency stochastic gravitational wave background from pulsar timing observations: Current limits and future prospects}",
    eprint = "astro-ph/0609013",
    archivePrefix = "arXiv",
    doi = "10.1086/508702",
    journal = "Astrophys. J.",
    volume = "653",
    pages = "1571--1576",
    year = "2006"
}

@article{EPTA:2011kjn,
    author = "van Haasteren, R. and others",
    collaboration = "EPTA",
    title = "{Placing limits on the stochastic gravitational-wave background using European Pulsar Timing Array data}",
    eprint = "1103.0576",
    archivePrefix = "arXiv",
    primaryClass = "astro-ph.CO",
    doi = "10.1111/j.1365-2966.2011.18613.x",
    journal = "Mon. Not. Roy. Astron. Soc.",
    volume = "414",
    number = "4",
    pages = "3117--3128",
    year = "2011",
    note = "[Erratum: Mon.Not.Roy.Astron.Soc. 425, 1597 (2012)]"
}

@article{Demorest:2012bv,
    author = "Demorest, P. B. and others",
    title = "{Limits on the Stochastic Gravitational Wave Background from the North American Nanohertz Observatory for Gravitational Waves}",
    eprint = "1201.6641",
    archivePrefix = "arXiv",
    primaryClass = "astro-ph.CO",
    doi = "10.1088/0004-637X/762/2/94",
    journal = "Astrophys. J.",
    volume = "762",
    pages = "94",
    year = "2013"
}

@article{Shannon:2013wma,
    author = "Shannon, R. M. and others",
    title = "{Gravitational-wave Limits from Pulsar Timing Constrain Supermassive Black Hole Evolution}",
    eprint = "1310.4569",
    archivePrefix = "arXiv",
    primaryClass = "astro-ph.CO",
    doi = "10.1126/science.1238012",
    journal = "Science",
    volume = "342",
    number = "6156",
    pages = "334--337",
    year = "2013"
}

@article{Shannon:2015ect,
    author = "Shannon, R. M. and others",
    title = "{Gravitational waves from binary supermassive black holes missing in pulsar observations}",
    eprint = "1509.07320",
    archivePrefix = "arXiv",
    primaryClass = "astro-ph.CO",
    doi = "10.1126/science.aab1910",
    journal = "Science",
    volume = "349",
    number = "6255",
    pages = "1522--1525",
    year = "2015"
}

@article{EPTA:2015qep,
    author = "Lentati, L. and others",
    collaboration = "EPTA",
    title = "{European Pulsar Timing Array Limits On An Isotropic Stochastic Gravitational-Wave Background}",
    eprint = "1504.03692",
    archivePrefix = "arXiv",
    primaryClass = "astro-ph.CO",
    doi = "10.1093/mnras/stv1538",
    journal = "Mon. Not. Roy. Astron. Soc.",
    volume = "453",
    number = "3",
    pages = "2576--2598",
    year = "2015"
}

@article{NANOGrav:2015aud,
    author = "Arzoumanian, Z. and others",
    collaboration = "NANOGrav",
    title = "{The NANOGrav Nine-year Data Set: Limits on the Isotropic Stochastic Gravitational Wave Background}",
    eprint = "1508.03024",
    archivePrefix = "arXiv",
    primaryClass = "astro-ph.GA",
    doi = "10.3847/0004-637X/821/1/13",
    journal = "Astrophys. J.",
    volume = "821",
    number = "1",
    pages = "13",
    year = "2016"
}

@article{Verbiest:2016vem,
    author = "Verbiest, J. P. W. and others",
    title = "{The International Pulsar Timing Array: First Data Release}",
    eprint = "1602.03640",
    archivePrefix = "arXiv",
    primaryClass = "astro-ph.IM",
    doi = "10.1093/mnras/stw347",
    journal = "Mon. Not. Roy. Astron. Soc.",
    volume = "458",
    number = "2",
    pages = "1267--1288",
    year = "2016"
}

@article{NANOGRAV:2018hou,
    author = "Arzoumanian, Z. and others",
    collaboration = "NANOGRAV",
    title = "{The NANOGrav 11-year Data Set: Pulsar-timing Constraints On The Stochastic Gravitational-wave Background}",
    eprint = "1801.02617",
    archivePrefix = "arXiv",
    primaryClass = "astro-ph.HE",
    doi = "10.3847/1538-4357/aabd3b",
    journal = "Astrophys. J.",
    volume = "859",
    number = "1",
    pages = "47",
    year = "2018"
}

@ARTICLE{2025ApJ...991L..19C,
       author = {{Chen}, Nianyi and {Di Matteo}, Tiziana and {Zhou}, Yihao and {Kelley}, Luke Zoltan and {Blecha}, Laura and {Ni}, Yueying and {Bird}, Simeon and {Yang}, Yanhui and {Croft}, Rupert},
        title = "{The Gravitational-wave Background from Massive Black Holes in the ASTRID Simulation}",
      journal = {Astrophys. J. Letters},
         year = 2025,
        month = sep,
       volume = {991},
       number = {1},
          eid = {L19},
        pages = {L19},
          doi = {10.3847/2041-8213/adefe2},
archivePrefix = {arXiv},
       eprint = {2502.01024},
 primaryClass = {astro-ph.GA},
       adsurl = {https://ui.adsabs.harvard.edu/abs/2025ApJ...991L..19C}
}

@ARTICLE{2018MNRAS.473.3410R,
       author = {{Ryu}, Taeho and {Perna}, Rosalba and {Haiman}, Zolt{\'a}n and {Ostriker}, Jeremiah P. and {Stone}, Nicholas C.},
        title = "{Interactions between multiple supermassive black holes in galactic nuclei: a solution to the final parsec problem}",
      journal = {Mon. Not. Roy. Ast. Soc.},
         year = 2018,
        month = jan,
       volume = {473},
       number = {3},
        pages = {3410-3433},
          doi = {10.1093/mnras/stx2524},
archivePrefix = {arXiv},
       eprint = {1709.06501},
 primaryClass = {astro-ph.GA},
       adsurl = {https://ui.adsabs.harvard.edu/abs/2018MNRAS.473.3410R}
}

@ARTICLE{2019ApJ...871...84M,
       author = {{Mu{\~n}oz}, Diego J. and {Miranda}, Ryan and {Lai}, Dong},
        title = "{Hydrodynamics of Circumbinary Accretion: Angular Momentum Transfer and Binary Orbital Evolution}",
      journal = {Astrophys. J.},
         year = 2019,
        month = jan,
       volume = {871},
       number = {1},
          eid = {84},
        pages = {84},
          doi = {10.3847/1538-4357/aaf867},
archivePrefix = {arXiv},
       eprint = {1810.04676},
 primaryClass = {astro-ph.HE},
       adsurl = {https://ui.adsabs.harvard.edu/abs/2019ApJ...871...84M}
}

@ARTICLE{2019ApJ...875...66M,
       author = {{Moody}, Mackenzie S.~L. and {Shi}, Ji-Ming and {Stone}, James M.},
        title = "{Hydrodynamic Torques in Circumbinary Accretion Disks}",
      journal = {Astrophys. J.},
         year = 2019,
        month = apr,
       volume = {875},
       number = {1},
          eid = {66},
        pages = {66},
          doi = {10.3847/1538-4357/ab09ee},
archivePrefix = {arXiv},
       eprint = {1903.00008},
 primaryClass = {astro-ph.HE},
       adsurl = {https://ui.adsabs.harvard.edu/abs/2019ApJ...875...66M}
}

@ARTICLE{2011MNRAS.415.3033R,
       author = {{Roedig}, C. and {Dotti}, M. and {Sesana}, A. and {Cuadra}, J. and {Colpi}, M.},
        title = "{Limiting eccentricity of subparsec massive black hole binaries surrounded by self-gravitating gas discs}",
      journal = {Mon. Not. Roy. Ast. Soc.},
         year = 2011,
        month = aug,
       volume = {415},
       number = {4},
        pages = {3033-3041},
          doi = {10.1111/j.1365-2966.2011.18927.x},
archivePrefix = {arXiv},
       eprint = {1104.3868},
 primaryClass = {astro-ph.CO},
       adsurl = {https://ui.adsabs.harvard.edu/abs/2011MNRAS.415.3033R}
}

@ARTICLE{2014ApJ...783..134F,
       author = {{Farris}, Brian D. and {Duffell}, Paul and {MacFadyen}, Andrew I. and {Haiman}, Zoltan},
        title = "{Binary Black Hole Accretion from a Circumbinary Disk: Gas Dynamics inside the Central Cavity}",
      journal = {Astrophys. J.},
         year = 2014,
        month = mar,
       volume = {783},
       number = {2},
          eid = {134},
        pages = {134},
          doi = {10.1088/0004-637X/783/2/134},
archivePrefix = {arXiv},
       eprint = {1310.0492},
 primaryClass = {astro-ph.HE},
       adsurl = {https://ui.adsabs.harvard.edu/abs/2014ApJ...783..134F}
}

@ARTICLE{2014MNRAS.439.3476R,
       author = {{Roedig}, Constanze and {Sesana}, Alberto},
        title = "{Migration of massive black hole binaries in self-gravitating discs: retrograde versus prograde}",
      journal = {Mon. Not. Roy. Ast. Soc.},
         year = 2014,
        month = apr,
       volume = {439},
       number = {4},
        pages = {3476-3489},
          doi = {10.1093/mnras/stu194},
archivePrefix = {arXiv},
       eprint = {1307.6283},
 primaryClass = {astro-ph.HE},
       adsurl = {https://ui.adsabs.harvard.edu/abs/2014MNRAS.439.3476R}
}

@ARTICLE{2022ApJ...929L..13F,
       author = {{Franchini}, Alessia and {Lupi}, Alessandro and {Sesana}, Alberto},
        title = "{Resolving Massive Black Hole Binary Evolution via Adaptive Particle Splitting}",
      journal = {Astrophys. J. Letters},
         year = 2022,
        month = apr,
       volume = {929},
       number = {1},
          eid = {L13},
        pages = {L13},
          doi = {10.3847/2041-8213/ac63a2},
archivePrefix = {arXiv},
       eprint = {2201.05619},
 primaryClass = {astro-ph.HE},
       adsurl = {https://ui.adsabs.harvard.edu/abs/2022ApJ...929L..13F}
}

@ARTICLE{2020A&A...641A..64H,
       author = {{Heath}, R.~M. and {Nixon}, C.~J.},
        title = "{On the orbital evolution of binaries with circumbinary discs}",
      journal = {Astronomy and Astrophysics},
         year = 2020,
        month = sep,
       volume = {641},
          eid = {A64},
        pages = {A64},
          doi = {10.1051/0004-6361/202038548},
archivePrefix = {arXiv},
       eprint = {2007.11592},
 primaryClass = {astro-ph.HE},
       adsurl = {https://ui.adsabs.harvard.edu/abs/2020A&A...641A..64H}
}

@ARTICLE{2021ApJ...914L..21D,
       author = {{D'Orazio}, Daniel J. and {Duffell}, Paul C.},
        title = "{Orbital Evolution of Equal-mass Eccentric Binaries due to a Gas Disk: Eccentric Inspirals and Circular Outspirals}",
      journal = {Astrophys. J. Letters},
         year = 2021,
        month = jun,
       volume = {914},
       number = {1},
          eid = {L21},
        pages = {L21},
          doi = {10.3847/2041-8213/ac0621},
archivePrefix = {arXiv},
       eprint = {2103.09251},
 primaryClass = {astro-ph.HE},
       adsurl = {https://ui.adsabs.harvard.edu/abs/2021ApJ...914L..21D}
}

@article{Duffell:2019uuk,
    author = "Duffell, Paul C. and D'Orazio, Daniel and Derdzinski, Andrea and Haiman, Zoltan and MacFadyen, Andrew and Rosen, Anna L. and Zrake, Jonathan",
    title = "{Circumbinary Disks: Accretion and Torque as a Function of Mass Ratio and Disk Viscosity}",
    eprint = "1911.05506",
    archivePrefix = "arXiv",
    primaryClass = "astro-ph.SR",
    doi = "10.3847/1538-4357/abab95",
    journal = "Astrophys. J.",
    volume = "901",
    number = "1",
    pages = "25",
    year = "2020"
}

@article{Duffell:2024fwy,
    author = "Duffell, Paul C. and others",
    title = "{The Santa Barbara Binary{\ensuremath{-}}disk Code Comparison}",
    eprint = "2402.13039",
    archivePrefix = "arXiv",
    primaryClass = "astro-ph.SR",
    doi = "10.3847/1538-4357/ad5a7e",
    journal = "Astrophys. J.",
    volume = "970",
    number = "2",
    pages = "156",
    year = "2024"
}

@ARTICLE{2015MNRAS.454L..66S,
       author = {{Sesana}, Alberto and {Khan}, Fazeel Mahmood},
        title = "{Scattering experiments meet N-body - I. A practical recipe for the evolution of massive black hole binaries in stellar environments}",
      journal = {Mon. Not. Roy. Ast. Soc.},
         year = 2015,
        month = nov,
       volume = {454},
       number = {1},
        pages = {L66-L70},
          doi = {10.1093/mnrasl/slv131},
archivePrefix = {arXiv},
       eprint = {1505.02062},
 primaryClass = {astro-ph.GA},
       adsurl = {https://ui.adsabs.harvard.edu/abs/2015MNRAS.454L..66S}
}

@ARTICLE{2015ApJ...810...49V,
       author = {{Vasiliev}, Eugene and {Antonini}, Fabio and {Merritt}, David},
        title = "{The Final-parsec Problem in the Collisionless Limit}",
      journal = {Astrophys. J.},
         year = 2015,
        month = sep,
       volume = {810},
       number = {1},
          eid = {49},
        pages = {49},
          doi = {10.1088/0004-637X/810/1/49},
archivePrefix = {arXiv},
       eprint = {1505.05480},
 primaryClass = {astro-ph.GA},
       adsurl = {https://ui.adsabs.harvard.edu/abs/2015ApJ...810...49V}
}

@ARTICLE{2017MNRAS.464.3131K,
       author = {{Kelley}, Luke Zoltan and {Blecha}, Laura and {Hernquist}, Lars},
        title = "{Massive black hole binary mergers in dynamical galactic environments}",
      journal = {Mon. Not. Roy. Ast. Soc.},
         year = 2017,
        month = jan,
       volume = {464},
       number = {3},
        pages = {3131-3157},
          doi = {10.1093/mnras/stw2452},
archivePrefix = {arXiv},
       eprint = {1606.01900},
 primaryClass = {astro-ph.HE},
       adsurl = {https://ui.adsabs.harvard.edu/abs/2017MNRAS.464.3131K}
}

@ARTICLE{2020MNRAS.495.4681I,
       author = {{Izquierdo-Villalba}, David and {Bonoli}, Silvia and {Dotti}, Massimo and {Sesana}, Alberto and {Rosas-Guevara}, Yetli and {Spinoso}, Daniele},
        title = "{From galactic nuclei to the halo outskirts: tracing supermassive black holes across cosmic history and environments}",
      journal = {Mon. Not. Roy. Ast. Soc.},
         year = 2020,
        month = jul,
       volume = {495},
       number = {4},
        pages = {4681-4706},
          doi = {10.1093/mnras/staa1399},
archivePrefix = {arXiv},
       eprint = {2001.10548},
 primaryClass = {astro-ph.GA},
       adsurl = {https://ui.adsabs.harvard.edu/abs/2020MNRAS.495.4681I}
}

@ARTICLE{2007ApJ...664..226L,
       author = {{Lauer}, Tod R. and {Gebhardt}, Karl and {Faber}, S.~M. and {Richstone}, Douglas and {Tremaine}, Scott and {Kormendy}, John and {Aller}, M.~C. and {Bender}, Ralf and {Dressler}, Alan and {Filippenko}, Alexei V. and {Green}, Richard and {Ho}, Luis C.},
        title = "{The Centers of Early-Type Galaxies with Hubble Space Telescope. VI. Bimodal Central Surface Brightness Profiles}",
      journal = {Astrophys. J.},
         year = 2007,
        month = jul,
       volume = {664},
       number = {1},
        pages = {226-256},
          doi = {10.1086/519229},
archivePrefix = {arXiv},
       eprint = {astro-ph/0609762},
 primaryClass = {astro-ph},
       adsurl = {https://ui.adsabs.harvard.edu/abs/2007ApJ...664..226L}
}

@article{2016PhRvL.116f1102A,
    author = "Abbott, B. P. and others",
    collaboration = "LIGO Scientific, Virgo",
    title = "{Observation of Gravitational Waves from a Binary Black Hole Merger}",
    eprint = "1602.03837",
    archivePrefix = "arXiv",
    primaryClass = "gr-qc",
    reportNumber = "LIGO-P150914",
    doi = "10.1103/PhysRevLett.116.061102",
    journal = "Phys. Rev. Lett.",
    volume = "116",
    number = "6",
    pages = "061102",
    year = "2016"
}

@article{LIGOScientific:2025slb,
    author = "Abac, A. G. and others",
    collaboration = "LIGO Scientific, VIRGO, KAGRA",
    title = "{GWTC-4.0: Updating the Gravitational-Wave Transient Catalog with Observations from the First Part of the Fourth LIGO-Virgo-KAGRA Observing Run}",
    eprint = "2508.18082",
    archivePrefix = "arXiv",
    primaryClass = "gr-qc",
    reportNumber = "LIGO-P2400386",
    month = "8",
    year = "2025"
}

@article{KAGRA:2021duu,
    author = "Abbott, R. and others",
    collaboration = "KAGRA, VIRGO, LIGO Scientific",
    title = "{Population of Merging Compact Binaries Inferred Using Gravitational Waves through GWTC-3}",
    eprint = "2111.03634",
    archivePrefix = "arXiv",
    primaryClass = "astro-ph.HE",
    reportNumber = "LIGO-P2100239 ; Data release: https://zenodo.org/record/5655785, LIGO-P2100239",
    doi = "10.1103/PhysRevX.13.011048",
    journal = "Phys. Rev. X",
    volume = "13",
    number = "1",
    pages = "011048",
    year = "2023"
}

@article{KAGRA:2025oiz,
    author = "Abac, A. G. and others",
    collaboration = "LIGO Scientific, Virgo, KAGRA",
    title = "{GW250114: Testing Hawking{\textquoteright}s Area Law and the Kerr Nature of Black Holes}",
    eprint = "2509.08054",
    archivePrefix = "arXiv",
    primaryClass = "gr-qc",
    reportNumber = "LIGO-P2500421",
    doi = "10.1103/kw5g-d732",
    journal = "Phys. Rev. Lett.",
    volume = "135",
    number = "11",
    pages = "111403",
    year = "2025"
}

@article{2017PhRvL.119p1101A,
    author = "Abbott, B. P. and others",
    collaboration = "LIGO Scientific, Virgo",
    title = "{GW170817: Observation of Gravitational Waves from a Binary Neutron Star Inspiral}",
    eprint = "1710.05832",
    archivePrefix = "arXiv",
    primaryClass = "gr-qc",
    reportNumber = "LIGO-P170817",
    doi = "10.1103/PhysRevLett.119.161101",
    journal = "Phys. Rev. Lett.",
    volume = "119",
    number = "16",
    pages = "161101",
    year = "2017"
}

@article{Lasky:2015lej,
    author = "Lasky, Paul D. and others",
    title = "{Gravitational-wave cosmology across 29 decades in frequency}",
    eprint = "1511.05994",
    archivePrefix = "arXiv",
    primaryClass = "astro-ph.CO",
    doi = "10.1103/PhysRevX.6.011035",
    journal = "Phys. Rev. X",
    volume = "6",
    number = "1",
    pages = "011035",
    year = "2016"
}

@article{LISA:2017pwj,
    author = "Amaro-Seoane, Pau and others",
    collaboration = "LISA",
    title = "{Laser Interferometer Space Antenna}",
    eprint = "1702.00786",
    archivePrefix = "arXiv",
    primaryClass = "astro-ph.IM",
    month = "2",
    year = "2017"
}

@article{LiteBIRD:2023iei,
    author = "Fuskeland, U. and others",
    collaboration = "LiteBIRD",
    title = "{Tensor-to-scalar ratio forecasts for extended LiteBIRD frequency configurations}",
    eprint = "2302.05228",
    archivePrefix = "arXiv",
    primaryClass = "astro-ph.CO",
    doi = "10.1051/0004-6361/202346155",
    journal = "Astron. Astrophys.",
    volume = "676",
    pages = "A42",
    year = "2023"
}

@article{Wolz:2023lzb,
    author = "Wolz, Kevin and others",
    title = "{The Simons Observatory: pipeline comparison and validation for large-scale B-modes}",
    eprint = "2302.04276",
    archivePrefix = "arXiv",
    primaryClass = "astro-ph.CO",
    doi = "10.1051/0004-6361/202346105",
    journal = "Astron. Astrophys.",
    volume = "686",
    pages = "A16",
    year = "2024"
}

@article{NANOGrav:2023gor,
    author = "Agazie, Gabriella and others",
    collaboration = "NANOGrav",
    title = "{The NANOGrav 15 yr Data Set: Evidence for a Gravitational-wave Background}",
    eprint = "2306.16213",
    archivePrefix = "arXiv",
    primaryClass = "astro-ph.HE",
    doi = "10.3847/2041-8213/acdac6",
    journal = "Astrophys. J. Lett.",
    volume = "951",
    number = "1",
    pages = "L8",
    year = "2023"
}

@article{Antoniadis:2023ott,
    author = "Antoniadis, J. and others",
    collaboration = "EPTA, InPTA:",
    title = "{The second data release from the European Pulsar Timing Array - III. Search for gravitational wave signals}",
    eprint = "2306.16214",
    archivePrefix = "arXiv",
    primaryClass = "astro-ph.HE",
    doi = "10.1051/0004-6361/202346844",
    journal = "Astron. Astrophys.",
    volume = "678",
    pages = "A50",
    year = "2023"
}

@article{Reardon:2023gzh,
    author = "Reardon, Daniel J. and others",
    title = "{Search for an Isotropic Gravitational-wave Background with the Parkes Pulsar Timing Array}",
    eprint = "2306.16215",
    archivePrefix = "arXiv",
    primaryClass = "astro-ph.HE",
    doi = "10.3847/2041-8213/acdd02",
    journal = "Astrophys. J. Lett.",
    volume = "951",
    number = "1",
    pages = "L6",
    year = "2023"
}

@article{Xu:2023wog,
    author = "Xu, Heng and others",
    title = "{Searching for the Nano-Hertz Stochastic Gravitational Wave Background with the Chinese Pulsar Timing Array Data Release I}",
    eprint = "2306.16216",
    archivePrefix = "arXiv",
    primaryClass = "astro-ph.HE",
    doi = "10.1088/1674-4527/acdfa5",
    journal = "Res. Astron. Astrophys.",
    volume = "23",
    number = "7",
    pages = "075024",
    year = "2023"
}

@article{Caprini:2018mtu,
    author = "Caprini, Chiara and Figueroa, Daniel G.",
    title = "{Cosmological Backgrounds of Gravitational Waves}",
    eprint = "1801.04268",
    archivePrefix = "arXiv",
    primaryClass = "astro-ph.CO",
    doi = "10.1088/1361-6382/aac608",
    journal = "Class. Quant. Grav.",
    volume = "35",
    number = "16",
    pages = "163001",
    year = "2018"
}

@article{Taylor:2021yjx,
    author = "Taylor, Stephen R.",
    title = "{The Nanohertz Gravitational Wave Astronomer}",
    eprint = "2105.13270",
    archivePrefix = "arXiv",
    primaryClass = "astro-ph.HE",
    month = "5",
    year = "2021"
}

@article{2018ASSL..457...95P,
    author = "Perrodin, Delphine and Sesana, Alberto",
    title = "{Radio pulsars: testing gravity and detecting gravitational waves}",
    eprint = "1709.02816",
    archivePrefix = "arXiv",
    primaryClass = "astro-ph.HE",
    doi = "10.1007/978-3-319-97616-7\_3",
    journal = "Astrophys. Space Sci. Libr.",
    volume = "457",
    pages = "95--148",
    year = "2018"
}

@book{Maggiore:2007ulw,
    author = "Maggiore, Michele",
    title = "{Gravitational Waves. Vol. 1: Theory and Experiments}",
    doi = "10.1093/acprof:oso/9780198570745.001.0001",
    isbn = "978-0-19-171766-6, 978-0-19-852074-0",
    publisher = "Oxford University Press",
    year = "2007"
}

@article{Planck:2018vyg,
    author = "Aghanim, N. and others",
    collaboration = "Planck",
    title = "{Planck 2018 results. VI. Cosmological parameters}",
    eprint = "1807.06209",
    archivePrefix = "arXiv",
    primaryClass = "astro-ph.CO",
    doi = "10.1051/0004-6361/201833910",
    journal = "Astron. Astrophys.",
    volume = "641",
    pages = "A6",
    year = "2020",
    note = "[Erratum: Astron.Astrophys. 652, C4 (2021)]"
}

@article{Damour:2000wa,
    author = "Damour, Thibault and Vilenkin, Alexander",
    title = "{Gravitational wave bursts from cosmic strings}",
    eprint = "gr-qc/0004075",
    archivePrefix = "arXiv",
    reportNumber = "IHES-P-00-32",
    doi = "10.1103/PhysRevLett.85.3761",
    journal = "Phys. Rev. Lett.",
    volume = "85",
    pages = "3761--3764",
    year = "2000"
}

@article{Sorbo:2011rz,
    author = "Sorbo, Lorenzo",
    title = "{Parity violation in the Cosmic Microwave Background from a pseudoscalar inflaton}",
    eprint = "1101.1525",
    archivePrefix = "arXiv",
    primaryClass = "astro-ph.CO",
    doi = "10.1088/1475-7516/2011/06/003",
    journal = "JCAP",
    volume = "06",
    pages = "003",
    year = "2011"
}

@article{2019MNRAS.483.1731L,
    author = "Lyubarsky, Yuri",
    title = "{Radio emission of the Crab and Crab-like pulsars}",
    eprint = "1811.11122",
    archivePrefix = "arXiv",
    primaryClass = "astro-ph.HE",
    doi = "10.1093/mnras/sty3233",
    journal = "Mon. Not. Roy. Astron. Soc.",
    volume = "483",
    number = "2",
    pages = "1731--1736",
    year = "2019"
}

@article{2013LRR....16....9Y,
    author = "Yunes, Nicol{\'a}s and Siemens, Xavier",
    title = "{Gravitational-Wave Tests of General Relativity with Ground-Based Detectors and Pulsar Timing-Arrays}",
    eprint = "1304.3473",
    archivePrefix = "arXiv",
    primaryClass = "gr-qc",
    doi = "10.12942/lrr-2013-9",
    journal = "Living Rev. Rel.",
    volume = "16",
    pages = "9",
    year = "2013"
}

@article{1983ApJ...265L..39H,
    author = "Hellings, R. w. and Downs, G. s.",
    title = "{UPPER LIMITS ON THE ISOTROPIC GRAVITATIONAL RADIATION BACKGROUND FROM PULSAR TIMING ANALYSIS}",
    doi = "10.1086/183954",
    journal = "Astrophys. J. Lett.",
    volume = "265",
    pages = "L39--L42",
    year = "1983"
}

@article{Cordes:2013iea,
    author = "Cordes, James M.",
    title = "{Limits to PTA sensitivity: spin stability and arrival time precision of millisecond pulsars}",
    doi = "10.1088/0264-9381/30/22/224002",
    journal = "Class. Quant. Grav.",
    volume = "30",
    pages = "224002",
    year = "2013"
}

@article{Shannon:2010bv,
    author = "Shannon, Ryan M. and Cordes, James M.",
    title = "{Assessing the Role of Spin Noise in the Precision Timing of Millisecond Pulsars}",
    eprint = "1010.4794",
    archivePrefix = "arXiv",
    primaryClass = "astro-ph.SR",
    doi = "10.1088/0004-637X/725/2/1607",
    journal = "Astrophys. J.",
    volume = "725",
    pages = "1607--1619",
    year = "2010"
}

@article{2016MNRAS.455.4339T,
    author = "Tiburzi, Caterina and Hobbs, George and Kerr, Matthew and Coles, William and Dai, Shi and Manchester, Richard and Possenti, Andrea and Shannon, Ryan and You, Xiaopeng",
    title = "{A study of spatial correlations in pulsar timing array data}",
    eprint = "1510.02363",
    archivePrefix = "arXiv",
    primaryClass = "astro-ph.IM",
    doi = "10.1093/mnras/stv2143",
    journal = "Mon. Not. Roy. Astron. Soc.",
    volume = "455",
    number = "4",
    pages = "4339--4350",
    year = "2016"
}

@article{Allen:1997ad,
    author = "Allen, Bruce and Romano, Joseph D.",
    title = "{Detecting a stochastic background of gravitational radiation: Signal processing strategies and sensitivities}",
    eprint = "gr-qc/9710117",
    archivePrefix = "arXiv",
    reportNumber = "WISC-MILW-97-TH-14",
    doi = "10.1103/PhysRevD.59.102001",
    journal = "Phys. Rev. D",
    volume = "59",
    pages = "102001",
    year = "1999"
}

@article{EPTA:2023sfo,
    primaryClass = "astro-ph.HE",
    doi = "10.1093/mnras/stz2857",
    journal = "Mon. Not. Roy. Astron. Soc.",
    volume = "490",
    number = "4",
    pages = "4666--4687",
    year = "2019"
}

@article{Rana:2025ano,
    author = "Rana, Prerna and others",
    title = "{The Indian Pulsar Timing Array Data Release 2: I. Dataset and Timing Analysis}",
    eprint = "2506.16769",
    archivePrefix = "arXiv",
    primaryClass = "astro-ph.IM",
    doi = "10.1017/pasa.2025.10066",
    month = "6",
    year = "2025"
}

@article{Miles:2022lkg,
    author = "Miles, Matthew T. and others",
    title = "{The MeerKAT Pulsar Timing Array: first data release}",
    eprint = "2212.04648",
    archivePrefix = "arXiv",
    primaryClass = "astro-ph.HE",
    doi = "10.1093/mnras/stac3644",
    journal = "Mon. Not. Roy. Astron. Soc.",
    volume = "519",
    number = "3",
    pages = "3976--3991",
    year = "2023"
}

@article{Perera:2019sca,
    author = "Perera, B. B. P. and others",
    title = "{The International Pulsar Timing Array: Second data release}",
    eprint = "1909.04534",
    archivePrefix = "arXiv",
    primaryClass = "astro-ph.HE",
    doi = "10.1093/mnras/stz2857",
    journal = "Mon. Not. Roy. Astron. Soc.",
    volume = "490",
    number = "4",
    pages = "4666--4687",
    year = "2019"
}

@ARTICLE{2024A&A...691A.145M,
       author = {{Maiolino}, Roberto and {Scholtz}, Jan and {Curtis-Lake}, Emma and {Carniani}, Stefano and {Baker}, William and {de Graaff}, Anna and {Tacchella}, Sandro and {{\"U}bler}, Hannah and {D'Eugenio}, Francesco and {Witstok}, Joris and {Curti}, Mirko and {Arribas}, Santiago and {Bunker}, Andrew J. and {Charlot}, St{\'e}phane and {Chevallard}, Jacopo and {Eisenstein}, Daniel J. and {Egami}, Eiichi and {Ji}, Zhiyuan and {Jones}, Gareth C. and {Lyu}, Jianwei and {Rawle}, Tim and {Robertson}, Brant and {Rujopakarn}, Wiphu and {Perna}, Michele and {Sun}, Fengwu and {Venturi}, Giacomo and {Williams}, Christina C. and {Willott}, Chris},
        title = "{JADES: The diverse population of infant black holes at 4 < z < 11: Merging, tiny, poor, but mighty}",
      journal = {Astronomy and Astrophysics},
         year = 2024,
        month = nov,
       volume = {691},
          eid = {A145},
        pages = {A145},
          doi = {10.1051/0004-6361/202347640},
archivePrefix = {arXiv},
       eprint = {2308.01230},
 primaryClass = {astro-ph.GA},
       adsurl = {https://ui.adsabs.harvard.edu/abs/2024A&A...691A.145M}
}

@ARTICLE{1963PhRv..131..435P,
       author = {{Peters}, P.~C. and {Mathews}, J.},
        title = "{Gravitational Radiation from Point Masses in a Keplerian Orbit}",
      journal = {Physical Review},
         year = 1963,
        month = jul,
       volume = {131},
       number = {1},
        pages = {435-440},
          doi = {10.1103/PhysRev.131.435},
       adsurl = {https://ui.adsabs.harvard.edu/abs/1963PhRv..131..435P}
}

@ARTICLE{2024PhRvD.110f3547I,
       author = {{Inomata}, Keisuke and {Kamionkowski}, Marc and {Toral}, Celia M. and {Taylor}, Stephen R.},
        title = "{Overlap reduction functions for pulsar timing arrays and astrometry}",
      journal = {Phys. Rev. D},
         year = 2024,
        month = sep,
       volume = {110},
       number = {6},
          eid = {063547},
        pages = {063547},
          doi = {10.1103/PhysRevD.110.063547},
archivePrefix = {arXiv},
       eprint = {2406.00096},
 primaryClass = {astro-ph.CO},
       adsurl = {https://ui.adsabs.harvard.edu/abs/2024PhRvD.110f3547I}
}

@ARTICLE{2024JCAP...05..030C,
       author = {{{\c{C}}al{\i}{\c{s}}kan}, Mesut and {Chen}, Yifan and {Dai}, Liang and {Kumar}, Neha Anil and {Stomberg}, Isak and {Xue}, Xiao},
        title = "{Dissecting the stochastic gravitational wave background with astrometry}",
      journal = {Jcap},
         year = 2024,
        month = may,
       volume = {2024},
       number = {5},
          eid = {030},
        pages = {030},
          doi = {10.1088/1475-7516/2024/05/030},
archivePrefix = {arXiv},
       eprint = {2312.03069},
 primaryClass = {gr-qc},
       adsurl = {https://ui.adsabs.harvard.edu/abs/2024JCAP...05..030C}
}

@article{Zic:2023gta,
    author = "Zic, Andrew and others",
    title = "{The Parkes Pulsar Timing Array third data release}",
    eprint = "2306.16230",
    archivePrefix = "arXiv",
    primaryClass = "astro-ph.HE",
    doi = "10.1017/pasa.2023.36",
    journal = "Publ. Astron. Soc. Austral.",
    volume = "40",
    pages = "e049",
    year = "2023"
}

@article{NANOGrav:2023hde,
    author = "Agazie, Gabriella and others",
    collaboration = "NANOGrav",
    title = "{The NANOGrav 15 yr Data Set: Observations and Timing of 68 Millisecond Pulsars}",
    eprint = "2306.16217",
    archivePrefix = "arXiv",
    primaryClass = "astro-ph.HE",
    doi = "10.3847/2041-8213/acda9a",
    journal = "Astrophys. J. Lett.",
    volume = "951",
    number = "1",
    pages = "L9",
    year = "2023"
}

@article{Chen:2025qqb,
    author = "Chen, Siyuan and others",
    title = "{The Chinese Pulsar Timing Array Data Release I - Single pulsar noise analysis}",
    eprint = "2506.04850",
    archivePrefix = "arXiv",
    primaryClass = "astro-ph.HE",
    doi = "10.1051/0004-6361/202452550",
    journal = "Astron. Astrophys.",
    volume = "699",
    pages = "A165",
    year = "2025"
}

@ARTICLE{2025arXiv250500797K,
       author = {{Kelley}, Luke Zoltan},
        title = "{Pulsar Timing Arrays}",
      journal = {arXiv e-prints},
         year = 2025,
        month = may,
          eid = {arXiv:2505.00797},
        pages = {arXiv:2505.00797},
          doi = {10.48550/arXiv.2505.00797},
archivePrefix = {arXiv},
       eprint = {2505.00797},
 primaryClass = {astro-ph.HE},
       adsurl = {https://ui.adsabs.harvard.edu/abs/2025arXiv250500797K}
}

@ARTICLE{2024arXiv240207571C,
       author = {{Colpi}, Monica and {Danzmann}, Karsten and {Hewitson}, Martin and {Holley-Bockelmann}, Kelly and {Jetzer}, Philippe and {Nelemans}, Gijs and {Petiteau}, Antoine and {Shoemaker}, David and {Sopuerta}, Carlos and {Stebbins}, Robin and {Tanvir}, Nial and {Ward}, Henry and {Weber}, William Joseph and {Thorpe}, Ira and {Daurskikh}, Anna and {Deep}, Atul and {Fern{\'a}ndez N{\'u}{\~n}ez}, Ignacio and {Garc{\'\i}a Marirrodriga}, C{\'e}sar and {Gehler}, Martin and {Halain}, Jean-Philippe and {Jennrich}, Oliver and {Lammers}, Uwe and {Larra{\~n}aga}, Jonan and {Lieser}, Maike and {L{\"u}tzgendorf}, Nora and {Martens}, Waldemar and {Mondin}, Linda and {Piris Ni{\~n}o}, Ana and {Amaro-Seoane}, Pau and {Arca Sedda}, Manuel and {Auclair}, Pierre and {Babak}, Stanislav and {Baghi}, Quentin and {Baibhav}, Vishal and {Baker}, Tessa and {Bayle}, Jean-Baptiste and {Berry}, Christopher and {Berti}, Emanuele and {Boileau}, Guillaume and {Bonetti}, Matteo and {Brito}, Richard and {Buscicchio}, Riccardo and {Calcagni}, Gianluca and {Capelo}, Pedro R. and {Caprini}, Chiara and {Caputo}, Andrea and {Castelli}, Eleonora and {Chen}, Hsin-Yu and {Chen}, Xian and {Chua}, Alvin and {Davies}, Gareth and {Derdzinski}, Andrea and {Domcke}, Valerie Fiona and {Doneva}, Daniela and {Dvorkin}, Irna and {Mar{\'\i}a Ezquiaga}, Jose and {Gair}, Jonathan and {Haiman}, Zoltan and {Harry}, Ian and {Hartwig}, Olaf and {Hees}, Aurelien and {Heffernan}, Anna and {Husa}, Sascha and {Izquierdo-Villalba}, David and {Karnesis}, Nikolaos and {Klein}, Antoine and {Korol}, Valeriya and {Korsakova}, Natalia and {Kupfer}, Thomas and {Laghi}, Danny and {Lamberts}, Astrid and {Larson}, Shane and {Le Jeune}, Maude and {Lewicki}, Marek and {Littenberg}, Tyson and {Madge}, Eric and {Mangiagli}, Alberto and {Marsat}, Sylvain and {Vilchez}, Ivan Martin and {Maselli}, Andrea and {Mathews}, Josh and {van de Meent}, Maarten and {Muratore}, Martina and {Nardini}, Germano and {Pani}, Paolo and {Peloso}, Marco and {Pieroni}, Mauro and {Pound}, Adam and {Quelquejay-Leclere}, Hippolyte and {Ricciardone}, Angelo and {Rossi}, Elena Maria and {Sartirana}, Andrea and {Savalle}, Etienne and {Sberna}, Laura and {Sesana}, Alberto and {Shoemaker}, Deirdre and {Slutsky}, Jacob and {Sotiriou}, Thomas and {Speri}, Lorenzo and {Staab}, Martin and {Steer}, Dani{\`e}le and {Tamanini}, Nicola and {Tasinato}, Gianmassimo and {Torrado}, Jesus and {Torres-Orjuela}, Alejandro and {Toubiana}, Alexandre and {Vallisneri}, Michele and {Vecchio}, Alberto and {Volonteri}, Marta and {Yagi}, Kent and {Zwick}, Lorenz},
}

@article{Stinebring:2013yza,
    author = "Stinebring, Dan",
    title = "{Effects of the interstellar medium on detection of low-frequency gravitational waves}",
    eprint = "1310.8316",
    archivePrefix = "arXiv",
    primaryClass = "astro-ph.HE",
    doi = "10.1088/0264-9381/30/22/224006",
    journal = "Class. Quant. Grav.",
    volume = "30",
    pages = "224006",
    year = "2013"
}

@article{2015MNRAS.451.2417R,
    author = "Rosado, Pablo A. and Sesana, Alberto and Gair, Jonathan",
    title = "{Expected properties of the first gravitational wave signal detected with pulsar timing arrays}",
    eprint = "1503.04803",
    archivePrefix = "arXiv",
    primaryClass = "astro-ph.HE",
    doi = "10.1093/mnras/stv1098",
    journal = "Mon. Not. Roy. Astron. Soc.",
    volume = "451",
    number = "3",
    pages = "2417--2433",
    year = "2015"
}

@article{2013CQGra..30v4015S,
    author = "Siemens, Xavier and Ellis, Justin and Jenet, Fredrick and Romano, Joseph D.",
    title = "{The stochastic background: scaling laws and time to detection for pulsar timing arrays}",
    eprint = "1305.3196",
    archivePrefix = "arXiv",
    primaryClass = "astro-ph.IM",
    doi = "10.1088/0264-9381/30/22/224015",
    journal = "Class. Quant. Grav.",
    volume = "30",
    pages = "224015",
    year = "2013"
}

@article{1996ApJ...465..566P,
    author = "Pyne, Ted and Gwinn, Carl R. and Birkinshaw, Mark and Eubanks, T. Marshall and Matsakis, Demetrios N.",
    title = "{Gravitational radiation and very long baseline interferometry}",
    eprint = "astro-ph/9507030",
    archivePrefix = "arXiv",
    doi = "10.1086/177443",
    journal = "Astrophys. J.",
    volume = "465",
    pages = "566--577",
    year = "1996"
}

@article{Book:2010pf,
    author = "Book, Laura G. and Flanagan, Eanna E.",
    title = "{Astrometric Effects of a Stochastic Gravitational Wave Background}",
    eprint = "1009.4192",
    archivePrefix = "arXiv",
    primaryClass = "astro-ph.CO",
    doi = "10.1103/PhysRevD.83.024024",
    journal = "Phys. Rev. D",
    volume = "83",
    pages = "024024",
    year = "2011"
}

@article{Gaia:2016zol,
    author = "Prusti, T. and others",
    collaboration = "Gaia",
    title = "{The Gaia Mission}",
    eprint = "1609.04153",
    archivePrefix = "arXiv",
    primaryClass = "astro-ph.IM",
    doi = "10.1051/0004-6361/201629272",
    journal = "Astron. Astrophys.",
    volume = "595",
    number = "Gaia Data Release 1",
    pages = "A1",
    year = "2016"
}

@article{Mihaylov:2018uqm,
    author = "Mihaylov, Deyan P. and Moore, Christopher J. and Gair, Jonathan R. and Lasenby, Anthony and Gilmore, Gerard",
    title = "{Astrometric Effects of Gravitational Wave Backgrounds with non-Einsteinian Polarizations}",
    eprint = "1804.00660",
    archivePrefix = "arXiv",
    primaryClass = "gr-qc",
    doi = "10.1103/PhysRevD.97.124058",
    journal = "Phys. Rev. D",
    volume = "97",
    number = "12",
    pages = "124058",
    year = "2018"
}

@article{Pardo:2023cag,
    author = "Pardo, Kris and Chang, Tzu-Ching and Dor{\'e}, Olivier and Wang, Yijun",
    title = "{Gravitational Wave Detection with Relative Astrometry using Roman's Galactic Bulge Time Domain Survey}",
    eprint = "2306.14968",
    archivePrefix = "arXiv",
    primaryClass = "astro-ph.GA",
    month = "6",
    year = "2023"
}

@article{Vaglio:2025tex,
    author = "Vaglio, Massimo and Falxa, Mikel and Mentasti, Giorgio and Renzini, Arianna I. and Kuntz, Adrien and Barausse, Enrico and Contaldi, Carlo and Sesana, Alberto",
    title = "{Searching for Gravitational Waves with Gaia and its Cross-Correlation with PTA: Absolute vs Relative Astrometry}",
    eprint = "2507.18593",
    archivePrefix = "arXiv",
    primaryClass = "gr-qc",
    reportNumber = "Imperial-TP-2025-CC-7",
    month = "7",
    year = "2025"
}

@article{Qin:2018yhy,
    author = "Qin, Wenzer and Boddy, Kimberly K. and Kamionkowski, Marc and Dai, Liang",
    title = "{Pulsar-timing arrays, astrometry, and gravitational waves}",
    eprint = "1810.02369",
    archivePrefix = "arXiv",
    primaryClass = "astro-ph.CO",
    doi = "10.1103/PhysRevD.99.063002",
    journal = "Phys. Rev. D",
    volume = "99",
    number = "6",
    pages = "063002",
    year = "2019"
}

@article{Cruz:2024diu,
    author = "Cruz, N. M. Jim{\'e}nez and Malhotra, Ameek and Tasinato, Gianmassimo and Zavala, Ivonne",
    title = "{Astrometry meets pulsar timing arrays: Synergies for gravitational wave detection}",
    eprint = "2412.14010",
    archivePrefix = "arXiv",
    primaryClass = "astro-ph.CO",
    doi = "10.1103/8k1p-pzcg",
    journal = "Phys. Rev. D",
    volume = "112",
    number = "8",
    pages = "083558",
    year = "2025"
}

@article{TianQin:2015yph,
    author = "Luo, Jun and others",
    collaboration = "TianQin",
    title = "{TianQin: a space-borne gravitational wave detector}",
    eprint = "1512.02076",
    archivePrefix = "arXiv",
    primaryClass = "astro-ph.IM",
    doi = "10.1088/0264-9381/33/3/035010",
    journal = "Class. Quant. Grav.",
    volume = "33",
    number = "3",
    pages = "035010",
    year = "2016"
}

@article{Hu:2017mde,
    author = "Hu, Wen-Rui and Wu, Yue-Liang",
    title = "{The Taiji Program in Space for gravitational wave physics and the nature of gravity}",
    doi = "10.1093/nsr/nwx116",
    journal = "Natl. Sci. Rev.",
    volume = "4",
    number = "5",
    pages = "685--686",
    year = "2017"
}

@article{Sesana:2019vho,
    author = "Sesana, Alberto and others",
    title = "{Unveiling the gravitational universe at $\mu$-Hz frequencies}",
    eprint = "1908.11391",
    archivePrefix = "arXiv",
    primaryClass = "astro-ph.IM",
    doi = "10.1007/s10686-021-09709-9",
    journal = "Exper. Astron.",
    volume = "51",
    number = "3",
    pages = "1333--1383",
    year = "2021"
}

@article{Hui:2012yp,
    author = "Hui, Lam and McWilliams, Sean T. and Yang, I-Sheng",
    title = "{Binary systems as resonance detectors for gravitational waves}",
    eprint = "1212.2623",
    archivePrefix = "arXiv",
    primaryClass = "gr-qc",
    doi = "10.1103/PhysRevD.87.084009",
    journal = "Phys. Rev. D",
    volume = "87",
    number = "8",
    pages = "084009",
    year = "2013"
}

@article{Foster:2025nzf,
    author = "Foster, Joshua W. and Blas, Diego and Bourgoin, Adrien and Hees, Aurelien and Herrero-Valea, M{\'\i}riam and Jenkins, Alexander C. and Xue, Xiao",
    title = "{Discovering $\mu$Hz gravitational waves and ultra-light dark matter with binary resonances}",
    eprint = "2504.15334",
    archivePrefix = "arXiv",
    primaryClass = "astro-ph.CO",
    reportNumber = "FERMILAB-PUB-25-0091-T",
    month = "4",
    year = "2025"
}

@article{Miles:2024seg,
    author = "Miles, Matthew T. and others",
    title = "{The MeerKAT Pulsar Timing Array: the first search for gravitational waves with the MeerKAT radio telescope}",
    eprint = "2412.01153",
    archivePrefix = "arXiv",
    primaryClass = "astro-ph.HE",
    doi = "10.1093/mnras/stae2571",
    journal = "Mon. Not. Roy. Astron. Soc.",
    volume = "536",
    number = "2",
    pages = "1489--1500",
    year = "2024"
}

@article{Agazie:2024jbf,
    author = "Agazie, Gabriella and others",
    title = "{The NANOGrav 15 yr Data Set: Looking for Signs of Discreteness in the Gravitational-wave Background}",
    eprint = "2404.07020",
    archivePrefix = "arXiv",
    primaryClass = "astro-ph.HE",
    doi = "10.3847/1538-4357/ad93d5",
    journal = "Astrophys. J.",
    volume = "978",
    number = "1",
    pages = "31",
    year = "2025"
}

@article{NANOGrav:2023tcn,
    author = "Agazie, Gabriella and others",
    collaboration = "NANOGrav",
    title = "{The NANOGrav 15 yr Data Set: Search for Anisotropy in the Gravitational-wave Background}",
    eprint = "2306.16221",
    archivePrefix = "arXiv",
    primaryClass = "astro-ph.HE",
    doi = "10.3847/2041-8213/acf4fd",
    journal = "Astrophys. J. Lett.",
    volume = "956",
    number = "1",
    pages = "L3",
    year = "2023"
}

@article{NANOGrav:2023pdq,
    author = "Agazie, Gabriella and others",
    collaboration = "NANOGrav",
    title = "{The NANOGrav 15 yr Data Set: Bayesian Limits on Gravitational Waves from Individual Supermassive Black Hole Binaries}",
    eprint = "2306.16222",
    archivePrefix = "arXiv",
    primaryClass = "astro-ph.HE",
    doi = "10.3847/2041-8213/ace18a",
    journal = "Astrophys. J. Lett.",
    volume = "951",
    number = "2",
    pages = "L50",
    year = "2023"
}

@article{EPTA:2023gyr,
    author = "Antoniadis, J. and others",
    collaboration = "EPTA, InPTA",
    title = "{The second data release from the European Pulsar Timing Array - V. Search for continuous gravitational wave signals}",
    eprint = "2306.16226",
    archivePrefix = "arXiv",
    primaryClass = "astro-ph.HE",
    doi = "10.1051/0004-6361/202348568",
    journal = "Astron. Astrophys.",
    volume = "690",
    pages = "A118",
    year = "2024"
}

@article{EPTA:2023xxk,
    author = "Antoniadis, J. and others",
    collaboration = "EPTA, InPTA",
    title = "{The second data release from the European Pulsar Timing Array - IV. Implications for massive black holes, dark matter, and the early Universe}",
    eprint = "2306.16227",
    archivePrefix = "arXiv",
    primaryClass = "astro-ph.CO",
    doi = "10.1051/0004-6361/202347433",
    journal = "Astron. Astrophys.",
    volume = "685",
    pages = "A94",
    year = "2024"
}

@article{NANOGrav:2023hvm,
    author = "Afzal, Adeela and others",
    collaboration = "NANOGrav",
    title = "{The NANOGrav 15 yr Data Set: Search for Signals from New Physics}",
    eprint = "2306.16219",
    archivePrefix = "arXiv",
    primaryClass = "astro-ph.HE",
    reportNumber = "FERMILAB-PUB-23-589-T",
    doi = "10.3847/2041-8213/acdc91",
    journal = "Astrophys. J. Lett.",
    volume = "951",
    number = "1",
    pages = "L11",
    year = "2023",
    note = "[Erratum: Astrophys.J.Lett. 971, L27 (2024), Erratum: Astrophys.J. 971, L27 (2024)]"
}

@article{NANOGrav:2023hfp,
    author = "Agazie, Gabriella and others",
    collaboration = "NANOGrav",
    title = "{The NANOGrav 15 yr Data Set: Constraints on Supermassive Black Hole Binaries from the Gravitational-wave Background}",
    eprint = "2306.16220",
    archivePrefix = "arXiv",
    primaryClass = "astro-ph.HE",
    doi = "10.3847/2041-8213/ace18b",
    journal = "Astrophys. J. Lett.",
    volume = "952",
    number = "2",
    pages = "L37",
    year = "2023"
}

@article{InternationalPulsarTimingArray:2023mzf,
    author = "Agazie, G. and others",
    collaboration = "International Pulsar Timing Array",
    title = "{Comparing Recent Pulsar Timing Array Results on the Nanohertz Stochastic Gravitational-wave Background}",
    eprint = "2309.00693",
    archivePrefix = "arXiv",
    primaryClass = "astro-ph.HE",
    doi = "10.3847/1538-4357/ad36be",
    journal = "Astrophys. J.",
    volume = "966",
    number = "1",
    pages = "105",
    year = "2024"
}

@article{Kormendy:2013dxa,
    author = "Kormendy, John and Ho, Luis C.",
    title = "{Coevolution (Or Not) of Supermassive Black Holes and Host Galaxies}",
    eprint = "1304.7762",
    archivePrefix = "arXiv",
    primaryClass = "astro-ph.CO",
    doi = "10.1146/annurev-astro-082708-101811",
    journal = "Ann. Rev. Astron. Astrophys.",
    volume = "51",
    pages = "511--653",
    year = "2013"
}

@article{2004PhRvD..69h2005B,
    author = "Barack, Leor and Cutler, Curt",
    title = "{LISA capture sources: Approximate waveforms, signal-to-noise ratios, and parameter estimation accuracy}",
    eprint = "gr-qc/0310125",
    archivePrefix = "arXiv",
    doi = "10.1103/PhysRevD.69.082005",
    journal = "Phys. Rev. D",
    volume = "69",
    pages = "082005",
    year = "2004"
}

@article{Rajagopal:1994zj,
    author = "Rajagopal, Mohan and Romani, Roger W.",
    title = "{Ultralow frequency gravitational radiation from massive black hole binaries}",
    eprint = "astro-ph/9412038",
    archivePrefix = "arXiv",
    doi = "10.1086/175813",
    journal = "Astrophys. J.",
    volume = "446",
    pages = "543--549",
    year = "1995"
}

@article{Jaffe:2002rt,
    author = "Jaffe, Andrew H. and Backer, Donald C.",
    title = "{Gravitational waves probe the coalescence rate of massive black hole binaries}",
    eprint = "astro-ph/0210148",
    archivePrefix = "arXiv",
    doi = "10.1086/345443",
    journal = "Astrophys. J.",
    volume = "583",
    pages = "616--631",
    year = "2003"
}

@article{Wyithe:2002ep,
    author = "Wyithe, J. Stuart B. and Loeb, Abraham",
    title = "{Low - frequency gravitational waves from massive black hole binaries: Predictions for LISA and pulsar timing arrays}",
    eprint = "astro-ph/0211556",
    archivePrefix = "arXiv",
    doi = "10.1086/375187",
    journal = "Astrophys. J.",
    volume = "590",
    pages = "691--706",
    year = "2003"
}

@article{Sesana:2004sp,
    author = "Sesana, Alberto and Haardt, Francesco and Madau, Piero and Volonteri, Marta",
    title = "{Low - frequency gravitational radiation from coalescing massive black hole binaries in hierarchical cosmologies}",
    eprint = "astro-ph/0401543",
    archivePrefix = "arXiv",
    doi = "10.1086/422185",
    journal = "Astrophys. J.",
    volume = "611",
    pages = "623--632",
    year = "2004"
}

@article{Phinney:2001di,
    author = "Phinney, E. S.",
    title = "{A Practical theorem on gravitational wave backgrounds}",
    eprint = "astro-ph/0108028",
    archivePrefix = "arXiv",
    month = "7",
    year = "2001"
}

@article{Sato-Polito:2023gym,
    author = "Sato-Polito, Gabriela and Zaldarriaga, Matias and Quataert, Eliot",
    title = "{Where are the supermassive black holes measured by PTAs?}",
    eprint = "2312.06756",
    archivePrefix = "arXiv",
    primaryClass = "astro-ph.CO",
    doi = "10.1103/PhysRevD.110.063020",
    journal = "Phys. Rev. D",
    volume = "110",
    number = "6",
    pages = "063020",
    year = "2024"
}

@article{Sesana:2008mz,
    author = "Sesana, Alberto and Vecchio, Alberto and Colacino, Carlo Nicola",
    title = "{The stochastic gravitational-wave background from massive black hole binary systems: implications for observations with Pulsar Timing Arrays}",
    eprint = "0804.4476",
    archivePrefix = "arXiv",
    primaryClass = "astro-ph",
    doi = "10.1111/j.1365-2966.2008.13682.x",
    journal = "Mon. Not. Roy. Astron. Soc.",
    volume = "390",
    pages = "192",
    year = "2008"
}

@article{1987thyg.book..330T,
    author = "Thorne, K. S.",
    title = "{GRAVITATIONAL RADIATION}",
    year = "1987"
}

@article{Haiman:2009te,
    author = "Haiman, Zolt{\'a}n and Haiman, Zoltan and Kocsis, Bence and Kocsis, Bence and Menou, Kristen and Menou, Kristen",
    title = "{The Population of Viscosity- and Gravitational Wave-Driven Supermassive Black Hole Binaries Among Luminous AGN}",
    eprint = "0904.1383",
    archivePrefix = "arXiv",
    primaryClass = "astro-ph.CO",
    doi = "10.1088/0004-637X/700/2/1952",
    journal = "Astrophys. J.",
    volume = "700",
    pages = "1952--1969",
    year = "2009",
    note = "[Erratum: Astrophys.J. 937, 129 (2022)]"
}

@article{Amaro-Seoane:2009ucl,
    author = "Amaro-Seoane, Pau and Sesana, Alberto and Hoffman, Loren and Benacquista, Matthew and Eichhorn, Christoph and Makino, Junichiro and Spurzem, Rainer",
    title = "{Triplets of supermassive black holes: Astrophysics, Gravitational Waves and Detection}",
    eprint = "0910.1587",
    archivePrefix = "arXiv",
    primaryClass = "astro-ph.CO",
    doi = "10.1111/j.1365-2966.2009.16104.x",
    journal = "Mon. Not. Roy. Astron. Soc.",
    volume = "402",
    pages = "2308",
    year = "2010"
}

@article{Boyle:2010rt,
    author = "Boyle, Latham and Pen, Ue-Li",
    title = "{Pulsar timing arrays as imaging gravitational wave telescopes: angular resolution and source (de)confusion}",
    eprint = "1010.4337",
    archivePrefix = "arXiv",
    primaryClass = "astro-ph.HE",
    doi = "10.1103/PhysRevD.86.124028",
    journal = "Phys. Rev. D",
    volume = "86",
    pages = "124028",
    year = "2012"
}

@article{Ferranti:2024jsh,
    author = "Ferranti, Irene and Shaifullah, Golam and Chalumeau, Aurelien and Sesana, Alberto",
    title = "{Separating deterministic and stochastic gravitational wave signals in realistic pulsar timing array datasets}",
    eprint = "2407.21105",
    archivePrefix = "arXiv",
    primaryClass = "astro-ph.HE",
    doi = "10.1051/0004-6361/202451809",
    journal = "Astron. Astrophys.",
    volume = "694",
    pages = "A194",
    year = "2025"
}

@article{Kocsis:2011ch,
    author = "Kocsis, Bence and Ray, Alak and Portegies Zwart, Simon",
    title = "{Mapping the Galactic Center with Gravitational Wave measurements using Pulsar Timing}",
    eprint = "1110.6172",
    archivePrefix = "arXiv",
    primaryClass = "astro-ph.GA",
    doi = "10.1088/0004-637X/752/1/67",
    journal = "Astrophys. J.",
    volume = "752",
    pages = "67",
    year = "2012"
}

@article{Chen:2025uzf,
    author = "Chen, Xian and V{\'a}zquez-Aceves, Ver{\'o}nica and Chen, Siyuan and Lee, Kejia and Guo, Yanjun and Liu, Kuo",
    title = "{Detecting intermediate-mass black holes using miniature pulsar timing arrays in globular clusters}",
    eprint = "2507.08201",
    archivePrefix = "arXiv",
    primaryClass = "astro-ph.HE",
    doi = "10.1103/nb5d-qzm8",
    journal = "Phys. Rev. Res.",
    volume = "7",
    number = "3",
    pages = "033300",
    year = "2025"
}

@article{Fastidio:2024crh,
    author = "Fastidio, Federica and Gualandris, Alessia and Sesana, Alberto and Bortolas, Elisa and Dehnen, Walter",
    title = "{Eccentricity evolution of PTA sources from cosmological initial conditions}",
    eprint = "2406.02710",
    archivePrefix = "arXiv",
    primaryClass = "astro-ph.GA",
    doi = "10.1093/mnras/stae1411",
    journal = "Mon. Not. Roy. Astron. Soc.",
    volume = "532",
    number = "1",
    pages = "295--304",
    year = "2024"
}

@article{Begelman:1980vb,
    author = "Begelman, M. C. and Blandford, R. D. and Rees, M. J.",
    title = "{Massive black hole binaries in active galactic nuclei}",
    doi = "10.1038/287307a0",
    journal = "Nature",
    volume = "287",
    pages = "307--309",
    year = "1980"
}

@article{Dosopoulou:2016hbg,
    author = "Dosopoulou, Fani and Antonini, Fabio",
    title = "{Dynamical friction and the evolution of Supermassive Black hole Binaries: the final hundred-parsec problem}",
    eprint = "1611.06573",
    archivePrefix = "arXiv",
    primaryClass = "astro-ph.GA",
    doi = "10.3847/1538-4357/aa6b58",
    journal = "Astrophys. J.",
    volume = "840",
    number = "1",
    pages = "31",
    year = "2017"
}

@article{Colpi:1999cm,
    author = "Colpi, Monica and Mayer, Lucio and Governato, Fabio",
    title = "{Dynamical friction and the evolution of satellites in virialized halos: the theory of linear response}",
    eprint = "astro-ph/9907088",
    archivePrefix = "arXiv",
    doi = "10.1086/307952",
    journal = "Astrophys. J.",
    volume = "525",
    pages = "720",
    year = "1999"
}

@article{McWilliams:2012an,
    author = "McWilliams, Sean T. and Ostriker, Jeremiah P. and Pretorius, Frans",
    title = "{Gravitational waves and stalled satellites from massive galaxy mergers at $z \leq 1$}",
    eprint = "1211.5377",
    archivePrefix = "arXiv",
    primaryClass = "astro-ph.CO",
    doi = "10.1088/0004-637X/789/2/156",
    journal = "Astrophys. J.",
    volume = "789",
    pages = "156",
    year = "2014"
}

@article{Izquierdo-Villalba:2021prf,
    author = "Izquierdo-Villalba, David and Sesana, Alberto and Bonoli, Silvia and Colpi, Monica",
    title = "{Massive black hole evolution models confronting the n-Hz amplitude of the stochastic gravitational wave background}",
    eprint = "2108.11671",
    archivePrefix = "arXiv",
    primaryClass = "astro-ph.GA",
    doi = "10.1093/mnras/stab3239",
    journal = "Mon. Not. Roy. Astron. Soc.",
    volume = "509",
    number = "3",
    pages = "3488--3503",
    year = "2021"
}

@article{Sesana:2010qb,
    author = "Sesana, Alberto",
    title = "{Self consistent model for the evolution of eccentric massive black hole binaries in stellar environments: implications for gravitational wave observations}",
    eprint = "1006.0730",
    archivePrefix = "arXiv",
    primaryClass = "astro-ph.CO",
    doi = "10.1088/0004-637X/719/1/851",
    journal = "Astrophys. J.",
    volume = "719",
    pages = "851--864",
    year = "2010"
}

@article{Quinlan:1996vp,
    author = "Quinlan, Gerald D.",
    title = "{The dynamical evolution of massive black hole binaries - I. hardening in a fixed stellar background}",
    eprint = "astro-ph/9601092",
    archivePrefix = "arXiv",
    reportNumber = "RUTGERS-ASTROPHYSICS-PREPRINT-SERIES-NO-187",
    doi = "10.1016/S1384-1076(96)00003-6",
    journal = "New Astron.",
    volume = "1",
    pages = "35--56",
    year = "1996"
}

@article{Sesana:2006xw,
    author = "Sesana, Alberto and Haardt, Francesco and Madau, Piero",
    title = "{Interaction of massive black hole binaries with their stellar environment. 1. Ejection of hypervelocity stars}",
    eprint = "astro-ph/0604299",
    archivePrefix = "arXiv",
    doi = "10.1086/507596",
    journal = "Astrophys. J.",
    volume = "651",
    pages = "392--400",
    year = "2006"
}

@article{Milosavljevic:2002bn,
    author = "Milosavljevic, Milos and Merritt, David",
    title = "{Long term evolution of massive black hole binaries}",
    eprint = "astro-ph/0212459",
    archivePrefix = "arXiv",
    doi = "10.1086/378086",
    journal = "Astrophys. J.",
    volume = "596",
    pages = "860",
    year = "2003"
}

@article{Preto:2011gu,
    author = "Preto, Miguel and Berentzen, Ingo and Berczik, Peter and Spurzem, Rainer",
    title = "{Fast coalescence of massive black hole binaries from mergers of galactic nuclei: implications for low-frequency gravitational-wave astrophysics}",
    eprint = "1102.4855",
    archivePrefix = "arXiv",
    primaryClass = "astro-ph.GA",
    doi = "10.1088/2041-8205/732/2/L26",
    journal = "Astrophys. J. Lett.",
    volume = "732",
    pages = "L26",
    year = "2011"
}

@article{Khan:2011gi,
    author = "Khan, Fazeel Mahmood and Just, Andreas and Merritt, David",
    title = "{Efficient Merger of Binary Supermassive Black Holes in Merging Galaxies}",
    eprint = "1103.0272",
    archivePrefix = "arXiv",
    primaryClass = "astro-ph.CO",
    doi = "10.1088/0004-637X/732/2/89",
    journal = "Astrophys. J.",
    volume = "732",
    pages = "89",
    year = "2011"
}

@article{Mihos:1995ri,
    author = "Mihos, J Christopher and Hernquist, Lars",
    title = "{Gasdynamics and starbursts in major mergers}",
    eprint = "astro-ph/9512099",
    archivePrefix = "arXiv",
    doi = "10.1086/177353",
    journal = "Astrophys. J.",
    volume = "464",
    pages = "641",
    year = "1996"
}

@article{Franchini:2021uiy,
    author = "Franchini, Alessia and Sesana, Alberto and Dotti, Massimo",
    title = "{Circumbinary disc self-gravity governing supermassive black hole binary mergers}",
    eprint = "2106.13253",
    archivePrefix = "arXiv",
    primaryClass = "astro-ph.HE",
    doi = "10.1093/mnras/stab2234",
    journal = "Mon. Not. Roy. Astron. Soc.",
    volume = "507",
    number = "1",
    pages = "1458--1467",
    year = "2021"
}

@article{Nixon:2010by,
    author = "Nixon, C. J. and Cossins, P. J. and King, A. R. and Pringle, J. E.",
    title = "{Retrograde Accretion and Merging Supermassive Black Holes}",
    eprint = "1011.1914",
    archivePrefix = "arXiv",
    primaryClass = "astro-ph.HE",
    doi = "10.1111/j.1365-2966.2010.17952.x",
    journal = "Mon. Not. Roy. Astron. Soc.",
    volume = "412",
    pages = "1591",
    year = "2011"
}

@article{Siwek:2023rlk,
    author = "Siwek, Magdalena and Weinberger, Rainer and Hernquist, Lars",
    title = "{Orbital evolution of binaries in circumbinary discs}",
    eprint = "2302.01785",
    archivePrefix = "arXiv",
    primaryClass = "astro-ph.HE",
    doi = "10.1093/mnras/stad1131",
    journal = "Mon. Not. Roy. Astron. Soc.",
    volume = "522",
    number = "2",
    pages = "2707--2717",
    year = "2023"
}

@article{Ivanov:1998qk,
    author = "Ivanov, P. B. and Papaloizou, J. C. B. and Polnarev, A. G.",
    title = "{The evolution of a supermassive binary caused by an accretion disc}",
    eprint = "astro-ph/9812198",
    archivePrefix = "arXiv",
    reportNumber = "TAC-1998-027",
    doi = "10.1046/j.1365-8711.1999.02623.x",
    journal = "Mon. Not. Roy. Astron. Soc.",
    volume = "307",
    pages = "79",
    year = "1999"
}

@article{2002ApJ...578..775B,
    author = "Blaes, Omer and Lee, Man Hoi and Socrates, Aristotle",
    title = "{The kozai mechanism and the evolution of binary supermassive black holes}",
    eprint = "astro-ph/0203370",
    archivePrefix = "arXiv",
    doi = "10.1086/342655",
    journal = "Astrophys. J.",
    volume = "578",
    pages = "775--786",
    year = "2002"
}

@article{Bonetti:2017lnj,
    author = "Bonetti, Matteo and Sesana, Alberto and Barausse, Enrico and Haardt, Francesco",
    title = "{Post-Newtonian evolution of massive black hole triplets in galactic nuclei {\textendash} III. A robust lower limit to the nHz stochastic background of gravitational waves}",
    eprint = "1709.06095",
    archivePrefix = "arXiv",
    primaryClass = "astro-ph.GA",
    doi = "10.1093/mnras/sty874",
    journal = "Mon. Not. Roy. Astron. Soc.",
    volume = "477",
    number = "2",
    pages = "2599--2612",
    year = "2018"
}

@article{Sayeb:2023vav,
    author = "Sayeb, Mohammad and Blecha, Laura and Kelley, Luke Zoltan",
    title = "{MBH binary intruders: triple systems from cosmological simulations}",
    eprint = "2311.18228",
    archivePrefix = "arXiv",
    primaryClass = "astro-ph.GA",
    month = "11",
    year = "2023"
}

@article{Bonetti:2017dan,
    author = "Bonetti, Matteo and Haardt, Francesco and Sesana, Alberto and Barausse, Enrico",
    title = "{Post-Newtonian evolution of massive black hole triplets in galactic nuclei {\textendash} II. Survey of the parameter space}",
    eprint = "1709.06088",
    archivePrefix = "arXiv",
    primaryClass = "astro-ph.GA",
    doi = "10.1093/mnras/sty896",
    journal = "Mon. Not. Roy. Astron. Soc.",
    volume = "477",
    number = "3",
    pages = "3910--3926",
    year = "2018"
}

@article{Enoki:2004ew,
    author = "Enoki, Motohiro and Inoue, Kaiki Taro and Nagashima, Masahiro and Sugiyama, Naoshi",
    title = "{Gravitational waves from supermassive black hole coalescence in a hierarchical galaxy formation model}",
    eprint = "astro-ph/0404389",
    archivePrefix = "arXiv",
    doi = "10.1086/424475",
    journal = "Astrophys. J.",
    volume = "615",
    pages = "19",
    year = "2004"
}

@article{Sesana:2008xk,
    author = "Sesana, A. and Vecchio, A. and Volonteri, M.",
    title = "{Gravitational waves from resolvable massive black hole binary systems and observations with Pulsar Timing Arrays}",
    eprint = "0809.3412",
    archivePrefix = "arXiv",
    primaryClass = "astro-ph",
    doi = "10.1111/j.1365-2966.2009.14499.x",
    journal = "Mon. Not. Roy. Astron. Soc.",
    volume = "394",
    pages = "2255",
    year = "2009"
}

@article{Ravi:2012bz,
    author = "Ravi, V. and Wyithe, J. S. B. and Hobbs, G. and Shannon, R. M. and Manchester, R. N. and Yardley, D. R. B. and Keith, M. J.",
    title = "{Does a 'stochastic' background of gravitational waves exist in the pulsar timing band?}",
    eprint = "1210.3854",
    archivePrefix = "arXiv",
    primaryClass = "astro-ph.CO",
    doi = "10.1088/0004-637X/761/2/84",
    journal = "Astrophys. J.",
    volume = "761",
    pages = "84",
    year = "2012"
}

@article{Sesana:2012ak,
    author = "Sesana, A.",
    title = "{Systematic investigation of the expected gravitational wave signal from supermassive black hole binaries in the pulsar timing band}",
    eprint = "1211.5375",
    archivePrefix = "arXiv",
    primaryClass = "astro-ph.CO",
    doi = "10.1093/mnrasl/slt034",
    journal = "Mon. Not. Roy. Astron. Soc.",
    volume = "433",
    pages = "1",
    year = "2013"
}

@article{Kulier:2013gda,
    author = "Kulier, Andrea and Ostriker, Jeremiah P. and Natarajan, Priyamvada and Lackner, Claire N. and Cen, Renyue",
    title = "{Understanding black hole mass assembly via accretion and mergers at late times in cosmological simulations}",
    eprint = "1307.3684",
    archivePrefix = "arXiv",
    primaryClass = "astro-ph.CO",
    doi = "10.1088/0004-637X/799/2/178",
    journal = "Astrophys. J.",
    volume = "799",
    number = "2",
    pages = "178",
    year = "2015"
}

@article{Ravi:2014nua,
    author = "Ravi, V. and Wyithe, J. S. B. and Shannon, R. M. and Hobbs, G.",
    title = "{Prospects for gravitational-wave detection and supermassive black hole astrophysics with pulsar timing arrays}",
    eprint = "1406.5297",
    archivePrefix = "arXiv",
    primaryClass = "astro-ph.CO",
    doi = "10.1093/mnras/stu2659",
    journal = "Mon. Not. Roy. Astron. Soc.",
    volume = "447",
    pages = "2772",
    year = "2015"
}

@article{Roebber:2015iva,
    author = "Roebber, Elinore and Holder, Gilbert and Holz, Daniel E. and Warren, Michael",
    title = "{Cosmic variance in the nanohertz gravitational wave background}",
    eprint = "1508.07336",
    archivePrefix = "arXiv",
    primaryClass = "astro-ph.CO",
    doi = "10.3847/0004-637X/819/2/163",
    journal = "Astrophys. J.",
    volume = "819",
    pages = "163",
    year = "2016"
}

@article{Sesana:2016yky,
    author = "Sesana, Alberto and Shankar, Francesco and Bernardi, Mariangela and Sheth, Ravi K.",
    title = "{Selection bias in dynamically measured supermassive black hole samples: consequences for pulsar timing arrays}",
    eprint = "1603.09348",
    archivePrefix = "arXiv",
    primaryClass = "astro-ph.GA",
    doi = "10.1093/mnrasl/slw139",
    journal = "Mon. Not. Roy. Astron. Soc.",
    volume = "463",
    number = "1",
    pages = "L6--L11",
    year = "2016"
}

@article{Rasskazov:2016jjk,
    author = "Rasskazov, Alexander and Merritt, David",
    title = "{Evolution Of Massive Black Hole Binaries In Rotating Stellar Nuclei: Implications For Gravitational Wave Detection}",
    eprint = "1606.07484",
    archivePrefix = "arXiv",
    primaryClass = "astro-ph.GA",
    doi = "10.1103/PhysRevD.95.084032",
    journal = "Phys. Rev. D",
    volume = "95",
    number = "8",
    pages = "084032",
    year = "2017"
}

@article{Kelley:2017lek,
    author = "Kelley, Luke Zoltan and Blecha, Laura and Hernquist, Lars and Sesana, Alberto and Taylor, Stephen R.",
    title = "{The Gravitational Wave Background from Massive Black Hole Binaries in Illustris: spectral features and time to detection with pulsar timing arrays}",
    eprint = "1702.02180",
    archivePrefix = "arXiv",
    primaryClass = "astro-ph.HE",
    doi = "10.1093/mnras/stx1638",
    journal = "Mon. Not. Roy. Astron. Soc.",
    volume = "471",
    number = "4",
    pages = "4508--4526",
    year = "2017"
}

@article{Dvorkin:2017vvm,
    author = "Dvorkin, Irina and Barausse, Enrico",
    title = "{The nightmare scenario: measuring the stochastic gravitational-wave background from stalling massive black-hole binaries with pulsar-timing arrays}",
    eprint = "1702.06964",
    archivePrefix = "arXiv",
    primaryClass = "astro-ph.GA",
    doi = "10.1093/mnras/stx1454",
    journal = "Mon. Not. Roy. Astron. Soc.",
    volume = "470",
    number = "4",
    pages = "4547--4556",
    year = "2017"
}

@article{Ryu:2018yhv,
    author = "Ryu, Taeho and Perna, Rosalba and Haiman, Zolt{\'a}n and Ostriker, Jeremiah P. and Stone, Nicholas C.",
    title = "{Interactions between multiple supermassive black holes in galactic nuclei: a solution to the final parsec problem}",
    eprint = "1709.06501",
    archivePrefix = "arXiv",
    doi = "10.1093/mnras/stx2524",
    journal = "Mon. Not. Roy. Astron. Soc.",
    volume = "473",
    number = "3",
    pages = "3410--3433",
    year = "2018"
}

@article{Chen:2018znx,
    author = "Chen, Siyuan and Sesana, Alberto and Conselice, Christopher J.",
    title = "{Constraining astrophysical observables of Galaxy and Supermassive Black Hole Binary Mergers using Pulsar Timing Arrays}",
    eprint = "1810.04184",
    archivePrefix = "arXiv",
    primaryClass = "astro-ph.GA",
    doi = "10.1093/mnras/stz1722",
    journal = "Mon. Not. Roy. Astron. Soc.",
    volume = "488",
    number = "1",
    pages = "401--418",
    year = "2019"
}

@article{Chen:2020qlp,
    author = "Chen, Yunfeng and Yu, Qingjuan and Lu, Youjun",
    title = "{Dynamical evolution of cosmic supermassive binary black holes and their gravitational wave radiation}",
    eprint = "2005.10818",
    archivePrefix = "arXiv",
    primaryClass = "astro-ph.HE",
    doi = "10.3847/1538-4357/ab9594",
    journal = "Astrophys. J.",
    volume = "897",
    number = "1",
    pages = "86",
    year = "2020"
}

@article{Siwek:2020adv,
    author = "Siwek, Magdalena S. and Kelley, Luke Zoltan and Hernquist, Lars",
    title = "{The effect of differential accretion on the Gravitational Wave Background and the present day MBH Binary population}",
    eprint = "2005.09010",
    archivePrefix = "arXiv",
    primaryClass = "astro-ph.GA",
    doi = "10.1093/mnras/staa2361",
    journal = "Mon. Not. Roy. Astron. Soc.",
    volume = "498",
    number = "1",
    pages = "537--547",
    year = "2020"
}

@article{Simon:2023dyi,
    author = "Simon, Joseph",
    title = "{Exploring Proxies for the Supermassive Black Hole Mass Function: Implications for Pulsar Timing Arrays}",
    eprint = "2306.01832",
    archivePrefix = "arXiv",
    primaryClass = "astro-ph.GA",
    doi = "10.3847/2041-8213/acd18e",
    journal = "Astrophys. J. Lett.",
    volume = "949",
    number = "2",
    pages = "L24",
    year = "2023"
}

@article{NANOGrav:2020bcs,
    author = "Arzoumanian, Zaven and others",
    collaboration = "NANOGrav",
    title = "{The NANOGrav 12.5 yr Data Set: Search for an Isotropic Stochastic Gravitational-wave Background}",
    eprint = "2009.04496",
    archivePrefix = "arXiv",
    primaryClass = "astro-ph.HE",
    doi = "10.3847/2041-8213/abd401",
    journal = "Astrophys. J. Lett.",
    volume = "905",
    number = "2",
    pages = "L34",
    year = "2020"
}

@article{Truant:2024aci,
    author = "Truant, Riccardo J. and Izquierdo-Villalba, David and Sesana, Alberto and Shaifullah, Golam Mohiuddin and Bonetti, Matteo",
    title = "{Resolving the nano-hertz gravitational wave sky: The detectability of eccentric binaries with PTA experiments}",
    eprint = "2407.12078",
    archivePrefix = "arXiv",
    primaryClass = "astro-ph.GA",
    doi = "10.1051/0004-6361/202451556",
    journal = "Astron. Astrophys.",
    volume = "694",
    pages = "A282",
    year = "2025"
}

@article{Gardiner:2025mwf,
    author = "Gardiner, Emiko C. and B{\'e}csy, Bence and Kelley, Luke Zoltan and Cornish, Neil J.",
    title = "{Characterizing Continuous Gravitational Waves from Supermassive Black Hole Binaries in Realistic Pulsar Timing Array Data}",
    eprint = "2502.16016",
    archivePrefix = "arXiv",
    primaryClass = "astro-ph.CO",
    doi = "10.3847/1538-4357/ade4c2",
    journal = "Astrophys. J.",
    volume = "988",
    number = "2",
    pages = "222",
    year = "2025"
}

@article{Sesana:2010mx,
    author = "Sesana, A. and Vecchio, A.",
    editor = "Marka, Zsuzsa and Marka, Szabolcs",
    title = "{Gravitational waves and pulsar timing: stochastic background, individual sources and parameter estimation}",
    eprint = "1001.3161",
    archivePrefix = "arXiv",
    primaryClass = "astro-ph.CO",
    doi = "10.1088/0264-9381/27/8/084016",
    journal = "Class. Quant. Grav.",
    volume = "27",
    pages = "084016",
    year = "2010"
}

@article{Burke-Spolaor:2013aba,
    author = "Burke-Spolaor, Sarah",
    title = "{Multi-messenger approaches to binary supermassive black holes in the {\textquoteleft}continuous-wave{\textquoteright} regime}",
    eprint = "1308.4408",
    archivePrefix = "arXiv",
    primaryClass = "astro-ph.CO",
    doi = "10.1088/0264-9381/30/22/224013",
    journal = "Class. Quant. Grav.",
    volume = "30",
    pages = "224013",
    year = "2013"
}

@article{Zhou:2025lsk,
    author = "Zhou, Yihao and Di Matteo, Tiziana and Chen, Nianyi and Kelley, Luke Zoltan and Blecha, Laura and Ni, Yueying and Bird, Simeon and Yang, Yanhui and Croft, Rupert",
    title = "{Central Cluster Galaxies: A Hot Spot for Detectable Gravitational Waves from Black Hole Mergers}",
    eprint = "2502.01845",
    archivePrefix = "arXiv",
    primaryClass = "astro-ph.GA",
    doi = "10.3847/2041-8213/adf101",
    journal = "Astrophys. J. Lett.",
    volume = "988",
    number = "2",
    pages = "L74",
    year = "2025"
}

@article{Wen:2002km,
    author = "Wen, Linqing",
    title = "{On the eccentricity distribution of coalescing black hole binaries driven by the Kozai mechanism in globular clusters}",
    eprint = "astro-ph/0211492",
    archivePrefix = "arXiv",
    doi = "10.1086/378794",
    journal = "Astrophys. J.",
    volume = "598",
    pages = "419--430",
    year = "2003"
}

@article{1991PhRvL..67.1486C,
    author = "Christodoulou, D.",
    title = "{Nonlinear nature of gravitation and gravitational wave experiments}",
    doi = "10.1103/PhysRevLett.67.1486",
    journal = "Phys. Rev. Lett.",
    volume = "67",
    pages = "1486--1489",
    year = "1991"
}

@article{Agazie:2025oug,
    author = "Agazie, Gabriella and others",
    title = "{The NANOGrav 15 yr Data Set: Search for Gravitational-wave Memory}",
    eprint = "2502.18599",
    archivePrefix = "arXiv",
    primaryClass = "gr-qc",
    doi = "10.3847/1538-4357/add874",
    journal = "Astrophys. J.",
    volume = "987",
    number = "1",
    pages = "5",
    year = "2025"
}

@article{Matt:2025bao,
    author = "Matt, Cayenne and others",
    title = "{Inferring Mbh-Mbulge Evolution from the Gravitational Wave Background}",
    eprint = "2508.18126",
    archivePrefix = "arXiv",
    primaryClass = "astro-ph.HE",
    month = "8",
    year = "2025"
}

@article{Liepold:2024woa,
    author = "Liepold, Emily R. and Ma, Chung-Pei",
    title = "{Big Galaxies and Big Black Holes: The Massive Ends of the Local Stellar and Black Hole Mass Functions and the Implications for Nanohertz Gravitational Waves}",
    eprint = "2407.14595",
    archivePrefix = "arXiv",
    primaryClass = "astro-ph.GA",
    doi = "10.3847/2041-8213/ad66b8",
    journal = "Astrophys. J. Lett.",
    volume = "971",
    number = "2",
    pages = "L29",
    year = "2024"
}

@article{Grishchuk:1974ny,
    author = "Grishchuk, L. P.",
    title = "{Amplification of gravitational waves in an isotropic universe}",
    journal = "Sov. Phys. JETP",
    volume = "40",
    number = "3",
    pages = "409--415",
    year = "1975"
}

@article{Starobinsky:1979ty,
    author = "Starobinsky, Alexei A.",
    editor = "Khalatnikov, I. M. and Mineev, V. P.",
    title = "{Spectrum of relict gravitational radiation and the early state of the universe}",
    journal = "JETP Lett.",
    volume = "30",
    pages = "682--685",
    year = "1979"
}

@article{Rubakov:1982df,
    author = "Rubakov, V. A. and Sazhin, M. V. and Veryaskin, A. V.",
    title = "{Graviton Creation in the Inflationary Universe and the Grand Unification Scale}",
    doi = "10.1016/0370-2693(82)90641-4",
    journal = "Phys. Lett. B",
    volume = "115",
    pages = "189--192",
    year = "1982"
}

@article{Fabbri:1983us,
    author = "Fabbri, R. and Pollock, M. d.",
    title = "{The Effect of Primordially Produced Gravitons upon the Anisotropy of the Cosmological Microwave Background Radiation}",
    doi = "10.1016/0370-2693(83)91322-9",
    journal = "Phys. Lett. B",
    volume = "125",
    pages = "445--448",
    year = "1983"
}

@article{Anber:2006xt,
    author = "Anber, Mohamed M. and Sorbo, Lorenzo",
    title = "{N-flationary magnetic fields}",
    eprint = "astro-ph/0606534",
    archivePrefix = "arXiv",
    doi = "10.1088/1475-7516/2006/10/018",
    journal = "JCAP",
    volume = "10",
    pages = "018",
    year = "2006"
}

@article{Barnaby:2011qe,
    author = "Barnaby, Neil and Pajer, Enrico and Peloso, Marco",
    title = "{Gauge Field Production in Axion Inflation: Consequences for Monodromy, non-Gaussianity in the CMB, and Gravitational Waves at Interferometers}",
    eprint = "1110.3327",
    archivePrefix = "arXiv",
    primaryClass = "astro-ph.CO",
    reportNumber = "UMN-TH-3017-11",
    doi = "10.1103/PhysRevD.85.023525",
    journal = "Phys. Rev. D",
    volume = "85",
    pages = "023525",
    year = "2012"
}

@article{Adshead:2013qp,
    author = "Adshead, Peter and Martinec, Emil and Wyman, Mark",
    title = "{Gauge fields and inflation: Chiral gravitational waves, fluctuations, and the Lyth bound}",
    eprint = "1301.2598",
    archivePrefix = "arXiv",
    primaryClass = "hep-th",
    doi = "10.1103/PhysRevD.88.021302",
    journal = "Phys. Rev. D",
    volume = "88",
    number = "2",
    pages = "021302",
    year = "2013"
}

@article{Namba:2015gja,
    author = "Namba, Ryo and Peloso, Marco and Shiraishi, Maresuke and Sorbo, Lorenzo and Unal, Caner",
    title = "{Scale-dependent gravitational waves from a rolling axion}",
    eprint = "1509.07521",
    archivePrefix = "arXiv",
    primaryClass = "astro-ph.CO",
    doi = "10.1088/1475-7516/2016/01/041",
    journal = "JCAP",
    volume = "01",
    pages = "041",
    year = "2016"
}

@article{Maleknejad:2016qjz,
    author = "Maleknejad, Azadeh",
    title = "{Axion Inflation with an SU(2) Gauge Field: Detectable Chiral Gravity Waves}",
    eprint = "1604.03327",
    archivePrefix = "arXiv",
    primaryClass = "hep-ph",
    doi = "10.1007/JHEP07(2016)104",
    journal = "JHEP",
    volume = "07",
    pages = "104",
    year = "2016"
}

@article{Dimastrogiovanni:2016fuu,
    author = "Dimastrogiovanni, Emanuela and Fasiello, Matteo and Fujita, Tomohiro",
    title = "{Primordial Gravitational Waves from Axion-Gauge Fields Dynamics}",
    eprint = "1608.04216",
    archivePrefix = "arXiv",
    primaryClass = "astro-ph.CO",
    doi = "10.1088/1475-7516/2017/01/019",
    journal = "JCAP",
    volume = "01",
    pages = "019",
    year = "2017"
}

@article{Domcke:2016bkh,
    author = "Domcke, Valerie and Pieroni, Mauro and Bin{\'e}truy, Pierre",
    title = "{Primordial gravitational waves for universality classes of pseudoscalar inflation}",
    eprint = "1603.01287",
    archivePrefix = "arXiv",
    primaryClass = "astro-ph.CO",
    doi = "10.1088/1475-7516/2016/06/031",
    journal = "JCAP",
    volume = "06",
    pages = "031",
    year = "2016"
}

@article{Guzzetti:2016mkm,
    author = "Guzzetti, M. C. and Bartolo, N. and Liguori, M. and Matarrese, S.",
    title = "{Gravitational waves from inflation}",
    eprint = "1605.01615",
    archivePrefix = "arXiv",
    primaryClass = "astro-ph.CO",
    doi = "10.1393/ncr/i2016-10127-1",
    journal = "Riv. Nuovo Cim.",
    volume = "39",
    number = "9",
    pages = "399--495",
    year = "2016"
}

@article{Bartolo:2016ami,
    author = "Bartolo, Nicola and others",
    title = "{Science with the space-based interferometer LISA. IV: Probing inflation with gravitational waves}",
    eprint = "1610.06481",
    archivePrefix = "arXiv",
    primaryClass = "astro-ph.CO",
    reportNumber = "ACFI-T16-19, UMN-TH-3608-16, CERN-TH-2016-222, KCL-PH-TH-2016-58, IFT-UAM-CSIC-16-104",
    doi = "10.1088/1475-7516/2016/12/026",
    journal = "JCAP",
    volume = "12",
    pages = "026",
    year = "2016"
}

@article{Garcia-Bellido:2017aan,
    author = "Garcia-Bellido, Juan and Peloso, Marco and Unal, Caner",
    title = "{Gravitational Wave signatures of inflationary models from Primordial Black Hole Dark Matter}",
    eprint = "1707.02441",
    archivePrefix = "arXiv",
    primaryClass = "astro-ph.CO",
    reportNumber = "IFT-UAM-CSIC-17-056, UMN-TH-3630-17, IFT--UAM-CSIC-17-056",
    doi = "10.1088/1475-7516/2017/09/013",
    journal = "JCAP",
    volume = "09",
    pages = "013",
    year = "2017"
}

@article{Garcia-Bellido:2023ser,
    author = "Garcia-Bellido, Juan and Papageorgiou, Alexandros and Peloso, Marco and Sorbo, Lorenzo",
    title = "{A flashing beacon in axion inflation: recurring bursts of gravitational waves in the strong backreaction regime}",
    eprint = "2303.13425",
    archivePrefix = "arXiv",
    primaryClass = "astro-ph.CO",
    doi = "10.1088/1475-7516/2024/01/034",
    journal = "JCAP",
    volume = "01",
    pages = "034",
    year = "2024"
}

@article{Figueroa:2023oxc,
    author = "Figueroa, Daniel G. and Lizarraga, Joanes and Urio, Ander and Urrestilla, Jon",
    title = "{Strong Backreaction Regime in Axion Inflation}",
    eprint = "2303.17436",
    archivePrefix = "arXiv",
    primaryClass = "astro-ph.CO",
    doi = "10.1103/PhysRevLett.131.151003",
    journal = "Phys. Rev. Lett.",
    volume = "131",
    number = "15",
    pages = "151003",
    year = "2023"
}

@article{Figueroa:2024rkr,
    author = "Figueroa, Daniel G. and Lizarraga, Joanes and Loayza, Nicol{\'a}s and Urio, Ander and Urrestilla, Jon",
    title = "{Nonlinear dynamics of axion inflation: A detailed lattice study}",
    eprint = "2411.16368",
    archivePrefix = "arXiv",
    primaryClass = "astro-ph.CO",
    doi = "10.1103/PhysRevD.111.063545",
    journal = "Phys. Rev. D",
    volume = "111",
    number = "6",
    pages = "063545",
    year = "2025"
}

@article{Sharma:2024nfu,
    author = "Sharma, Ramkishor and Brandenburg, Axel and Subramanian, Kandaswamy and Vikman, Alexander",
    title = "{Lattice simulations of axion-U(1) inflation: gravitational waves, magnetic fields, and scalar statistics}",
    eprint = "2411.04854",
    archivePrefix = "arXiv",
    primaryClass = "astro-ph.CO",
    reportNumber = "NORDITA-2024-040",
    doi = "10.1088/1475-7516/2025/05/079",
    journal = "JCAP",
    volume = "05",
    pages = "079",
    year = "2025"
}

@article{Matarrese:1992rp,
    author = "Matarrese, Sabino and Pantano, Ornella and Saez, Diego",
    title = "{A General relativistic approach to the nonlinear evolution of collisionless matter}",
    reportNumber = "DFPD-92-A-39",
    doi = "10.1103/PhysRevD.47.1311",
    journal = "Phys. Rev. D",
    volume = "47",
    pages = "1311--1323",
    year = "1993"
}

@article{Matarrese:1993zf,
    author = "Matarrese, Sabino and Pantano, Ornella and Saez, Diego",
    title = "{General relativistic dynamics of irrotational dust: Cosmological implications}",
    eprint = "astro-ph/9310036",
    archivePrefix = "arXiv",
    reportNumber = "DFPD-93-A-67",
    doi = "10.1103/PhysRevLett.72.320",
    journal = "Phys. Rev. Lett.",
    volume = "72",
    pages = "320--323",
    year = "1994"
}

@article{Matarrese:1997ay,
    author = "Matarrese, Sabino and Mollerach, Silvia and Bruni, Marco",
    title = "{Second order perturbations of the Einstein-de Sitter universe}",
    eprint = "astro-ph/9707278",
    archivePrefix = "arXiv",
    reportNumber = "SISSA-83-97-A",
    doi = "10.1103/PhysRevD.58.043504",
    journal = "Phys. Rev. D",
    volume = "58",
    pages = "043504",
    year = "1998"
}

@article{Nakamura:2004rm,
    author = "Nakamura, Kouji",
    title = "{Second-order gauge invariant cosmological perturbation theory: Einstein equations in terms of gauge invariant variables}",
    eprint = "gr-qc/0605108",
    archivePrefix = "arXiv",
    doi = "10.1143/PTP.117.17",
    journal = "Prog. Theor. Phys.",
    volume = "117",
    pages = "17--74",
    year = "2007"
}

@article{Ananda:2006af,
    author = "Ananda, Kishore N. and Clarkson, Chris and Wands, David",
    title = "{The Cosmological gravitational wave background from primordial density perturbations}",
    eprint = "gr-qc/0612013",
    archivePrefix = "arXiv",
    doi = "10.1103/PhysRevD.75.123518",
    journal = "Phys. Rev. D",
    volume = "75",
    pages = "123518",
    year = "2007"
}

@article{Baumann:2007zm,
    author = "Baumann, Daniel and Steinhardt, Paul J. and Takahashi, Keitaro and Ichiki, Kiyotomo",
    title = "{Gravitational Wave Spectrum Induced by Primordial Scalar Perturbations}",
    eprint = "hep-th/0703290",
    archivePrefix = "arXiv",
    doi = "10.1103/PhysRevD.76.084019",
    journal = "Phys. Rev. D",
    volume = "76",
    pages = "084019",
    year = "2007"
}

@article{Unal:2018yaa,
    author = "Unal, Caner",
    title = "{Imprints of Primordial Non-Gaussianity on Gravitational Wave Spectrum}",
    eprint = "1811.09151",
    archivePrefix = "arXiv",
    primaryClass = "astro-ph.CO",
    doi = "10.1103/PhysRevD.99.041301",
    journal = "Phys. Rev. D",
    volume = "99",
    number = "4",
    pages = "041301",
    year = "2019"
}

@article{Adshead:2021hnm,
    author = "Adshead, Peter and Lozanov, Kaloian D. and Weiner, Zachary J.",
    title = "{Non-Gaussianity and the induced gravitational wave background}",
    eprint = "2105.01659",
    archivePrefix = "arXiv",
    primaryClass = "astro-ph.CO",
    doi = "10.1088/1475-7516/2021/10/080",
    journal = "JCAP",
    volume = "10",
    pages = "080",
    year = "2021"
}

@article{Domenech:2021ztg,
    author = "Dom{\`e}nech, Guillem",
    title = "{Scalar Induced Gravitational Waves Review}",
    eprint = "2109.01398",
    archivePrefix = "arXiv",
    primaryClass = "gr-qc",
    doi = "10.3390/universe7110398",
    journal = "Universe",
    volume = "7",
    number = "11",
    pages = "398",
    year = "2021"
}

@article{Easther:2006gt,
    author = "Easther, Richard and Lim, Eugene A.",
    title = "{Stochastic gravitational wave production after inflation}",
    eprint = "astro-ph/0601617",
    archivePrefix = "arXiv",
    doi = "10.1088/1475-7516/2006/04/010",
    journal = "JCAP",
    volume = "04",
    pages = "010",
    year = "2006"
}

@article{GarciaBellido:2007dg,
    author = "Garcia-Bellido, Juan and Figueroa, Daniel G.",
    title = "{A stochastic background of gravitational waves from hybrid preheating}",
    eprint = "astro-ph/0701014",
    archivePrefix = "arXiv",
    reportNumber = "IFT-UAM-CSIC-06-46",
    doi = "10.1103/PhysRevLett.98.061302",
    journal = "Phys. Rev. Lett.",
    volume = "98",
    pages = "061302",
    year = "2007"
}

@article{GarciaBellido:2007af,
    author = "Garcia-Bellido, Juan and Figueroa, Daniel G. and Sastre, Alfonso",
    title = "{A Gravitational Wave Background from Reheating after Hybrid Inflation}",
    eprint = "0707.0839",
    archivePrefix = "arXiv",
    primaryClass = "hep-ph",
    reportNumber = "IFT-UAM-CSIC-07-38",
    doi = "10.1103/PhysRevD.77.043517",
    journal = "Phys. Rev. D",
    volume = "77",
    pages = "043517",
    year = "2008"
}

@article{Dufaux:2007pt,
    author = "Dufaux, Jean Francois and Bergman, Amanda and Felder, Gary N. and Kofman, Lev and Uzan, Jean-Philippe",
    title = "{Theory and Numerics of Gravitational Waves from Preheating after Inflation}",
    eprint = "0707.0875",
    archivePrefix = "arXiv",
    primaryClass = "astro-ph",
    doi = "10.1103/PhysRevD.76.123517",
    journal = "Phys. Rev. D",
    volume = "76",
    pages = "123517",
    year = "2007"
}

@article{Dufaux:2008dn,
    author = "Dufaux, Jean-Francois and Felder, Gary and Kofman, Lev and Navros, Olga",
    title = "{Gravity Waves from Tachyonic Preheating after Hybrid Inflation}",
    eprint = "0812.2917",
    archivePrefix = "arXiv",
    primaryClass = "astro-ph",
    reportNumber = "FTUAM-08-25, IFT-UAM-CSIC-08-90",
    doi = "10.1088/1475-7516/2009/03/001",
    journal = "JCAP",
    volume = "03",
    pages = "001",
    year = "2009"
}

@article{Dufaux:2010cf,
    author = "Dufaux, Jean-Francois and Figueroa, Daniel G. and Garcia-Bellido, Juan",
    title = "{Gravitational Waves from Abelian Gauge Fields and Cosmic Strings at Preheating}",
    eprint = "1006.0217",
    archivePrefix = "arXiv",
    primaryClass = "astro-ph.CO",
    reportNumber = "IFT-UAM-CSIC-10-38, CERN-PH-TH-2010-121",
    doi = "10.1103/PhysRevD.82.083518",
    journal = "Phys. Rev. D",
    volume = "82",
    pages = "083518",
    year = "2010"
}

@article{Bethke:2013aba,
    author = "Bethke, Laura and Figueroa, Daniel G. and Rajantie, Arttu",
    title = "{Anisotropies in the Gravitational Wave Background from Preheating}",
    eprint = "1304.2657",
    archivePrefix = "arXiv",
    primaryClass = "astro-ph.CO",
    reportNumber = "IMPERIAL-TP-2013-LB-1",
    doi = "10.1103/PhysRevLett.111.011301",
    journal = "Phys. Rev. Lett.",
    volume = "111",
    number = "1",
    pages = "011301",
    year = "2013"
}

@article{Bethke:2013vca,
    author = "Bethke, Laura and Figueroa, Daniel G. and Rajantie, Arttu",
    title = "{On the Anisotropy of the Gravitational Wave Background from Massless Preheating}",
    eprint = "1309.1148",
    archivePrefix = "arXiv",
    primaryClass = "astro-ph.CO",
    doi = "10.1088/1475-7516/2014/06/047",
    journal = "JCAP",
    volume = "06",
    pages = "047",
    year = "2014"
}

@article{Figueroa:2017vfa,
    author = "Figueroa, Daniel G. and Torrenti, Francisco",
    title = "{Gravitational wave production from preheating: parameter dependence}",
    eprint = "1707.04533",
    archivePrefix = "arXiv",
    primaryClass = "astro-ph.CO",
    reportNumber = "CERN-TH-2017-152, IFT-UAM-CSIC-17-069",
    doi = "10.1088/1475-7516/2017/10/057",
    journal = "JCAP",
    volume = "10",
    pages = "057",
    year = "2017"
}

@article{Adshead:2019igv,
    author = "Adshead, Peter and Giblin, John T. and Pieroni, Mauro and Weiner, Zachary J.",
    title = "{Constraining Axion Inflation with Gravitational Waves across 29 Decades in Frequency}",
    eprint = "1909.12843",
    archivePrefix = "arXiv",
    primaryClass = "astro-ph.CO",
    doi = "10.1103/PhysRevLett.124.171301",
    journal = "Phys. Rev. Lett.",
    volume = "124",
    number = "17",
    pages = "17",
    year = "2020"
}

@article{Cosme:2022htl,
    author = "Cosme, Catarina and Figueroa, Daniel G. and Loayza, Nicolas",
    title = "{Gravitational wave production from preheating with trilinear interactions}",
    eprint = "2206.14721",
    archivePrefix = "arXiv",
    primaryClass = "astro-ph.CO",
    doi = "10.1088/1475-7516/2023/05/023",
    journal = "JCAP",
    volume = "05",
    pages = "023",
    year = "2023"
}

@article{Antusch:2025ewc,
    author = "Antusch, Stefan and Marschall, Kenneth and Torrenti, Francisco",
    title = "{Equation of state during (p)reheating with trilinear interactions}",
    eprint = "2507.13465",
    archivePrefix = "arXiv",
    primaryClass = "astro-ph.CO",
    doi = "10.1088/1475-7516/2025/11/002",
    journal = "JCAP",
    volume = "11",
    pages = "002",
    year = "2025"
}

@article{Giovannini:1998bp,
    author = "Giovannini, Massimo",
    title = "{Gravitational waves constraints on postinflationary phases stiffer than radiation}",
    eprint = "hep-ph/9806329",
    archivePrefix = "arXiv",
    doi = "10.1103/PhysRevD.58.083504",
    journal = "Phys. Rev. D",
    volume = "58",
    pages = "083504",
    year = "1998"
}

@article{Giovannini:1999bh,
    author = "Giovannini, Massimo",
    title = "{Production and detection of relic gravitons in quintessential inflationary models}",
    eprint = "astro-ph/9903004",
    archivePrefix = "arXiv",
    reportNumber = "TUPT-01-99",
    doi = "10.1103/PhysRevD.60.123511",
    journal = "Phys. Rev. D",
    volume = "60",
    pages = "123511",
    year = "1999"
}

@article{Boyle:2007zx,
    author = "Boyle, Latham A. and Buonanno, Alessandra",
    title = "{Relating gravitational wave constraints from primordial nucleosynthesis, pulsar timing, laser interferometers, and the CMB: Implications for the early Universe}",
    eprint = "0708.2279",
    archivePrefix = "arXiv",
    primaryClass = "astro-ph",
    doi = "10.1103/PhysRevD.78.043531",
    journal = "Phys. Rev. D",
    volume = "78",
    pages = "043531",
    year = "2008"
}

@article{Li:2016mmc,
    author = "Li, Bohua and Shapiro, Paul R. and Rindler-Daller, Tanja",
    title = "{Bose-Einstein-condensed scalar field dark matter and the gravitational wave background from inflation: new cosmological constraints and its detectability by LIGO}",
    eprint = "1611.07961",
    archivePrefix = "arXiv",
    primaryClass = "astro-ph.CO",
    doi = "10.1103/PhysRevD.96.063505",
    journal = "Phys. Rev. D",
    volume = "96",
    number = "6",
    pages = "063505",
    year = "2017"
}

@article{Li:2021htg,
    author = "Li, Bohua and Shapiro, Paul R.",
    title = "{Precision cosmology and the stiff-amplified gravitational-wave background from inflation: NANOGrav, Advanced LIGO-Virgo and the Hubble tension}",
    eprint = "2107.12229",
    archivePrefix = "arXiv",
    primaryClass = "astro-ph.CO",
    doi = "10.1088/1475-7516/2021/10/024",
    journal = "JCAP",
    volume = "10",
    pages = "024",
    year = "2021"
}

@article{Figueroa:2018twl,
    author = "Figueroa, Daniel G. and Tanin, Erwin H.",
    title = "{Inconsistency of an inflationary sector coupled only to Einstein gravity}",
    eprint = "1811.04093",
    archivePrefix = "arXiv",
    primaryClass = "astro-ph.CO",
    doi = "10.1088/1475-7516/2019/10/050",
    journal = "JCAP",
    volume = "10",
    pages = "050",
    year = "2019"
}

@article{Figueroa:2019paj,
    author = "Figueroa, Daniel G. and Tanin, Erwin H.",
    title = "{Ability of LIGO and LISA to probe the equation of state of the early Universe}",
    eprint = "1905.11960",
    archivePrefix = "arXiv",
    primaryClass = "astro-ph.CO",
    doi = "10.1088/1475-7516/2019/08/011",
    journal = "JCAP",
    volume = "08",
    pages = "011",
    year = "2019"
}

@article{Gouttenoire:2021wzu,
    author = "Gouttenoire, Yann and Servant, G{\'e}raldine and Simakachorn, Peera",
    title = "{Revealing the Primordial Irreducible Inflationary Gravitational-Wave Background with a Spinning Peccei-Quinn Axion}",
    eprint = "2108.10328",
    archivePrefix = "arXiv",
    primaryClass = "hep-ph",
    reportNumber = "DESY 21-126",
    month = "8",
    year = "2021"
}

@article{Co:2021lkc,
    author = "Co, Raymond T. and Dunsky, David and Fernandez, Nicolas and Ghalsasi, Akshay and Hall, Lawrence J. and Harigaya, Keisuke and Shelton, Jessie",
    title = "{Gravitational wave and CMB probes of axion kination}",
    eprint = "2108.09299",
    archivePrefix = "arXiv",
    primaryClass = "hep-ph",
    reportNumber = "UMN-TH-4023/21, FTPI-MINN-21-15, CERN-TH-2021-124",
    doi = "10.1007/JHEP09(2022)116",
    journal = "JHEP",
    volume = "09",
    pages = "116",
    year = "2022"
}

@article{Gouttenoire:2021jhk,
    author = "Gouttenoire, Yann and Servant, Geraldine and Simakachorn, Peera",
    title = "{Kination cosmology from scalar fields and gravitational-wave signatures}",
    eprint = "2111.01150",
    archivePrefix = "arXiv",
    primaryClass = "hep-ph",
    reportNumber = "DESY 21-134",
    month = "11",
    year = "2021"
}

@article{Oikonomou:2023qfz,
    author = "Oikonomou, V. K.",
    title = "{Flat energy spectrum of primordial gravitational waves versus peaks and the NANOGrav 2023 observation}",
    eprint = "2306.17351",
    archivePrefix = "arXiv",
    primaryClass = "astro-ph.CO",
    doi = "10.1103/PhysRevD.108.043516",
    journal = "Phys. Rev. D",
    volume = "108",
    number = "4",
    pages = "043516",
    year = "2023"
}

@article{Ghiglieri:2015nfa,
    author = "Ghiglieri, J. and Laine, M.",
    title = "{Gravitational wave background from Standard Model physics: Qualitative features}",
    eprint = "1504.02569",
    archivePrefix = "arXiv",
    primaryClass = "hep-ph",
    doi = "10.1088/1475-7516/2015/07/022",
    journal = "JCAP",
    volume = "07",
    pages = "022",
    year = "2015"
}

@article{Ghiglieri:2020mhm,
    author = "Ghiglieri, J. and Jackson, G. and Laine, M. and Zhu, Y.",
    title = "{Gravitational wave background from Standard Model physics: Complete leading order}",
    eprint = "2004.11392",
    archivePrefix = "arXiv",
    primaryClass = "hep-ph",
    doi = "10.1007/JHEP07(2020)092",
    journal = "JHEP",
    volume = "07",
    pages = "092",
    year = "2020"
}

@article{Ringwald:2020ist,
    author = {Ringwald, Andreas and Sch{\"u}tte-Engel, Jan and Tamarit, Carlos},
    title = "{Gravitational Waves as a Big Bang Thermometer}",
    eprint = "2011.04731",
    archivePrefix = "arXiv",
    primaryClass = "hep-ph",
    reportNumber = "DESY 20-187, DESY-20-187, TUM-HEP-1293-20",
    doi = "10.1088/1475-7516/2021/03/054",
    journal = "JCAP",
    volume = "03",
    pages = "054",
    year = "2021"
}

@article{Ghiglieri:2022rfp,
    author = {Ghiglieri, Jacopo and Sch{\"u}tte-Engel, Jan and Speranza, Enrico},
    title = "{Freezing-in gravitational waves}",
    eprint = "2211.16513",
    archivePrefix = "arXiv",
    primaryClass = "hep-ph",
    doi = "10.1103/PhysRevD.109.023538",
    journal = "Phys. Rev. D",
    volume = "109",
    number = "2",
    pages = "023538",
    year = "2024"
}

@article{Zhou:2013tsa,
    author = "Zhou, Shuang-Yong and Copeland, Edmund J. and Easther, Richard and Finkel, Hal and Mou, Zong-Gang and Saffin, Paul M.",
    title = "{Gravitational Waves from Oscillon Preheating}",
    eprint = "1304.6094",
    archivePrefix = "arXiv",
    primaryClass = "astro-ph.CO",
    doi = "10.1007/JHEP10(2013)026",
    journal = "JHEP",
    volume = "10",
    pages = "026",
    year = "2013"
}

@article{Antusch:2016con,
    author = "Antusch, Stefan and Cefala, Francesco and Orani, Stefano",
    title = "{Gravitational waves from oscillons after inflation}",
    eprint = "1607.01314",
    archivePrefix = "arXiv",
    primaryClass = "astro-ph.CO",
    doi = "10.1103/PhysRevLett.118.011303",
    journal = "Phys. Rev. Lett.",
    volume = "118",
    number = "1",
    pages = "011303",
    year = "2017",
    note = "[Erratum: Phys.Rev.Lett. 120, 219901 (2018)]"
}

@article{Antusch:2017vga,
    author = "Antusch, Stefan and Cefala, Francesco and Orani, Stefano",
    title = "{What can we learn from the stochastic gravitational wave background produced by oscillons?}",
    eprint = "1712.03231",
    archivePrefix = "arXiv",
    primaryClass = "astro-ph.CO",
    doi = "10.1088/1475-7516/2018/03/032",
    journal = "JCAP",
    volume = "03",
    pages = "032",
    year = "2018"
}

@article{Liu:2017hua,
    author = "Liu, Jing and Guo, Zong-Kuan and Cai, Rong-Gen and Shiu, Gary",
    title = "{Gravitational Waves from Oscillons with Cuspy Potentials}",
    eprint = "1707.09841",
    archivePrefix = "arXiv",
    primaryClass = "astro-ph.CO",
    doi = "10.1103/PhysRevLett.120.031301",
    journal = "Phys. Rev. Lett.",
    volume = "120",
    number = "3",
    pages = "031301",
    year = "2018"
}

@article{Amin:2018xfe,
    author = "Amin, Mustafa A. and Braden, Jonathan and Copeland, Edmund J. and Giblin, John T. and Solorio, Christian and Weiner, Zachary J. and Zhou, Shuang-Yong",
    title = "{Gravitational waves from asymmetric oscillon dynamics?}",
    eprint = "1803.08047",
    archivePrefix = "arXiv",
    primaryClass = "astro-ph.CO",
    doi = "10.1103/PhysRevD.98.024040",
    journal = "Phys. Rev. D",
    volume = "98",
    pages = "024040",
    year = "2018"
}

@article{Kamionkowski:1993fg,
    author = "Kamionkowski, Marc and Kosowsky, Arthur and Turner, Michael S.",
    title = "{Gravitational radiation from first order phase transitions}",
    eprint = "astro-ph/9310044",
    archivePrefix = "arXiv",
    reportNumber = "IASSNS-HEP-93-44, FERMILAB-PUB-93-235-A",
    doi = "10.1103/PhysRevD.49.2837",
    journal = "Phys. Rev. D",
    volume = "49",
    pages = "2837--2851",
    year = "1994"
}

@article{Caprini:2007xq,
    author = "Caprini, Chiara and Durrer, Ruth and Servant, Geraldine",
    title = "{Gravitational wave generation from bubble collisions in first-order phase transitions: An analytic approach}",
    eprint = "0711.2593",
    archivePrefix = "arXiv",
    primaryClass = "astro-ph",
    reportNumber = "CERN-PH-TH-2007-206, SACLAY-T07-142",
    doi = "10.1103/PhysRevD.77.124015",
    journal = "Phys. Rev. D",
    volume = "77",
    pages = "124015",
    year = "2008"
}

@article{Huber:2008hg,
    author = "Huber, Stephan J. and Konstandin, Thomas",
    title = "{Gravitational Wave Production by Collisions: More Bubbles}",
    eprint = "0806.1828",
    archivePrefix = "arXiv",
    primaryClass = "hep-ph",
    doi = "10.1088/1475-7516/2008/09/022",
    journal = "JCAP",
    volume = "09",
    pages = "022",
    year = "2008"
}

@article{Hindmarsh:2013xza,
    author = "Hindmarsh, Mark and Huber, Stephan J. and Rummukainen, Kari and Weir, David J.",
    title = "{Gravitational waves from the sound of a first order phase transition}",
    eprint = "1304.2433",
    archivePrefix = "arXiv",
    primaryClass = "hep-ph",
    reportNumber = "HIP-2013-07-TH",
    doi = "10.1103/PhysRevLett.112.041301",
    journal = "Phys. Rev. Lett.",
    volume = "112",
    pages = "041301",
    year = "2014"
}

@article{Hindmarsh:2015qta,
    author = "Hindmarsh, Mark and Huber, Stephan J. and Rummukainen, Kari and Weir, David J.",
    title = "{Numerical simulations of acoustically generated gravitational waves at a first order phase transition}",
    eprint = "1504.03291",
    archivePrefix = "arXiv",
    primaryClass = "astro-ph.CO",
    reportNumber = "HIP-2015-13-TH",
    doi = "10.1103/PhysRevD.92.123009",
    journal = "Phys. Rev. D",
    volume = "92",
    number = "12",
    pages = "123009",
    year = "2015"
}

@article{Caprini:2015zlo,
    author = "Caprini, Chiara and others",
    title = "{Science with the space-based interferometer eLISA. II: Gravitational waves from cosmological phase transitions}",
    eprint = "1512.06239",
    archivePrefix = "arXiv",
    primaryClass = "astro-ph.CO",
    reportNumber = "DESY-15-246",
    doi = "10.1088/1475-7516/2016/04/001",
    journal = "JCAP",
    volume = "04",
    pages = "001",
    year = "2016"
}

@article{Hindmarsh:2017gnf,
    author = "Hindmarsh, Mark and Huber, Stephan J. and Rummukainen, Kari and Weir, David J.",
    title = "{Shape of the acoustic gravitational wave power spectrum from a first order phase transition}",
    eprint = "1704.05871",
    archivePrefix = "arXiv",
    primaryClass = "astro-ph.CO",
    reportNumber = "HIP-2017-02-TH, HIP-2017-02/TH",
    doi = "10.1103/PhysRevD.96.103520",
    journal = "Phys. Rev. D",
    volume = "96",
    number = "10",
    pages = "103520",
    year = "2017",
    note = "[Erratum: Phys.Rev.D 101, 089902 (2020)]"
}

@article{Cutting:2018tjt,
    author = "Cutting, Daniel and Hindmarsh, Mark and Weir, David J.",
    title = "{Gravitational waves from vacuum first-order phase transitions: from the envelope to the lattice}",
    eprint = "1802.05712",
    archivePrefix = "arXiv",
    primaryClass = "astro-ph.CO",
    reportNumber = "HIP-2018-4-TH",
    doi = "10.1103/PhysRevD.97.123513",
    journal = "Phys. Rev. D",
    volume = "97",
    number = "12",
    pages = "123513",
    year = "2018"
}

@article{Cutting:2019zws,
    author = "Cutting, Daniel and Hindmarsh, Mark and Weir, David J.",
    title = "{Vorticity, kinetic energy, and suppressed gravitational wave production in strong first order phase transitions}",
    eprint = "1906.00480",
    archivePrefix = "arXiv",
    primaryClass = "hep-ph",
    reportNumber = "HIP-2019-15/TH",
    doi = "10.1103/PhysRevLett.125.021302",
    journal = "Phys. Rev. Lett.",
    volume = "125",
    number = "2",
    pages = "021302",
    year = "2020"
}

@article{Pol:2019yex,
    author = "Roper Pol, Alberto and Mandal, Sayan and Brandenburg, Axel and Kahniashvili, Tina and Kosowsky, Arthur",
    title = "{Numerical simulations of gravitational waves from early-universe turbulence}",
    eprint = "1903.08585",
    archivePrefix = "arXiv",
    primaryClass = "astro-ph.CO",
    reportNumber = "NORDITA-2019-024",
    doi = "10.1103/PhysRevD.102.083512",
    journal = "Phys. Rev. D",
    volume = "102",
    number = "8",
    pages = "083512",
    year = "2020"
}

@article{Caprini:2019egz,
    author = "Caprini, Chiara and others",
    title = "{Detecting gravitational waves from cosmological phase transitions with LISA: an update}",
    eprint = "1910.13125",
    archivePrefix = "arXiv",
    primaryClass = "astro-ph.CO",
    reportNumber = "DESY-19-159, IPPP/19/27, HIP-2019-14/TH, MITP/19-066, IFT-UAM/CSIC-19-139",
    doi = "10.1088/1475-7516/2020/03/024",
    journal = "JCAP",
    volume = "03",
    pages = "024",
    year = "2020"
}

@article{Cutting:2020nla,
    author = "Cutting, Daniel and Escartin, Elba Granados and Hindmarsh, Mark and Weir, David J.",
    title = "{Gravitational waves from vacuum first order phase transitions II: from thin to thick walls}",
    eprint = "2005.13537",
    archivePrefix = "arXiv",
    primaryClass = "astro-ph.CO",
    reportNumber = "HIP-2020-13/TH",
    doi = "10.1103/PhysRevD.103.023531",
    journal = "Phys. Rev. D",
    volume = "103",
    number = "2",
    pages = "023531",
    year = "2021"
}

@article{Han:2023olf,
    author = "Han, Chengcheng and Xie, Ke-Pan and Yang, Jin Min and Zhang, Mengchao",
    title = "{Self-interacting dark matter implied by nano-Hertz gravitational waves}",
    eprint = "2306.16966",
    archivePrefix = "arXiv",
    primaryClass = "hep-ph",
    doi = "10.1103/PhysRevD.109.115025",
    journal = "Phys. Rev. D",
    volume = "109",
    number = "11",
    pages = "115025",
    year = "2024"
}

@article{Ashoorioon:2022raz,
    author = "Ashoorioon, Amjad and Rezazadeh, Kazem and Rostami, Abasalt",
    title = "{NANOGrav signal from the end of inflation and the LIGO mass and heavier primordial black holes}",
    eprint = "2202.01131",
    archivePrefix = "arXiv",
    primaryClass = "astro-ph.CO",
    reportNumber = "IPM/P-2022/06",
    doi = "10.1016/j.physletb.2022.137542",
    journal = "Phys. Lett. B",
    volume = "835",
    pages = "137542",
    year = "2022"
}

@article{Athron:2023mer,
    author = "Athron, Peter and Fowlie, Andrew and Lu, Chih-Ting and Morris, Lachlan and Wu, Lei and Wu, Yongcheng and Xu, Zhongxiu",
    title = "{Can Supercooled Phase Transitions Explain the Gravitational Wave Background Observed by Pulsar Timing Arrays?}",
    eprint = "2306.17239",
    archivePrefix = "arXiv",
    primaryClass = "hep-ph",
    doi = "10.1103/PhysRevLett.132.221001",
    journal = "Phys. Rev. Lett.",
    volume = "132",
    number = "22",
    pages = "221001",
    year = "2024"
}

@article{Li:2023yaj,
    author = "Li, Yao-Yu and Zhang, Chi and Wang, Ziwei and Cui, Ming-Yang and Tsai, Yue-Lin Sming and Yuan, Qiang and Fan, Yi-Zhong",
    title = "{Primordial magnetic field as a common solution of nanohertz gravitational waves and the Hubble tension}",
    eprint = "2306.17124",
    archivePrefix = "arXiv",
    primaryClass = "astro-ph.HE",
    doi = "10.1103/PhysRevD.109.043538",
    journal = "Phys. Rev. D",
    volume = "109",
    number = "4",
    pages = "043538",
    year = "2024"
}

@article{Vachaspati:1984gt,
    author = "Vachaspati, Tanmay and Vilenkin, Alexander",
    title = "{Gravitational Radiation from Cosmic Strings}",
    reportNumber = "HUTP-84/A065",
    doi = "10.1103/PhysRevD.31.3052",
    journal = "Phys. Rev. D",
    volume = "31",
    pages = "3052",
    year = "1985"
}

@article{Sakellariadou:1990ne,
    author = "Sakellariadou, M.",
    title = "{Gravitational waves emitted from infinite strings}",
    doi = "10.1103/PhysRevD.42.354",
    journal = "Phys. Rev. D",
    volume = "42",
    pages = "354--360",
    year = "1990",
    note = "[Erratum: Phys.Rev.D 43, 4150 (1991)]"
}

@article{Damour:2001bk,
    author = "Damour, Thibault and Vilenkin, Alexander",
    title = "{Gravitational wave bursts from cusps and kinks on cosmic strings}",
    eprint = "gr-qc/0104026",
    archivePrefix = "arXiv",
    reportNumber = "IHES-P-01-15",
    doi = "10.1103/PhysRevD.64.064008",
    journal = "Phys. Rev. D",
    volume = "64",
    pages = "064008",
    year = "2001"
}

@article{Damour:2004kw,
    author = "Damour, Thibault and Vilenkin, Alexander",
    title = "{Gravitational radiation from cosmic (super)strings: Bursts, stochastic background, and observational windows}",
    eprint = "hep-th/0410222",
    archivePrefix = "arXiv",
    doi = "10.1103/PhysRevD.71.063510",
    journal = "Phys. Rev. D",
    volume = "71",
    pages = "063510",
    year = "2005"
}

@article{Figueroa:2012kw,
    author = "Figueroa, Daniel G. and Hindmarsh, Mark and Urrestilla, Jon",
    title = "{Exact Scale-Invariant Background of Gravitational Waves from Cosmic Defects}",
    eprint = "1212.5458",
    archivePrefix = "arXiv",
    primaryClass = "astro-ph.CO",
    doi = "10.1103/PhysRevLett.110.101302",
    journal = "Phys. Rev. Lett.",
    volume = "110",
    number = "10",
    pages = "101302",
    year = "2013"
}

@article{Hiramatsu:2013qaa,
    author = "Hiramatsu, Takashi and Kawasaki, Masahiro and Saikawa, Ken'ichi",
    title = "{On the estimation of gravitational wave spectrum from cosmic domain walls}",
    eprint = "1309.5001",
    archivePrefix = "arXiv",
    primaryClass = "astro-ph.CO",
    reportNumber = "ICRR-REPORT-659-2013-8, IPMU13-0182, YITP-13-87",
    doi = "10.1088/1475-7516/2014/02/031",
    journal = "JCAP",
    volume = "02",
    pages = "031",
    year = "2014"
}

@article{Blanco-Pillado:2017oxo,
    author = "Blanco-Pillado, Jose J. and Olum, Ken D.",
    title = "{Stochastic gravitational wave background from smoothed cosmic string loops}",
    eprint = "1709.02693",
    archivePrefix = "arXiv",
    primaryClass = "astro-ph.CO",
    doi = "10.1103/PhysRevD.96.104046",
    journal = "Phys. Rev. D",
    volume = "96",
    number = "10",
    pages = "104046",
    year = "2017"
}

@article{Auclair:2019wcv,
    author = "Auclair, Pierre and others",
    title = "{Probing the gravitational wave background from cosmic strings with LISA}",
    eprint = "1909.00819",
    archivePrefix = "arXiv",
    primaryClass = "astro-ph.CO",
    doi = "10.1088/1475-7516/2020/04/034",
    journal = "JCAP",
    volume = "04",
    pages = "034",
    year = "2020"
}

@article{Gouttenoire:2019kij,
    author = "Gouttenoire, Yann and Servant, G{\'e}raldine and Simakachorn, Peera",
    title = "{Beyond the Standard Models with Cosmic Strings}",
    eprint = "1912.02569",
    archivePrefix = "arXiv",
    primaryClass = "hep-ph",
    reportNumber = "DESY-19-204",
    doi = "10.1088/1475-7516/2020/07/032",
    journal = "JCAP",
    volume = "07",
    pages = "032",
    year = "2020"
}

@article{Figueroa:2020lvo,
    author = "Figueroa, Daniel G. and Hindmarsh, Mark and Lizarraga, Joanes and Urrestilla, Jon",
    title = "{Irreducible background of gravitational waves from a cosmic defect network: update and comparison of numerical techniques}",
    eprint = "2007.03337",
    archivePrefix = "arXiv",
    primaryClass = "astro-ph.CO",
    doi = "10.1103/PhysRevD.102.103516",
    journal = "Phys. Rev. D",
    volume = "102",
    number = "10",
    pages = "103516",
    year = "2020"
}

@article{Gorghetto:2021fsn,
    author = "Gorghetto, Marco and Hardy, Edward and Nicolaescu, Horia",
    title = "{Observing invisible axions with gravitational waves}",
    eprint = "2101.11007",
    archivePrefix = "arXiv",
    primaryClass = "hep-ph",
    doi = "10.1088/1475-7516/2021/06/034",
    journal = "JCAP",
    volume = "06",
    pages = "034",
    year = "2021"
}

@article{Chang:2021afa,
    author = "Chang, Chia-Feng and Cui, Yanou",
    title = "{Gravitational waves from global cosmic strings and cosmic archaeology}",
    eprint = "2106.09746",
    archivePrefix = "arXiv",
    primaryClass = "hep-ph",
    doi = "10.1007/JHEP03(2022)114",
    journal = "JHEP",
    volume = "03",
    pages = "114",
    year = "2022"
}

@article{Yamada:2022aax,
    author = "Yamada, Masaki and Yonekura, Kazuya",
    title = "{Cosmic F- and D-strings from pure Yang{\textendash}Mills theory}",
    eprint = "2204.13125",
    archivePrefix = "arXiv",
    primaryClass = "hep-th",
    reportNumber = "TU-1153",
    doi = "10.1016/j.physletb.2023.137724",
    journal = "Phys. Lett. B",
    volume = "838",
    pages = "137724",
    year = "2023"
}

@article{Yamada:2022imq,
    author = "Yamada, Masaki and Yonekura, Kazuya",
    title = "{Cosmic strings from pure Yang{\textendash}Mills theory}",
    eprint = "2204.13123",
    archivePrefix = "arXiv",
    primaryClass = "hep-th",
    reportNumber = "TU-1152",
    doi = "10.1103/PhysRevD.106.123515",
    journal = "Phys. Rev. D",
    volume = "106",
    number = "12",
    pages = "123515",
    year = "2022"
}

@article{Kitajima:2023cek,
    author = "Kitajima, Naoya and Lee, Junseok and Murai, Kai and Takahashi, Fuminobu and Yin, Wen",
    title = "{Gravitational waves from domain wall collapse, and application to nanohertz signals with QCD-coupled axions}",
    eprint = "2306.17146",
    archivePrefix = "arXiv",
    primaryClass = "hep-ph",
    reportNumber = "TU-1198",
    doi = "10.1016/j.physletb.2024.138586",
    journal = "Phys. Lett. B",
    volume = "851",
    pages = "138586",
    year = "2024"
}

@article{Baeza-Ballesteros:2023say,
    author = "Baeza-Ballesteros, Jorge and Copeland, Edmund J. and Figueroa, Daniel G. and Lizarraga, Joanes",
    title = "{Gravitational wave emission from a cosmic string loop: Global case}",
    eprint = "2308.08456",
    archivePrefix = "arXiv",
    primaryClass = "astro-ph.CO",
    doi = "10.1103/PhysRevD.110.043522",
    journal = "Phys. Rev. D",
    volume = "110",
    number = "4",
    pages = "043522",
    year = "2024"
}

@article{Dankovsky:2024ipq,
    author = "Dankovsky, I. and Ramazanov, S. and Babichev, E. and Gorbunov, D. and Vikman, A.",
    title = "{Numerical analysis of melting domain walls and their gravitational waves}",
    eprint = "2410.21971",
    archivePrefix = "arXiv",
    primaryClass = "hep-ph",
    doi = "10.1088/1475-7516/2025/02/064",
    journal = "JCAP",
    volume = "02",
    pages = "064",
    year = "2025"
}

@article{Ferreira:2024eru,
    author = "Ferreira, Ricardo Z. and Notari, Alessio and Pujol{\`a}s, Oriol and Rompineve, Fabrizio",
    title = "{Collapsing domain wall networks: impact on pulsar timing arrays and primordial black holes}",
    eprint = "2401.14331",
    archivePrefix = "arXiv",
    primaryClass = "astro-ph.CO",
    reportNumber = "CERN-TH-2024-020",
    doi = "10.1088/1475-7516/2024/06/020",
    journal = "JCAP",
    volume = "06",
    pages = "020",
    year = "2024"
}

@article{Baeza-Ballesteros:2024otj,
    author = "Baeza-Ballesteros, Jorge and Copeland, Edmund J. and Figueroa, Daniel G. and Lizarraga, Joanes",
    title = "{Particle and gravitational wave emission by local string loops: Lattice calculation}",
    eprint = "2408.02364",
    archivePrefix = "arXiv",
    primaryClass = "astro-ph.CO",
    doi = "10.1103/ym9p-scmw",
    journal = "Phys. Rev. D",
    volume = "112",
    number = "4",
    pages = "043540",
    year = "2025"
}

@article{Notari:2025kqq,
    author = "Notari, Alessio and Rompineve, Fabrizio and Torrenti, Francisco",
    title = "{The spectrum of gravitational waves from annihilating domain walls}",
    eprint = "2504.03636",
    archivePrefix = "arXiv",
    primaryClass = "astro-ph.CO",
    doi = "10.1088/1475-7516/2025/07/049",
    journal = "JCAP",
    volume = "07",
    pages = "049",
    year = "2025"
}

@article{Babichev:2025stm,
    author = "Babichev, E. and Dankovsky, I. and Gorbunov, D. and Ramazanov, S. and Vikman, A.",
    title = "{Biased domain walls: faster annihilation, weaker gravitational waves}",
    eprint = "2504.07902",
    archivePrefix = "arXiv",
    primaryClass = "hep-ph",
    doi = "10.1088/1475-7516/2025/10/103",
    journal = "JCAP",
    volume = "10",
    pages = "103",
    year = "2025"
}

@article{Figueroa:2023zhu,
    author = "Figueroa, Daniel G. and Pieroni, Mauro and Ricciardone, Angelo and Simakachorn, Peera",
    title = "{Cosmological Background Interpretation of Pulsar Timing Array Data}",
    eprint = "2307.02399",
    archivePrefix = "arXiv",
    primaryClass = "astro-ph.CO",
    reportNumber = "CERN-TH-2023-132",
    doi = "10.1103/PhysRevLett.132.171002",
    journal = "Phys. Rev. Lett.",
    volume = "132",
    number = "17",
    pages = "171002",
    year = "2024"
}

@article{Ellis:2023oxs,
    author = {Ellis, John and Fairbairn, Malcolm and Franciolini, Gabriele and H{\"u}tsi, Gert and Iovino, Antonio and Lewicki, Marek and Raidal, Martti and Urrutia, Juan and Vaskonen, Ville and Veerm{\"a}e, Hardi},
    title = "{What is the source of the PTA GW signal?}",
    eprint = "2308.08546",
    archivePrefix = "arXiv",
    primaryClass = "astro-ph.CO",
    reportNumber = "KCL-PH-TH/2023-43, CERN-TH-2023-153, AION-REPORT/2023-08",
    doi = "10.1103/PhysRevD.109.023522",
    journal = "Phys. Rev. D",
    volume = "109",
    number = "2",
    pages = "023522",
    year = "2024"
}

@article{Tristram:2021tvh,
    author = "Tristram, M. and others",
    title = "{Improved limits on the tensor-to-scalar ratio using BICEP and Planck data}",
    eprint = "2112.07961",
    archivePrefix = "arXiv",
    primaryClass = "astro-ph.CO",
    doi = "10.1103/PhysRevD.105.083524",
    journal = "Phys. Rev. D",
    volume = "105",
    number = "8",
    pages = "083524",
    year = "2022"
}

@article{Planck:2018jri,
    author = "Akrami, Y. and others",
    collaboration = "Planck",
    title = "{Planck 2018 results. X. Constraints on inflation}",
    eprint = "1807.06211",
    archivePrefix = "arXiv",
    primaryClass = "astro-ph.CO",
    doi = "10.1051/0004-6361/201833887",
    journal = "Astron. Astrophys.",
    volume = "641",
    pages = "A10",
    year = "2020"
}

@article{Kuroyanagi:2014nba,
    author = "Kuroyanagi, Sachiko and Takahashi, Tomo and Yokoyama, Shuichiro",
    title = "{Blue-tilted Tensor Spectrum and Thermal History of the Universe}",
    eprint = "1407.4785",
    archivePrefix = "arXiv",
    primaryClass = "astro-ph.CO",
    reportNumber = "ICRR-REPORT-686-2014-12",
    doi = "10.1088/1475-7516/2015/02/003",
    journal = "JCAP",
    volume = "02",
    pages = "003",
    year = "2015"
}

@article{Kuroyanagi:2020sfw,
    author = "Kuroyanagi, Sachiko and Takahashi, Tomo and Yokoyama, Shuichiro",
    title = "{Blue-tilted inflationary tensor spectrum and reheating in the light of NANOGrav results}",
    eprint = "2011.03323",
    archivePrefix = "arXiv",
    primaryClass = "astro-ph.CO",
    doi = "10.1088/1475-7516/2021/01/071",
    journal = "JCAP",
    volume = "01",
    pages = "071",
    year = "2021"
}

@article{Vagnozzi:2023lwo,
    author = "Vagnozzi, Sunny",
    title = "{Inflationary interpretation of the stochastic gravitational wave background signal detected by pulsar timing array experiments}",
    eprint = "2306.16912",
    archivePrefix = "arXiv",
    primaryClass = "astro-ph.CO",
    doi = "10.1016/j.jheap.2023.07.001",
    journal = "JHEAp",
    volume = "39",
    pages = "81--98",
    year = "2023"
}

@article{Unal:2023srk,
    author = "Unal, Caner and Papageorgiou, Alexandros and Obata, Ippei",
    title = "{Axion-gauge dynamics during inflation as the origin of pulsar timing array signals and primordial black holes}",
    eprint = "2307.02322",
    archivePrefix = "arXiv",
    primaryClass = "astro-ph.CO",
    doi = "10.1016/j.physletb.2024.138873",
    journal = "Phys. Lett. B",
    volume = "856",
    pages = "138873",
    year = "2024"
}

@article{Li:2025udi,
    author = "Li, Bohua and Meyers, Joel and Shapiro, Paul R.",
    title = "{Multimodality in the Search for New Physics in Pulsar Timing Data and the Case of Kination-amplified Gravitational-wave Background from Inflation}",
    eprint = "2503.18937",
    archivePrefix = "arXiv",
    primaryClass = "astro-ph.CO",
    doi = "10.3847/1538-4357/adcc14",
    journal = "Astrophys. J.",
    volume = "985",
    number = "1",
    pages = "117",
    year = "2025"
}

@article{Tomita:1975kj,
    author = "Tomita, Kenji",
    title = "{Evolution of Irregularities in a Chaotic Early Universe}",
    reportNumber = "RRK 75-3",
    doi = "10.1143/PTP.54.730",
    journal = "Prog. Theor. Phys.",
    volume = "54",
    pages = "730",
    year = "1975"
}

@article{Acquaviva:2002ud,
    author = "Acquaviva, Viviana and Bartolo, Nicola and Matarrese, Sabino and Riotto, Antonio",
    title = "{Second order cosmological perturbations from inflation}",
    eprint = "astro-ph/0209156",
    archivePrefix = "arXiv",
    reportNumber = "DFPD-A-02-21",
    doi = "10.1016/S0550-3213(03)00550-9",
    journal = "Nucl. Phys. B",
    volume = "667",
    pages = "119--148",
    year = "2003"
}

@article{Mollerach:2003nq,
    author = "Mollerach, Silvia and Harari, Diego and Matarrese, Sabino",
    title = "{CMB polarization from secondary vector and tensor modes}",
    eprint = "astro-ph/0310711",
    archivePrefix = "arXiv",
    doi = "10.1103/PhysRevD.69.063002",
    journal = "Phys. Rev. D",
    volume = "69",
    pages = "063002",
    year = "2004"
}

@article{Franciolini:2023pbf,
    author = "Franciolini, Gabriele and Iovino, Junior., Antonio and Vaskonen, Ville and Veermae, Hardi",
    title = "{Recent Gravitational Wave Observation by Pulsar Timing Arrays and Primordial Black Holes: The Importance of Non-Gaussianities}",
    eprint = "2306.17149",
    archivePrefix = "arXiv",
    primaryClass = "astro-ph.CO",
    doi = "10.1103/PhysRevLett.131.201401",
    journal = "Phys. Rev. Lett.",
    volume = "131",
    number = "20",
    pages = "201401",
    year = "2023"
}

@article{Franciolini:2023wjm,
    author = "Franciolini, Gabriele and Racco, Davide and Rompineve, Fabrizio",
    title = "{Footprints of the QCD Crossover on Cosmological Gravitational Waves at Pulsar Timing Arrays}",
    eprint = "2306.17136",
    archivePrefix = "arXiv",
    primaryClass = "astro-ph.CO",
    reportNumber = "CERN-TH-2023-080",
    doi = "10.1103/PhysRevLett.132.081001",
    journal = "Phys. Rev. Lett.",
    volume = "132",
    number = "8",
    pages = "081001",
    year = "2024",
    note = "[Erratum: Phys.Rev.Lett. 133, 189901 (2024)]"
}

@article{Inomata:2023zup,
    author = "Inomata, Keisuke and Kohri, Kazunori and Terada, Takahiro",
    title = "{Detected stochastic gravitational waves and subsolar-mass primordial black holes}",
    eprint = "2306.17834",
    archivePrefix = "arXiv",
    primaryClass = "astro-ph.CO",
    reportNumber = "KEK-TH-2535, KEK-Cosmo-0317, KEK-QUP-2023-0016, CTPU-PTC-23-28",
    doi = "10.1103/PhysRevD.109.063506",
    journal = "Phys. Rev. D",
    volume = "109",
    number = "6",
    pages = "063506",
    year = "2024"
}

@article{Ebadi:2023xhq,
    author = "Ebadi, Reza and Kumar, Soubhik and McCune, Amara and Tai, Hanwen and Wang, Lian-Tao",
    title = "{Gravitational waves from stochastic scalar fluctuations}",
    eprint = "2307.01248",
    archivePrefix = "arXiv",
    primaryClass = "astro-ph.CO",
    doi = "10.1103/PhysRevD.109.083519",
    journal = "Phys. Rev. D",
    volume = "109",
    number = "8",
    pages = "083519",
    year = "2024"
}

@article{Yi:2023mbm,
    author = "Yi, Zhu and Gao, Qing and Gong, Yungui and Wang, Yue and Zhang, Fengge",
    title = "{Scalar induced gravitational waves in light of Pulsar Timing Array data}",
    eprint = "2307.02467",
    archivePrefix = "arXiv",
    primaryClass = "gr-qc",
    doi = "10.1007/s11433-023-2266-1",
    journal = "Sci. China Phys. Mech. Astron.",
    volume = "66",
    number = "12",
    pages = "120404",
    year = "2023"
}

@article{Firouzjahi:2023lzg,
    author = "Firouzjahi, Hassan and Talebian, Alireza",
    title = "{Induced gravitational waves from ultra slow-roll inflation and pulsar timing arrays observations}",
    eprint = "2307.03164",
    archivePrefix = "arXiv",
    primaryClass = "gr-qc",
    doi = "10.1088/1475-7516/2023/10/032",
    journal = "JCAP",
    volume = "10",
    pages = "032",
    year = "2023"
}

@article{You:2023rmn,
    author = "You, Zhi-Qiang and Yi, Zhu and Wu, You",
    title = "{Constraints on primordial curvature power spectrum with pulsar timing arrays}",
    eprint = "2307.04419",
    archivePrefix = "arXiv",
    primaryClass = "gr-qc",
    doi = "10.1088/1475-7516/2023/11/065",
    journal = "JCAP",
    volume = "11",
    pages = "065",
    year = "2023"
}

@article{Balaji:2023ehk,
    author = "Balaji, Shyam and Dom{\`e}nech, Guillem and Franciolini, Gabriele",
    title = "{Scalar-induced gravitational wave interpretation of PTA data: the role of scalar fluctuation propagation speed}",
    eprint = "2307.08552",
    archivePrefix = "arXiv",
    primaryClass = "gr-qc",
    doi = "10.1088/1475-7516/2023/10/041",
    journal = "JCAP",
    volume = "10",
    pages = "041",
    year = "2023"
}

@article{Zhao:2023joc,
    author = "Zhu, Qing-Hua and Zhao, Zhi-Chao and Wang, Sai and Zhang, Xin",
    title = "{Unraveling the early universe{\textquoteright}s equation of state and primordial black hole production with PTA, BBN, and CMB observations*}",
    eprint = "2307.13574",
    archivePrefix = "arXiv",
    primaryClass = "astro-ph.CO",
    doi = "10.1088/1674-1137/ad79d5",
    journal = "Chin. Phys. C",
    volume = "48",
    number = "12",
    pages = "125105",
    year = "2024"
}

@article{Yi:2023tdk,
    author = "Yi, Zhu and You, Zhi-Qiang and Wu, You",
    title = "{Model-independent reconstruction of the primordial curvature power spectrum from PTA data}",
    eprint = "2308.05632",
    archivePrefix = "arXiv",
    primaryClass = "astro-ph.CO",
    doi = "10.1088/1475-7516/2024/01/066",
    journal = "JCAP",
    volume = "01",
    pages = "066",
    year = "2024"
}
  %\addbibresource{autoArxiv.bib,manualArxiv.bib}
  %\bibliographystyle{ieeetr}
\bibliographystyle{h-physrev4}
%\bibliographystyle{ar-style2}
%\end{multicols}

\end{document}